\documentclass[11pt]{article}
\usepackage{mymacros}

\title{q-Askey Deformations of Double-Scaled SYK}
\author[a,b]{Sergio E. Aguilar-Gutierrez,\orcidlink{0000-0003-0308-0061}}
\author[b]{Trivko Kukolj,\orcidlink{0000-0001-7163-8798}}
\author[b]{Josef Seitz\orcidlink{0009-0009-0066-3553}}

\affiliation[a]{Qubits and Spacetime Unit, Okinawa Institute of Science and Technology Graduate University,\footnote{\begin{CJK}{UTF8}{min}沖縄科学技術大学院大学\end{CJK}} 1919-1 Tancha, Onna, Okinawa 904 0495, Japan}
\affiliation[b]{Department of Particle Physics and Astrophysics, Weizmann Institute of Science, \\ Rehovot 7610001, Israel}

\emailAdd{sergio.ernesto.aguilar@gmail.com}
\emailAdd{trivko.kukolj@weizmann.ac.il}
\emailAdd{josef-emanuel.seitz@weizmann.ac.il}

\abstract{We construct families of deformations of the double-scaled SYK (DSSYK) model and investigate their bulk interpretation. We introduce microscopic deformations of the SYK model which, after ensemble-averaging and in the double-scaling limit, are described by a transfer matrix encoding the recurrence relations of basic orthogonal polynomials in the q-Askey scheme. For certain families of deformations in the semiclassical limit at finite temperature, the chord number (encoding Krylov complexity) corresponds to the length of an Einstein-Rosen bridge connecting an End-Of-The-World brane to an anti-de Sitter asymptotic boundary. By increasing one of the deformation parameters, the models eventually exhibit discrete energy levels, signaling a new geometric transition in sine dilaton gravity.
Via the SYK-Schur duality, Krylov complexity also admits a representation-theoretic interpretation as the spread of the SU(2) spin in the index of an $\mathcal{N}=2$ SU(2) gauge theory. We study the operator algebras of the deformed theories. The algebras can be type II$_1$ or type I$_\infty$ factors, depending on the operators that are included. The entanglement entropy between the type II$_1$ algebras for a pure state manifests as an extremal surface through the Ryu-Takayanagi formula. We discuss connections between our results and the emergence of baby universes in the bulk.}

\begin{document}

\maketitle
\section{Introduction}
Deformations in the holographic correspondence are crucial to extend the bulk/boundary dictionary beyond the constraining requirements of conformal field theories (CFT), which are a small subset in the space of quantum field theories, and beyond spacetimes with asymptotic anti-de Sitter (AdS) boundaries. Some notable examples arise in the context of finite cutoff holography \cite{McGough:2016lol}. In this work, we are interested in describing deformations of the boundary theory that describe spacetimes with End-Of-The-World (ETW) branes in a well-controlled setting in lower dimensions. We are particularly interested in deforming the duality between the SYK model \cite{kitaevTalks1,kitaevTalks2,kitaevTalks3,Sachdev_1993} in the double-scaling limit \cite{Cotler:2016fpe,Erdos:2014zgc,Berkooz:2018qkz,Berkooz:2018jqr}, called the DSSYK model (see \cite{Berkooz:2024lgq} for a recent review), and sine dilaton gravity \cite{Blommaert:2023opb,Blommaert:2024whf,Blommaert:2024ymv,Blommaert:2025avl,Blommaert:2025eps}.\footnote{Other bulk dual proposals include stretched horizon holography in two-dimensional de Sitter (dS$_2$) space \cite{Susskind:2021esx,Susskind:2022bia,Susskind:2023hnj,Lin:2022nss,Rahman:2022jsf,Rahman:2023pgt,Rahman:2024iiu,Rahman:2024vyg,Sekino:2025bsc,Miyashita:2025rpt,Aguilar-Gutierrez:2025otq,Aguilar-Gutierrez:2026ogo}, and (Schwarzschild-)dS$_3$ space \cite{Narovlansky:2023lfz,Verlinde:2024znh,Verlinde:2024zrh,Narovlansky:2025tpb,Blommaert:2025eps,Tietto:2025oxn,Aguilar-Gutierrez:2024nau,Marini:2026zjk} which are expected to be intricately related to sine dilaton gravity \cite{Blommaert:2025eps,Aguilar-Gutierrez:2026ogo}. See other interesting proposals in e.g.\cite{Milekhin:2023bjv,Milekhin:2024vbb,Narovlansky:2025tpb,Okuyama:2025hsd,Yuan:2024utc,Gubankova:2025gbx,Ahn:2025exp}.}

In the double-scaling limit, the number of fermions and the all-to-all interactions in the Hamiltonian both diverge, but a ratio between them is fixed. In this regime, the DSSYK model is controlled by a statistical description which can be solved through chord diagram methods \cite{Berkooz:2018jqr,Berkooz:2018qkz}. Chord diagrams are a graphical representation of expectation values involving the Wick contraction between Majorana fermions, where a Hilbert space trace is represented by a circle, whose periodicity is associated with the inverse temperature of the system, as we briefly review in Sec.~\ref{ssec:DSSYK review}. In this way, moments of the Hamiltonian in the microscopic system can be equivalently evaluated in terms of a transfer matrix acting on acting on the Hilbert space of an auxiliary quantum mechanical system describing the recurrence relations of summations of chord diagrams (see \cite{Berkooz:2024lgq} for a review).

In particular, the connection between the chord Hamiltonian of the DSSYK model and ETW branes in Jackiw–Teitelboim (JT) \cite{JACKIW1985343,TEITELBOIM198341} gravity
 (reviewed in \cite{Mertens:2022irh}) and sine dilaton gravity has been pointed out by \cite{Okuyama:2023byh,Blommaert:2025avl,Cui:2025sgy,Cao:2025pir,Aguilar-Gutierrez:2025hty}. There are different types of slicings in spacetime. When the slicing is such that it connects the ETW brane to an asymptotic boundary, the corresponding chord Hamiltonians have been associated with Askey-Wilson polynomials \cite{askey1985some} in the q-Askey scheme \cite{askey1985some,koekoek1994askey,koekoek2010hypergeometric}, specifically the generalised q-Hermite polynomials, and Al-Salam Chihara polynomials. Similar to the DSSYK model, there is a notion of a chord Hilbert space with a complete chord basis parametrised by a discrete set of numbers. There are additional continuous parameters in the corresponding chord Hamiltonians, which have been associated with the brane tensions in sine dilaton gravity in the few known examples. In particular, by connecting or splitting ETW branes in the bulk, \cite{Cui:2025sgy} has shown that one can recover correlation functions in the DSSYK model with an arbitrary number of matter chord operators, providing a precise realization of the holographic correspondence with matter. However, so far, only the first two members in the q-Askey scheme have been very concretely related to a bulk theory.

More recently, building on the SYK-Schur duality proposal by Gaiotto-Verlinde \cite{Gaiotto:2024kze}, \cite{Berkooz:2025ydg} showed there is an intricate connection between a four-dimensional $\mathcal{N}=2$ SU(2) gauge theory (originally solved by \cite{Seiberg:1994aj,Seiberg:1994rs}, and reviewed in \cite{Tachikawa:2013kta}) with different numbers of hypermultiplets in terms of auxiliary chord theories. In particular, it was found that the number of hypermultiplets determines the chord Hamiltonian, which encodes a recurrence relation in the q-Askey scheme for basic hypergeometric orthogonal polynomials \cite{askey1985some}. Therefore, the wavefunctions that solve the eigenvalue problem are given by the Askey-Wilson polynomials.\footnote{A possible generalisation of this result would be introducing a deformation of the DSSYK transfer matrix whose wavefunctions can be solved in terms Askey-Wilson functions \cite{stokman2002expansion}.} For certain specific cases, the chord auxiliary theories in the q-Askey scheme have been attributed to ETW branes in previous literature \cite{Okuyama:2023byh,Rajgadia:2026ask,Cao:2025pir,Aguilar-Gutierrez:2025hty}. Other generalizations with $\mathcal{N}=2$ SU(N) theories were investigated in \cite{Lewis:2025qjq}. However, the precise bulk interpretation of the different chord theories beyond those in the previous literature has not been carefully understood to date. In particular, one of the members in the q-Askey scheme is associated with the SU(2) theory with an adjoint hypermultiplet, which we refer to as an ``adjoint deformation'' throughout this work. It has special properties that suggest it can describe random matrix theory for a specific point in its parameter space. It was suggested in \cite{Berkooz:2025ydg} that the chord auxiliary theory corresponding to the SU(2) theory with an adjoint hypermultiplet could be interpreted in terms of an entangled pair of ETW branes. However, a rigorous justification is still missing. Moreover, the classification of chord theories in the q-Askey scheme ends at a finite order \cite{Berkooz:2025ydg}. This mathematical fact has not received a chord-diagrammatic or bulk explanation to date. In contrast, from the SU(2) gauge theory picture, adding more hypermultiplets leads to theories with Landau poles.

Recently, some examples of chord Hamiltonians describing an ETW brane in the bulk have been interpreted by deforming the SYK model in \cite{Cao:2025pir,Berkooz:2025ydg,Rajgadia:2026ask}. From the microscopic Hamiltonian, one can deduce the respective chord rules after performing annealed ensemble averaging in the double-scaling limit. However, to date, the microscopic approach has only been implemented for the chord theory associated with generalised q-Hermite polynomials, based on operators that are well-adapted to probe properties of the SYK in the TFD state \cite{Cao:2025pir} or KM states \cite{Rajgadia:2026ask}. A more systematic construction for chord theories in the q-Askey scheme, which may also probe beyond TFD or KM states, is still missing.
It is thus natural to ask: 
\begin{quote}
    \emph{What is the microscopic theory for each of the chord theories in the q-Askey scheme, and what is their bulk interpretation?}
\end{quote}

\noindent In this paper, we construct the microscopic theory for each of the chord theories explicitly, as a double-scaling limit of a deformed SYK Hamiltonian. We then investigate the bulk interpretation of these microscopic Hamiltonians in two ways. Both rely on the semiclassical limit of the theory, where the combination $q^n$ is fixed, $n$ being the chord number.

On the one hand, one can build an ordered basis, called the Krylov basis, which minimises the cost function of a given evolving state, known as Krylov complexity \cite{Parker:2018yvk,Balasubramanian:2022tpr} (see recent reviews by \cite{Baiguera:2025dkc,Nandy:2024evd,Rabinovici:2025otw}).\footnote{Recent work on Krylov operator complexity for deformed SYK models has been investigated by \cite{Chapman:2024pdw}.} This has found multiple applications in the literature, and it has been argued to be a concrete measure of quantum chaos (see \cite{Baggioli:2024wbz,Alishahiha:2024vbf,Erdmenger:2023wjg,Balasubramanian:2022tpr,Camargo:2026szl} among others). Recent developments show that Krylov complexity of the Hartle-Hawking (HH) state in the DSSYK model and some of its deformations can be precisely matched to the geodesic length between the Dirichlet boundaries of an AdS black hole \cite{Rabinovici:2023yex} 
(several extensions have appeared in \cite{Ambrosini:2024sre,Ambrosini:2025hvo,Heller:2024ldz,Heller:2025ddj,Xu:2024gfm,Bhattacharjee:2022ave,Fu:2025kkh,Aguilar-Gutierrez:2024nau,Aguilar-Gutierrez:2025hty,Aguilar-Gutierrez:2025mxf,Aguilar-Gutierrez:2025pqp,Aguilar-Gutierrez:2025sqh,Aguilar-Gutierrez:2026jjv,Aguilar-Gutierrez:2026ogo,Miyaji:2025ucp}). 
Similarly, \cite{Aguilar-Gutierrez:2025hty} showed that in the case of Al-Salam Chihara polynomials, Krylov spread complexity holographically corresponds to the geodesic length between a timelike ETW brane with Neumann boundary conditions and an asymptotic AdS$_2$ cutoff surface with Dirichlet boundary conditions. This allows for a precise bulk-boundary holographic dictionary, as the ETW brane parameters, such as the tension and ``fake'' temperature \cite{Lin:2023trc}, are encoded in the geodesic length. For this reason, Krylov spread complexity in the HH state may probe the properties of the chord Hamiltonians in the q-Askey scheme more generally, as we show.

On the other hand, we also study the bulk interpretation via algebraic methods. This is motivated by pioneering work on operator algebras in holographic systems \cite{Papadodimas:2013jku,Jefferson:2018ksk,Leutheusser:2021frk,Leutheusser:2022bgi,Witten:2021unn} and more recent developments for rigorously defining generalised entropies in quantum gravity \cite{Witten:2021unn,Chandrasekaran:2022cip,Chandrasekaran:2022eqq,Jensen:2023yxy,Kudler-Flam:2023qfl} among others (see \cite{Witten:2018zxz,Sorce:2023fdx,Casini:2022rlv,Liu:2025krl} for reviews). The algebraic formulation is particularly well-suited to study the DSSYK model \cite{Xu:2024hoc,Rajgadia:2026ask} (see also \cite{Cao:2025pir}) and its deformations put forward in this work. When studying the large $N$ limit of a quantum mechanical system, the matrix elements of a given operator evaluated in a sequence of states that depend on $N$ might not converge in the norm topology of the operator algebra as $N\rightarrow\infty$ even if the matrix elements themselves do. So that closure of the algebra is imposed under weak operator topology instead to define the von Neumann algebra.

 In the case of the DSSYK model, this leads to different types of algebras depending on what operators are retained in the large N limit \cite{Xu:2024hoc,Rajgadia:2026ask}, and where the matrix elements are evaluated in the chord Hilbert space. The latter describes states encoding correlation functions of the microscopic SYK model in the double-scaling limit with annealed ensemble averaging. This is understood through a Gelfand-Naimark-Segal \cite{gelfand1943imbedding,Segal1947IrreducibleRO} (GNS) construction of the Hilbert space, which captures how SYK operators in the double-scaling limit act on a special state that generates all chord states. One of the advantages of the algebraic formulation is that it allows more clarity on how to connect information of the microscopic system to its GNS representation, which is often seen from the type classification of the algebra. In particular, the generators of the algebra may be restricted to probe only thermodynamic properties of the state of the microscopic system. In this case, the algebras are type II or type III in the classification, which only have access to mixed density matrices in the system. In other cases where the generators can probe the quantum state of the full system, they are said to be type I, which can describe pure states. This has improved our understanding of the emergence of spacetime from the boundary theory algebra in holographic systems \cite{Gesteau:2024rpt,Gesteau:2023rrx,Ouseph:2023juq,Lashkari:2024lkt,Jefferson:2018ksk,Leutheusser:2021frk,Leutheusser:2022bgi,Penington:2023dql,Kolchmeyer:2023gwa} (among others). In particular, for systems described by a type II$_1$ algebra, such as the DSSYK model \cite{Lin:2022rbf,Xu:2024hoc} (restricted to operators that only probe its thermal properties), one can find a rigorous notion of entanglement entropy between a subsystem algebra and its commutant \cite{Aguilar-Gutierrez:2025otq,Tang:2024xgg,Aguilar-Gutierrez:2026ogo}. The algebraic entanglement entropy manifests in the bulk, as predicted by the Ryu-Takayanagi (RT) formula \cite{Ryu:2006bv,Ryu:2006ef} (reviewed in \cite{Nishioka:2009un,Rangamani:2016dms,Chen:2019lcd,Harlow:2014yka}, and extended in \cite{Hubeny:2007xt,Faulkner:2013ana,Engelhardt:2014gca}). A central outcome of this work is that the bulk manifestation of algebraic entanglement entropy continues to hold after deforming the SYK model, in the regime where the RT formula is valid.

\subsection{Results and outline}
In Sec.~\ref{sec:micro models}, we define q-Askey deformations of the SYK model, and we study their ensemble averaging description in terms of chord Hamiltonians arising from an auxiliary quantum-mechanical system with Askey-Wilson polynomials as wavefunctions. In this scheme, the label $n_F$ indicates the corresponding member in the hierarchy,\footnote{We employ the same notation as in \cite{Berkooz:2025ydg}, where $n_F$ denoted the number of half-hypermultiplets in the corresponding SU(2) gauge theory, which does not play a major role in our work.} where $n_F=8$ denotes the most general type, the Askey-Wilson polynomials, while $n_F=0$ is the undeformed case, the q-Hermite polynomials, which describe the chord Hamiltonian of the DSSYK model. In our notation, the families of basic hypergeometric orthogonal polynomials are given by $n_F=0,2,4,6,8$. 

{The families of microscopic deformations are constructed in two ways, which are based on the KM and TFD states respectively, and by PETS more generally. In both cases, let us consider $2N$ Majorana fermions, where $\{\psi_i,\psi_j\}=2\delta_{ij}$. The SYK model is described by
\begin{eqnarray}\label{eq:H syk aain}
    \hH_{\rm SYK}&\equiv\rmi^{p/2}\sum_{i_1\dots i_{p}}J_{i_1\dots i_p}\prod_{j=1}^{p}\psi_{i_j}~,
\end{eqnarray}
with $1\leq i_1\leq i_2\leq\dots\leq i_{p}\leq 2N$, while $J_{i_1\dots i_p}$ are random coupling constants obeying a Gaussian distribution \eqref{eq:J coupling}.}

{In our first approach, based on KM states, we introduce the following family of SYK deformations,
\begin{equation}\label{eq:def_1}
    \hH_{\rm SYK}^{(n_F)}=\frac{1}{2}\qty(\hH_{\rm def}{\rm b}_{n_F}(\hH_{\rm int})+{\rm b}_{n_F}(\hH_{\rm int})\hH^\dagger_{\rm def})+{\rm a}_{n_F}\qty(\hH_{\rm int})~,
\end{equation}
where ${\rm a}_{n_F}(\hH_{\rm int})$ and ${\rm b}_{n_F}(\hH_{\rm int})$ are Hermitian operators that depend on up to four deformation parameters, and whose explicit form is shown in Sec.~\ref{sec:micro models} (see \eqref{eq:targetedeq}). Meanwhile, $\hH_{\rm int}$ (introduced in \cite{Rajgadia:2026ask} and {further studied in App.}~\ref{sapp:H int H def DS}) and $\hH_{\rm def}$ are the following deformation operators:
\begin{subequations}\label{eq:part_2}
    \begin{align}
    \hH_{\rm int}&\equiv\qty{\hat{M},\hat{M}^\dagger}~,\quad \hat{M}=\rmi^{w/2}\sum_{i_1,i_2,\dots, i_{w}}R_{i_1\dots i_{w}}\prod_{j=1}^{w}\frac{1}{2}(\hat{\psi}_{2i_j-1}+\rmi \hat{\psi}_{2i_j})~,\label{seq:H int}\\
    \hH_{\rm def}&\equiv\hH_{\rm SYK}+\qty[\hH_{\rm SYK},~\frac{\rmi}{4p}\sum_{i=1}^N\qty[\hat{\psi}_{2i_j-1},~\hat{\psi}_{2i_j}]]~,\label{seq:H def}
    \end{align}
\end{subequations}
Here $1\leq i_1\leq i_2\leq\dots\leq i_w\leq N$, while $R_{I''}\equiv R_{i_1,\dots i_{w}}$ are complex-valued random Gaussian-distributed couplings; $[x,y]_a\equiv xy-a~yx$; $\hat{M}$ and $\hat{M}^\dagger$ are composite SYK operators that change the spin of the KM states (Sec.~\ref{ssec:DSSYK review} contains more details). {In particular, when all the deformation parameters vanish, $b_{n_F}=1$ and $a_{n_F}=0$, recovering the undeformed SYK model}. We selected the specific form of \eqref{eq:def_1} for the transfer matrix of the corresponding auxiliary system in the double-scaling limit to encode the recurrence relation of the Askey-Wilson polynomials, which are at the top of the hierarchy in the q-Askey scheme.}

{In the second approach, we bipartition the $2N$ Majorana fermions evenly between two subsystems, $L/R$, denoted as $\psi_i^{(L)}$ and $\psi^{(R)}_i$ for $i=1,\dots, N$. We will focus on deforming one of the subsystems, $R$. The deformed microscopic theories are generally described by:
\begin{equation}
    \hH_{\rm SYK'}^{(n_F)}=\frac{1}{2}\qty(\hH_{\rm def'}{\rm b}_{n_F}(\hH_{\rm int'})+{\rm b}_{n_F}(\hH_{\rm int'})\hH^\dagger_{\rm def'})+{\rm a}_{n_F}\qty(\hH_{\rm int'})~,
\end{equation}
where $\hH_{\rm int'}$ and $\hH_{\rm def'}$ are deformation operators of the form 
\begin{subequations}\label{eq:part_1}
    \begin{align}\label{seq:H intp}
    \hH_{\rm int'}&\equiv\begin{pmatrix}
        N\\s
    \end{pmatrix}^{-1}\rmi^{s}\sum_{i_1\dots i_{s}}\prod_{j=1}^s\psi^{(L)}_{i_j}\psi^{(R)}_{i_j} ~,\\\label{seq:H defp}
    \hH_{\rm def'}&\equiv\hH_{\rm SYK}^{(R)}+\qty[\hH_{\rm SYK}^{(R)},~\frac{\rmi}{p}\sum_{j=1}^N\psi_{j}^{(L)}\psi_{j}^{(R)}]~,
    \end{align}
\end{subequations}
where $\hH_{\rm SYK}^{(R)}$ corresponds to \eqref{eq:H syk aain} with $\psi_i\rightarrow\psi_i^{(R)}$; $\hH_{\rm int'}$ is a generalization of the Maldacena-Qi interaction term \cite{Maldacena:2018lmt} that can generate traversable wormholes in the bulk dual. By formulating the microscopic theories in terms of $\psi_i^{(L/R)}$, they are naturally described by partially entangled thermal states (PETS) \cite{Goel:2018ubv,Goel:2023svz}, where the subsystems $L/R$ are defined by the number of operator insertions, and they can take different temperatures.}

After the annealed ensemble averaging in the double-scaling limit, the above deformations encode the recurrence relations of the Askey-Wilson polynomials, which describe wavefunctions for a fixed energy. Crucially, there are two types of energy spectra that the wavefunctions belong to, which depend on the magnitude of the deformation parameters. The most-studied solutions in the DSSYK literature are associated with the continuous energy spectrum. However, the different families of orthogonal polynomials in the q-Askey scheme also admit wavefunctions with a discrete energy spectrum, which have been less explored to date. The bound states arise through an analytic continuation in the orthogonality relations of the Askey-Wilson polynomials when the magnitude of the deformation parameters exceeds a critical value. Intuitively, they correspond to bound states in the Schrödinger equation in the low-energy limit of the deformed theories. They are associated with $\hH_{\rm int}$ in \eqref{eq:part_1}. Beyond the low-energy limit, the discrete spectrum contributes additively to the full partition function alongside the continuous spectrum. At the semiclassical level, the discrete spectrum contribution vanishes, as its effect is non-perturbative in $\lambda$. By increasing the magnitude of the deformation one can then transition between the two regimes. We provide a bulk interpretation for this behaviour below.

In particular, we analyse the triple-scaling limits of the deformed chord theories. We find that the $n_F=2, 4, 6, 8$ all recover the expected Schwarzian dynamics in JT gravity with a timelike ETW brane with Neumann boundary conditions, which shows that the semiclassical description at very low temperatures \cite{Berkooz:2018qkz} describes JT gravity with ETW branes, or just JT gravity. In the PETS construction, the very low temperature limit means that we study the leading order non-trivial expansion of the deformed SYK Hamiltonian in terms of one of the independent temperature parameters, while the one for the complement subsystem can take an independent value. We discuss an alternative realization of the family of PETS, which makes a closer connection with Kourkoulou-Maldacena (KM) states \cite{Kourkoulou:2017zaj}.

\begin{figure}
    \centering
    \subfloat[]{\includegraphics[height=0.29\linewidth]{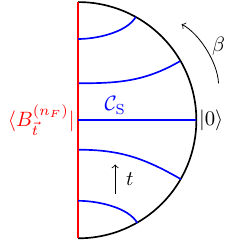}\label{sfig:ETW_nF2}}\hspace{0.4cm}\subfloat[]{\includegraphics[height=0.29\linewidth]{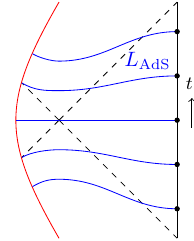}}\hspace{0.4cm}\subfloat[]{\includegraphics[height=0.28\linewidth]{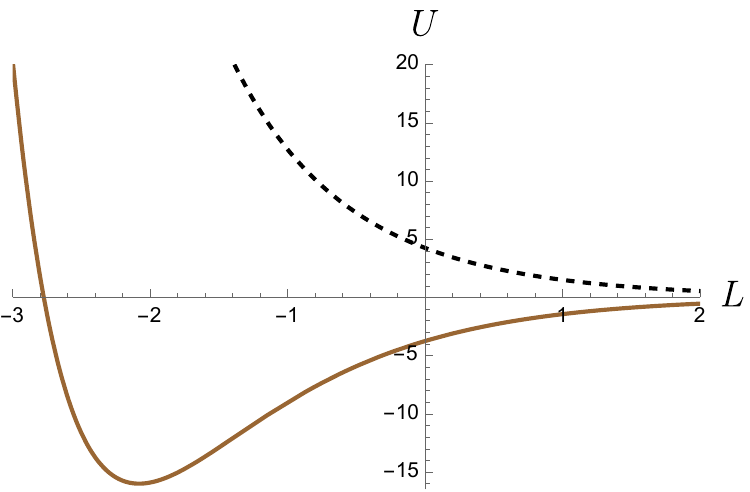}}
    \caption{Interpretation of the q-Askey deformed theories from different perspectives. (a) Chord-diagrammatic representation, where $\ket{B^{(n_F)}_{\vec{t}}}$ \eqref{eq:ETW op} is the chord state associated with ETW branes in the dual bulk theory. (b) Corresponding Lorentzian bulk interpretation for the $n_F=2,~4,~6$ cases. Krylov complexity in the boundary theory determines a parameter $L$ \eqref{eq:length ETW wormhole} (blue) which in the bulk corresponds to the geodesic length $L_{\rm AdS}$ joining one asymptotic boundary of an AdS$_2$ black hole at anchoring points (black) to an ETW brane (red). For the most general type of deformation (Sec.~\ref{ssec:explicit microscopic}), we expect a more intricate bulk geometry. (c) In the low energy limit, most of the q-Askey deformations are described by a Schrödinger equation with a Morse potential $U(L)=\nu\rme^{-L}+\rme^{-2L}/4$. This admits scattering states when $\nu>0$ (dashed black line) and bound states (solid brown line) otherwise.}
    \label{fig:CS_nF2}
\end{figure}

In Sec.~\ref{sec:Krylov Askey} we discuss Krylov complexity with the HH state as a reference in the Lanczos algorithm, using the matterless sector of the deformed theory. We investigate the bulk interpretation and show that the Krylov complexity of the HH state corresponds to the minimal geodesic length connecting an AdS$_2$ boundary to an ETW brane, whose details depend on the parameters of the deformation.
We then study the semiclassical limit of the deformed theories at finite temperature. We find that the family of deformations in the $n_F=2,~4,~6$ classification still describes single-sided ETW branes in the bulk. The representation of these cases is displayed in The boundary and bulk representations of the system are shown in Fig.~\ref{fig:CS_nF2}.

Meanwhile, the bulk interpretation of the $n_F=8$ and adjoint theories are, in general, more involved. The exact bulk dual of the $n_F=2,~4$ deformed theories at the quantum gravity level is known in sine dilaton gravity; while that for the $n_F=6,~8$ and adjoint cases are not yet known; however, our results suggest that they would correspond to extensions of the known cases in sine dilaton gravity with ETW branes. Modifications involving two bulk asymptotic boundaries are necessary for the adjoint case; it might describe a spacetime where a pair of ETW branes is glued with Israel junction conditions.
We show that the Krylov spread complexity of the deformed theories is not necessarily convex, unlike in chaotic quantum mechanical systems at infinite $N$, which linearly at late times and faster than linear at earlier times \cite{Alishahiha:2024vbf,Nandy:2024evd,Rabinovici:2025otw,Baiguera:2025dkc}. This indicates a different behaviour with respect to the undeformed DSSYK (which is maximally \cite{Maldacena:2015waa} chaotic \cite{aleiner1996divergence,Rozenbaum:2016mmv} with respect to the ``fake temperature'' \cite{Lin:2023trc}). This is primarily due to the deformation of the SYK model being a function of an operator bilinear in the Majorana fermions. We associate the intermediate chaotic behaviour with the specific terms of the deformation in the microscopic models, where one includes matter operators that change the spin of the PETS. In the double-scaling limit, this corresponds to dressed integrable operators. 

Interestingly, there is a particular limit where the adjoint case describes RMT in the auxiliary quantum mechanical system. The Krylov complexity in this specific regime of the adjoint case can be evaluated exactly. The adjoint case also allows for a very simple connection between the deformed chord theory and the $\mathcal{N}=2$ SU(2) with matter hypermultiplets description under the Schur-SYK duality proposed by Gaiotto-Verlinde \cite{Gaiotto:2024kze}: Krylov complexity is the rate of representation-theoretic spread of the SU(2) spin in the index. As an aside, we provide an interpretation of Krylov complexity in terms of the quantum group of the DSSYK model and its deformations. In the specific $n_F=4$ case with a magnetic field, Krylov complexity corresponds to the radial spread of the wavefunction of a particle of the quantum disk (see \cite{Berkooz:2018jqr,Berkooz:2022mfk,Blommaert:2023opb, Almheiri:2024ayc,Belaey:2025ijg, vanderHeijden:2025zkr, Schouten:2025tvn}).

Our results on Krylov complexity also clarify the bulk interpretation of the discrete energy spectrum solutions for the deformed chord Hamiltonians, which describe bound state excitations, when investigating the semiclassical limit beyond the low energies, which is associated with sine dilaton gravity (even in the quantum regime \cite{Blommaert:2024ymv}). We find that while the Krylov complexity of the HH state is still identified with a bulk geodesic length, the black hole ADM energy becomes discretised, and the time coordinate undergoes a Wick rotation. We interpret these solutions as arising from a new contour in sine dilaton gravity with ETW branes, describing an effective AdS$_2$ black hole geometry \cite{Blommaert:2024ymv}. The new contour includes a Wick rotation for the time relative to the continuum spectrum solutions. Hence, as one of the deformation parameters increases, there occurs a transition between the Lorentzian AdS$_2$ black hole describing the continuum energy solution and a Euclidean one whose ADM energy is quantised. {The sum over all the discrete spectrum contributions in the partition function corresponds to different energy eigenstates of a Wick-rotated AdS$_2$ black hole, together with the continuous spectrum associated with an AdS$_2$ black hole in sine dilaton gravity.\footnote{The simultaneous existence of the discrete and continuous states might signal a spacetime superposition \cite{Giacomini:2021gei,Giacomini:2020ahk} in the bulk.} The bulk origin of the black hole saddle-point solution associated with the discrete spectrum remains an open problem. However, at least in the low-energy regime, the discrete spectrum is a consequence of generating sufficiently negative tension in the ETW brane to produce states with negative length, and it has equivalent effects as opening an eternal traversable wormhole due to negative energy matter in AdS$_2$.} 

We also relate our results to Cauchy slice holography \cite{Araujo-Regado:2022gvw,Araujo-Regado:2022jpj,Soni:2024aop,Khan:2025gld,Araujo-Regado:2025elv}, where one finds a transition between Euclidean and Lorentzian geometries by increasing the T$\overline{\text{T}}$ deformation parameter \cite{Smirnov:2016lqw,Cavaglia:2016oda} in higher-dimensional finite cutoff holography \cite{Hartman:2018tkw,Taylor:2018xcy}. Thus, the ETW brane parameter in sine dilaton gravity, which is controlled by the deformation parameter in our setting, acts analogously to the radial cutoff in Cauchy slice holography. In our case, the energy spectrum varies discretely along the flow induced by the deformation parameter instead of continuously for T$\overline{\text{T}}$ deformations.

In Sec.~\ref{sec:OP algebras}, we discuss the operator algebras of the deformed theories, particularly, we identify that operators encoding thermal information of the system are part of a type II$_1$ von Neumann factor. We denote this as the q-Askey double-scaled algebra, which has the form
\begin{equation}\label{eq:part 2}
    \mathcal{A}_{\rm DS}^{(n_F)}\equiv\langle \hH_{n_F}^{(R)},~\hmO_\Delta^{(R)}\rangle''~,
\end{equation}
where $\hH_{n_F}^{(R)}$ and $\hmO_\Delta^{(R)}$ are the chord Hamiltonians and matter operators in the auxiliary chord theory {and $\langle~,~\rangle$ denotes the algebra generated by the corresponding operators. $''$ is the bicommutant, which makes $\mathcal{A}_{\rm DS}^{(n_F)}$ closed under weak operator topology}. The repeated action of $\hH_{n_F}^{(R)}$ and $\hmO_\Delta^{(R)}$ on an appropriate tracial state generates a GNS Hilbert space, under which there is a closed action of the double-commutant $\mathcal{A}_{\rm DS}^{(n_F)}$.

We also explore the geometric meaning of the entanglement entropy between the type II$_1$ algebras. Once we include an operator that probes the purity of the global state of the system into the previous algebra, corresponding to $\hH_{\rm int}$ in \eqref{eq:part_1} in the auxiliary chord theory, one recovers a type I$_\infty$ algebra. We study the operator algebra of the deformed theories and the entanglement entropy between the algebras for a given pure state. We emphasise the connection between modular inclusions and the Schwarzian limit of the theory. The algebras are constructed from the deformed Hamiltonians and DSSYK matter operators acting on the chord Hilbert space of the deformed theory, which we call the q-Askey double-scaled algebra. We define an isomorphism map between the double-scaled algebras of the undeformed DSSYK model and that of q-Askey deformations (following a similar discussion as \cite{Cao:2025pir}). The resulting algebras are type II$_1$ factors. Meanwhile, when we attach the total chord number operator to the algebra, it becomes type I$_\infty$ factor, which indicates there is access to the purity of the full system prepared in a PETS. Our results agree with the interpretation of the chord number as being dual to a geodesic length in the bulk, which has access to the quantum state of the bulk theory for a given Cauchy slice. Given that the Krylov state complexity is the expectation value of the chord number, this also extends the bulk interpretation of our results, including matter. Meanwhile, the type II$_1$ factor indicates one has access only to thermal information, encoded by a density matrix belonging to the operator algebra in terms of the deformed Hamiltonian and matter operator. Therefore, one does not have access to the purity of the state used to compute the expectation value of ensemble-averaged operators. We apply our results to study entanglement entropy between the q-Askey double-scaled algebras for a given pure state. We then obtain a precise match to the extremal area in the bulk, which corresponds to the dilaton in the Schwarzian limit of the deformed DSSYK models. This discussion crucially relies on the homology constraint in the extremal area surface (which is just a point) with the addition of an ETW brane, which provides a concrete realization of holographic entanglement entropy. The results are consistent with the interpretation of JT gravity as the s-wave reduction of a BTZ black hole. In the higher-dimensional setting, there are thermal and connected RT surfaces (between the anchoring boundary surface and the ETW brane). The thermal phase becomes trivial in the s-wave reduction, while the connected surface gives a non-trivial matching.

Finally, Sec.~\ref{sec:disc} contains a summary of our work and future directions. {Importantly, we discuss how to study the emission of baby universes through the Haking-Page transition based on the q-Askey deformation of a two-copy DSSYK model. The q-Askey deformation terms used in this work are equivalent to generalised Maldacena-Qi \cite{Maldacena:2018lmt} interaction terms that can be used to generate baby universes. The discrete spectrum states in this work provide an addition to existing literature on this process beyond JT gravity.} 
We also include pedagogical appendices and technical details supporting the main text, including a glossary of our notation in App~\ref{app:nomenclature}.

\section{Microscopic q-Askey deformations}\label{sec:micro models}
{In this section, we introduce deformations of the SYK model whose annealed ensemble-averaged moments are described in terms of recurrence relations of the basic hypergeometric orthogonal polynomials in the q-Askey scheme. This serves as our definition for q-Askey deformations of the SYK model. We construct their chord Hilbert space, and we study their triple-scaling limit. We also show that there is some degree of universality in our results: the ensemble average expectation values can be taken in a class of PETS, where the details of the microscopic states modify certain observables in the ensemble-averaged theory.} 

\paragraph{Outline}In Sec.~\ref{ssec:DSSYK review} we review the double-scaling limit of the SYK model; its two-copy formalism \cite{Lin:2022rbf}, {and provide new results regarding PETS in the double scaling limit}. In Sec.~\ref{ssec:def qAskey def} we define q-Askey deformations of the DSSYK model, as well as their chord Hilbert space. Sec.~\ref{eq:partitil func} then discusses the discrete and continuous energy spectrum contributions to the partition function. In Secs.~\ref{ssec:KM state deformations} and \ref{ssec:explicit microscopic} we present the explicit microscopic deformations of the SYK model based on KM states and TFD states respectively. In Sec.~\ref{ssec:triple scaling limit} we investigate the triple-scaling limit of the deformed theories, which describe either JT gravity with ETW branes or pure JT gravity.

\subsection{The DSSYK model and partially entangled thermal states}\label{ssec:DSSYK review}
We first review how to evaluate observables in the DSSYK model, and we also provide new results at the end of the section on how to evaluate expectation values in PETS \cite{Goel:2018ubv,Goel:2023svz}.

Consider the SYK model, consisting of $N$ Majorana fermions $\hat{\psi}_{1\leq i\leq N}$, $\qty{\hat{\psi}_i,~\hat{\psi}_j}=2\delta_{ij}$, with $p$-local all-to-all interactions:
\begin{equation}\label{eqn:H-SYK}
    \hH_{\rm SYK}\equiv \rmi^{p/2}\sum_{1\leq i_1\leq\dots \leq i_p\leq N}J_I\Psi_I~,\quad I\equiv \qty{i_1\dots i_{p}}~,
\end{equation}
where we denote
\begin{equation}\label{eq:string Psi}
    \Psi_I\equiv\hat{\psi}_{i_1}\dots\hat{\psi}_{i_p}~,\quad J_{I}\equiv J_{i_1\dots i_p}\in\mathbb{R}~.\\
\end{equation}
We also consider operators in the same universality class as \eqref{eqn:H-SYK} (which we refer to as \textit{SYK operators}):
\begin{subequations}\label{eq:def SYK matter ops}
\begin{align}
    \hat{\mathcal{O}}_{\rm SYK}&\equiv \rmi^{p'/2}\sum_{1\leq\dots\leq p'\leq N}O_{I'}\Psi_{I'}~,\quad I'\equiv\qty{i_1\dots i_{p'}}~,\\
    &O_{I'}\equiv O_{i_1\dots i_{p'}}\in\mathbb{R}~,\qquad \Delta\equiv p'/p~.\label{eq:Delta def}
\end{align}
\end{subequations}
In both cases, we assume Gaussian distributions for the random couplings:
\begin{subequations}\label{eq:J coupling}
\begin{align}
\overline{J_I}&=0~,\quad \overline{J_{I_1}J_{I_2}}=J^2\delta_{I_1I_2}\begin{pmatrix}
N\\
p
\end{pmatrix}^{-1}~,\\
\overline{O_{I'}}&=0~,\quad \overline{O_{I'_1}O_{I'_2}}=O^2\delta_{I'_1I'_2}\begin{pmatrix}
N\\
p'
\end{pmatrix}^{-1}~,
\end{align}
\end{subequations}
where $J$ and $O$ are arbitrary real constants, and the normalizations are chosen so that $\Tr[\mathbb{1}]=1$. We will set $J=1$ and $O=1$ for convenience, which can be restored in the expressions by dimensional analysis.

\begin{figure}
    \centering
    \includegraphics[width=0.3\linewidth]{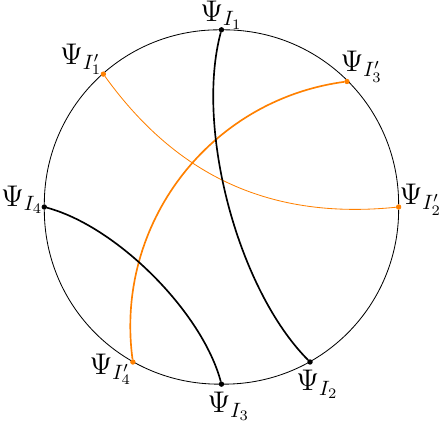}
    \caption{Example of a chord diagram to evaluate \eqref{eq:BNS exp val} with SYK Hamiltonians (black) and SYK operator insertions (orange) as $\eval{\expval{(\hH_{\rm SYK}\hmO_{\rm SYK})^4}}_{\rm DS}$. Each node in the thermal circle represents a string of Majorana fermions, which are pairwise Wick contracted between each other due to \eqref{eq:J coupling}.}
    \label{fig:examplechord}
\end{figure}

A remarkable simplification occurs when studying the double-scaling limit
\begin{equation}\label{eq:DS limit}
{\rm DS}~:\quad N~,~~p\rightarrow\infty~,\quad q\equiv\rme^{-p^2/N}~{\rm fixed}~.
\end{equation}
It has been shown in \cite{Berkooz:2018qkz} that observables of the type
\begin{equation}\label{eq:BNS exp val}
\expval{P(\hH_{\rm SYK},\hat{\mathcal{O}}_{\rm SYK})}\equiv\overline{\Tr(P(\hH_{\rm SYK},\hat{\mathcal{O}}_{\rm SYK}))}~,
\end{equation}
with $P$ being a polynomial function of the operators in the argument. \eqref{eq:BNS exp val} can be computed by an effective quantum system described by chord rules. {This description arises by representing the trace in \eqref{eq:BNS exp val} as a thermal circle and including the SYK operators as nodes. Due to the Gaussian distributions of the couplings $J_I$ and $O_{I'}$ in \eqref{eq:J coupling}, the nodes are pairwise connected, leading to Wick contraction between the strings of Majorana fermions. An example is displayed in Fig.~\ref{fig:examplechord}.}
 Concretely, expectation values of polynomial functions of the Hamiltonian moments with SYK matter operators take the form:
\begin{subequations}\label{eq:eq_exp_values}
\begin{align}
\eval{\expval{P(\hH_{\rm SYK},\hat{\mathcal{O}}_{\rm SYK})}}_{\rm DS}&=\bra{\Omega}P(\hat{a}+\hat{a}^\dagger,~\hat{b}+\hat{b}^\dagger)\ket{\Omega}~,
\end{align}
\end{subequations}
where DS denotes \eqref{eq:DS limit}, and $\ket{\Omega}$ is the zero-chord number state in the auxiliary chord space of the undeformed theory. The latter is given as:
\begin{equation}\label{eq:H space nF=0}
\mathcal{H}_0={\rm span}\qty{\ket{n}~|~~ n\in\mathbb{Z}_{\geq0}}={\rm L}^2(\theta\in[0,\pi],\rmd\mu(\theta))~,
\end{equation}
so that we denote $\ket{\Omega}\equiv\ket{n=0}_{n_F=0}$. Operators acting on the Hilbert space obey the relations \cite{Berkooz:2018jqr}:
\begin{equation}\label{eq:H SYK0}
    \hH_{n_F=0}\ket{n}=\ket{n+1}+[n]_{q^2}\ket{n-1}~,
\end{equation}
with $[n]_{q^2}\equiv \frac{1-q^{2n}}{1-q^2}$. The different creation and annihilation operators in \eqref{eq:eq_exp_values} obey the commutation relations:
\begin{equation}\label{eq:q-oscillator}
    [\hat{a},~\hat{a}^\dagger]_{q^2}\equiv \hat{a} \hat{a}^\dagger - q^2 \hat{a}^\dagger \hat{a} =1~,\quad [\hat{b},~\hat{b}^\dagger]=1~.
\end{equation}
The system is isomorphic to a q-deformed harmonic oscillator, with creation and annihilation operators changing the number of open chords, $n$, {in the \emph{monic} Krylov basis representation \cite{Muck:2022xfc} as:}
\begin{equation}\label{eq:a a dagger}
    {\rm Monic:}~~\hat{a}^\dagger\ket{n}=\ket{n+1}~,\quad \hat{a}\ket{n}={[n]_{q^2}}\ket{n-1}~.
\end{equation}
In terms of the above basis, one can then represent the Hamiltonian in the auxiliary description as:
\begin{equation}\label{eq:zero particle H}
    \hH_{n_F=0}\equiv \hat{a}^\dagger+\hat{a}=\rme^{-\rmi \hat{p}}+\rme^{\rmi \hat{p}}[\hat{n}]_{q^2}~,
\end{equation}
where we define
\begin{equation}\label{eq: Oscillators 1}
    \rme^{\pm\rmi \hat{p}}\ket{n}=\ket{n\mp1}~,\quad \hat{n}\ket{n}=n\ket{n}~.
\end{equation}
We call $\hat{n}$ the chord number operator; it has the following chord algebra relations:
\begin{equation}\label{eq:property n}
    [\hat{n},~\hat{a}^\dagger]=\hat{a}^\dagger~,\quad [\hat{n},~\hat{a}]=-\hat{a}~.
\end{equation}
One can straightforwardly implement a change of basis where the Hamiltonian is symmetric, which is the \emph{orthogonal} Krylov basis, {where the raising and lowering operators act in the chord number basis as}
\begin{equation}\label{eq:a a dagger ortho}
    {\rm Orthogonal:}~~\hat{a}^\dagger\ket{n}=\sqrt{[n+1]_{q^2}}\ket{n+1}~,\quad \hat{a}\ket{n}=\sqrt{[n]_{q^2}}\ket{n-1}~,
\end{equation}
{which preserves the algebra in \eqref{eq:property n}.}

Next, we review an extension of the chord Hilbert space with matter introduced in \cite{Lin:2022rbf,Lin:2023trc}. This relies on realizing that chord operators of the DSSYK model form a pair of double-scaled algebras \cite{Lin:2022rbf,Xu:2024hoc} corresponding to two commuting left/right subsystems (or copies) of the theory, defined by: 
\begin{equation}\label{eq:algebra}
\mathcal{A}_{L/R}=\langle\hH_{L/R},~\hat{\mathcal{O}}_\Delta^{{(L/R)}}\rangle''~,    
\end{equation}
where we use the same notation as in \eqref{eq:part 2}. This algebra of operators allows us to generate the GNS Hilbert space by acting with polynomial functions of the operators on the cyclic separating state, which is $\ket{\Omega}$ \cite{Xu:2024hoc}, which results in the chord Hilbert space with matter \cite{Lin:2022rbf}:
\begin{equation}\label{eq:states notation matter}
\mathcal{H}_m\equiv{\rm span}\qty{\ket{\Delta_1,\dots,\Delta_m;n_0,n_1,\dots,n_m}}~,
\end{equation}
where $\Delta_{1\leq i\leq m}$ are the conformal dimensions of the matter chord operators
\begin{equation}\label{eq:reference pets}
\qty{\hat{\mathcal{O}}_{\Delta_{1}},\dots,~\hat{\mathcal{O}}_{\Delta_{m}}}~,  \quad m\in\mathbb{Z}_{>0}~;
\end{equation}
while $n_0$ in \eqref{eq:states notation matter} is the number of DSSYK chords (called $H$-chords) to the left of all matter chords, $n_1$ is the number between the first two particles, all the way up to the number of chords between all the $m$ particles. The inner product in $\mathcal{H}_m$ was developed in \cite{Lin:2023trc,Xu:2024hoc}. This is used to define the full chord Hilbert space containing all possible operator insertions with fixed conformal dimensions $\Delta_{1\leq i\leq m}$:
\begin{equation}\label{eq:H ffull}
    \mathcal{H}_{\rm chord}=\bigoplus_{m=0}^\infty\mathcal{H}_m~.
\end{equation}
The dynamics in the model are controlled by the chord Hamiltonian with matter \cite{Lin:2022rbf}
\begin{equation}\label{eqn:hH_LR}
    \hH_{L/R}=\hat{a}^\dagger_{L/R}+\hat{a}_{L/R}~,
\end{equation}
where, in the monic representation \cite{Muck:2022xfc}
\begin{subequations}\label{eq:two_sided_DSSYK_H}
    \begin{align}
\hat{a}_L^\dagger&=\hat{a}_0^\dagger~,\quad \hat{a}_L=\sum_{i=0}^m\hat{\alpha}_i[\hat{n}_i]_{q^2}~q^{2\hat{n}_i^<}~,\quad \hat{n}_i^<=\sum_{j=0}^{i-1}\qty(\hat{n}_j+\Delta_{j+1})~,\label{eq:HLmultiple}\\
    \hat{a}_R^\dagger&=\hat{a}_m^\dagger~,\quad \hat{a}_R=\sum_{i=0}^m\hat{\alpha}_i[\hat{n}_i]_{q^2}~q^{2\hat{n}_i^>}~,\quad\hat{n}_i^>=\sum_{j={i+1}}^{m}\qty(\hat{n}_j+\Delta_{j})~,\label{eq:HRmultiple}
\end{align}
\end{subequations}
which was obtained by \cite{Lin:2022rbf} by considering different types of crossings between Hamiltonian and matter chords once either type of chord is inserted or removed from the chord diagram.
The creation and annihilation operators act on the chord basis as:
\begin{subequations}\label{eq:Fock Hm}
    \begin{align}\label{eq:Fock Hm 1}
    \hat{a}^\dagger_{i}\ket{\Delta_1,\dots,\Delta_m;n_0,\dots n_i,\dots, n_m}&=\ket{\Delta_1,\dots,\Delta_m;n_0,\dots, n_i+1,\dots n_m}\\\label{eq:Fock Hm 2}
    \hat{\alpha}_{i}\ket{\Delta_1,\dots,\Delta_m;n_0,\dots n_i,\dots, n_m}&=\ket{\Delta_1,\dots,\Delta_m;n_0,\dots, n_i-1,\dots n_m}~.
\end{align}
\end{subequations}
We also define the total chord number, which counts the number of $H$- and matter-chords as:
\begin{equation}\label{eq:total chord number}
    \hat{N}\equiv \sum_{i=0}^m\hat{n}_i+\Delta_i~.
\end{equation}
Once this operator is included in the double-scaled algebra, it becomes a type I$_\infty$, the so-called chord algebra \cite{Lin:2023trc}.

\paragraph{Two-copy preparation}
In the following, we study a pair of decoupled physical SYK models, with Hilbert space:
\begin{equation}\label{eq:SYK D}
    \mH_{\rm SYK^2}\equiv\mH_{\rm SYK}\otimes\mH_{\rm SYK}~,
\end{equation}
where:
\begin{equation}
    \mathcal{H}_{\rm SYK}={\rm span}\qty{\ket{E_n}}~,\quad \bra{E_n}\ket{E_m}=\delta_{nm}~,
\end{equation}
where we denote its eigenstates of the SYK Hamiltonian as $\ket{E_n}$:
\begin{equation}\label{eq:SYK H}
    \hH_{\rm SYK}\ket{E_n}=E_n\ket{E_n}~.
\end{equation}
We similarly describe SYK matter operators \eqref{eq:def SYK matter ops} acting on either copy by:
\begin{equation}\label{eq:O left}
    \hat{\mathcal{O}}_{\rm SYK}^{(L)}\equiv\hat{\mathcal{O}}_{\rm SYK}\otimes\mathbb{1}~,\quad  \hat{\mathcal{O}}_{\rm SYK}^{(R)}\equiv\mathbb{1}\otimes\hat{\mathcal{O}}_{\rm SYK}~.
\end{equation}
General states in \eqref{eq:SYK D} in the presence of matter operators \eqref{eq:O left} can be written as:
\begin{equation}\label{eq:SYK double states}
    \ket{\psi_{\rm SYK}}\equiv \sum_{nm}{\psi_{nm}\ket{E_n}\otimes\ket{E_m}}~.
\end{equation}
A simple case is when we impose an equal energy constraint in the full Hilbert space \eqref{eq:SYK D}:
\begin{subequations}
\begin{align}
    &\left(\hH_{\rm SYK}^{(L)}-\hH_{\rm SYK}^{(R)}\right)\ket{\psi_{\rm SYK}}=0~,\label{eq:states constraint}\\
    \hH_{\rm SYK}^{(L)}&\equiv \hH_{\rm SYK}\otimes \mathbb{1}~,\quad \hH_{\rm SYK}^{(R)}\equiv \mathbb{1}\otimes\hH_{\rm SYK}~,\label{eq:HLR SYK}
\end{align}
\end{subequations}
whose double-scaling limit was explored in \cite{Narovlansky:2023lfz}. The states of the form \eqref{eq:states constraint} are TFD-like, and they can be written as \eqref{eq:SYK double states} with $\psi_{nm}=f(E_n)\delta_{nm}$, with $f$ a function of $E_n$. In particular, for $f(E_n)=1$, we simply recover the infinite temperature TFD:
\begin{equation}\label{eq:inf temp TFD}
    \ket{\rm TFD_\infty}\equiv\sum_n\ket{E_n}\otimes\ket{E_n}~.
\end{equation}
A special set of states \eqref{eq:SYK double states}, called PETS \cite{Goel:2018ubv}, can be constructed by acting with matter operators and the SYK Hamiltonians on the TFD state 
\begin{equation}\label{eq:other_PETS}
\ket{\psi_{\rm PETS}}\equiv\rme^{-\frac{\beta_L}{2}\hH_{\rm SYK}^{(L)}-\frac{\beta_R}{2}\hH_{\rm SYK}^{(R)}}\hmO_{\rm SYK}^{(L)}\ket{\rm TFD_{\infty}}~,
\end{equation}
where $\hmO_{\rm SYK}^{(L)}$ and $\ket{\rm TFD_{\infty}}$ are defined in \eqref{eq:O left} and \eqref{eq:inf temp TFD} respectively. The norm of \eqref{eq:other_PETS} defines a partition function with matter insertions, which will be discussed in Sec.~\ref{ssec:explicit microscopic}.

As noticed by \cite{Lin:2022rbf}, in this formulation, taking traces over the full Hilbert space of a single SYK model is equivalent to evaluating expectation values with respect to the TFD state for operators acting on only one of the subsystems of the two-copy SYK system, i.e.:
\begin{equation}
\expval{P\qty(\hH_{\rm SYK}^{(L/R)},~\hat{\mathcal{O}}_{\rm SYK}^{(L/R)})}\equiv\overline{\bra{\rm TFD_\infty}P\qty(\hH_{\rm SYK}^{(L/R)},~\hat{\mathcal{O}}_{\rm SYK}^{(L/R)})\ket{\rm TFD_\infty}}~,
\end{equation}
where $P$ represents a polynomial function. {Note that the correlation functions of generic two-sided operators do not need to converge in the large $N$ limit, but those acting on one-sided Hilbert space do, and are determined by chord rules. The same occurs in the set of deformations that we investigate in the remainder of the section.}

As found by Lin \cite{Lin:2022rbf}, one recovers the chord diagram relations as if one evaluated traces in the single SYK model \eqref{eq:BNS exp val}, namely:
\begin{equation}\label{eq:double scaling L}
    \expval{P\qty(\hH_{\rm SYK}^{(L/R)},~\hat{\mathcal{O}}_{\rm SYK}^{(L/R)})}\underset{\rm DS}{\rightarrow}\bra{\Omega}P\qty(\hH_{L/R},\hat{\mathcal{O}}_\Delta^{(L/R)})\ket{\Omega}~,
\end{equation}
where $\hH_{L/R},~\hat{\mathcal{O}}_\Delta^{(L/R)}$ are the operators acting on the extended Hilbert space with matter in \eqref{eq:algebra}. 

\paragraph{Additional PETS in the double-scaling limit}
{We will now discuss an extension of the above results using a family of PETS which is different from \eqref{eq:other_PETS}, and that was first defined in \cite{Goel:2018ubv}, and whose double-scaling limit is developed in this section and Apps.~\ref{app:more PETS}, \ref{app:PETS extension algebra}.\footnote{See \cite{Rajgadia:2026ask} for related recent developments about KM states in the double-scaled context.}}

To set up the stage, we consider a two-copy SYK model or one of its q-Askey deformations. We bipartition $4N$ Majorana fermions, where the $L/R$ subsystems have an equal number of fermions each, denoted $\hat{\psi}^{(L)}_{i}$ and $\hat{\psi}^{(R)}_i$ with $1\leq i\leq 2N$, which generates a Hilbert space $\mathcal{H}_L\otimes\mathcal{H}_R$. For this type of system, \cite{Goel:2018ubv} (see App.~A) constructed a class of PETS given by,
\begin{equation}\label{eq:def PETS}
\ket{s}_{\rm PETS}=\rme^{-\beta_L\hH_L}\otimes\rme^{-\beta_L\hH_L}\ket{s}
\end{equation}                                    
where $\ket{s}$ are eigenstates of the spin operator
\begin{equation}\label{eq:spin op}
\qty(\hat{S}^{(\varepsilon)}_k-s^{(\varepsilon)}_k)\ket{s}=0~,\quad \hat{S}^{(\varepsilon)}_k\equiv \rmi\hat{\psi}^{(\varepsilon)}_{2k-1}\hat{\psi}^{(\varepsilon)}_{2k}~,\quad \varepsilon=L,~R~.
\end{equation}
Depending on how precisely the $\hat{\psi}^{(\varepsilon)}_i$ are bipartitioned, one recovers different entangled states. For instance, we construct the TFD state by taking $\hat{\psi}_i^{(L)}$ and $\hat{\psi}_i^{(R)}$ as either $\hat{\psi}_{2k}$ and $\hat{\psi}_{2k-1}$ respectively, with $k\in\mathbb{Z}_{\geq0}$ and $s^{(L)}_k=s^{(R)}_k=1$. Meanwhile, one recovers a tensor product of thermal pure states (i.e.~KM states \cite{Kourkoulou:2017zaj}) when $\hat{\psi}^{(L)}=\qty{\hat{\psi}_{2k-1},~\hat{\psi}_{2k}}_{k=1}^{N}$ and similarly for $\hat{\psi}^{(R)}$. 
\begin{figure}
    \centering
    \subfloat[]{\includegraphics[height=0.32\linewidth]{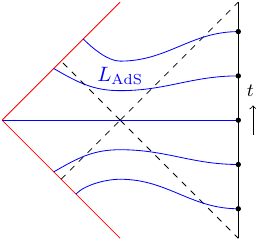}\label{sfig:CS4}}\hspace{0.5cm}\subfloat[]{\includegraphics[height=0.32\linewidth]{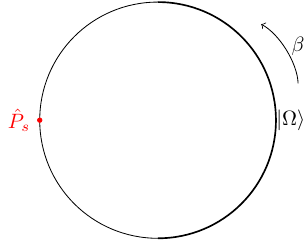}\label{sfig:ETW_nF4}}
    \caption{(a) Lorentzian Bulk interpretation of the KM state. The ETW brane corresponds to a shockwave propagating outside the black hole (red null rays) generated by inserting heavy operators in the thermal circle (red dot), and we display the minimal length geodesics (blue) joining one asymptotic boundary (black) of an AdS$_2$ black hole to a pair of ETW branes (red). (b) Chord diagram of the KM state, where $\hat{P}_s$ represents a projection on the KM state used in the Hamiltonian moments \eqref{eq:H moments KM}.}
    \label{fig:CS_nF4}
    \end{figure}
The values $s^{(\varepsilon)}_k$ in each subsystem can be fixed independently. The tensor product of states can be visualised from Fig.~\ref{fig:CS_nF4} as a two-copy identical system.\footnote{The operator size \eqref{eq:rescaled op size} for the pair of KM states corresponds to an effective coupling between SYK systems \cite{Sasieta:2025vck}.} For more general bipartitions, one can interpolate between the thermal and pure states, as presented in Fig.~\ref{fig:PETS}.

Note that the different bipartitions correspond to different numbers of nearest neighbor fermion pairings in each of the two subsystems (i.e.~for $\hat{\psi}^{(\varepsilon)}_{2k-1}$ and $\hat{\psi}^{(\varepsilon)}_{2k}$) that defines the spin operators $\hat{S}_k^{(\varepsilon)}$. 
Let us denote the total number of nearest neighbor pairings in each subsystem ($L/R$) by $n_{\rm tot}$, which interpolates between $n_{\rm tot}=0$ for the TFD state, and $n_{\rm tot}=N$ for a product of thermal pure states. Depending on the number of pairings, one can construct a set of creation and annihilation operators
\begin{equation}
\hat{c}^{(\varepsilon)}_k=\frac{1}{2}\qty(\hat{\psi}^{(\varepsilon)}_{2k-1}+\rmi\hat{\psi}^{(\varepsilon)}_{2k})~,\quad \hat{c}^{(\varepsilon)}_k{}^\dagger=\frac{1}{2}\qty(\hat{\psi}^{(\varepsilon)}_{2k-1}-\rmi\hat{\psi}^{(\varepsilon)}_{2k})~,
\end{equation}
which change the spin of the $\ket{s}$ states in \eqref{eq:spin op} due to the commutation relations
\begin{equation}
[\hat{S}^{(\varepsilon)}_j,~\hat{c}^{(\varepsilon)}_i]=2\delta_{ij}\hat{c}^{(\varepsilon)}_j{}^\dagger~,\quad [\hat{S}^{(\varepsilon)}_j,~\hat{c}^{(\varepsilon)}_i{}^\dagger]=2\delta_{ij}\hat{c}^{(\varepsilon)}_j~;
\end{equation}
other commutators vanish.

The creation/annihilation operators lead to a natural set of matter operators acting on the PETS \eqref{eq:def PETS}
\begin{align}\label{eq:def M Mdagger}
\hat{M}_{\varepsilon}&\equiv\rmi^{w/2}\sum_{I''}R_{I''}{C}^{(\varepsilon)}_{I''}~,\quad\hat{M}_{\varepsilon}^\dagger\equiv(-\rmi)^{w/2}\sum_{I''}R^*_{I''}{C}^{(\varepsilon)}_I{}^\dagger~,
\end{align}
where $I''=\qty{i_1,\dots,i_{w}}$ with $1\leq i_1\leq\cdots\leq i_{w}\leq n_{\rm tot}$; and we defined
\begin{equation}\label{eq:C upper}
    \begin{aligned}
        {C}^{(\varepsilon)}_I&\equiv \hat{c}^{(\varepsilon)}_{i_1}\dots \hat{c}^{(\varepsilon)}_{i_{w}}~,&R_I&\equiv R_{i_1,\dots,i_{w}}\\
    {C}^{(\varepsilon)}_I{}^\dagger&\equiv \hat{c}^{(\varepsilon)}_{i_w}{}^\dagger\dots \hat{c}^{(\varepsilon)}_{i_{1}}{}^\dagger~, &R_I^*&\equiv R_{i_1,\dots,i_{w}}^*~.
    \end{aligned}
\end{equation}
The random couplings $R_{i_1,\dots,i_{w}}$ and $R_{i_1,\dots,i_{w}}^*$ can be chosen to be Gaussian-distributed
\begin{equation}\label{eq:gaussian dist R}
\overline{R_{i_1,\dots,i_{w}}}=\overline{R_{i_1,\dots,i_{w}}^*}=0~,\quad \overline{R_{i_1,\dots,i_{w}}R_{i_1,\dots,i_{w}}^*}=R^2\begin{pmatrix}
n_{\rm tot}\\
w
\end{pmatrix}^{-1}
\end{equation}
where $R\in\mathbb{R}$ is a constant, which we will take $R=1$. In the following, we are interested in computing in the double-scaling limit
\begin{equation}\label{eq:full polynomial expression}
\begin{aligned}
    &\overline{\bra{s}_{\rm PETS}P(\hH^{(R)}_{\rm SYK},\hat{M}_R,\hat{M}_R^\dagger)\ket{s}_{\rm PETS}}=\overline{\Tr[\ket{s}_{\rm PETS}\bra{s}_{\rm PETS}P(\hH^{(R)}_{\rm SYK},\hat{M}_R,\hat{M}_R^\dagger)]}
\end{aligned}
\end{equation}
 with $P(\cdot)$ a polynomial function, $\hH_{\rm SYK}^{(R)}$ the SYK Hamiltonian acting on the right subsystem of the bipartition, and $\Tr$ indicating the Hilbert space trace over $\mathcal{H}_L\otimes\mathcal{H}_R$. Let us take $n_{\rm tot}\equiv \mathbb{k} N$, where $\mathbb{k}$ depends on the specific partition that we are interested in; $\mathbb{k}=0$ for the TFD state, while $\mathbb{k}=1$ for the product of pure thermal states. We carry out the corresponding analysis of the chord rules with PETS for \eqref{eq:full polynomial expression} in App.~\ref{app:more PETS}. The different values of $\mathbb{k}\in[0,~1]$ modify the penalty factors for chords involving intersections between $\hH_{\rm SYK}^{(R)}$ with $\hat{M}_R$ contracted with $\hat{M}_R^\dagger$. It shows that there are several ways of recovering the same chord auxiliary system, albeit with modifications in the observables, which are sensitive to the same details (e.g.~the number of fermionic pairings) of the underlying microscopic theory. This agrees with and extends the arguments put forward for KM states in \cite{Rajgadia:2026ask}.

In particular, to make connection with \cite{Rajgadia:2026ask}, let us focus in the $\mathbb{k}=1$ case, 
\begin{equation}
\ket{s}_{\rm PETS}\bra{s}_{\rm PETS}=\ket{s}_L\otimes\ket{s}_R\bra{s}_L\otimes\bra{s}_R~.
\end{equation}
Decomposing the polynomial function \eqref{eq:full polynomial expression} into a product of factors on the L/R subsystems, we evaluate \eqref{eq:full polynomial expression} in terms of chord diagrams in the L/R subsystems in App.~\ref{app:more PETS}. This reproduces the results in \cite{Rajgadia:2026ask}, which uses the KM state in a single SYK model to evaluate expectation values of SYK operators instead of the full trace in \eqref{eq:BNS exp val}, resulting in additional matter chord operators which modify the spin of the KM states. The KM state approach of \cite{Rajgadia:2026ask} and the one based on the TFD state in this subsection lead to similar physical conclusions as when using a more generic PETS. In either case, by keeping operators in the algebra of observables that have access to the purity of the PETS, we can recover a type I$_\infty$ algebra (see Sec.~\ref{ssec:qA chord algebra}). Meanwhile, if one only uses operators that have access to thermal information of the system (such as the Hamiltonian and chord matter operators acting on one of the subsystems), by tracing out one of the subsystems, it will be a mixed state rather than a pure one (see App.~\ref{app:PETS extension algebra} for more details), which is in agreement with \cite{Rajgadia:2026ask}. 

{We emphasize that the $\mathbb{k}=1$ limit above describes the product of two pure thermal states, each describing a different SYK system. This allows for clearer bulk interpretation of the SYK deformations in this work, which are introduced in Sec.~\ref{ssec:KM state deformations}.}

\subsection{q-Askey deformations and their Hilbert space}\label{ssec:def qAskey def}
{In this section, we discuss the general properties of the \emph{q-Askey deformations}, which are defined by deforming the SYK model in such a way that the auxiliary quantum mechanical system that reproduces annealed ensemble average observables of the microscopic theory in the double scaling limit encodes the recurrence relation of the orthogonal polynomials in the q-Askey scheme. We are mainly interested in the theory after ensemble averaging. Meanwhile, the explicit microscopic deformations of the SYK model are discussed in Sec.~\ref{ssec:KM state deformations} and \ref{ssec:explicit microscopic}, which are based on KM and TFD states, as well as PETS interpolating between the two. These constructions are motivated by recent works \cite{Cao:2025pir,Rajgadia:2026ask} in the $n_F=2$ case.}

{Based on the above definition of q-Askey deformations, the annealed ensemble-averaged moments of the deformed SYK Hamiltonian, denoted $\hH_{\rm SYK}^{n_F}$, in the double-scaling limit correspond to a chord Hamiltonian $\hH_{n_F}$ in the auxiliary quantum mechanical system:
\begin{equation}\label{eq:Hamiltonia moments}
    \lim_{\rm DS}\overline{\bra{\psi}P\qty(\hH_{\rm SYK}^{n_F})\ket{\psi}}=\bra{0}P\qty(\hH_{n_F})\ket{0}~,
\end{equation}
where, $\ket{\psi}$ represents a state in the SYK model (in Sec.~\ref{ssec:KM state deformations} and \ref{ssec:explicit microscopic}).} $P$ is a generic polynomial function, and $\ket{0}$ is the empty chord state in the auxiliary system, defined in the same way as $\ket{\Omega}$ in the $n_F=0$ case. By repeated action of $\hH_{n_F}$ on $\ket{0}$, one can generate a basis $\ket{n}$ in the Hilbert space of the deformed theory without matter:
\begin{equation}\label{eq:H space nF}
    \mathcal{H}_{n_F}\equiv{\rm span}\qty{\ket{n}|n\in\mathbb{Z}_{\geq0}}={\rm L}^2(\theta\in[0,\pi],~\rmd \mu_{n_F}(\theta))
\end{equation}
where $\mu_{n_F}(\theta)$ is the measure in the energy basis, which we specify below; and the elements of the basis are defined recursively through the relation:
\begin{equation}\label{eq:recurrence states q Askey}
    \hH_{n_F}\ket{n}=a_n\ket{n}+b_{n+1}\ket{n+1}+c_n\ket{n-1}~,
\end{equation}
where $a_n$, $b_n$, $c_n$ are real coefficients, also specified below. Let:
\begin{equation}\label{eq:cont spectrum}
    \hH_{n_F}\ket{\theta}_{n_F}=E(\theta)\ket{\theta}_{n_F}~,\quad E(\theta)=\frac{2\cos\theta}{\sqrt{1-q^2}}~,\quad \theta\in[0,~\pi]
\end{equation}
represent the parametrization of the continuous energy spectrum of the Hamiltonian.

Note that in \eqref{eq:Hamiltonia moments} we also allow for matter operator insertions. Similar to the DSSYK model with matter, we define a corresponding extended chord Hilbert space in the same way as \eqref{eq:H ffull}. The main difference with respect to the DSSYK model is that the action of the chord Hamiltonian onto the extended chord number basis $\ket{\Delta_1,\dots,\Delta_m;n_0,\dots n_m}$ takes a different form.

{Due to the definition of q-Askey deformations, we require that the recurrence relation in the auxiliary chord Hamiltonian \eqref{eq:recurrence states q Askey} for the continuous spectrum \eqref{eq:cont spectrum} takes the form:}
\begin{equation}
    2x~ \psi_n(x)=a_n\psi_n(x)+b_{n+1}\psi_{n+1}(x)+c_{n}\psi_{n-1}(x)~,
\end{equation}
where $\psi_n(\cos\theta)\equiv {}_{n_F}\bra{\theta}\ket{n}$, and its solutions are:
\begin{equation}\label{eq:psi n P n}
    \psi_n(x)=g_nP_n(x)
\end{equation}
with $g_n$ normalization coefficients encoding chord diagram summations. {The explicit form of $g_n$ will be specified in Sec.~\ref{sec:Krylov Askey} since which is not relevant for the present discussion}. On the other hand, $P_n(x)$ in \eqref{eq:psi n P n} are the \emph{Askey-Wilson polynomials}, which satisfy an orthogonality relation \cite{koekoek2010hypergeometric}:\footnote{A brief review of additional mathematical properties of basic hypergeometric orthogonal polynomials can be found in App.~B of \cite{Berkooz:2025ydg}.}
\begin{subequations}\label{eq:recurrence discrete}
    \begin{align}
       \label{eq:orthogonality relation} h_n\delta_{nm}
       &=\begin{cases}
            \int _{-1}^1\rmd x~w(x)P_n(x)P_m(x)~,&{\rm Cond_1=True}\\
            \int _{-1}^1\rmd x~w(x)P_n(x)P_m(x)+\sum_{l=0}^{l_{\rm max}} w_lP_n(x_l)P_m(x_l)~,&{\rm Cond_2=True}
        \end{cases}\\[5pt]
        h_n&\equiv\frac{(t_1t_2t_3t_4;q^2)_\infty(1-t_1t_2t_3t_4q^{2n-2})(q^2,t_1t_2,t_1t_3,t_1t_4,t_2t_3,t_2t_4,t_3t_4;q^2)_n}{(q^2,t_1t_2,t_1t_3,t_1t_4,t_2t_3,t_2t_4,t_3t_4;q^2)_\infty(1-t_1t_2t_3t_4q^{4n-2})(t_1t_2t_3t_4;q^2)_n}~,\label{eq:hn expl}\\[5pt]
        w(\cos\theta)&\equiv\frac{(\rme^{\pm2\rmi\theta};q^2)_\infty}{2\pi\sin\theta (t_1\rme^{\pm\rmi\theta},t_2\rme^{\pm\rmi\theta},t_3\rme^{\pm\rmi\theta},t_4\rme^{\pm\rmi\theta};q^2)_\infty}~,
    \end{align}
\end{subequations}
where $P_0(x)=1$, and $t_{1\leq i\leq 4}$ are parameters of the polynomial, $w_l$ is a discrete weight function that we specify below, as well as the bound $l_{\rm max}$. 

``$\rm Cond_1$'' stands for the following condition on the parameters $t_i$:
\begin{flalign}\label{eqn:Cond-continuous}
    |t_i|\leq 1~,~~t_1t_2,~t_1t_3,~t_1t_4,~t_2t_3,~t_2t_4,~t_3t_4\neq 1~,
\end{flalign}
while ``$\rm Cond_2$'' stands for:
\begin{equation}\label{eq:dissrete cond 1}
\exists\alpha\in \{t_1,t_2,t_3,t_4\}~,~~\abs{\alpha}>1~,~~ t_1t_2,~t_1t_3,~t_1t_4,~t_2t_3,~t_2t_4,~t_3t_4~\in\qty{{z\in\mathbb{C}}|~\abs{z}\leq 1}~.
\end{equation}
Now, we can represent $w_l$ and $l_{\rm max}$ in \eqref{eq:orthogonality relation}. Let $\alpha\in \{t_1,t_2,t_3,t_4\}$ be the parameter that obeys \eqref{eq:dissrete cond 1}, then \cite{koekoek2010hypergeometric}:
\begin{subequations}
    \begin{align}
l&\in[0,~l_{\rm max}]~,\quad l_{\rm max}\equiv \left\lfloor\frac{1}{2\lambda}\log\abs{\alpha}\right\rfloor~,\label{eq:ceiling}\\
\label{eq:wl expl}
    w_l&\equiv \qty(\frac{q^2}{t_1t_2t_3t_4})^l\frac{(1-\alpha^2)^{-1}(\alpha^{-2};q^2)_\infty(1-\alpha^2q^{4l})\prod_{i=1}^4(\alpha t_i;q^2)_l}{(q^2;q^2)_\infty(q^2;q^2)_l\prod_{t_{1\leq i\leq 4}\neq\alpha}(\alpha t_i,\alpha^{-1}t_i;q^2)_\infty(\alpha t_i^{-1} q^2;q^2)_l}~.
\end{align}
\end{subequations}
In general, basic hypergeometric orthogonal polynomials can be represented in terms of basic hypergeometric series \cite{koekoek2010hypergeometric}
\begin{flalign}\label{eqn:hypergeometric}
    _r\phi_s\bigg({\genfrac{}{}{0pt}{}{a_1,...,a_r}{b_1,...,b_s}}; q^2,z\bigg) = 
    \sum_{k=0}^\infty \frac{(a_1,...a_r;q^2)_k}{(b_1,...b_s;q^2)_k}(-1)^{(1+s-r)k} q^{2(1+s-r)\binom{k}{2}} 
    \frac{z^k}{(q^2;q^2)_k}.
\end{flalign}
The values $x_l$ in \eqref{eq:recurrence discrete} satisfy \cite{koekoek2010hypergeometric}:
\begin{equation}\label{eq:xl}
    x_l=\frac{1}{2}\qty(\alpha q^{2l}+\frac{1}{\alpha q^{2l}})~, \quad \text{for}~~ \alpha\in\qty{t_{1\leq i\leq 4}}~~{\rm such~that}~~\abs{\alpha}\geq1~,
\end{equation}
where $l$ depends on the specific value of $\alpha$ by \eqref{eq:ceiling}:
\begin{equation}\label{eq:dissrete cond 2}
   l\in\mathbb{Z}_{\geq0}\quad \text{such that}\quad 1<\abs{\alpha q^{2l}}\leq \abs{\alpha}~.
\end{equation}
Note that the label $n_F$ in \eqref{eq:recurrence states q Askey} corresponds to the classification of the corresponding polynomial in the q-Askey scheme. The different elements in this hierarchy correspond to special cases of the Askey-Wilson polynomials where some of the parameters $t_i$ in the orthogonality relation \eqref{eq:recurrence discrete} take special values, as seen in Tab.~\ref{tab:qaskey}. 
\begin{table}[t!]
\centering
\caption{Classification of the different basic hypergeometric orthogonal polynomials into families labeled by $n_F$ in the q-Askey scheme.}\label{tab:qaskey}
\begin{tabular}{|c|c|c|}
\hline
Label & Parameter values & Polynomial family \\
\hline
$n_F=0$ & $t_1=t_2=t_3=t_4=0$ & q-Hermite polynomials \\
\hline
$n_F=2$ & $t_2=t_3=t_4=0$ & generalised q-Hermite polynomials  \\
\hline
$n_F=4$ & $t_3=t_4=0$ & Al-Salam-Chihara polynomials \\
\hline
$n_F=6$ & $t_4=0$  & Continuous dual q-Hahn polynomials \\
\hline
$n_F=8$ & No restrictions on $t_i$. & Askey-Wilson polynomials \\
\hline
$n_F=Adj.$ & $t_1=-t_3$, $t_2=-t_4=q t_1$ & Continuous q-ultraspherical polynomials \\
\hline
\end{tabular}
\end{table}

\subsection{Partition function: Discrete and continuous contributions}\label{eq:partitil func}
We note that the orthogonality relation for the general $n_F$ case \eqref{eq:recurrence discrete}, for $n=m=0$, serves to define the partition function of the $\hH_{n_F}$ system
\begin{equation}\label{eq:parition}
Z_{n_F}(\beta)\equiv \bra{0}\rme^{-\beta\hH_{n_F}}\ket{0}~.
\end{equation}
Below, we will analyze (i) the continuous and (ii) discrete energy spectrum contributions to the partition function, as well as its semiclassical limit.

\paragraph{ETW brane state}
Restricting to the case where \eqref{eqn:Cond-continuous} is  satisfied, we can evaluate the partition function \eqref{eq:parition} in the zero-particle $n_F=0$ DSSYK model \eqref{eq:zero particle H}, and the asymptotic boundary described by the chord DSSYK model in the HH state, namely,
\begin{equation}\label{eq:mu nF first}
\begin{aligned}
    \bra{0}\rme^{-\beta\hH_{n_F}}\ket{0}&=\int _0^\pi \rmd\theta~\mu_{n_F}(\theta)~\rme^{-\beta E(\theta)}\\
    &\equiv\bra{(B^{(n_F)}_{\vec{t}})^{1/2}}\rme^{-\beta\hH_{n_F=0}}\ket{(B^{(n_F)}_{\vec{t}})^{1/2}}=\bra{B^{(n_F)}_{\vec{t}}}\rme^{-\beta\hH_{n_F=0}}\ket{\Omega}~,
\end{aligned}
\end{equation}
where we have defined:
\begin{subequations}
    \begin{align}
 \ket{(B^{(n_F)}_{\vec{t}})^{1/2}}&\equiv\qty(\hat{B}^{(n_F)}_{\vec{t}})^{1/2}\ket{\Omega}~,\qquad \ket{B^{(n_F)}_{\vec{t}}}\equiv\hat{B}^{(n_F)}_{\vec{t}}\ket{\Omega}~,\label{eq:ETW brane state general ext}\\
 \hat{B}^{(n_F)}_{\vec{t}}&\equiv\frac{\prod_{i\leq j\leq4}(t_it_j;q^2)_\infty}{(t_1\rme^{\pm\rmi\theta(\hH_{n_F=0})},\dots,t_4\rme^{\pm\rmi\theta(\hH_{n_F=0})};q)_\infty(t_1t_2t_3t_4;q^2)_\infty}~,\label{eq:ETW op}\\
 \mu_{n_F=8}(\theta)&=\bra{\Omega}\hat{B}^{(n_F)}_{\vec{t}}\ket{\theta}\equiv\frac{(\rme^{\pm2\rmi \theta};q^2)_\infty\prod_{i<j<4}(t_it_j;q^2)_\infty}{2\pi(t_1\rme^{\pm \rmi \theta},\dots,t_{4}\rme^{\pm\rmi \theta};q^2)_\infty(t_1t_2t_3t_4;q^2)_\infty}~.
    \end{align}
\end{subequations}
Here $\theta(E)\equiv \cos^{-1}\qty(\sqrt{1-q^2}E/2)$ and we selected the normalization of the measure $\mu_{n_F}(\theta)$ such that
\begin{eqnarray}
    \bra{(B_{\vec{t}}^{(n_R)})^{1/2}}\ket{(B_{\vec{t}}^{(n_R)})^{1/2}}=1~.
\end{eqnarray}
In later sections, $\ket{(B_{\vec{t}}^{(n_F)})^{1/2}}$ will be associated with the state of an ETW brane.\footnote{$\ket{B_{\vec{t}}^{(n_F=4)}}$ is a generalised q-coherent state \cite{Watanabe:2025rwp}. We discuss an extension for the $n_F=8$-case in App.~\ref{app:extended q coherent}.}

\paragraph{Semiclassical evaluation}
The semiclassical partition function of the boundary theory encodes information about a topological expansion in the bulk partition function, which equals the boundary one. For this reason, we now consider the semiclassical limit ($\lambda\rightarrow0$ while other parameters are fixed) of the partition function \eqref{eq:parition} when either \eqref{eqn:Cond-continuous} or \eqref{eq:dissrete cond 1} are satisfied, to learn about what the bulk interpretation of the two type of spectra may be. 

We first focus on the continuous case \eqref{eq:mu nF first}. We find (see App.~\ref{app:Ent branes}) that
\begin{equation}\label{eq:semiclassical cont Z}
Z_{\rm cont}(\beta)\equiv\int _0^\pi \rmd\theta~\mu_{n_F}(\theta)~\rme^{-\beta E(\theta)}\underset{\lambda\rightarrow0}{\rightarrow}\rme^{S_{n_F}(\theta)-\beta E(\theta)}
\end{equation}
where $S_{n_F}(\theta)$ appears in \eqref{eq:thermal ETW}. This limit in the partition function can be interpreted in the bulk theory simply as a black hole saddle in sine dilaton gravity \cite{Blommaert:2025avl,Aguilar-Gutierrez:2025hty}.

We now analyze the contribution to the partition function of the discrete spectrum, when \eqref{eq:dissrete cond 1} is satisfied. Inserting the factor $\rme^{-\beta\hH_{n_F}}$ into the resolution of the identity in energy basis \eqref{eq:recurrence discrete}, we find that \eqref{eq:mu nF first} is modified to
\begin{eqnarray}\label{full partition function}
     \bra{0}\rme^{-\beta\hH_{n_F}}\ket{0}&=\int _0^\pi \rmd\theta~\mu_{n_F}(\theta)~\rme^{-\beta E(\theta)}+\sum_{l=0}^{l_{\rm max}}\tilde{w}_l~\rme^{-{\beta x_l}/{\sqrt{1-q^2}}}~,
\end{eqnarray}
where $\tilde{w}_l\equiv w_l/h_0$ with $h_0$ and $w_l$ appearing in \eqref{eq:hn expl} and \eqref{eq:wl expl} respectively, while $l_{\rm max}$ \eqref{eq:ceiling} and $x_l$ \eqref{eq:discrete spectrum}.

We carry out the semiclassical evaluation of the partition function \eqref{full partition function} in App.~\ref{app:semiclassical Z}. By adding the continuous and discrete contributions in \eqref{full partition function} with \eqref{eq:semiclassical cont Z} and \eqref{eq:semiclassical disc S}, we find that
\begin{equation}\label{eq:semiclassical disc S2}
Z(\beta)=Z_{\rm cont}(\beta)+\sum_{l=0}^{l_{\rm max}}\rme^{S_l-\beta x_l}~,
\end{equation}
where $S_l$ appears in \eqref{eq:semiclassical disc S}, $x_l$ in \eqref{eq:discrete spectrum}. However, the analysis reveals that the summation in \eqref{eq:semiclassical disc S2} vanishes at leading order in the semiclassical limit. This means that the contribution of the discrete spectrum modes to the partition function is only present at the non-perturbative level in $\lambda$, which is consistent with our later observations in Sec.~\ref{ssec:discrete spectrum}.

The total partition function indicates that when the discrete spectrum condition is satisfied \eqref{eq:dissrete cond 1}, there is an additive contribution to the partition function that does not affect the continuous contribution. Interpreted in terms of a dual bulk partition function, \eqref{eq:semiclassical disc S2} suggests that the semiclassical limit retains the black hole associated with the continuous spectrum in sine dilaton gravity, and beyond the semiclassical limit, there is a pair of disk topologies at the same inverse temperature $\beta$ for a given value of $l$. In Sec.~\ref{ssec:discrete spectrum}, we will associate the discrete contributions with AdS$_2$ black holes in sine dilaton gravity, which are Wick rotated with respect to the continuous spectrum case.

{In the rest of the section, we describe the microscopic models that give rise to the auxiliary system described by $\hH_{n_F}$ based on KM states, and its PETS generalization.}

\subsection{Microscopic models from Kourkoulou-Maldacena states}\label{ssec:KM state deformations}
{We now formulate the microscopic deformations of the SYK model based on KM states \cite{Kourkoulou:2017zaj} introduced in Sec.~\ref{ssec:DSSYK review}. This choice of states allows for a clearer bulk understanding of the deformations in terms of single-sided black holes with ETW branes, while the PETS generalization describe two-sided black holes (Sec.~\ref{ssec:explicit microscopic}).}

{Using the results in \cite{Rajgadia:2026ask}, with the intermediate details explained in App.~\ref{sapp:H int H def DS}, we can express similarly to \eqref{eq:Hamiltonia moments} that
\begin{equation}\label{eq:H moments KM}
    \lim_{\rm DS}\overline{\bra{\rm s}P\qty(\hH_{\rm SYK},~\qty{\hat{M},~\hat{M}^\dagger})\ket{\rm s}}=\bra{0}P(\hH_{n_F=0},q^{\Delta_w\hat{n}}\ket{0}~,
\end{equation}
with $\ket{s}$ in \eqref{eq:spin op}; $\qty{x,y}\equiv xy+yx$; $\Delta_w\equiv w/p$; we omitted the label $\varepsilon$ since we work on a one-copy system; DS indicates the double-scaling limit; $\hH_{n_F=0}$ is the DSSYK chord transfer matrix \eqref{eq:H SYK0}, and $\hat{n}$ is the chord number operator counting $H$ chords. Using \eqref{eq:H moments KM}, we construct new families of deformation of the SYK model with a clear double-scaling limit below.}

\paragraph{\texorpdfstring{$n_F=2$}{} q-Askey deformation}
{Before explaining the most general case (associated with Askey-Wilson polynomials, when $n_F=8$ in the notation of Tab.~\ref{tab:qaskey}), we analyze the simplest family of deformations of the DSSYK model. This deformation of the DSSYK model was first discussed in \cite{Rajgadia:2026ask} (7.2)\footnote{{See App.~\ref{sapp:H int H def DS} for discussion about the double-scaling limit.}}
\begin{equation}
\hH_{\rm SYK}^{(n_F=2)}=\hH_{\rm SYK}+\frac{t_1}{\sqrt{1-q^2}}\hH_{\rm int}~,\quad \hH_{\rm int}\equiv\qty{\hat{M},~\hat{M}^\dagger}.
\end{equation}
In the ensemble-averaged theory, one has
\begin{equation}\label{eq:H moments KM2}
    \lim_{\rm DS}\overline{\bra{\rm s}\qty(\hH_{\rm SYK}^{(n_F=2)})^k\ket{\rm s}}=\bra{0}\hH_{n_F=2}^k\ket{0}~,
\end{equation}
with
\begin{eqnarray}\label{eq:def transfer}
    \hH_{n_F=2}\equiv \hat{a}+\hat{a}^\dagger+\frac{t_1q^{\Delta_w\hat{n}}}{\sqrt{1-q^2}}~.
\end{eqnarray}
Here the creation $\hat{a}^\dagger$ and annihilation operators act on the corresponding chord Hilbert space span $\qty{\ket{n}}$ by \eqref{eq:a a dagger}. To match the Hamiltonian eigenvalue problem to the recurrence relation of generalised q-Hermite polynomials (see Sec.~\ref{ssec:explicit microscopic}), we choose $\Delta_w=2$. Note, however, that the double-scaled Hamiltonian \eqref{eq:def transfer} is also valid for $\Delta_w\neq 2$.}

\paragraph{\texorpdfstring{$n_F=8$}{} deformation}
{One can similarly construct the Hamiltonian deformations in the most general case ($n_F=8$) that encodes Askey-Wilson polynomials. Explicitly, let us define
\begin{equation}\label{eq:H nf8 again}
\begin{aligned}
    \hH_{\rm SYK}^{(n_F=8)}&\equiv\frac{1}{2}\qty(\hH_{\rm def}{\rm b}_{n_F}(\hH_{\rm int})+{\rm b}_{n_F}(\hH_{\rm int})\hH^\dagger_{\rm def})+\frac{1}{\sqrt{1-q^2}}{\rm a}_{n_F}\qty(\hH_{\rm int})~,\\
    \hH_{\rm def}&\equiv\hH_{\rm SYK}+\frac{\rmi}{4p}\qty[\hH_{\rm SYK},\qty[\hat{\psi}_{2j-1},~\hat{\psi}_{2j}]]~,
\end{aligned}
\end{equation}
where
\begin{equation}\label{eq:targetedeq}
\begin{aligned}
{\rm a}_{n_F=8}(x)\equiv &t_1^{-1}\qty(1-\sum_{k_1,k_2=0}^\infty \frac{(x^2r)^{k_1+k_2}}{q^{2k_2}}(1-t_1t_2x)(1-t_1t_3x)(1-t_1t_4x)\qty(1-q^{-2}{r x} ))\\&+t_1\qty(1-\sum_{k_1,k_2=0}^\infty \frac{(x^2r)^{k_1+k_2}}{q^{2k_1+4k_2}}(1-x)(1-t_2t_3xq^{-2})(1-t_2t_4xq^{-2})(1-t_3t_4xq^{-2}))~,\\
    {\rm b}_{n_F=8}(x)\equiv &\qty(1-\sum_{k=1}^\infty \frac{(2k-3)!!}{2^k\,k!}(q^{-4}t_it_j\,x)^k)\qty(\prod_{1\leq i<j\leq 4}\qty(1-\sum_{m=1}^\infty \frac{(2m-3)!!}{2^m\,m!}(q^{-2}t_it_j\,x)^m))\\
&\times\sum_{m=0}^\infty q^{-4}r\, x^{2m}\sum_{n=0}^\infty\frac{(2n)!}{4^n(n!)^2}(q^{-6}r\,x^2)^n\sum_{j=0}^\infty\frac{(2j)!}{4^j(j!)^2}(q^{-2}r\,x^2)^j~.
    \end{aligned}
\end{equation}
with $r\equiv t_1t_2t_3t_4$. By applying \eqref{eq:H moments KM}, we find the following chord Hamiltonian satisfying \eqref{eq:Hamiltonia moments} (see App.~\ref{sapp:H int H def DS} for the derivation),
\begin{flalign}\begin{aligned}
\label{eq:micro nF8}
    \hH_{n_F=8}=b_{n_F=8}(q^{2\hat{n}})\ha^\dagger +\ha \,b_{n_F=8}(q^{2\hat{n}})+\frac{a_{n_F=8}(q^{2\hat{n}})}{\sqrt{1-q^2}}~.
\end{aligned}
\end{flalign}
\begin{equation}\label{eq:a b nf8}
\begin{aligned}
a_{n_F=8}(x)\equiv &t_1+t_1^{-1}-\frac{(1-t_1t_2x)(1-t_1t_3x)(1-t_1t_4x)(1-t_1t_2t_3t_4 q^{-2}x)}{t_1(1-t_1t_2t_3t_4q^{-2}x^2)(1-t_1t_2t_3t_4x^{2})}\\& \qquad\qquad
    -\frac{t_1(1-x)(1-t_2t_3xq^{-2})(1-t_2t_4xq^{-2})(1-t_3t_4xq^{-2})}{(1-t_1t_2t_3t_4x^2q^{-4})(1-t_1t_2t_3t_4x^2q^{-2})}~,\\
    b_{n_F=8}(x)\equiv &\frac{1}{(1-t_1t_2t_3t_4q^{-4}x^2)}\sqrt{\frac{\prod_{1\leq i<j\leq 4}(1-t_it_jq^{-2}x)(1-t_1t_2t_3t_4xq^{-4})}{(1-t_1t_2t_3t_4x^2q^{-6})(1-t_1t_2t_3t_4x^2q^{-2})}}~,
    \end{aligned}
\end{equation}
{where the creation and annihilation operators in the chord Hamiltonian \eqref{eq:micro nF8} act as in \eqref{eq:a a dagger ortho}.} Note that we represent the above transfer matrix in the chord number basis $\ket{n}$, as in the definition \eqref{eq:recurrence states q Askey}. Given that the factor $q^{2n}\leq1$ and in general we require $ t_1 t_2 t_3 t_4<1$ for both the continuous \eqref{eqn:Cond-continuous} and discrete \eqref{eq:dissrete cond 1} parametrisations of the corresponding Askey-Wilson polynomials, then terms such as $\lim_{\tilde{N}\rightarrow\infty}(t_1t_2t_3t_4 q^{2n})^{\tilde{N}}\rightarrow 0$. For these reasons, the geometric series terms in ${\rm a}_{n_F=8}$ and ${\rm b}_{n_F=8}$ \eqref{eq:targetedeq} results in the denominators of ${a}_{n_F=8}$ and ${b}_{n_F=8}$ \eqref{eq:a b nf8}.}

The eigenstates for general $t_i$ are given in terms of Askey-Wilson polynomials \eqref{eq:recurrence discrete}, which are carefully analysed in Sec.~\ref{ssec:Krylov nF8}. In particular, similar to the $n_F=2$ case, the corresponding partition function \eqref{eq:mu nF first} for the deformed theory \eqref{eq:micro nF8} is represented as a chord diagram in Fig.~\ref{fig:CS_nF2} (a). 

\begin{figure}
    \centering
    \subfloat[]{\includegraphics[width=0.45\linewidth]{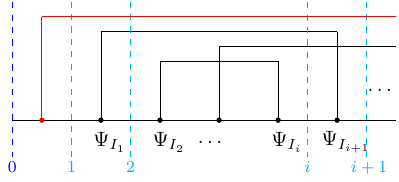}}\hspace{0.5cm}\subfloat[]{\includegraphics[width=0.45\linewidth]{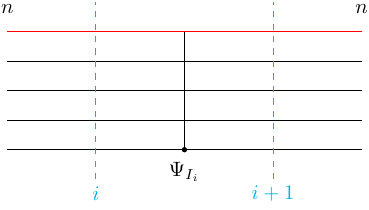}}\\
\subfloat[]{\includegraphics[width=0.45\linewidth]{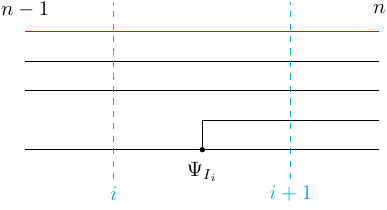}}\hspace{0.5cm}\subfloat[]{\includegraphics[width=0.45\linewidth]{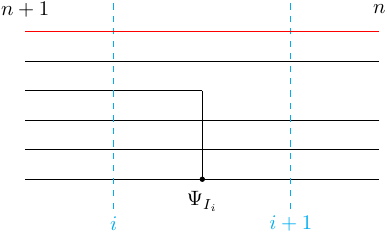}}
    \caption{{(a) Slicing of the chord diagram in Fig.~\ref{fig:CS_nF2}, including H-chords as in Fig.~\ref{fig:examplechord}, into a line. The vertices (cyan dashed lines) represent locations between the insertion of the $\Psi_I$ strings (black or red dots), which are contracted with each other (depicted by the black or red solid lines, respectively). In (b) an open chord can intercept an effective ETW brane term (red solid line) with penalty factor $a_{n_F}(q^{2n})/\sqrt{1-q^2}$. We also depict two processes where one ends with $n$ chords at vertex $i$, by either creating (c) or annihilating (d) an open chord between the vertices $i$ and $i+1$.}}
    \label{fig:OpenCloseChords}
    \end{figure}

{To visualize the chord diagram in more detail, it is convenient to perform a change of chord number basis from an orthogonal one where the Hamiltonian takes the symmetric and manifestly Hermitian form in \eqref{eq:micro H nF8}, to the monic basis \cite{Muck:2022xfc}, where it takes the form
\begin{flalign}\begin{aligned}\label{eq:HnF8 micro monic}
&\hH_{n_F=8}=\ha^\dagger b_{n_F=8}(q^{2\hat{n}})+\ha+\frac{a_{n_F=8}(q^{2\hat{n}})}{\sqrt{1-q^2}}~.
\end{aligned}
\end{flalign}
where the creation and annihilation operators appear in \eqref{eq:a a dagger} and the Hamiltonian is no longer symmetric, but still Hermitian in terms of the chord inner product \cite{Lin:2022rbf,Aguilar-Gutierrez:2025mxf}}. Let $v_n^{(i)}$ be the sum over all chord diagrams with $n$ open chords at a given position $i$ between consecutive $\Psi_{I}$ nodes. As in \cite{Berkooz:2018qkz,Berkooz:2018jqr}, an open chord between any two indices $i$ and $i+1$ may close (intercepting all possible $n-1$ open chords), or a new open chord may be created (without intercepting others). However, an open chord between $i$ and $i+1$ that intercepts the ETW brane state produces a penalty factor $a_{n_F}(q^{\Delta_w n})/\sqrt{1-q^2}$. Then by considering the sum over all diagrams with $n$ open chords at vertex $i+1$, we have that
\begin{itemize}
\item Given a vertex $i+1$ with $n$ open chord before it, there could have been $n-1$ or $n+1$ open chords at vertex $i$. In the latter, the penalty factor for intercepting open chords $\hH_{n_F}$ is $q^{\Delta_w\#}b_{n_F}(q^{2n})$, where $\#$ represents $\hH$ interceptions.
\item If the open chord $i+1$ intercepts the ETW brane state, it does not modify the number of open chords, and its penalty factor is $a_{n_F}(q^{\Delta_w n})/\sqrt{1-q^2}$.
\end{itemize}
The sum over the chord diagrams with n open chords at the $i$-th vertex $v^{(i)}_n$ then obeys
\begin{equation}\label{eq:chord diagram recurrence}
v^{(i+1)}_n=\sum_{j=1}^nq^{2j}b_{n_F}(q^{\Delta_w n})v^{(i)}_{n+1}+v^{(i)}_{n-1}+\frac{a_{n_F}(q^{\Delta_w n})}{\sqrt{1-q^2}}v_{n-1}^{(i)}~,
\end{equation}
so that, similar to the original argument by \cite{Berkooz:2018qkz,Berkooz:2018jqr} in the undeformed DSSYK, $v^{(i)}_n=\hH_{n_F}^i\ket{n}$ corresponds to \eqref{eq:HnF8 micro monic} when $\Delta_w=2$. The chord diagram representation of the partition function \eqref{eq:mu nF first} for the deformed theory \eqref{eq:micro nF2} is displayed in Fig.~\ref{fig:CS_nF2} (a), where the corresponding state $\ket{B_{\vec{t}}^{(n_F)}}$ is defined in \eqref{eq:mu nF first}. We illustrate the chord rules for the deformed DSSYK model in Fig.~\ref{fig:OpenCloseChords}.
\paragraph{Summary}
{The previous results apply more generally when we interpolate between the TFD and KM states the family of PETS in \eqref{eq:other_PETS} and \eqref{eq:def PETS}. The different q-Askey deformed SYK Hamiltonians in Sec.~\ref{ssec:KM state deformations} and \ref{ssec:explicit microscopic} can be seen as different state-dependent representations. As shown in App.~\ref{app:more PETS}, the chord diagram rules depend on the factor $\mathbb{k}\in[0,1]$, which indicates the emergence of different operator algebras in the large $N$ limit, due to different weak operator topologies. However, since the factor $\mathbb{k}$ appears in a product with the factor $\Delta_w$, one can reproduce the same chord Hamiltonian for all the members in the q-Askey scheme in this work in the double-scaling limit. The difference between the microscopic theories is relevant when evaluating appropriate observables, which are discussed in App.~\ref{app: moremicroscopic}.}

\subsection{Microscopic models from partially entangled thermal states}\label{ssec:explicit microscopic}
\begin{figure}
    \centering
    \subfloat[]{\includegraphics[width=0.55\linewidth]{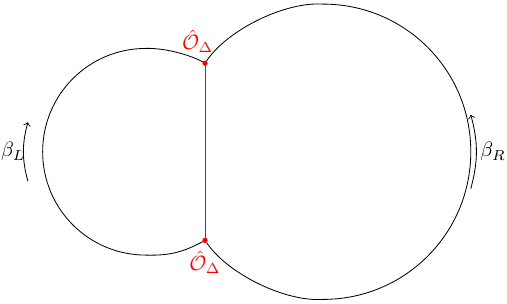}}\hfill\subfloat[]{\includegraphics[width=0.3\linewidth]{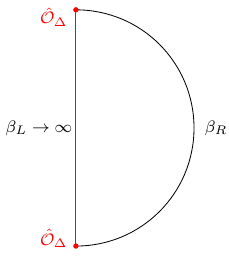}}
    \caption{{Representation of the two-point function \eqref{eq:partition one particle} resulting in a PETS \cite{Goel:2018ubv}. (a) General case where the inverse temperature $\beta_{L/R}$ of each subsystem in the boundary is finite. (b) In the $\beta_L\rightarrow\infty$ limit in \eqref{eq:partition one particle}, the left asymptotic boundary is sent to infinity. The matter worldline in the bulk acts as an effective ETW brane  \cite{Goel:2018ubv}. While the left system is described by its ground state, the effective ETW brane remains in a (generalised) q-coherent state (App.~\ref{app:extended q coherent}). The Euclidean evolution on the right-hand side is generated by the right SYK Hamiltonian $\hH_{\rm SYK}^{(R)}$ \eqref{eq:HLR SYK} or its deformation $\hH_{\rm SYK}^{(R)}\rightarrow \hH_{\rm SYK}^{n_F}$ \eqref{eq:Hamiltonia moments}. Different choices of the Hamiltonian correspond to different slicings in the bulk \cite{Blommaert:2025avl}.}}
    \label{fig:PETS}
\end{figure}

{In this section, we are interested in describing an AdS$_2$ black hole with an ETW brane in a PETS preparation of state with matter operators that act in a similar way to the ETW brane in a single-sided black hole.} Specifically, consider
\begin{equation}\label{eq:partition one particle}
    Z_\Delta(\beta_L,\beta_R)\equiv\overline{\bra{\rm TFD_\infty}\hat{\mathcal{O}}^{(R)}_{\rm SYK}\rme^{-\beta_R\hH_{\rm SYK}^{(R)}}\hat{\mathcal{O}}^{(R)}_{\rm SYK}\rme^{-\beta_L\hH_{\rm SYK}^{(R)}}\ket{\rm TFD_\infty}}~,
\end{equation}
where we take the zero-temperature limit of one of the boundary subsystems relative to the matter chord while keeping the temperature for the complement fixed. {This preparation of state describes the same dynamics as for an eternal single-sided AdS$_2$ black hole with an ETW brane \cite{Goel:2023svz}, as illustrated in Fig.~\ref{fig:PETS}.}

While $\hH_{\rm SYK}^{(L)}$ is irrelevant in this limit, the Hamiltonian of the remaining boundary can describe different dynamics. For instance, when the Hamiltonian takes the same form as if there were no ETW brane, it captures the evolution of observables that are gravitationally dressed to the asymptotic boundary (such as two-sided geodesic lengths dressed to the boundaries) {which are not affected by the matter insertion in the PETS.} Meanwhile, the dynamics generated by the Hamiltonian are modified when we choose to describe a slicing that does involve both the remaining boundary and the {would-be} ETW brane \cite{Blommaert:2025avl}. This motivates us to search for deformations of the SYK model that describe the dynamics expected from introducing ETW branes in the bulk using PETS \cite{Goel:2018ubv}.

For concreteness, consider $2N$ Majorana fermions $\psi_L$ and $\psi_R$  evenly distributed in subsystems $L$ and $R$. The Hilbert space is $\mathcal{H}_L\otimes \mathcal{H}_R$. Let us denote $\hH_{\rm SYK'}^{n_F}$ as a deformation of the SYK model, which we specify below. Similar to Sec.~\ref{ssec:KM state deformations}, while the derivations for the $n_F=8$ case suffice to describe all the families of deformations in this work, for clarity in the presentation we illustrate the strategy to build the deformations of interest in the SYK model for the $n_F=2$ case, which we later generalise for $n_F=8$, and we also illustrate the adjoint deformation, as it has a different construction from the microscopic perspective. In App.~\ref{app: moremicroscopic} we illustrate as well the special cases $n_F=4$ and $6$ for completeness.

\paragraph{\texorpdfstring{$n_F=2$}{} q-Askey deformation}
Let us consider the two-copy SYK system to study the dynamics of the system illustrated in Fig.~\ref{fig:PETS}. We describe the Euclidean evolution on the right boundary side relative to the operator insertion by a term:
\begin{equation}\label{eq:doubled H nF}
\hH_{\rm SYK'}^{(n_F=2)}\equiv \hH_{\rm SYK}^{(R)}+\frac{t_1}{1-q^2}\hH_{\rm int'}~,
\end{equation}
where $\hH_{\rm SYK}^{(R)}$ appears in \eqref{eq:HLR SYK}, and we included an interaction term based on \cite{Lin:2022rbf},
\begin{eqnarray}\label{eq:H int}
   \hH_{\rm int'}\equiv \rmi^{s}\begin{pmatrix}
    N\\s
\end{pmatrix}^{-1}\sum_{I_s}\Psi_{I_s}^{(L)}\Psi_{I_s}^{(R)}~,\quad I_s\equiv i_1\dots i_s~.
\end{eqnarray}
It was realised in \cite{Lin:2022rbf} that the interaction term is related to the total chord number operator \eqref{eq:total chord number} through
\begin{equation}\label{eq:relation with chord number}
    \lim_{\rm DS}\overline{\bra{\rm TFD_\infty}P\qty(\hH_{\rm int'},~\hH_{\rm SYK}^{(L/R)},~\hmO_{\rm SYK}^{(L/R)})\ket{\rm TFD_\infty}}=\bra{\Omega}P\qty(q^{\Delta_s\hat{N}},\ha_{L/R}+\ha^\dagger_{L/R},~\hmO_{\Delta}^{(L/R)})\ket{\Omega}~,
\end{equation}
where $P$ is a polynomial function, and $\Delta_s\equiv s/p$, where $p$ is the number of all-to-all interactions in the undeformed SYK Hamiltonian \eqref{eqn:H-SYK}. 

Meanwhile, the specific case $s=1$ (instead of $s\rightarrow\infty$ used in \eqref{eq:relation with chord number}) is associated with a rescaled version of the operator size \cite{Lin:2022rbf,Lin:2023trc,Cao:2025pir},\footnote{The operator size itself corresponds to $p~\hat{s}$.}
\begin{equation}\label{eq:rescaled op size}
\hat{s}\equiv\frac{1}{p}\sum_{j=1}^N\qty(1+\rmi ~\psi^{(L)}_{j}\psi^{(R)}_{j})~,   
\end{equation}
which plays the role of the total chord number in the double scaling limit, as first pointed out by \cite{Lin:2022rbf}.

From now on, we will set $\Delta_s=2$, unless stated otherwise, to derive a connection with the q-Askey scheme. Using the identification between microscopic and chord operators in \eqref{eq:relation with chord number}, it follows that the chord Hamiltonian corresponding to the deformed theory \eqref{eq:doubled H nF} is:
\begin{equation}\label{eq:H SYK2}
    \begin{aligned}
\hH_{n_F=2}=\hat{a}_R+\hat{a}^\dagger_R+\frac{t_1}{\sqrt{1-q^2}}q^{2\hat{N}}~.
    \end{aligned}
\end{equation}
where the action of the creation/annihilation operators with matter insertions appears in \eqref{eq:Fock Hm 1}, and the total chord number $\hat{N}$ is defined in \eqref{eq:total chord number}. In particular, when $\Delta=0$ in Fig.~\ref{fig:PETS} we have that
\begin{equation}\label{eq:micro nF2}
    \eval{\hH_{n_F=2}}_{\Delta=0}=\ha+\ha^\dagger+\frac{t_1}{\sqrt{1-q^2}}q^{2\hat{n}}~,
\end{equation}
where $\hat{n}$ is the chord number operator, and $\ha$, $\ha^\dagger$ the annhilation and creation operators in the zero-particle sector \eqref{eq: Oscillators 1}. Note that the operator $q^{2\hat{n}}$ is bounded due to the spectrum of  $\hat{n}$ being positive. This implies that the spectrum of the Hamiltonian \eqref{eq:micro nF2} is bounded.

\paragraph{Comparison to the other approaches}The chord Hamiltonian \eqref{eq:micro nF2} has been investigated in a different setting, originally by \cite{Okuyama:2023byh}, and more recent microscopic realizations have been investigated by \cite{Cao:2025pir,Rajgadia:2026ask}. In particular, we discuss an alternative construction of the chord Hamiltonian \eqref{eq:micro nF2} in Sec.~\ref{ssec:KM state deformations} based on \cite{Rajgadia:2026ask}, which we compare with our approach. Our procedure extends the previous results by incorporating the effect of matter SYK operators into the auxiliary chord theory \eqref{eq:H SYK2}. Moreover, while our derivation relies on a hermitian Hamiltonian, one could relax this condition.\footnote{For instance, a closely connected deformation of the SYK model to the one presented here would be to implement a replacement in \eqref{eq:doubled H nF}
\begin{equation}\label{eq:MQ}
 t_1\hH_{\rm int'}\rightarrow\mu\hat{s}~, 
\end{equation}
with $\mu$ some constant (which can be taken as pure imaginary). The term \eqref{eq:MQ} corresponds to a Maldacena-Qi interaction \cite{Maldacena:2018lmt} recently studied within the DSSYK model context by \cite{Marini:2026zjk}.}

Below, we discuss other types of deformations whose auxiliary system description corresponds to other families of basic orthogonal polynomials in the q-Askey scheme.

\paragraph{\texorpdfstring{$n_F=8$}{} deformation}
Next, we introduce the SYK Hamiltonian deformation:
\begin{flalign}\label{eq:micro H nF8}
\begin{aligned}
\hH_{\rm SYK'}^{(n_F)}&\equiv\frac{1}{2}\qty(\hH_{\rm def'}{\rm b}_{n_F}(\hH_{\rm int'})+{\rm b}_{n_F}(\hH_{\rm int'})\hH^\dagger_{\rm def'})+\frac{{\rm a}_{n_F}\qty(\hH_{\rm int'})}{\sqrt{1-q^2}}~,\\
    \hH_{\rm def'}&\equiv \hH_{\rm SYK}^{(R)}+\qty[\hH_{\rm SYK}^{(R)},~\hat{s}]~,
\end{aligned}\end{flalign}
where $\hH_{\rm int'}$ and $\hat{s}$ are defined in \eqref{eq:H int}, \eqref{eq:rescaled op size} respectively, with ${\rm a}_{n_F=8}$ and ${\rm b}_{n_F=8}$ as in \eqref{eq:targetedeq}. Using \eqref{eq:relation with chord number}, we then recover the following chord Hamiltonian\footnote{Note that $\hH_{\rm def'}$ in the deformed SYK Hamiltonian \eqref{eq:micro H nF8} corresponds to the factor $2\ha^\dagger$ by the same algebraic argument in App. \ref{sapp:H int H def DS} (in particular, see \eqref{eq:Double scaling H def}) since $\hat{s}$ corresponds to $\hat{n}$ in the chord diagram.}
\begin{flalign}\begin{aligned}\label{eq:H SYK8}
\hH_{n_F=8}= \hat{a}_R\,b_{n_F=8}(q^{2\hat{N}})+b_{n_F=8}(q^{2\hat{N}})\hat{a}^\dagger_R+\frac{a_{n_F=8}(q^{2\hat{N}})}{\sqrt{1-q^2}}~,
\end{aligned}
\end{flalign}
where $a_{n_F=8}$ and $b_{n_F=8}$ appear in \eqref{eq:a b nf8}; {$\hat{a}_R$ and $\hat{a}_R^\dagger$ act in the orthogonal representation \eqref{eq:a a dagger ortho};} and when $\Delta=0$, the above expression becomes \eqref{eq:micro nF8}.

\paragraph{Adjoint deformation}
Motivated by recent work \cite{Berkooz:2025ydg}, we wish to describe an entangled quantum system describing PETS on a two-sided bulk geometry at the semiclassical level. For this reason, we construct the following microscopic theory:
\begin{equation}\label{eq:Adj micro H}
\begin{aligned}
\hH_{\rm SYK'}^{(\rm Adj)}=&\frac{1}{2}\qty(b_{\rm Adj}(\hH_{\rm int})\hH_{\rm def''}+\hH_{\rm def''}^\dagger b_{\rm Adj}(\hH_{\rm int}))+\frac{t_1+t_2}{\sqrt{1-q^2}}\hH_{\rm int'}~,
\end{aligned}
\end{equation}
where
\begin{equation}
\begin{aligned}
b_{\rm Adj}(x)&\equiv\qty(1-\sum_{n=1}^\infty\frac{(2n-3)!!}{2^n\,n!}\qty(\chi q^{-2}x)^m)\qty(\sum_{n=0}\frac{(2n)!}{4^n(n!)^2}(\chi q^{-2} x)^n)\qty(\sum_{k=0}\frac{(2k)!}{4^k(k!)^2}(\chi\,x)^k)~,\\
\hH_{\rm def''}&\equiv \frac{1}{2}\qty(\hH_{\rm SYK}^{(L)}+\hH_{\rm SYK}^{(R)}+\qty[\hH_{\rm SYK}^{(L)}+\hH_{\rm SYK}^{(R)},~\hat{s}])~.
\end{aligned}
\end{equation}
The Hamiltonian moments of \eqref{eq:Adj micro H} are encoded after ensemble averaging in the double-scaling limit by the chord Hamiltonian. Similar to the previous cases, one finds that in the case without matter, the chord version of \eqref{eq:Adj micro H} is:
\begin{equation}\label{eq:H SYKAdj}
\begin{aligned}
    \hH_{\rm Adj}=\sqrt{\frac{1-\chi^2 q^{2\hat{n}-2}}{(1-\chi q^{2\hat{n}-2})(1-\chi q^{2\hat{n}})}}\ha^\dagger+\ha\sqrt{\frac{1-\chi^2 q^{2\hat{n}-2}}{(1-\chi q^{2\hat{n}-2})(1-\chi q^{2\hat{n}})}}~.
\end{aligned}
\end{equation}
In particular, when $\chi=q^2$, 
the corresponding auxiliary theory encodes an $SU(2)$ random matrix theory (RMT)\footnote{We leave a numerical analysis of the level spacing of the Hamiltonian for future studies; our observations regarding RMT are restricted to the chord description of \eqref{eq:Adj micro H} when $\chi=q^2$ in this work.}, {as we discuss in Sec. \ref{ssec:Krylov nF Adj}.}

\subsection{Triple-scaling limits}\label{ssec:triple scaling limit}
We are interested in the low-energy behaviour of the auxiliary quantum mechanical systems described by the deformations introduced in this section. Therefore, we implement different expansions of chord Hamiltonians in terms of effective length variables describing emergent geodesic lengths in the dual bulk theory, which serves as a connection to the next section.

\paragraph{The case of $n_F=0$}
We begin by discussing the Schwarzian regime, also called triple-scaling limit for the regular DSSYK chord Hamiltonian, which involves different regularizations compared to the triple-scaling limits of the $n_F=2,~4,~6$ and $8$ cases.

Let us define the limits:
\begin{align}\label{eq:def IR UV triple scaling limit}
\rme^{-L}&\equiv\frac{q^{2 n}}{4\lambda^{2}}\quad\text{and}\quad {k}\equiv\frac{\theta}{\lambda}\quad\text{are fixed as }\lambda\rightarrow0~,
\end{align}
{where $L$ is the corresponding eigenvalue of:
\begin{align}\label{eq:dS length}
    \hat{L}&\equiv 2\lambda\hat{n}+\log(2\lambda^2)\mathbb{1}~.
\end{align}
This leads to:
\begin{subequations}\label{eq:triple scaling IR}
\begin{align}
\frac{2-\sqrt{1-q^2}\hH}{4\lambda^2}= \hat{P}^2+\rme^{-{\hat{L}}}+\mathcal{O}\qty(\lambda^2)~,
    \end{align}
\end{subequations}
where $[\hat{P},~\hat{L}]=2\lambda\rmi$. The resulting rescaled Hamiltonian with the zero point energy subtraction above is isomorphically dual to the ADM Hamiltonian of JT gravity at leading order in the genus expansion \cite{Harlow:2018tqv}. There is an associated limit where one performs an expansion of the energy spectrum around $\theta\simeq \pi$, which is associated with dS$_2$ space \cite{Heller:2025ddj,Aguilar-Gutierrez:2025otq,Aguilar-Gutierrez:2026ogo,Blommaert:2024whf}. In their picture, the chord number would be proportional to the (timelike) geodesic length connecting anchoring regions at $\mathcal{I}^\pm$, where the chord Hamiltonian takes the role of a spatial displacement operator along $\mathcal{I}^\pm$.

\paragraph{The cases $n_F=2,~4,~6,~8$}
We are now interested in the Schwarzian regime of the deformations of the SYK model. It was previously realised in the $n_F=2$ case in \cite{Okuyama:2023byh} that one can recover the Morse potential in JT gravity \cite{Gao:2021uro} by setting the deformation parameter $t=q^{2\nu+1}$ in the corresponding chord Hamiltonian, which encodes the recurrence relations for the generalised q-Hermite polynomials \eqref{eq:micro nF2}. 

We start from a general ansatz
\begin{eqnarray}\label{eq:constants}
    \frac{q^{2n}}{2\lambda}\equiv\rme^{-\tilde{L}}~,~~t_{i}=a_i q^{\Delta_i}~,\quad i=1,\dots,~4~,
\end{eqnarray}
where $\tilde{L}$, $a_i$, $\Delta_i$ are fixed parameters as $\lambda\rightarrow0$, which we seek to determine. The expansion of \eqref{eq:micro nF8} with \eqref{eq:constants} gives:
\begin{eqnarray}
    2-\sqrt{1-q^2}\hH_{n_F=8}\rightarrow2\lambda\prod_{i=1}^4(1-a_i)\rme^{-L}+\mathcal{O}(\lambda^2)~.
\end{eqnarray}
The $\lambda^2$ term contains the canonical conjugate momenta of the parameter $L$. Thus, to recover non-trivial dynamics, we require that at least one of the parameters $a_i=1$ in \eqref{eq:constants}. Without loss of generality, we select $a_1=1$ and perform the relabeling:
\begin{equation}\label{eq:triple scaling}
    \Delta_1\equiv 2\tilde{\nu}+1~,\quad a_{2}=a~,\quad a_{3}=b~,\quad a_{4}=c ~,\quad \text{as}\quad\lambda\rightarrow0~,
\end{equation}
where $\tilde{L}$, $\tilde{\nu}$, $a$, $b$, $c$ are $\mathcal{O}(\lambda^0)$ as $\lambda\rightarrow0$. One can then easily recover the Schwarzian limit of the other hierarchy levels $n_F=2,~4$ and $6$ in the q-Askey scheme by simply setting $a=b=c=0$, $b=c=0$, or $c=0$ respectively.

By applying the ansatz in \eqref{eq:constants} with \eqref{eq:triple scaling} for the corresponding $n_F=8$ chord Hamiltonian \eqref{eq:micro nF8}, as explained in the previous subsection, we recover a triple-scaled Hamiltonian of the same form as in the previous cases:
\begin{equation}\label{eq:IR H one}
    \hH_{\rm IR}\equiv\frac{2-\sqrt{1-q^2}\hH_{n_F=8}}{4\lambda^2}=-\partial_{\tilde{L}}^2+A_1(a,b,c)~\tilde{\nu}~\rme^{-\tilde{L}}+\frac{A_2(a,b,c)}{4}\rme^{-2\tilde{L}}+\mathcal{O}(\lambda^2)~,
\end{equation}
where:
\begin{flalign}
\begin{aligned}
    A_1&=(1 - a)  (1 - b) (1 - c) ~,\\
    A_2&=(a-1)^2 (b-1)^2 c^2+(a-1)^2 (b-1)^2-2 c \left(-2 \left(a^2+1\right) b+(a-1)^2 (1+b^2)\right)~.
\end{aligned}
\end{flalign}
Note crucially that $\Delta_{2\leq i\leq 4}$ in \eqref{eq:constants} do not contribute when talking the above limit. This means that $a$, $b$, $c$ a relabeling of three of the deformation parameters $t_{i}$ which do not to scale with $\lambda$ in the semiclassical regime; while one of $t_i$ scales as a factor $q^{\Delta}$ with $\Delta$ fixed as $\lambda\rightarrow0$. Thus, the above relations show that there are conditions on the scaling of the different deformation parameters for them to generate a semiclassical Hamiltonian of the form \eqref{eq:IR H one}. One can see from this expression that if one were to fix the scaling of at least two of the deformation parameters to be $q^{\Delta_i}$ (i.e.~$a_i=1$ for at least two parameters in the ansatz \eqref{eq:constants}), then the corresponding Hamiltonian takes the form of JT gravity without branes \cite{Harlow:2018tqv}. This can be interpreted from the fact that the set of deformation parameters $t_i$ determines the quantum mechanical system \eqref{eq:micro H nF8}, which in turn determine the bulk dual theory. The scaling for the deformation parameters in $\lambda$ as $\lambda\rightarrow0$ thus corresponds to different bulk interpretations. It would be interesting to investigate this further in the $n_F=8$ case. 

Let us also fix:
\begin{equation}
k=\frac{\theta}{2\lambda}~,\quad\text{as}~~   \lambda\rightarrow0,
\end{equation}
so that at leading order in the semiclassical limit, the recurrence relation for $n_F=8$ deformation becomes a Schrodinger equation for the wavefunction in the triple-scaling limit:
\begin{equation}
    \Psi_k(\tilde{L})\equiv \eval{{}_{n_F}\bra{\theta}\ket{n}}_{\rm TS}~,
\end{equation}
where TS denotes the triple-scaling limit defined in \eqref{eq:constants} and \eqref{eq:triple scaling}. Specifically, the wavefunctions associated with the continuous part of the deformed SYK spectrum obey:
\begin{equation}\label{eq:Schrodinger with Morse}
   \qty(\partial_{\tilde{L}}^2+k^2-A_1(a,b,c)~\tilde{\nu}~\rme^{-\tilde{L}}-\frac{A_2(a,b,c)}{4}\rme^{-2\tilde{L}}) \Psi_k(\tilde{L})=0~,
\end{equation}
which describes JT gravity with a Morse potential \cite{Gao:2021uro}. Note that $A_2$ in the previous equation can be absorbed into a shift of the canonical variable $L$. Under the redefinition:
\begin{equation}\label{eq:ETW brane dictionary substitution}
    \tilde{L}\rightarrow L\equiv \tilde{L}-\frac{1}{2}\log A_2(a,b,c)~,\quad \nu\equiv A_{1}(a,b,c)~\tilde{\nu}/\sqrt{A_2(a,b,c)}~, 
\end{equation}
equation \eqref{eq:Schrodinger with Morse} takes the standard form corresponding to JT gravity with an ETW brane \cite{Gao:2021uro}:
\begin{equation}\label{eq:morse potential}
    \qty(\partial_{L}^2+k^2-U(L))\Psi_k(L)=0~,\quad U(L)\equiv \nu~\rme^{-L}+\frac{1}{4}\rme^{-2L}~.
\end{equation}
The precise connection between the different q-Askey deformed chord theories with ETW branes beyond the low-energy limit is described in Sec.~\ref{sec:Krylov Askey}.

Similarly, based on the conditions \eqref{eq:dissrete cond 1} and \eqref{eq:dissrete cond 2}, the deformed theory has a discrete part of the spectrum (parametrised by $x_l$ in \eqref{eq:xl}) when any of the parameters $\abs{a}$, $\abs{b}$, $\abs{c}>1$. In this case, by keeping the zero-point subtracted energy finite as $\lambda\rightarrow0$,
\begin{eqnarray}\label{eq:discrete zero point subtracted}
k_l^2=\frac{x_l-2~{\rm sign}(\alpha)}{\sqrt{1-q^2}}~,\quad \abs{\alpha}>1~,
\end{eqnarray}
one obtains a Schrodinger equation analogous to \eqref{eq:Schrodinger with Morse},\footnote{Equation \eqref{eq:discrete states schrodinger} describes bound states of the Morse potential, as realised by \cite{Rajgadia:2026ask} in the $n_F=2$ case.}
\begin{equation}\label{eq:discrete states schrodinger}
    \qty(\partial_{L}^2-k_l^2-\nu~\rme^{-L}-\frac{1}{4}\rme^{-2L}) \Psi_l(L)=0~,
\end{equation}
where $L$ and $\nu$ are defined in \eqref{eq:morse potential}.

The wavefunctions solving the
Schrodinger equation in \eqref{eq:morse potential} are Whittaker functions
\begin{equation}
\begin{aligned}
    \Psi_{k}(L)=e^{L/2}W_{-\nu,\rmi k}(e^{-L})~,\quad\Psi_{l}(L)=e^{L/2}W_{-\nu,\rmi k_l}(e^{-L})~,
\end{aligned}
\end{equation}
which are normalised as
\begin{equation}
\int _{-\infty}^{\infty}\dd l\,\psi_{k_{1},\nu}(l)\psi_{k_{2},\nu}(l)=\frac{\delta(k_{1}-k_{2})}{\rho_{\nu}(k_{1})},\quad\rho_{\nu}(k)=\frac{\Gamma(\nu+\frac{1}{2}\pm ik)}{2\pi\Gamma(\pm2ik)}.
\end{equation}
In particular, in the case $n_F=2$, then $\abs{t_1}>1$ means $\nu\leq -1/2$; while the relation is more involved in general $n_F$, where $\nu$ is related to other $t_i$ through \eqref{eq:ETW brane dictionary substitution}.

The bulk interpretation of the discrete part of the spectrum is explained in Sec.~\ref{ssec:discrete spectrum}. In particular, while the IR limit indicates that the $n_F=8$ corresponds to an AdS$_2$ black hole with an ETW brane in the bulk, this picture changes at a different semiclassical regime (discussed in Sec.~\ref{sec:Krylov Askey}) where we keep temperatures finite\footnote{In contrast, the Schwarzian case corresponds to a very low temperature limit \cite{Berkooz:2018qkz}.}. In this regime, the $n_F=8$ case is dual to a bulk theory whose evolution goes beyond that describing ETW branes in an AdS$_2$ black hole background.\footnote{Similar to \cite{Heller:2025ddj,Aguilar-Gutierrez:2025otq,Aguilar-Gutierrez:2026ogo,Blommaert:2024whf} one could study the regime $\theta\simeq\pi$ with appropriate canonical variables to describe dS$_2$ space with a spacelike ETW brane that removes one of the boundaries at  $\mathcal{I}^+$ or $\mathcal{I}^-$. We leave a detailed analysis of this interpretation for future work.}

\paragraph{The adjoint case}
While the adjoint deformation corresponds to a special case of the $n_F=8$ system, one should note that the previous triple-scaling limits that recover the Morse potential all had in common that as $\lambda\rightarrow0$ we let $t_1=q^{2\Delta+1}$ with $\Delta\sim\mathcal{O}(1)$ and $t_{2\leq i\leq 4}$ held fixed. In contrast, the limit for the adjoint case corresponds to setting $t_1=-t_3=q\sqrt{\chi}$ and $t_2=-t_4=\sqrt{\chi}$. If we set $\chi\sim\mathcal{O}(1)$ as $\lambda\rightarrow0$, to study its classical limit, then it is outside the range of cases where we recover a Morse potential in the IR limit for the $n_F=8$ theory. This indicates that the resulting system describes a Liouvillian theory different from the one with an ETW brane obtained in the Schwarzian regime of the other $n_F=2$ to $8$ cases. 

To appreciate the difference between this case and the ETW brane limits taken in the previous cases, one may implement the scaling:
\begin{equation}\label{eq:opt 1}
    \chi=b q^2~,\quad q^{2n}=4\lambda^2\rme^{-L}~,
\end{equation}
with $b\sim\mathcal{O}(1)$ as $\lambda\rightarrow0$. The chord Hamiltonian in this limit can be expressed as:
\begin{equation}\label{eq:Liouville from Adj}
  \frac{2-\sqrt{1-q^2}\hH_{\rm Adj}}{4\lambda^2}=-\partial_{L}^2+\rme^{-L}+\mathcal{O}(\lambda^{2})~,
\end{equation}
where we defined $L\equiv\tilde{L}-2\log\abs{1-b}$. Thus, by absorbing the regularization parameter $\log\abs{1-b}$ in the canonical variable $L$, this limit reproduces the exact same Liouvillian-like dynamics of JT gravity without ETW branes \eqref{eq:H nF0}. 

Note that in contrast, if we implement: 
\begin{equation}\label{eq:opt 2}
    \chi=b q^2~,\quad q^{2n}=2\lambda\rme^{-L}~,
\end{equation}
then:
\begin{equation}\label{eq:adj H opt 2}
   \sqrt{1-q^2}\hH_{\rm Adj}-2=4\lambda^2\partial_L^2+\mathcal{O}(\lambda^{4})~,
\end{equation}
which corresponds to the canonical Hamiltonian of a free non-relativistic particle. For this reason, we associate the triple-scaling of the Adj.~case with the conditions stated in \eqref{eq:opt 1}. Note that the difference between the undeformed (i.e.~$n_F=0$) theory and the adjoint case is a constant $\mathcal{O}(1)$ regularization scale in the canonical length parameter $L$ by a factor $(1-b)$ in comparison to the triple-scaling limit introduced in \eqref{eq:def IR UV triple scaling limit} and \eqref{eq:opt 1} instead of \eqref{eq:opt 2} which was implemented in the previous cases associated to ETW branes. The physical interpretation is that the $\lambda^2$ regularization is for two-sided geodesics in the bulk, while $\lambda$ is for one-sided ones, which are less singular due to the ETW brane where the geodesic curve ends. Thus, the above Schwarzian limit of the adjoint deformation corresponds to AdS$_2$ black hole(s) with a total of two asymptotic boundaries. Note that there could exist ETW branes in the bulk that are glued together through Israel junction conditions \cite{Israel:1966rt} (which has been recently studied in sine dilaton gravity by \cite{Cui:2025sgy}). This would be compatible with the bulk interpretation in \cite{Berkooz:2025ydg}, which associates the adjoint case to a pair of entangled branes (which are not necessarily coupled). We study this perspective in App.~\ref{app:Ent branes}.

The result \eqref{eq:Liouville from Adj} highlights that in general, the q-Askey deformed theories can have several Schwarzian regimes, which depend on the scaling of the deformation parameters with $\lambda$ as we take $\lambda\rightarrow0$. This can be seen in the adjoint case being a special case of the $n_F=8$ deformation. The latter thus has Schwarzian limits, corresponding to \eqref{eq:triple scaling} and \eqref{eq:opt 1}. This indicates that, in general, a family of deformed SYK models can have different bulk interpretations depending on the specific parameters of the deformation.

\section{Interpreting q-Askey deformations from Krylov complexity}\label{sec:Krylov Askey}
In this section, we are interested in learning about the properties of the q-Askey deformed chord theories in Sec.~\ref{sec:micro models}, and particularly their semiclassical limit (i.e.~taking $\lambda n$ fixed as $\lambda\rightarrow0$ where $n$ indicates an eigenvalue of the chord number) with finite temperature effects. The probe that we mainly deal with is Krylov state complexity \cite{Balasubramanian:2022tpr}. This probe has successfully entered the holographic dictionary of the DSSYK model \cite{Rabinovici:2023yex}, where it is identified with the length of an Einstein-Rosen bridge in an AdS$_2$ black hole \cite{Rabinovici:2023yex}.

\paragraph{Outline}In Sec.~\ref{ssec:review Krylov} we review Krylov state complexity. In Sec.~\ref{ssec:semiclassical limit}, we explain how to evaluate Krylov state complexity in the semiclassical limit for q-Askey deformed theories with the discrete or continuous energy spectrum, and how the results can be used to evaluate two-point correlation functions for the deformed theories. Sec.~\ref{ssec:Krylov nF2} and \ref{ssec:Krylov nF4} discuss the computations for the $n_F=2$ and $4$ cases. In Sec.~\ref{ssec:quantum disk} we provide a quantum disk interpretation for Krylov complexity based on the quantum group of the DSSYK model and its deformations. Sec.~\ref{ssec:Krylov nF6} focuses on the $n_F=6$ case. Up to this point, the Krylov complexity has the same analytic structure which can be interpreted in terms of ETW branes in the bulk, as discussed in Sec.~\ref{ssec:sine dilaton gravity int}. Sec.~\ref{ssec:discrete spectrum} focuses on the bulk interpretation for the discrete spectrum solutions. Sec.~\ref{ssec:Krylov nF8} and \ref{ssec:Krylov nF Adj} analyse the $n_F=8$ and adjoint cases. In Sec.~\ref{ssec:RMT} we discuss a special limit of the adjoint deformation, which we identify with an emergent SU(2) RMT, and we also relate Krylov complexity to the Schur/SYK duality \cite{Gaiotto:2024kze}.

For completeness, we also include the evaluation of Krylov complexity of the HH state in the $n_F=0$ case in App.~\ref{app:Krylov nF0}.

\subsection{Review of Krylov complexity}\label{ssec:review Krylov}
To define Krylov state complexity (also called spread complexity \cite{Balasubramanian:2022tpr}), we consider an ordered basis:
\begin{equation}
    \qty{\ket{B_0},\ket{B_1},\dots,\ket{B_n},\dots}~,
\end{equation}
so that one can define a cost function with respect to a reference state $\ket{\psi(\tau)}$ as:
\begin{equation}\label{eq:cost}
    \mathcal{C}\equiv\sum_{n=0}^\infty c_n\frac{\abs{\bra{B_n}\ket{\psi(\tau)}}^2}{\bra{\psi(\tau)}\ket{\psi(\tau)}}~,
\end{equation}
where $\qty{c_n\in\mathbb{R}}$ is a monotonically increasing sequence, and we allow for the reference state to evolve in complex-valued time $\tau$ as \cite{Erdmenger:2023wjg}:
\begin{equation}\label{eq:liouv}
    \ket{\psi(\tau)}\equiv\rme^{-\tau\hat{H}_{n_F}}\ket{\psi(0)}~,
\end{equation}
where $\tau\equiv\frac{\beta}{2}+\rmi t$, with $\beta$ and $t\in \mathbb{R}$, leading to standard Schrödinger evolution when $\beta=0$, and $\hat{H}_{n_F}$ is the corresponding generator of evolution with respect to the parameter $\tau$. State complexity in \cite{Balasubramanian:2022tpr} is defined by minimizing the cost function \eqref{eq:cost} over all possible basis sets, which occurs when $\qty{\ket{B_n}}$ is the Krylov basis $\qty{\ket{n}}$ where $\ket{0}=\ket{\psi(\tau=0)}$ is the initial state in the basis, which we may choose as the zero-chord state $\ket{\Omega}$. The other elements in the Krylov basis are obtained recursively through the Lanczos algorithm in the form:
\begin{align}
    \ket{{n+1}}&=\frac{1}{b_{n+1}}\qty(\sqrt{1-q^2}\hat{H}_{n_F}\ket{n}-b_n\ket{{n-1}}-a_n\ket{n})~,\label{eq:lanczos new}
\end{align}
where we have rescaled the Hamiltonian by a constant factor $\sqrt{1-q^2}$, and the algorithm is initialised by the condition $b_0=0$, $f_{0}(x)=1$; and $b_n$ and $a_n$ are the Lanczos coefficients, which are determined from \eqref{eq:lanczos new} with the condition $\bra{n}\ket{m}=\delta_{nm}$, and $a_{0}=\sqrt{1-q^2}\bra{0}\hat{H}_{n_F}\ket{0}$. We are interested in Hamiltonians with a spectrum:
\begin{equation}\label{eq:explicit E}
    \hat{H}_{n_F}\ket{\theta}=E(\theta)\ket{\theta}~,
\end{equation}
where $E(\theta)$ appears in \eqref{eq:cont spectrum}.

The Hamiltonian can then be represented in the Krylov basis as:
\begin{equation}
    \begin{aligned}
        \hat{H}_{n_F}=&\hat{A}+\hat{A}^\dagger+\frac{a_{\hat{n}}}{\sqrt{1-q^2}}\\
        =&\frac{1}{\sqrt{1-q^2}}\qty(b_{\hat{n}}\rme^{-\rmi \hat{p}}+\rme^{\rmi \hat{p}}b_{\hat{n}}+a_{\hat{n}})~,
    \end{aligned}
\end{equation}
where the ladder operators are $\hat{A}\equiv \rme^{\rmi\hat{p}}b_{\hat{n}}/\sqrt{1-q^2}$, $\hat{A}^\dagger\equiv b_{\hat{n}}\rme^{-\rmi\hat{p}}/\sqrt{1-q^2}$ and $\hat{a}_{\hat{n}}$ are defined to act in the chord basis as:
\begin{equation}\label{eq:lanczos special}
    \rme^{\mp\rmi \hat{p}}\ket{n}=\ket{{n\pm 1}}~,\quad b_{\hat{n}}\ket{n}=b_{n}\ket{n}~,\quad a_{\hat{n}}\ket{n}=a_{n}\ket{n}~.
\end{equation}
One can then use the above algorithm to find the explicit coefficients in terms of Hamiltonian moments (see e.g.~\cite{Caputa:2023vyr}):
\begin{equation}\label{eq:Lanczos coeff}
\begin{aligned}
    a_0&=m_1~,\quad b_1=m_2-m_1^2~,\quad a_1 = \frac{m_3+m_1^3-2m_1m_2}{m_2-m_1^2}~,\\
    b_2&=\frac{m_2m_4-m_2^3 - m^2_3 - m^2_1m_4 - 2m_1^2m_2m_4 }{(m_2-m^2_1)^2}
\end{aligned}
\end{equation}
where
\begin{equation}
    m_n=(-1)^n\eval{\pdv[n]{\mathcal{Z}(\beta)}{\beta}}_{\beta=0}~,\quad \mathcal{Z}(\beta)\equiv\frac{\bra{\psi(t=0)}\rme^{-\beta\hat{H}_{n_F}}\ket{\psi(t=0)}}{\bra{\psi(t=0)}\ket{\psi(t=0)}}~.
\end{equation}
Then, based on \eqref{eq:cost} with $\qty{B_n}=\qty{\ket{n}}$ and $c_n=n$, Krylov spread complexity is simply defined as:
\begin{equation}\label{eq:spread Complexity}
    \mathcal{C}_{\rm S}(t)\equiv2\frac{\bra{\psi(t)}\hat{n}\ket{\psi(t)}}{\bra{\psi(t)}\ket{\psi(t)}}~,\quad \hat{n}\equiv \sum_{n=0}^\infty n\ket{n}\bra{n}~,
\end{equation}
{which means that Krylov complexity of the HH state exactly corresponds to the chord number for the deformed theories. This finding is modified when adding matter contributions in the Hamiltonian, even when $n_F=0$ \cite{Ambrosini:2024sre,Ambrosini:2025hvo,Aguilar-Gutierrez:2025pqp}.}

In the following, we are interested in evaluating \eqref{eq:spread Complexity} for the HH state of the q-Askey deformed DSSYK model.

\subsection{Semiclassical limit}\label{ssec:semiclassical limit}
After deducing the form of the Lanczos coefficients, one can work in the semiclassical limit, where $\ell\equiv\lambda \mathcal{C}_{\rm S}$ is fixed as $\lambda\rightarrow0$, to {write the path integral formulation of the theory as} (see relevant discussion in \cite{Blommaert:2024ymv,Aguilar-Gutierrez:2025pqp,Aguilar-Gutierrez:2025mxf,Aguilar-Gutierrez:2025hty,Aguilar-Gutierrez:2025sqh,Aguilar-Gutierrez:2026ogo,Bossi:2024ffa}):
\begin{subequations}
\begin{align}
    Z&=\int[\rmd \ell][\rmd p]\exp[\int\rmd\tau\qty(\frac{\rmi}{\sqrt{2\lambda}}p\partial_\tau\ell-H_{n_F})]~,\\
    H_{n_F}&=\frac{1}{\sqrt{2\lambda}}\qty(2b(\ell)\cos{p}+a(\ell))~,\label{eq:classical H}
    \end{align}
\end{subequations}
with $a(\ell)=\lim_{\lambda\rightarrow0}a_n$, $b(\ell)=\lim_{\lambda\rightarrow0}b_n$. One then finds the equations of motion:
\begin{subequations}\label{eq:EOM krylov basis def}
\begin{align}
    \frac{1}{\sqrt{2\lambda}}\dv{\ell}{t}&=-\frac{2}{\sqrt{2\lambda}}\sin p~b(\ell),\label{eq:first EOM}\\
    \frac{1}{\sqrt{2\lambda}}\dv{p}{t}&=-\frac{1}{\sqrt{2\lambda}}\qty(2\cos p~b'(\ell)+a'(\ell))~,\\
    \frac{1}{2\lambda}\dv[2]{\ell}{t}&={\frac{1}{\lambda}}\qty(\dv{\ell}\qty(b(\ell)^2)+\dv{a(\ell)}{\ell}{b(\ell)\cos p})\label{eq:2nd ODE}\\
    &={\frac{1}{\lambda}\qty(\dv{\ell}\qty(b(\ell)^2)+\dv{a(\ell)}{\ell}\qty(\cos\theta-\frac{a(\ell)}{2}))}~.\nonumber
\end{align}
\end{subequations}
These differential equations are supplemented by the initial conditions:
\begin{equation}\label{eq:initial condt}
   \ell(t=0)=\ell_0~, \quad \eval{\dv{\ell}{t}}_{t=0}=0~,
\end{equation}
where $\ell_0$ is a constant, determined by conservation of energy in the classical Hamiltonian. Note that the second condition follows from:
\begin{equation}\label{eq: derivative 1st cond}
\begin{aligned}
    &\eval{\dv{t}\bra{\psi(\tau)}\hat{n}\ket{\psi(\tau)}}_{t=0}=
    \bra{\Omega}\rme^{-\frac{\beta}{2}\hH_{n_F}}[\hH_{n_F},\hat{n}]\rme^{-\frac{\beta}{2}\hH_{n_F}}\ket{\Omega}\\
    &=2\sum_{n=0}^\infty n\qty(\bra{\Omega}\rme^{-\frac{\beta}{2}\hH_{n_F}}\ket{n}\bra{n}\dv{\rme^{-\frac{\beta}{2}\hH_{n_F}}}{\beta}\ket{\Omega}-\bra{\Omega}\dv{\rme^{-\frac{\beta}{2}\hH_{n_F}}}{\beta}\ket{n}\bra{n}\rme^{-\frac{\beta}{2}\hH_{n_F}}\ket{\Omega})\\
    &=2\sum_{n=0}^\infty n\qty(\psi_n(\beta)\psi'_n(\beta)-\psi_n(\beta)\psi'_n(\beta))=0~,
\end{aligned}
\end{equation}
where we expanded the chord number as $\hat{n}\equiv\sum_nn\ket{n}\bra{n}$ and we defined $\psi_n(\beta)\equiv\bra{\Omega}\rme^{-\frac{\beta}{2}\hH_{n_F}}\ket{n}$, which is real as long as $\hH_{n_F}$ is Hermitian, and $\psi'_n(\beta)=\rmd \psi_n(\beta)/\rmd \beta$. This leads to the second initial condition in \eqref{eq:initial condt}.

The initial value $\ell_0$ is determined by combining the second condition in \eqref{eq:initial condt} together with \eqref{eq:first EOM}, implying $p(t=0)=0$. Thus, the Hamiltonian \eqref{eq:classical H} with conserved energy eigenvalue $E=2\cos\theta/\sqrt{2\lambda}$ evaluates to:
\begin{equation}\label{eq:initial length}
   2\cos\theta=2b(\ell_0)+a(\ell_0)~.
\end{equation}
Note that while we illustrate the expressions with $\theta\in[0,~\pi]$, the exact same arguments apply for the discrete solutions, where one simply changes $\theta\rightarrow\rmi x_\ell$ with $\ell\in \mathbb{Z}_{\geq0}$ obeying the conditions \eqref{eq:dissrete cond 1}, \eqref{eq:dissrete cond 2}.

To solve the equation of motion \eqref{eq:2nd ODE}, we apply a substitution $u=\dv{\ell}{t}$, so that \eqref{eq:2nd ODE} can be expressed as
\begin{equation}
u\dv{u}{\ell}=2\dv{\ell}\qty(b(\ell)^2+\cos\theta~a(\ell)-\frac{a(\ell)^2}{4})~.
\end{equation}
This can be straightforwardly integrated using the initial condition \eqref{eq:initial length} as
\begin{equation}\label{eq:v2/4}
    \frac{u^2}{4}=b(\ell)^2+\cos\theta~a(\ell)-\frac{a(\ell)^2}{4}-\cos^2\theta~.
\end{equation}
This conserved equation of motion can be integrated, as
\begin{equation}\label{eq:special eq}
    t=\frac{1}{2}\int _{\ell_0}^{\ell(t)}\frac{\rmd\ell'}{\sqrt{b(\ell')^2+a(\ell')\qty(\cos\theta-a(\ell')/4)-\cos^2\theta}}~.
\end{equation}
One can then evaluate the Krylov complexity by substituting the explicit forms of $b(\ell)$, $a(\ell)$ and the initial value $\ell_0$ in \eqref{eq:initial length}. The characteristic behaviour of Krylov complexity for the different $n_F$ deformations is illustrated in Fig.~\ref{fig:nF}.\footnote{{It would be interesting to investigate if the variations of Krylov complexity with respect to the q-Askey deformation parameters can be used to extend the first law of spread complexity in \cite{Balasubramanian:2025xkj} with additional chemical potentials associated with the deformation parameters.}}
\begin{figure}
    \centering
\subfloat[]{\includegraphics[width=0.5\linewidth]{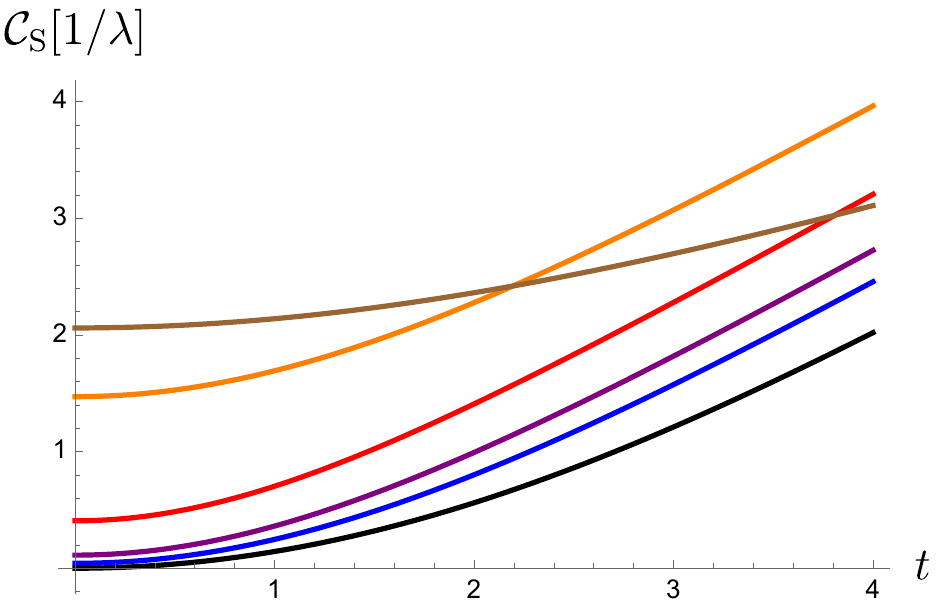}}\subfloat[]{\includegraphics[width=0.5\linewidth]{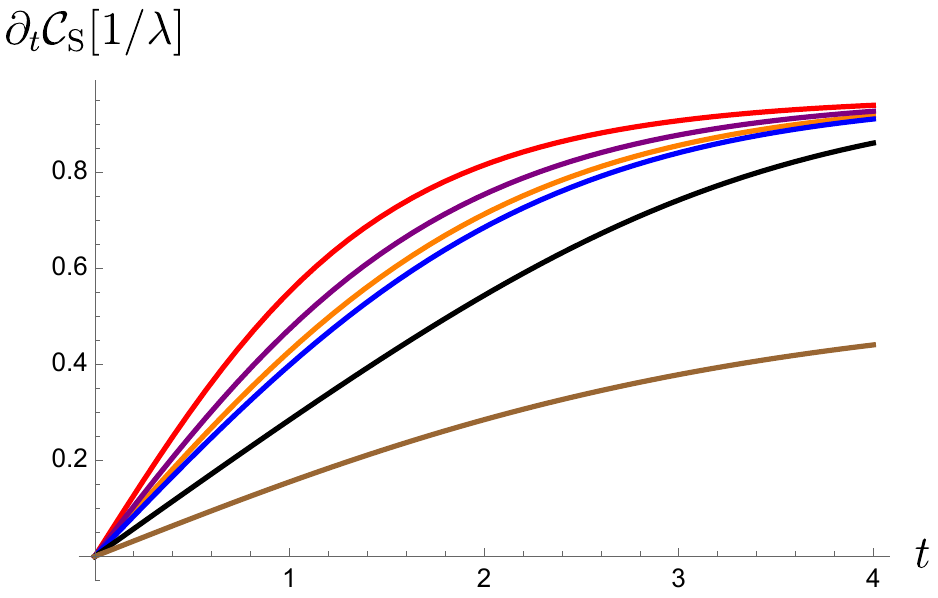}}
    \caption{(a) Semiclassical complexity $\mathcal{C}_{\rm S}$ \eqref{eq:Krylov HH state nF} and (b) its rate of growth for the HH state of the $n_F=0$ (orange), $2$ (red, $t_1=0.9$), $4$ (purple, $t_1=0.9$, $t_2=0.5$), $6$ (blue, $t_1=0.9$, $t_2=0.5$, $t_3=0.3$), $8$ (black, $t_1=0.9$, $t_2=0.5$, $t_3=0.3$, $t_4=0.5$) and the adjoint (brown, $\chi=0.5$) deformations of the DSSYK model. All cases are displayed for $\theta=0.5$, and we have rescaled Krylov complexity in units 1/$\lambda$, while $t$ is dimensionless since we set $J=1$ in \eqref{eq:J coupling}.}
    \label{fig:nF}
\end{figure}

In case of the $n_F=6$ q-Askey deformations (and its special cases, i.e. $n_F=2,~4$), the differential equation of interest \eqref{eq:2nd ODE} takes the form:
\begin{equation}\label{eq:ETW brane ODE}
    \dv[2]{\ell}{t}=a_1\rme^{-\ell}+a_2\rme^{-2\ell}~.
\end{equation}
The solution to the differential equation with initial conditions \eqref{eq:initial condt}, \eqref{eq:initial length}, gives:
\begin{equation}\label{eq:Krylov HH state nF}
    \mathcal{C}_{\rm S}(t)=\frac{\ell(t)}{\lambda}=\frac{1}{\lambda}\log\frac{a_1+(a_1+a_2\rme^{-\ell_0})\cosh(t)}{2a_1\rme^{-\ell_0}+a_2\rme^{-2\ell_0}}~.
\end{equation}
The above solution for the semiclassical Krylov complexity of the HH state is precisely of the form of ETW branes in the dual bulk theory, as we will see below. {In the solutions below, there is an overall factor rescaling the time in \eqref{eq:Krylov HH state nF} as $t\rightarrow2\sin\theta t$}. Note that:\footnote{On the other hand, the Boltzmann temperature at the semiclassical limit in the microcanonical ensemble of the undeformed DSSYK is given by \cite{Goel:2023svz}
\begin{eqnarray}\label{eq:Bolt temp}
    \beta(\theta)\equiv\frac{\pi-2\theta}{\sin\theta}~.
\end{eqnarray}
Note that the very low temperature limit in both \eqref{eq:Bolt temp} and \eqref{eq:fake temper} corresponds to $\beta\sim\mathcal{O}(\lambda)$ as $\lambda\rightarrow0$.}
\begin{equation}\label{eq:fake temper}
    \beta_{\rm fake}\equiv\frac{\pi}{2\sin\theta}~,
\end{equation}
is the fake temperature \cite{Lin:2023trc}, which is the thermal scale responsible for the decay of semiclassical correlation functions, as captured by the saddle point solutions to the path integral of the DSSYK model \cite{Aguilar-Gutierrez:2025pqp}. This naturally generalises to our arguments when there is an ETW brane. This is consistent with the interpretation of  Krylov spread complexity of the HH state \eqref{eq:Krylov HH state nF} as being dual to a geodesic length (captured by the corresponding two-point correlation \eqref{eq:correlation function} in the semiclassical limit \cite{Berkooz:2022fso}). We will show this more explicitly in Sec.~\ref{ssec:sine dilaton gravity int}.

\paragraph{Two-point correlation functions}\label{ssec:two-point functions} 
In the semiclassical limit, one can perform the computations of two-point functions from the Krylov spread complexity:
\begin{equation}\label{eq:correlation function}
   G_2(t)\equiv\frac{ \bra{0}\rme^{-\tau^*\hH_{n_F}}q^{2\Delta\hat{n}}\rme^{-\tau^*\hH_{n_F}}\ket{0}}{\bra{0}\rme^{-\beta\hH_{n_F}}\ket{0}}\underset{\rm DS}{\rightarrow}\rme^{-\Delta\ell(t)}~,
\end{equation}
where $\tau=\beta/2+\rmi t$, $\ell(t)=\lambda\mathcal{C}_{\rm S}(t)$ (with $\rme^{-\tau\hH_{n_F}}\ket{0}$ as initial state), and we use that as $\lambda\rightarrow0$ quantum fluctuations are suppressed, i.e.:
\begin{equation}
    \lim_{\lambda\rightarrow0}\expval{f(\hat{n})}=f(\expval{\hat{n}})~.
\end{equation}
We discuss the evaluation of correlation function for all $n_F$ at finite $\lambda$ in App. \ref{app:more two-point functions} using the known results for the asymmetric Fourier kernel for Al-Salam Chihara polynomials \cite{askey1996general}. However, there are significant simplifications in the semiclassical limit that allow us to solve the saddle points of the chord path integral, as explained in this section.

\subsection{\texorpdfstring{The $n_F=2$}{} case: generalised q-Hermite polynomials}\label{ssec:Krylov nF2}
Consider the $n_F=2$ deformed chord Hamiltonian \eqref{eq:H SYK2} 
in the symmetric form:
\begin{equation}\label{eq:H nF2}
    \begin{aligned}
        \hH_{n_F=2}\ket{n}=\sqrt{[n+1]_{q^2}}\ket{{n+1}}+\sqrt{[n]_{q^2}}\ket{{n-1}}+\frac{t_1}{\sqrt{1-q^2}}q^{2n}\ket{n}~,
    \end{aligned}
\end{equation}
which defines an inner product between the eigenstates of the Hamiltonian and the chord number states
\begin{equation}
    \bra{n}\ket{\theta}_{n_F=2} = \frac{H_n(\cos\theta;t_1|q^2)}{\sqrt{(q^2,q^2)_n}}~,
\end{equation}
where the generalised q-Hermite polynomials shown above are defined as
\begin{flalign}
    H_n(\cos\theta;t_1|q^2)\equiv e^{in\theta} 
    \,_2\phi_0\bigg(\genfrac{}{}{0pt}{}{q^{-2n}, t_1 e^{i\theta}}{-}; q^2, q^{2n}e^{-2i\theta} \bigg)~,
\end{flalign}
and $_r\phi_s$ was defined in \eqref{eqn:hypergeometric}. Note that the same relations are valid for the discrete energy spectrum solutions \eqref{eq:discrete states schrodinger}, where we need to perform the continuation $\theta\rightarrow\rmi x_l$.

After rescaling the Hamiltonian \eqref{eq:H nF2} by the overall factor $\sqrt{1-q^2}$, the corresponding Lanczos coefficients in the semiclassical limit are then:
\begin{equation}
    b(\ell)=\sqrt{1-\rme^{-\ell}}~,\quad a(\ell)=t_1\rme^{-\ell}~.
\end{equation}
Solving \eqref{eq:EOM krylov basis def} with \eqref{eq:initial length} we get \eqref{eq:spread Complexity} in the continuum approximation as (see Fig.~\ref{fig:nF}):
\begin{equation}\label{eq:womrhole XY}
    \begin{aligned}
        \mathcal{C}_{\rm S}(t)=\frac{\ell(t)}{\lambda}=\log\qty(\frac{ 1-t_1\cos
   \theta+\sqrt{t_1^2-2 t_1
   \cos \theta +1}~\cosh(2t\sin\theta)}{2 \sin^2\theta})~.
    \end{aligned}
\end{equation}
This corresponds to a renormalised geodesic length in JT gravity with an ETW brane \cite{Gao:2021uro,Okuyama:2023byh}, which we discuss later in Sec. \ref{ssec:sine dilaton gravity int}.

\paragraph{Growth and convexity}
Using the analytic expression \eqref{eq:womrhole XY}, it is straightforward to examine how the Krylov complexity (or its time derivative, the growth rate) depends on $t_1$:
\begin{flalign}
    \partial_{t_1}\dot{\mathcal{C}}_{\rm S}(t)=\frac{2\sin\theta\sinh(2t\sin\theta)~t_1\sin^2\theta}{\lambda~T_1(\theta)(1-t_1\cos\theta+T_1(\theta)\cosh(2t\sin\theta))^2}
\end{flalign}
Where $T_1(\theta)=\sqrt{t_1^2-2t_1\cos\theta+1}$. For $t>0$, $0<\theta<\pi$ and $t_1>0$ every factor on the right-hand side is positive, $\partial_{t_1}\dot{\mathcal{C}}_{\rm S}(t)>0$, i.e. as $t_1$ increases, the $n_F=2$ complexity grows faster; see Fig. \ref{fig:nf2_2}.
\begin{figure}[t!]
    \centering
    \subfloat[]{\includegraphics[width=0.5\linewidth]{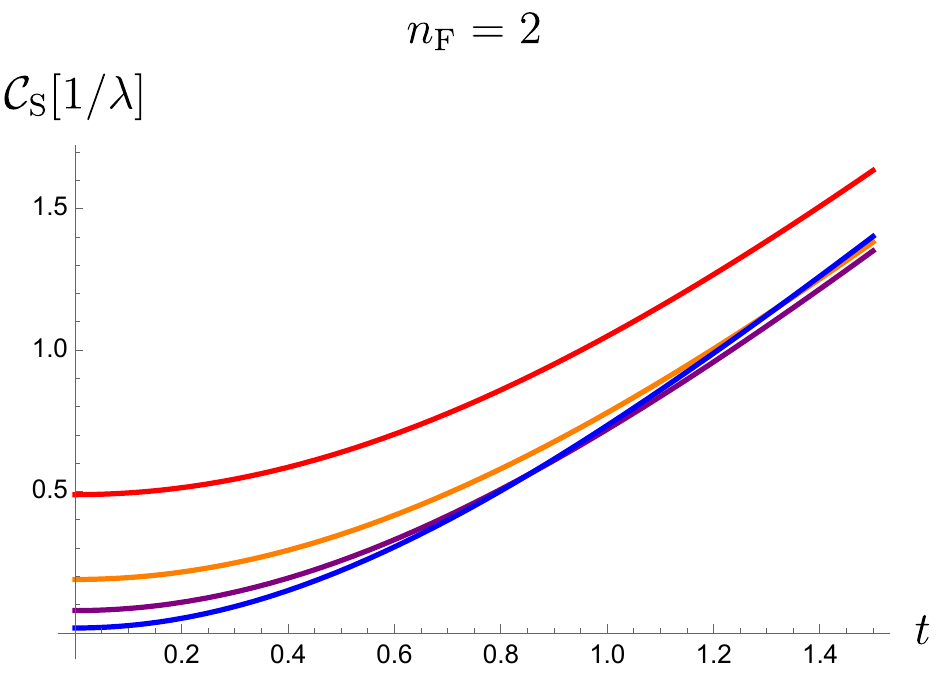}}\subfloat[]{\includegraphics[width=0.5\linewidth]{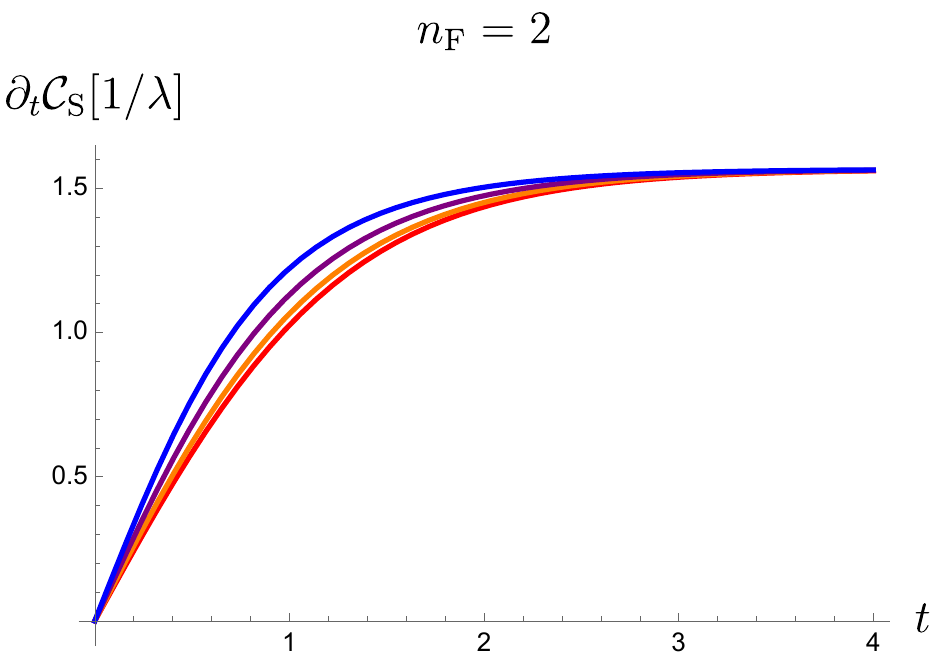}}
    \caption{(a) Krylov complexity and (b) its rate of growth for the $n_F=2$ case. The rate of growth $\rmd\mathcal{C}_{\rm S}(t)/\rmd t$ increases at intermediate times, as the value of $t_1$ is increased, while the initial value of $\mathcal{C}_{\rm S}$ increases with decreasing $t_1$. The different values of the deformation parameter: $t_1=0$ (red), $0.5$ (orange), $0.75$ (purple), $1$ (blue).}
    \label{fig:nf2_2}
\end{figure}
{Meanwhile, at very late times, the $\cosh(2t\sin\theta)^2$ in the denominator makes $\partial_{t_1}\mathcal{C}_{\rm S}\simeq 0$.} In the triple-scaling limit, this is like decreasing $\Delta$, so the effective barrier becomes shallower. 

One also finds that $\mathcal{C}_{\rm S}$ is convex for all allowed values of $t_1$. We can see this from the bulk equation of motion \eqref{eq:ETW brane ODE}:
\begin{equation}
    \dv[2]{\ell}{t}=2(1-t_1\cos\theta)\rme^{-\ell}+t_1^2\rme^{-2\ell}~,
\end{equation}
since we require $\abs{t_1}\leq1$ then $\dv[2]{\ell}{t}>0$ always. In particular, when $t_1\rightarrow1$, the $\rme^{-\ell}$ term is more suppressed in comparison with the $\rme^{-2\ell}$ term; where the latter is suppressed at late times since $\ell$ is monotonically increasing. 

\subsection{\texorpdfstring{The $n_F=4$}{} case: Al-Salam-Chihara polynomials}\label{ssec:Krylov nF4}
Let us consider 
the $n_F=4$ chord Hamiltonian \eqref{eq:H SYK8} with $t_3=t_4=0$ in its symmetric form:
\begin{equation}\label{eq:H nF4}
\begin{aligned}
    \hH_{n_F=4}\ket{n}=&\sqrt{[n+1]_{q^2}}\sqrt{1-t_1t_2 q^{2n}}\ket{{n+1}}+\sqrt{[n]_{q^2}}\sqrt{1-t_1t_2 q^{2(n-1)}}\ket{{n-1}}\\
    &+\frac{t_1+t_2}{\sqrt{1-{q^2}}}q^{2n}\ket{n}~,
\end{aligned}
\end{equation}
which is solved by 
\begin{flalign}
    \bra{n}\ket{\theta}_{n_F=4} =\frac{(t_1t_2;q)_\infty}{2}\frac{Q_n(\cos\theta;t_1,t_2|q^2)}{ \sqrt{(q^2,t_1t_2;q^2)_n}}~.
\end{flalign}
where we chose $\bra{0}\ket{\theta}_{n_F=2}=1$, and we denote the Al-Salam Chihara polynomials as
\begin{flalign}\label{eqn:ASC}
    Q_n(\cos\theta;t_1,t_2|q^2) = \frac{(t_1t_2;q^2)_n}{t_1^n} 
    \,_3 \phi_2\bigg({\genfrac{}{}{0pt}{}{q^{-2n},t_1e^{\pm i\theta}}{t_1t_2, 0}; q^2,q^2}\bigg)~.
\end{flalign}
We identify the Lanczos coefficients of the rescaled Hamiltonian in the semiclassical limit as:
\begin{equation}
    b(\ell)=\sqrt{1-\rme^{-\ell}}\sqrt{1-t_1t_2\rme^{-\ell}}~,\quad a(\ell)=(t_1+t_2)\rme^{-\ell}~.
\end{equation}
One finds the initial condition:
\begin{equation}
    \rme^{-\ell_0}=\frac{2 \sin ^2\theta}{1+t_1 t_2-\cos
   \theta  (t_1+t_2)+\sqrt{\prod_{i=1}^2\left(t_i^2-2 t_i \cos (\theta
   )+1\right)}}
\end{equation}
which allows to evaluate Krylov spread complexity in the semiclassical approximation as \eqref{eq:spread Complexity}  (see Fig.~\ref{fig:nF}):
\begin{equation}\label{eq:CS_nF4}
    \mathcal{C}_{\rm S}(t)=\frac{1}{\lambda}\log\qty(\frac{1+t_1t_2-(t_1+t_2)\cos
   \theta+\sqrt{\prod_{i=1,2}(t_i^2-2 t_i
   \cos \theta +1)}~\cosh(2\sin(\theta)~t)}{2\sin^2\theta})~.
\end{equation}
This corresponds to a geodesic length connecting the asymptotic boundary to an ETW brane in sine dilaton gravity \cite{Blommaert:2025avl,Cui:2025sgy}; which we discuss until Sec.~\ref{ssec:sine dilaton gravity int}. In particular, the case where $t_1=-t_2$, investigated in \cite{Xu:2024hoc,Aguilar-Gutierrez:2025hty}, corresponds to two ETW branes that are glued to each other through Israel junction conditions \cite{Israel:1966rt} (Fig.~\ref{fig:Ent_branes}) in JT gravity \cite{Engelhardt:2022qts} (equivalently in sine dilaton gravity with appropriate Weyl rescaling \cite{Blommaert:2025avl}).

\paragraph{Growth and convexity}
Similarly to the $n_F=2$ case, let us define $T_i(\theta)=\sqrt{t_i^2-2t_i\cos\theta+1}$ for $i=1,2$. From \eqref{eq:CS_nF4}, we find:
\begin{flalign}
    \partial_{t_i} \dot{\mathcal{C}}_{\rm S}(t) = 
    \frac{1}{\lambda}
    \frac{2\sin^3\theta T_j(\theta) (t_j-t_i)\sinh(2t\sin\theta)}{T_i(\theta)(1+t_1t_2-(t_1+t_2)\cos\theta+T_1(\theta)T_2(\theta)\cosh(2t\sin\theta))^2}
\end{flalign}
Where $i,j\in\{1,2\}, i\neq j$. For real $t_{1,2}$ we find:
\begin{flalign}
    \partial_{t_1} \dot{\mathcal{C}}_{\rm S}(t) \lessgtr 0 \iff t_1 \lessgtr t_2 \quad \text{and} \quad 
    \partial_{t_2} \dot{\mathcal{C}}_{\rm S}(t) \lessgtr 0 \iff t_2 \lessgtr t_1 
    \label{eqn:t1t2-growth}
\end{flalign}
Therefore, for fixed mean $\frac{t_1+t_2}{2}$, the $n_F=4$ Krylov complexity grows faster as the difference between the two ETW brane parameters is increased.

For $t_1=t_2$:
\begin{equation}\label{eq:nF_specia_case}
    \eval{\mathcal{C}_{\rm S}(t)}_{t_1=t_2}=\frac{2}{\lambda}\log(\frac{\cosh(2\sin\theta t/2)}{2\sin^2\theta})+\frac{2}{\lambda}\log( 2\sin\theta\sqrt{t_1^2-2t_1\cos\theta+1})~.
\end{equation}
In this case, any dependence on $t_i$ decouples, furthermore, from \eqref{eqn:t1t2-growth} it follows that $\mathcal{C}_{\rm S}$ has the slowest growth rate in this case, whereas splitting the two ETW brane parameters only increases the rate of growth, although at late enough times, they all converge to the same asymptotic rate of growth due to $1/\cosh^2(2\sin\theta~t)$ term in \eqref{eq:nF_specia_case}, as displayed in Fig.~\ref{fig:nF4_2}. 
\begin{figure}
    \centering
    \subfloat[]{\includegraphics[width=0.5\linewidth]{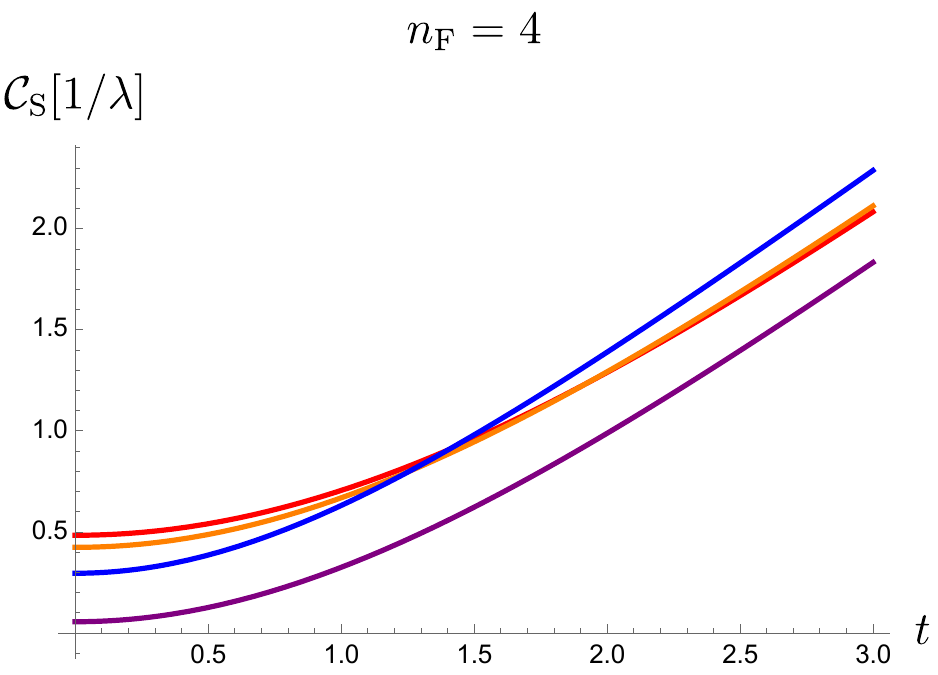}}\subfloat[]{\includegraphics[width=0.5\linewidth]{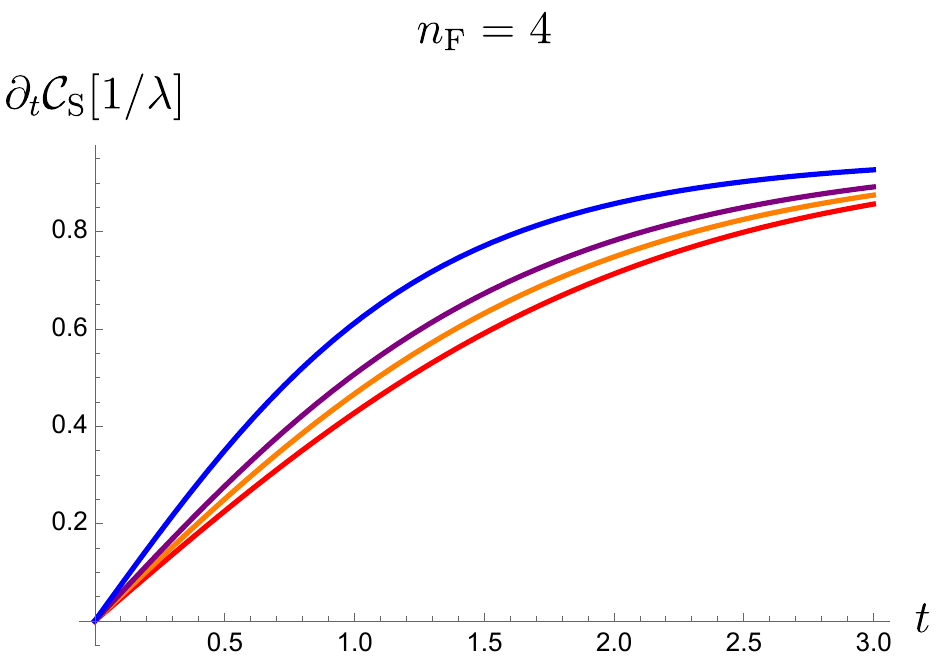}}
    \caption{(a) $\mathcal{C}_{\rm S}$ and (b) $\rmd\mathcal{C}_{\rm S}/\rmd t$ for different values of $\abs{t_2-t_1}$: $t_1=t_2=0.5$ (red); $t_1=0.75$, $t_2=0.25$ (orange); $t_1=1$, $t_2=0.5$ (purple); $t_1=1$, $t_2=0$ (blue). While the Krylov complexity grows faster as $\abs{t_1-t_2}$, at late enough times, they rate of growth of $\mathcal{C}_{\rm S}$ always converges to the same asymptotic value.}
    \label{fig:nF4_2}
\end{figure}
Moreover, the solution \eqref{eq:nF_specia_case} takes the same form as the $n_F=0$ case (up to a constant shift and time dilatation by a factor $1/2$, see App.~\ref{app:Krylov nF0}), so that the ETW brane parameter $t_1$ is just an overall additive constant, indicating that the system may be effectively a two-sided brane system.\footnote{If we used the DSSYK embedding in \eqref{eq:m=2 case} the constraints for $t_1=t_2=T$ would imply that
\begin{equation}
    \frac{1-q^{2n_1}}{2}=\sqrt{-q^{2n_1+2}}~,
\end{equation}
which would require that $n_1=\rmi \lambda\pi+\mathbb{R}$. This is not allowed since we require $n_1\in\mathbb{Z}_{\geq0}$ for the chord number.}

Another interesting regime to consider is $t_1=t_2^*=|t_1|e^{i\gamma}$. As was shown in \cite{Berkooz:2025ydg}, in this regime, the Hamiltonian moments of DSSYK with two chord reservoirs admit real values. It is interesting to examine how the growth of $\mathcal{C}_{\rm S}$ depends on the phase $\gamma$. By inserting $t_1=t_2^*=|t_1|e^{i\gamma}$ into \eqref{eq:CS_nF4}, we find:
\begin{flalign}\label{eqn:Cs-growth-gamma}
    \dot{\mathcal{C}_{\rm S}}(t,\gamma) = \frac{2}{\lambda}\frac{2\sin\theta\sinh(2t\sin\theta)\sqrt{G(\gamma)^2-4|t_1|^2\sin^2\gamma\sin^2\theta}/\lambda}{G(\gamma)+\sqrt{G(\gamma)^2-4|t_1|^2\sin^2\gamma\sin^2\theta}\cosh(2t\sin\theta)}
\end{flalign}
where:
\begin{flalign}
    G(\gamma)=1+|t_1|^2-2|t_1|\cos\theta\cos\gamma
\end{flalign}
We thus find that the growth rate of Krylov complexity depends non-monotonically on $\gamma$. While the growth is maximal for $\gamma=0,\pi$, by extremizing \eqref{eqn:Cs-growth-gamma}, we find a minimum at:
\begin{flalign}\label{eq:critical val nF4}
    \cos\gamma_*=\frac{2|t_1|\cos\theta}{1+|t_1|^2}
\end{flalign}
We display different growth rates in Fig.~\ref{fig:nf4_3}. 
\begin{figure}
    \centering
    \subfloat{\includegraphics[width=0.5\linewidth]{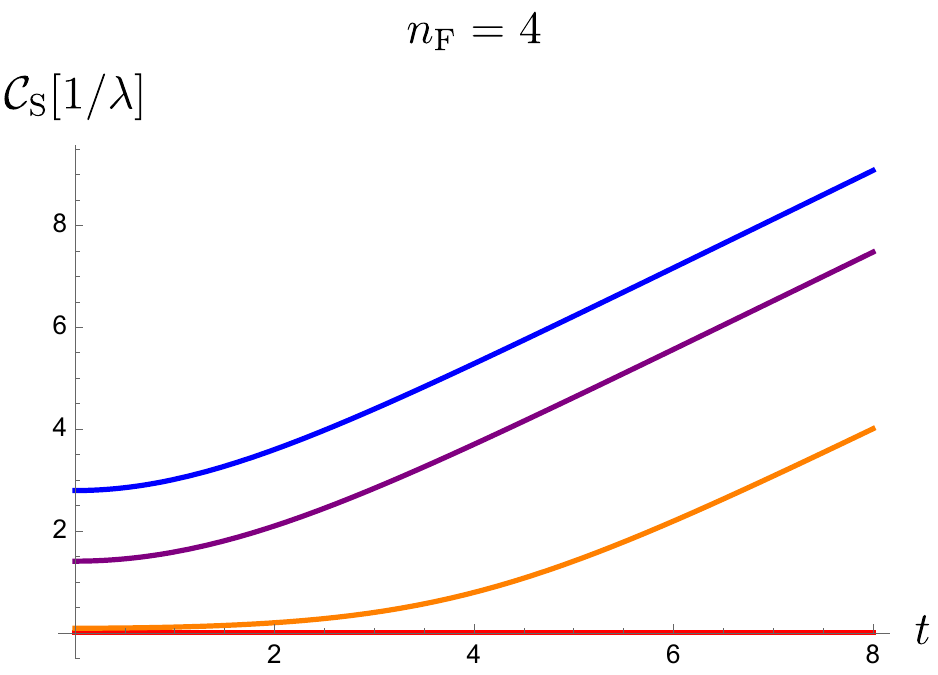}}\subfloat{\includegraphics[width=0.5\linewidth]{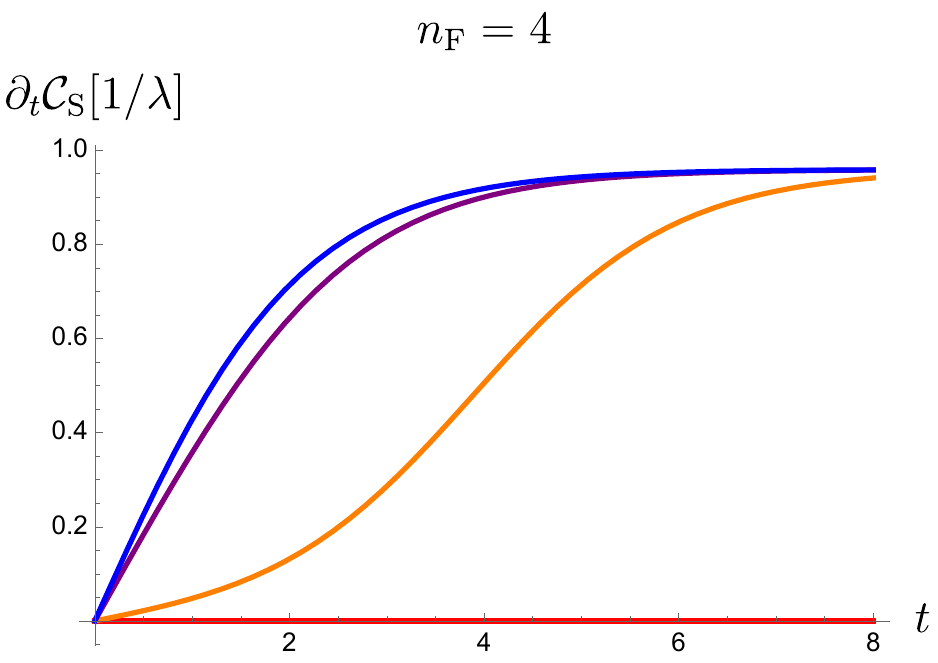}}
    \caption{We display the evolution of (a) spread complexity and (b) its growth rate in the $n_F=4$ case with $t_1=t_2^*=\abs{t_1}\rme^{\rmi \gamma}$. We fix $\abs{t_1}=1$ and we vary $\gamma=\theta$(=0.5) (red), $\pi/6$ (orange), $\pi/3$ (purple), $\pi$ (blue). The critical value for $\gamma$ in \eqref{eq:critical val nF4} generates the minimum rate of growth (red line, see \eqref{eq:explain flat line}), while $\gamma=\pi$ generates the maximum one.}
    \label{fig:nf4_3}
\end{figure}
We are particularly interested in the critical value in \eqref{eq:critical val nF4} where the spread complexity in \eqref{eq:nF_specia_case} becomes:
\begin{equation}\label{eq:explain flat line}
    \lambda\mathcal{C}_{\rm S}(t)=\log \left(\frac{\left| \abs{t_1}^2-1\right|  \sqrt{\abs{t_1}^4-2 \abs{t_1}^2 \cos (2 \theta )+1} \cosh (2 t
   \sin \theta)+\abs{t_1}^4-2 \abs{t_1}^2 \cos (2 \theta )+1}{2\sin ^2\theta   \left(\abs{t_1}^2+1\right)}\right)~.
\end{equation}
{When $\abs{t_1}=1$, corresponding to $\gamma_*=\theta+2n\pi$, $n\in\mathbb{Z}$,} then $\mathcal{C}_{\rm S}(t)=$const; so there is no evolution at all for the wormhole length. This can be understood from the Schur half-index
with $n$ fundamental Wilson line insertions \cite{Berkooz:2025ydg}:
\begin{equation}
    \mu_{n_F=4}(\theta)=\frac{({q^2},\rme^{\pm2\rmi\theta};{q^2})_\infty}{2\pi(\abs{t_1}\rme^{\pm\rmi\gamma}\rme^{\pm\rmi\theta};{q^2})_\infty}~,
\end{equation}
so that in the case $\abs{t_1}=1$ and $\gamma=\theta$ the measure for the would-be Al Salam-Chihara polynomials trivialises, indicating there is no longer a complete basis spanning the corresponding chord Hilbert space. The same observation carries on for the bulk Hilbert space of sine dilaton gravity with the ETW brane in the semi-open channel of \cite{Blommaert:2025avl}. 

Thus, our results show that there are different regimes where $C_S$ is not a convex function, unlike the behaviour of Krylov state complexity for the HH state of chaotic systems at infinite $N$ \cite{Alishahiha:2024vbf,Nandy:2024evd,Rabinovici:2025otw,Baiguera:2025dkc}. This indicates that the deformation of the SYK Hamiltonian \eqref{eq:H micro nF4} modifies the chaotic nature of the system. As shown in Sec.~\ref{ssec:KM state deformations}, the deformation in the Hamiltonian is closely related to integrable operators in the double-scaling limit \cite{Berkooz:2024evs,Berkooz:2024ofm,Almheiri:2024xtw,Aguilar-Gutierrez:2026jjv}. This is consistent with our microscopic derivation of the Hamiltonians in Sec.~\ref{ssec:explicit microscopic} since the {interaction term \eqref{eq:H int} is bilinear in the $L/R$ Majorana fermions, as in integrable deformations \cite{Nandy:2024evd}}.\footnote{It would be interesting to check if semiclassical crossed four-point functions have a Lyapunov exponent to confirm our observations. This could be carried out with the deformed DSSYK Hamiltonians with matter in Sec.~\ref{ssec:explicit microscopic} by studying the corresponding saddle point solutions in the path integral of the theories, as described in \cite{Aguilar-Gutierrez:2025pqp,Aguilar-Gutierrez:2025mxf,Aguilar-Gutierrez:2026ogo,Aguilar-Gutierrez:2025hty}. } 

The variation in the growth rate of Krylov complexity with $\gamma$, has a natural interpretation in terms of the quantum disk \cite{Lin:2022rbf,Berkooz:2022mfk,Berkooz:2025ydg,Almheiri:2024ayc} as we explain below.

\subsection{The quantum disk and radial dynamics in Krylov complexity}\label{ssec:quantum disk}
As pointed out in \cite{Berkooz:2025ydg}, the $n_F=4$ case has an equivalent physical description in terms of a particle moving on a $q$-deformed hyperbolic space, referred to as the ``quantum disk" \cite{Almheiri:2024ayc,vaksman2010quantumboundedsymmetricdomains} (see also \cite{Lin:2023trc,Blommaert:2023opb,Aguilar-Gutierrez:2025mxf}). Concretely, this space can be understood as a noncommutative deformation of the Poincaré disk, where the algebra of functions is replaced by operators obeying the $q$-oscillator algebra in \eqref{eq:q-oscillator}. The holomorphic and antiholomorphic coordinates\footnote{As usual, noncommutativity is related to discreteness: the coordinates do not commute, and admit discrete eigenvalues as operators.} on the quantum disk are identified with the ladder operators as:
\begin{flalign}
    z = \sqrt{1-q^2}\,a^\dagger, \quad z^* = \sqrt{1-q^2}\,a
\end{flalign}
The radial operator that measures the distance of the particle from the center $z=z^*=0$ is given by:
\begin{flalign}
    \hat{r}^2 = z z^* = (1-q^2)a^\dagger a
\end{flalign}
The isometry group is deformed from $SL(2,\mathbb{R})$ to the quantum group $\mathcal{U}_q(sl(2))$.\footnote{A brief review of quantum groups targeted to this system can be found in App.~A of \cite{Berkooz:2025ydg}.} The particle effectively transforms as a spinor under this symmetry, and its dynamics are governed by the transfer matrix \eqref{eq:H nF4}. As the algebra of noncommutative coordinates is naturally represented as $q$-harmonic oscillator, the Hilbert space of the particle is naturally represented as the Hilbert space of the $q$-harmonic oscillator:
\begin{flalign}
    \mathcal{H}_{QD} = \text{Span}\{ \ket{n},n\in\mathbb{Z}_{\geq 0}\}
\end{flalign}
By definition \eqref{eq:spread Complexity}, the Krylov complexity is therefore: 
\begin{flalign}
    \mathcal{C}_S(t) = \bra{\psi(t)}\hat{n}\ket{\psi(t)} = \langle \hat{n} \rangle_t~,
\end{flalign}
where $t$ is the time associated with the transfer matrix evolution. For the case of $n_F=4$, this encodes the physical propagation time on the quantum disk, since the transfer matrix corresponds to the Hamiltonian of the particle on this space \cite{Berkooz:2025ydg}. To relate the Krylov complexity to a physical quantity in this system, note that the eigenvalues of the radial operator are:
\begin{flalign}
    \hat{r}^2\ket{n} = (1-q^2)a^\dagger a \ket{n} = (1-q^{2\hat{n}})\ket{n} 
\end{flalign}
Consider now the triple-scaling limit $\lambda\to 0$, with $\lambda n=\text{fixed}$. By expanding $q=e^{-\lambda}$, we find:
\begin{flalign}
    \hat{r}^2 = (1-q^{2\hat{n}}) \to 2\lambda \hat{n} \quad\implies\quad \langle \hat{r}^2\rangle_t \to 2\lambda \mathcal{C}_S(t)
\end{flalign}
Therefore, at intermediate times (when $n\lesssim \lambda^{-1}$), the Krylov complexity describes the radial displacement of the particle from the center of the quantum disk. Furthermore, the growth of Krylov complexity describes the radial spreading of the particle wavefunction in this space. 

For the quantum disk, the parameters $t_{1,2}$ are:
\begin{equation}
    t_{1,2} = q e^{\pm i \gamma}. 
\end{equation} 
Here $\gamma$ encodes the spin of the particle (or equivalently, the magnetic field). Since the Krylov complexity growth is sensitive to $\gamma$, we see that the rate of spreading is influenced by the spin of the particle. For the case of `imaginary' spin, this question has been studied on the classical hyperbolic disk by \cite{Kitaev:2018wpr}. We expect to recover their results in the triple-scaling limit, which requires further treatment since their formulation is given as a statement on an $SL(2,R)/U(1)$ coset, in contrast to ours.

\subsection{\texorpdfstring{The $n_F=6$}{} case: continuous dual q-Hahn polynomials}\label{ssec:Krylov nF6}
Consider the recurrence relation encoded in the $n_F=6$ case \eqref{eq:H SYK8} with $t_4=0$ in its orthogonal form:
\begin{equation}\begin{aligned}\label{eq:H nF6}
    \hH_{n_F=6}\ket{n}=&\sqrt{[n+1]_{{q^2}}}\sqrt{\prod_{1\leq i<j\leq 3}(1-t_it_jq^{2n})} \ket{{n+1}}\\&
    + \sqrt{[n]_{{q^2}}}\sqrt{\prod_{1\leq i<j\leq 3}(1-t_it_jq^{2n-2})} \ket{{n-1}}\\&
    + \bigg(\frac{t_1+t_2+t_3}{\sqrt{1-{q^2}}}q^{2n}+\frac{t_1t_2t_3}{\sqrt{1-{q^2}}}(q^{2n-2}-q^{4n}-q^{4n-2})\bigg) \ket{n}~,
\end{aligned}\end{equation}
which, including as initial condition that $\bra{0}\ket{\theta}_{n_F=6}=1$, is solved by the wavefunctions
\begin{flalign}\begin{aligned}
    &\bra{n}\ket{\theta}_{n_F=6}=\prod_{1\leq i<j\leq 3} (t_it_j;q^2)_n^{-\frac{1}{2}{}}\frac{S_n(\cos\theta;t_1,t_2,t_3|q^2)}{\sqrt{ (q^2;q^2)_n}},\label{eqn:qHahn}
\end{aligned}\end{flalign}
where we introduced the continuous dual q-Hahn polynomials
\begin{flalign}
    S_n(\cos\theta;t_1,t_2,t_3|q^2) = 
    \frac{(t_1t_2,t_1t_3;q^2)_n}{t_1^n} \,_3\phi_2\bigg({\genfrac{}{}{0pt}{}{q^{-2n},t_1e^{\pm i\theta}}{t_1t_2,t_1t_3};q^2,q^2}\bigg)~.
\end{flalign}
We identify the Lanczos coefficients of the rescaled Hamiltonian in the continuum limit as:
\begin{subequations}
\begin{align}
     b(\ell)&=\sqrt{1-\rme^{-\ell}}\sqrt{\prod_{1\leq i<j\leq 3}(1-t_it_j\rme^{-\ell})} ~,\\ a(\ell)&=\qty(t_1+t_2+t_3)\rme^{-\ell}+\qty(t_1t_2t_3)(\rme^{-\ell}-2\rme^{-2\ell})~.
\end{align}
\end{subequations}
The initial condition \eqref{eq:initial condt} becomes:
\begin{equation}
\begin{aligned}
    \rme^{-\ell_0}=&\frac{2\sin^2\theta}{1+\prod_{i\neq j}t_it_j-(t_1 t_2
   t_3+\sum_{i=1}^3t_i)\cos \theta+\sqrt{\prod_{i=1}^3\left(t_i^2-2
   t_i \cos\theta+1\right)}}~.
\end{aligned}
\end{equation}
We then find Krylov spread complexity from \eqref{eq:spread Complexity}: 
\begin{equation}\label{eq:CS nF6}
    \lambda\mathcal{C}_{\rm S}(t)=\log\frac{1+\prod_{i\neq j}t_i t_j-\left(t_2 t_3 t_1+\sum_{i=1}^3t_i\right) \cos \theta +\sqrt{\prod_{i=1}^3\left(1+t_i^2-2 t_i \cos \theta
   \right)} \cosh (2 t \sin \theta)}{2\sin^2\theta}~,
\end{equation}
which we illustrate in Fig.~\ref{fig:nF}. The above expression is particularly useful to match the parameters in the DSSYK model to those in the bulk, as we describe below.

\subsection{Bulk interpretation of Krylov complexity}\label{ssec:sine dilaton gravity int}
In this subsection, we investigate the bulk interpretation of Krylov complexity in the semiclassical limit at finite temperature for the q-Askey deformed Hamiltonians.

Consider JT gravity with a timelike ETW brane with Neumann boundary conditions \cite{Gao:2021uro}:
{\small
\begin{equation}\label{eq:total JT action ETW}
\begin{aligned}
    &I_{\rm total}=I_{\rm JT}+\frac{1}{\kappa^2}\int _{\mathcal{S}}(\Phi_BK- m)~,\\[3pt]
    &I_{\rm JT}=-\frac{\Phi_0}{16\pi G_N}\qty(\int _{\mathcal{M}}\sqrt{g}\mathcal{R}+2\int _{\partial\mathcal{M}}\sqrt{h}K)-\frac{1}{16\pi G_N}\qty(\int _{\mathcal{M}}\sqrt{g}\Phi(\mathcal{R}+2)+2\int _{\partial\mathcal{M}}\sqrt{h}\Phi_B(K-1))~,
\end{aligned}
\end{equation}
}
where $\mathcal{M}$ is the spacetime manifold,  $\Phi_0$ is a constant, $\Phi_B=\eval{\Phi}_{\partial\mathcal{M}}$, $\mathcal{S}$ the worldvolume of the ETW brane, and $\nu$ the ETW brane mass/tension. The on-shell solutions describe an AdS$_2$ black hole with an ETW brane,
\begin{equation}\label{eq:metric AdS}
\begin{aligned}
      \rmd s^2&\equiv g_{\mu\nu}\rmd x^\mu\rmd x^\nu=\frac{-\rmd T^2+\rmd\sigma^2}{\sin^2\sigma}=-(\Phi^2-\Phi^2_h)\rmd t^2+\frac{\rmd\Phi^2}{\Phi^2-\Phi^2_h}~,\\
       \Phi&=\Phi_h\frac{\cos T}{\sin\sigma}~,
\end{aligned}
\end{equation}
where the ETW brane trajectory is given by 
\begin{equation}\label{eq:jafferis}
    \cos\sigma_{\rm brane}=\frac{m}{\sqrt{r_h^2+m^2}}\cos(T_{\rm brane})~,
\end{equation}
where we used the ETW brane tension parameter in \eqref{eq:total JT action ETW}.

The wormhole length between the asymptotic boundary $\partial\mathcal{M}$ located at $\Phi=\Phi_B\gg1$ and the ETW brane is \cite{Gao:2021uro,Blommaert:2025avl},\footnote{{One can ask if JT gravity with a particle insertion in the $\beta_L\rightarrow\infty$ limit described by \eqref{fig:PETS} reproduces the expected length with the ETW brane in \eqref{eq:length ETW wormhole}. The relevant length of compare with our calculations is one connecting the $R$ asymptotic AdS boundary to the matter insertion. This one-sided length can be calculated with the standard formula for lengths in AdS space
\begin{equation}
\cosh(d_{1,2})=\frac{\cos (T_1 -T_2) - \sin (\sigma_1)\sin (\sigma_2)}{\cos \sigma_1 \cos \sigma_2}~,
\end{equation}
where $ d_{1,2}$ is the geodesic distance between two points with coordinates $(\sigma_{1/2},~T_{1/2})$ in global AdS coordinates \eqref{eq:metric AdS}. While the two-sided length is determined by SL(2,$\mathbf{R}$) isometries of AdS; the one-sided length depends on the coordinates of the particle’s worldline. Given that when $\beta_L\to\infty$ there is no $L$ horizon; the worldline of an ETW brane is a compatible trajectory, and it agrees with our boundary result.}}
\begin{equation}\label{eq:length ETW wormhole}
L_{\rm AdS}\equiv\int _{\mathcal{S}}^{\partial\mathcal{M}}\rmd\xi\sqrt{g_{\mu\nu}\dv{x^\mu}{\xi}\dv{x^\nu}{\xi}}=\log\frac{m+\sqrt{m^2+\Phi_h^2}\cosh(2\Phi_h t)}{m+\sqrt{m^2+\Phi_h^2}}+L_0~,
\end{equation}
where $L_0$ is the scheme-dependent initial length of the wormhole (since its renormalised length to the one-sided asymptotic boundary), which we illustrate in Fig.~\ref{fig:CS_nF2}.                                                                                                                                

The wormhole length thus matches Krylov state complexity for the HH state in the $n_F=2$, $4$ and $6$ q-Askey deformations of the DSSYK model in \eqref{eq:Krylov HH state nF}. In particular, we identify a holographic dictionary relating the parameters of the bulk and boundary theories
\begin{subequations}
    \begin{align}
         m&=1+\prod_{i\neq j}t_i t_j-\left(t_2 t_3 t_1+\sum_{i=1}^3t_i\right) \cos \theta~,\quad \Phi_h=\sin\theta~,\\
   L_0&=\log\frac{1+\prod_{i\neq j}t_i t_j-\left(t_2 t_3 t_1+\sum_{i=1}^3t_i\right) \cos \theta+\sqrt{\prod_{i=1}^3(1+t_i^2-2 t_i \cos \theta)}}{2\sin^2\theta}~,
    \end{align}
\end{subequations}
where we used the explicit parameters $a_{1/2}$ for the $n_F=6$ in \eqref{eq:CS nF6}.

The three different parameters $t_{i=1,2,3}$ hint that there could be three ETW branes in the bulk, as argued by \cite{Berkooz:2025ydg} by expressing the different parameters in the partition function of the $n_F=6$ theory. This aligns with sine dilaton gravity in the $n_F=4$ case containing two ETW brane tension parameters (see Sec.~\ref{ssec:sine dilaton gravity int}). We expect that the additional deformation parameter in the chord Hamiltonian \eqref{eq:H nF6} corresponds to an additional bulk parameter in the sine dilaton gravity action displayed in the next subsection.

In particular, in the $n_F=4$ case, implementing the canonical quantization of sine dilaton gravity in the semiopen channel as in \cite{Blommaert:2025avl}, relates the wormhole length seen as an operator acting on the physical bulk Hilbert space $L_{\rm AdS}\rightarrow \hat{L}_{\rm AdS}$ with the chord number in the transfer matrix of the Al-Salam Chihara wavefunctions \eqref{eq:H nF4} as \cite{Aguilar-Gutierrez:2025hty}
\begin{eqnarray}\label{eq:ismophsm}
    \hat{L}_{\rm AdS}~~\text{isomorphic~to}~~\lambda\hat{n}~.
\end{eqnarray}
where the isomorphism refers to an equality between matrix elements of the operators in \eqref{eq:ismophsm} in the corresponding boundary/physical bulk Hilbert space. This means in particular that the expectation value in the HH state,
\begin{eqnarray}
    \mathcal{C}_{\rm S}=\bra{\psi_{\rm HH}}\hat{L}_{\rm AdS}\ket{\psi_{\rm HH}}~,
\end{eqnarray}
where $\ket{\psi_{\rm HH}}$ is the bulk HH state in sine dilaton gravity with ETW branes in \cite{Blommaert:2025avl}, while $\mathcal{C}_{\rm S}$ is the spread complexity of the Hartle-Hawking state. Thus, our findings show there is an exact equivalence, beyond the $\lambda\rightarrow0$ limit, between the wormhole length in sine dilaton gravity with an ETW brane in the semiopen channel of \cite{Blommaert:2025avl} and Krylov complexity in a specific state.

On the other hand, as we will see in the reminder of this work, the $n_F=8$ and, in particular, the adjoint deformation does not reproduce \eqref{eq:length ETW wormhole} at generic boundary times when the temperature is finite, while in the very low temperature limit, the ETW brane is a good description for the $n_F=8$ case, as shown in Sec.~\ref{ssec:semiclassical limit}. The very low temperature limit is also connected to late time limits, where a single ETW brane is still a good approximation (see Sec.~\ref{ssec:Krylov nF8} and \ref{ssec:Krylov nF Adj}). This means that in the two-sided description of the SYK states considered in this work (Fig.~\ref{fig:PETS}), we take the exact zero temperature limit for one of the sides, while for the other side, we look at first order corrections at very low temperatures, which capture the Schwarzian dynamics in the bulk.

\subsection{On the discrete spectrum and its bulk interpretation}\label{ssec:discrete spectrum}
In this subsection, we examine the discrete energy spectrum of the different $n_F\geq2$ theories described in Sec.~\ref{ssec:def qAskey def}, and we particularly discuss their bulk interpretation based on the previous results in this section.

\paragraph{Discrete energy spectrum}
First, we stress the discreteness in the spectrum of the deformed chord Hamiltonian for $\abs{t_i}\geq 1$ is a mathematical property of the Askey-Wilson polynomials, once the conditions \eqref{eq:dissrete cond 1} and \eqref{eq:dissrete cond 2} are satisfied. {Microscopically, this might indicate that the density of states of the deformed SYK system without ensemble averaging undergoes resonances once one of the deformation parameters is large enough. We leave a detailed investigation of this phenomenon for future studies. We will focus on the ensemble-averaged description below.}

From the completeness relation that includes the continuous and discrete spectra in \eqref{eq:recurrence discrete} and the explicit form of the parametrization $x_l$ \eqref{eq:xl} in the Askey-Wilson polynomials, the eigenstates $\ket{\theta_l}$ and the discrete energy spectrum $E(\theta_l)$ of the Hamiltonian \eqref{eq:recurrence states q Askey} take the form
\begin{subequations}\label{eq:energies discrete}
\begin{align}
\hH_{n_F}\ket{\theta_l}&=E(\theta_l)\ket{\theta_l}~,\\
    E(\theta_l)&=\frac{2}{\sqrt{1-q^2}}\cosh\theta_l~,\quad\theta_l\equiv\log\alpha-2\lambda l~,\label{eq:discrete spectrum}
\end{align}
\end{subequations}
where, as mentioned below \eqref{eq:dissrete cond 2}, $\alpha$ is any $t_i\in\mathbb{R}$ obeying $\abs{t_i}\geq1$ with \eqref{eq:dissrete cond 1}, and $l$ is  \eqref{eq:dissrete cond 2}. Importantly, there always exist bound states when $\abs{\alpha}>1$, $\lambda\geq 0$, and the energy spectrum is bounded from both below and above as seen from \eqref{eq:discrete spectrum} given the range of $l$ in \eqref{eq:ceiling} , which gives the bounds 
    {\begin{align}\label{eq:bounds discrete levels}
    \frac{\sqrt{1-q^2}}{2}E(\theta_l)\in\begin{cases}
        (1,~\frac{1}{2}\qty(\alpha+\alpha^{-1})]~,&\alpha>1~,\\
        [\frac{1}{2}(\alpha+\alpha^{-1}),~-1)~,&\alpha<-1~.
    \end{cases}
    \end{align}}
{Note that there is no overlap between the ranges of the discrete and continuous energy states; they complement each other, as we illustrate in Fig.~\ref{fig:spectrum_alpha}.}
\begin{figure}
    \centering
    \subfloat[]{\includegraphics[width=0.48\linewidth]{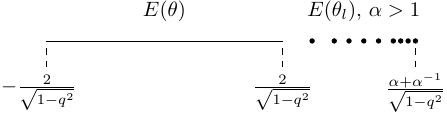}}\hfill\subfloat[]{\includegraphics[width=0.48\linewidth]{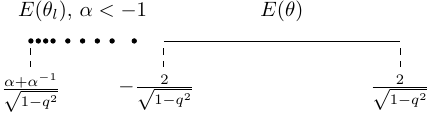}}
    \caption{Discrete ($E(\theta_l)$, \eqref{eq:discrete spectrum}) and continuous ($E(\theta)$, \eqref{eq:cont spectrum}) energy spectrum contributions for (a) $\alpha>1$ and (b) $\alpha<-1$.}
    \label{fig:spectrum_alpha}
\end{figure}
Furthermore, one sees from the above expressions that the $l_{\rm max}$ bound \eqref{eq:ceiling} determining the lower or upper bounds in the spectrum $E(\theta_l)$ \eqref{eq:discrete spectrum} for $\alpha>1$ or $\alpha<-1$ respectively is a non-perturbative parameter in $\lambda$. Indeed, one cannot reproduce the bounds in the spectrum \eqref{eq:bounds discrete levels} at any finite order in perturbation theory, as seen explicitly from \eqref{eq:discrete spectrum}. For instance, at first order quantum correction (i.e.~$\mathcal{O}(\lambda)$) to the energy spectrum is  
\begin{equation}
\sqrt{1-q^2}E(\theta_l)=\frac{1}{2}\qty(\alpha^{-1}+\alpha)-(\alpha^2-1)l\lambda/\alpha+\mathcal{O}(\lambda^2)~,
\end{equation}
in contrast to \eqref{eq:bounds discrete levels}. This connects with the observation in Sec.~\ref{eq:partitil func} that the {contribution of the discrete energy states to the classical partition function of the deformed theories vanishes, reflecting the quantum nature of the discrete energy states.} Holographically, this indicates the discrete spectrum is associated to a corresponding bulk process that is non-perturbative in terms of $G_N$. 

To recover a more concrete bulk interpretation, let us study the Krylov complexity in the $n_F=6$ case \eqref{eq:CS nF6} for the discrete spectrum states, which can be described by,\footnote{The periodic behavior of the chord number hints towards a possible connection to quantum scar states \cite{Milekhin:2023was,Liska:2022vrd,Caputa:2022zsr,turner2018quantum,bernien2017probing,Chakraborty:2025bhk}, which is work in progress.}
\begin{align}\label{eq:chord number ETW brane discr}
    &\lambda\mathcal{C}_{\rm S}(\tau)= \nonumber\\
    &\log\frac{1+\prod_{i\neq j}t_i t_j-\left(t_2 t_3 t_1+\sum_{i=1}^3t_i\right) \cosh \theta_l +\sqrt{\prod_{i=1}^3\left(1+t_i^2-2 t_i \cosh \theta_l
   \right)} \cosh (2 (\rmi\tau) \sinh \theta_l)}{2\sinh^2\theta_l}~,
\end{align}
which we illustrate in Fig.~\ref{fig:disc contribution},
\begin{figure}
\centering
\includegraphics[width=0.48\linewidth]{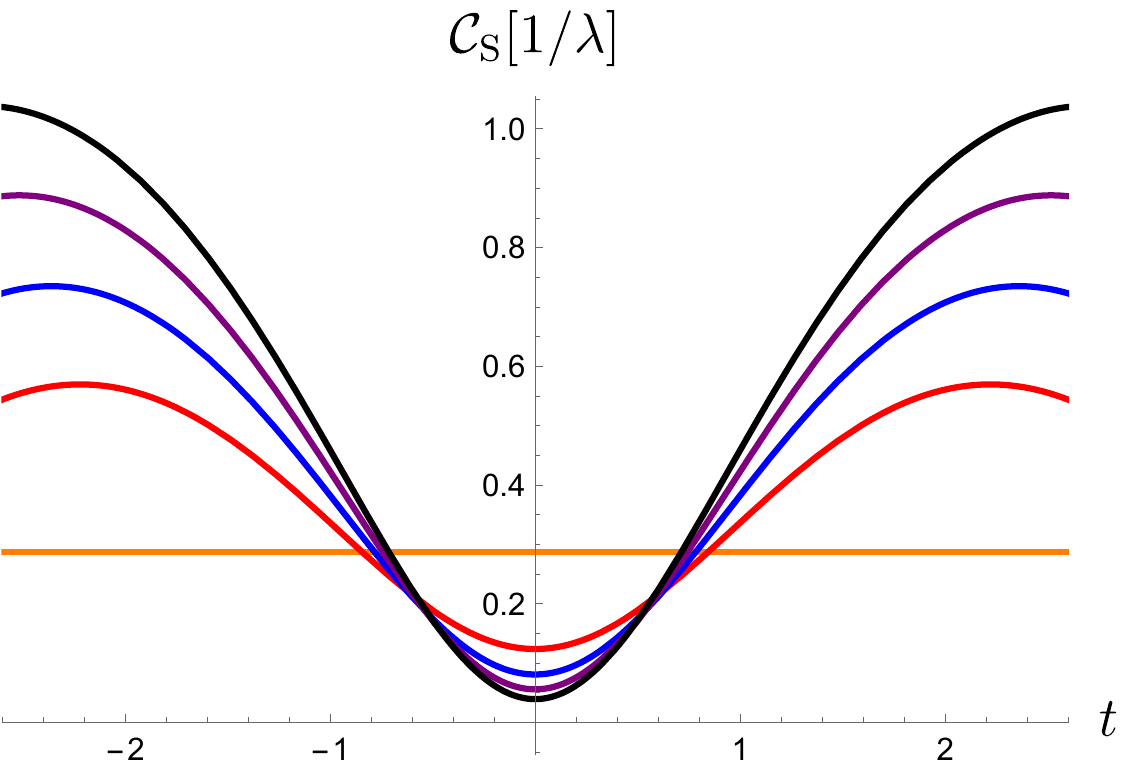}
\caption{{Krylov complexity in the semiclassical limit for the discrete spectrum states \eqref{eq:discrete spectrum}. The complexity is now periodic in time, and it is bounded. This agrees with the Morse potential in JT gravity, where the energy of a classical particle is converted between potential and kinetic terms.}}
\label{fig:disc contribution}
\end{figure}
We now seek to interpret \eqref{eq:chord number ETW brane discr} based on JT gravity and sine dilaton gravity.

\paragraph{Schwarzian interpretation and JT gravity}
To gain more intuition about the discrete nature of the spectrum, it is useful to study the Schwarzian limit of the q-Askey deformations following Sec.~\ref{ssec:triple scaling limit}. We seek to clarify the low energy description of the periodicity of Krylov complexity in the HH state \eqref{eq:chord number ETW brane discr} for the discrete spectrum states \eqref{eq:discrete spectrum}. {One can note that the Krylov complexity in \eqref{eq:chord number ETW brane discr} matches the geodesic length between an ETW brane with tension $m$ and an asymptotic boundary of global AdS$_2$ \cite{Gao:2021uro} (2.27):
\begin{eqnarray}
    L_{\rm AdS}\equiv\int _{\mathcal{S}}^{\partial\mathcal{M}}\rmd\xi\sqrt{g_{\mu\nu}\dv{x^\mu}{\xi}\dv{x^\nu}{\xi}}=\log\frac{m+\sqrt{m^2+\Phi_h^2}\cos(2\Phi_h t_g)}{m+\sqrt{m^2+\Phi_h^2}}+L_0~,
\end{eqnarray}
where we introduced global AdS$_2$ coordinates
\begin{eqnarray}\label{eq:global AdS}
    \rmd s^2=-(4\Phi_h^2+\rho^2)\rmd t_g^2+\frac{\rmd\rho^2}{\rho^2+4\Phi_h^2}~,
\end{eqnarray}
where $\tan T=\cot t_g$ and $\rho=-2\Phi_h\cot\sigma$ in terms of the compact global coordinates in \eqref{eq:metric AdS}, where $\Phi_h$ in \eqref{eq:global AdS} is a constant. Then, this means that the ETW brane removes the spacetime region associated with the AdS$_2$ black hole horizon, as illustrated in Fig.~\ref{fig:ETW AdS}, which occurs when the ETW brane tension is sufficiently negative, or equivalently, there is global negative energy matter \cite{AALO}.} In particular, for the $n_F=2$ case, it was previously realised by \cite{Rajgadia:2026ask} that the condition $\abs{t_1}>1$ (and other $t_i=0$ in \eqref{eq:dissrete cond 1}) corresponds $\nu<-1/2$ and that $l_{\rm max}$ in \eqref{eq:wl expl} when $n_F=2$ corresponds also corresponds to an upper bound associated to wavefunctions in the Morse potential. Then, \eqref{eq:chord number ETW brane discr} can be interpreted in the bulk as the geodesic length of an eternal traversable wormhole \cite{Maldacena:2018lmt}, as there is no black hole horizon when the brane tension is sufficiently negative (corresponding to $\nu<-1/2$ in the Morse potential). Similar to our findings for complexity, \cite{Maldacena:2018lmt} argued that the Lorentzian boundary-to-boundary two-point correlation functions in the eternal traversable wormhole are described in global AdS$_2$ coordinates.

{Next, t}he Schwarzian regime needs special treatment with respect to \eqref{eq:energies discrete} since we promote discrete variables to continuous ones. {This regime is conveniently described in the boundary particle formalism (reviewed in \cite{Mertens:2022irh}).
Considering the discrete energy states, the equation of motion of the boundary particle \eqref{eq:discrete states schrodinger} can be expressed as
\begin{eqnarray}
   - k^2_l=P^2+U(L)~,
\end{eqnarray}
where $U(L)$ is the Morse potential \eqref{eq:morse potential}; $k_l$ the zero-point subtracted energy which was defined in \eqref{eq:discrete zero point subtracted}; and $P$ represents the conjugate momentum to $L$, as in \eqref{eq:dS length}.
Let us first identify a local minimum of the Morse potential \eqref{eq:morse potential}, which appears when
\begin{equation}
L_{\rm min}=-\log(-2\nu)~,\quad U(L_{\rm min})=-\nu^2~.
\end{equation}
It follows that there exist negative energy states when $\nu<0 $ and negative length states only when $\nu\leq- 1/2$.\footnote{In higher-dimensional settings, negative tension branes lead to instabilities \cite{Marolf:2001ne}.} Thus, the discreteness of the spectrum persists in the low-energy limit. Next, we identify the points where the boundary particle only posses potential energy. This determines the maximum extend of a particle in the Morse potential $L_{\rm max}$ (where $U(L_{\rm max})=-k_l^2$) corresponding to
\begin{eqnarray}\label{eq:max energy}
    L_{\rm max}=
       \log{\frac{-\nu+\sqrt{\nu^2-k_l^2}}{2k_l^2}}~.
\end{eqnarray}
where one can note that $\lim_{k_l\rightarrow0}L_{\rm max}\rightarrow\infty$. There exist a second solution where the boundary particle loses all its kinetic energy, 
\begin{eqnarray}\label{eq:ETW brane distance}
    L_{\rm brane}=\log\frac{-\nu-\sqrt{\nu^2-k_l^2}}{2k_l^2}
\end{eqnarray}
which remains finite even when $k_l\rightarrow0$. \eqref{eq:ETW brane distance} corresponds to the location where the particle reaches the worldline particle.}

We display the Morse potential for scattering and bound states in Fig.~\ref{fig:Morse_potential}.
\begin{figure}
    \centering
    \subfloat[]{\includegraphics[width=0.485\linewidth]{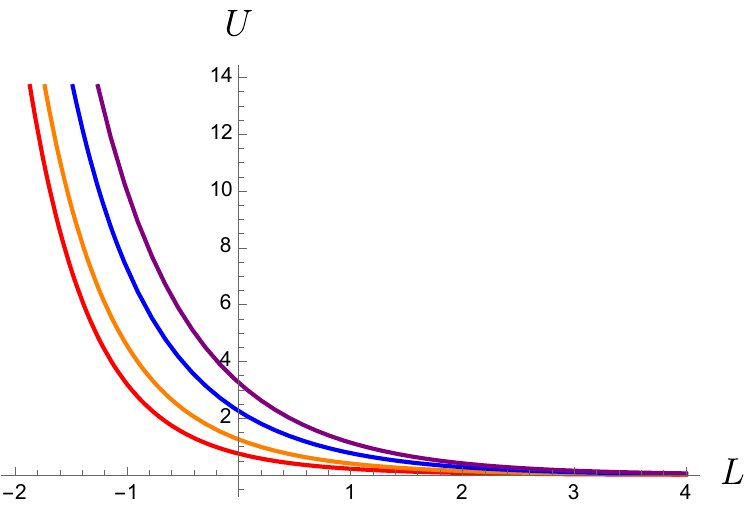}}\hfill\subfloat[]{\includegraphics[width=0.485\linewidth]{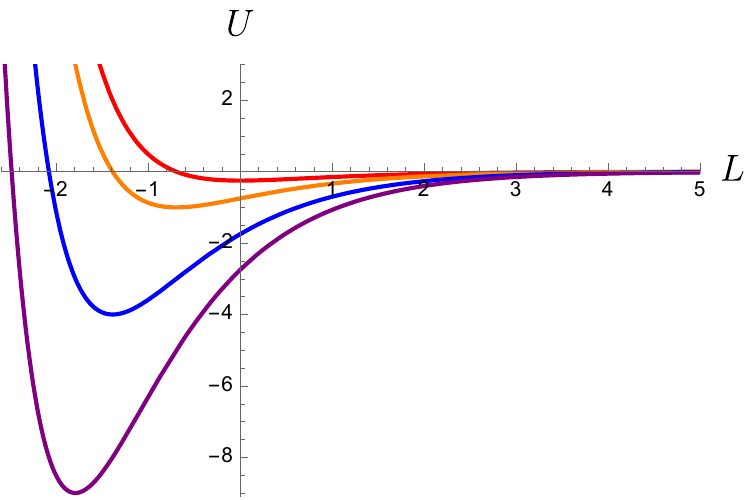}}
    \caption{Morse potential $U(L)=\nu\rme^{-L}+\rme^{-2L}/4$ \eqref{eq:morse potential} for (a) scattering states (with $\nu>0$) and (b) bound states (with $\nu<0$). The chosen parameters in the plots are $\abs{\nu}=1/2$ (red), $1$ (orange), $2$ (blue) and $4$ (purple). $L_{\rm AdS}$ is a periodic geodesic length connecting the asymptotic boundary to the ETW brane.}
    \label{fig:Morse_potential}
\end{figure}
{It can be seen from Fig.~\ref{fig:Morse_potential} that the geodesic length of the boundary particle can be negative when the brane tension is negative. This mean that the particle, starting at $L=L_{\rm max}$ moves \eqref{eq:max energy} in a timelike trajectory before hitting the ETW brane \eqref{eq:ETW brane distance}. Note that the expectation value of the chord number itself is positive definite when $\abs{\alpha}>1$, as we found in \eqref{eq:chord number ETW brane discr} (displayed in Fig.~\ref{fig:disc contribution}). In contrast, the geodesic length that becomes negative in \eqref{eq:morse potential} corresponds to a regularization of the chord number, given by $\tilde{L}$ in \eqref{eq:constants} which involves a subtraction by a $\log(2\lambda)$ term.}

{More generally, the same Morse potential with bound states can be generated by adding negative global energy matter (a SL$(2,\mathbb{R})$ charge \cite{Lin:2019qwu}) in JT gravity, resulting in eternal traversable wormholes \cite{Maldacena:2018lmt}. The negative global energy matter can be obtained in different ways, such as from the Casimir energy of AdS space with a bulk matter field and periodic boundary conditions \cite{Maldacena:2018lmt}, among others \cite{AALO,Garcia-Garcia:2020ttf,Gao:2016bin}. The geometric representations of this configuration in Euclidean and Lorentzian signatures are represented in Fig.~\ref{fig:ETW AdS}.
\begin{figure}
    \centering
    \subfloat[]{\includegraphics[width=0.35\textwidth]{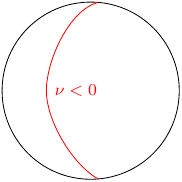}}\hspace{1.5cm}\subfloat[]{\includegraphics[width=0.25\textwidth]{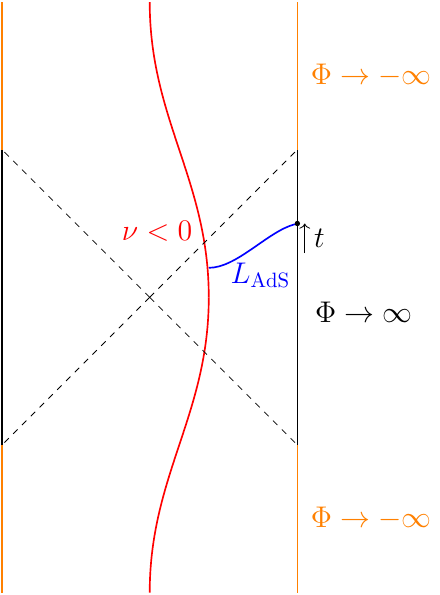}}
    \caption{{(a) Euclidean and (b) Lorentzian signature geometries representing an eternal traversable wormhole in AdS$_2$ space \cite{Maldacena:2018lmt} (black circle in (a), black rectangle in (b)) with negative global energy matter (red solid curve) corresponding to the negative tension ETW brane (Fig.~\ref{fig:Morse_potential}). {In the latter case, the worldline of the ETW brane moving along the trajectory \eqref{eq:jafferis} has a periodic profile in the AdS$_2$ eternal traversable wormhole, where the periodicity of the trajectory is encoded in the total chord number \eqref{eq:chord number ETW brane discr} corresponding to the geodesic length illustrated in the figure.}}}
    \label{fig:ETW AdS}
\end{figure}
In particular, the profile of the matter worldline is periodic when its global energy SL$(2,\mathbb{R})$ charge is negative, where the time period measured by the geodesic length connecting the asymptotic boundary to the ETW brane is $\pi/\sinh\theta$, which follows from the Krylov complexity \eqref{eq:chord number ETW brane discr}.}

In addition, negative and positive energy matter contributions in the Morse potential can be used to emit baby universes in JT gravity,\footnote{Recent interest in formulating closed and baby universes in holographic settings has been catalysed by \cite{Abdalla:2025gzn,Harlow:2025pvj} and different puzzles regarding Euclidean wormhole saddle points \cite{Antonini:2023hdh,Antonini:2024mci,Antonini:2025ioh,Engelhardt:2025vsp,Higginbotham:2025clp,Higginbotham:2025dvf,Liu:2025cml,Gesteau:2025obm,Kudler-Flam:2025cki,Mori:2025jej,Belin:2025ako}.} as we discuss in Sec.~\ref{ssec:open problems}.

\paragraph{Sine dilaton gravity interpretation}
We now analyse \eqref{eq:chord number ETW brane discr} in terms of sine dilaton gravity with ETW branes with Neumann boundary conditions, which was first investigated by \cite{Blommaert:2025avl}. The proposal for the bulk theory is
\begin{equation}\label{eq:SDG with ETW brane}
\begin{aligned}
    I_{\rm SDG}=-\frac{1}{2\kappa^2}\bigg(&\int _{\mathcal{M}}\rmd^2x\sqrt{g}\qty(\Phi\mathcal{R}+2\sin\Phi)+2\int _{\partial\mathcal{M}}\rmd x\sqrt{h}\qty(\Phi_{b} K)\\
    &-\zeta\int _{\rm brane}\sqrt{h}~\rme^{-\rmi\Phi_B}-\overline{\zeta}\int _{\rm brane}\sqrt{h}~\rme^{\rmi\Phi_B}\bigg)~,
\end{aligned}
\end{equation}
where $\zeta$, $\overline{\zeta}$ are constant parameters which play the role of the ETW brane tensions in JT gravity. It was shown in \cite{Blommaert:2025avl} that the canonical quantization of \eqref{eq:SDG with ETW brane} leads to an ADM Hamiltonian which is isomorphically dual to the $n_F=4$ chord Hamiltonian \eqref{eq:H nF4}, where the deformation parameters $t_1$ and $t_2$ are related to the brane tensions as
\begin{equation}
    t_1=\rme^{{\rm arcsinh}\bar\zeta+{\rm arcsinh}\zeta}~,\quad t_2=-\rme^{{\rm arcsinh}\bar\zeta-{\rm arcsinh}\zeta}~.
\end{equation}

For instance, the different special cases in Sec.~\ref{ssec:Krylov nF4}, namely  \eqref{eq:nF_specia_case} and \eqref{eqn:Cs-growth-gamma}, correspond to, \footnote{In the latter case, the maximum rate of complexity growth $\gamma=0,~\pi$ is achieved when $\abs{\zeta}=1$ \eqref{eq:critical val nF4} for the complexity to grow at its slowest rate, the corresponding brane tension is
\begin{equation}
    \zeta_{\rm crit}=\rmi\frac{2q^\Delta\cos\theta}{1+q^\Delta}~.
\end{equation}}

\begin{subequations}
    \begin{align}
       t_1=t_2~:&~~\zeta=\rmi~,~~\bar\zeta=-\frac{\rmi}{2t_1}(t_1^2+1)~, \\
       t_1=t_2^*=\abs{t_1}\rme^{\rmi \gamma}~:&~~\bar\zeta=\frac{\abs{t_1}+\abs{t_1}^{-1}}{2\rmi}~,\quad \zeta=\rmi\cos\gamma
    \end{align}
\end{subequations}
The background geometry is
\begin{equation}\label{eq:metric SD}
    \rmd s^2=F(\Phi)\rmd \tau^2+\frac{\rmd \Phi^2}{F(\Phi)}~,\quad F(\Phi)=\cosh \theta_l-\cos\Phi~,
\end{equation}
where $\Phi$ is the dilaton, so that the metric is parameterised in a radial gauge. 

According to \cite{Blommaert:2025avl}, the bulk interpretation of the chord number corresponds to the geodesic length connecting the asymptotic boundaries of an AdS$_2$ black hole, as described by \eqref{eq:length ETW wormhole}. This geometry is obtained by Weyl-rescaling the metric and choosing the contour in the complex plane for $\Phi\in \mathbb{C}$ appropriately. More specifically, the solutions where the black hole radius in sine dilaton gravity \cite{Blommaert:2024ymv,Blommaert:2024whf,Blommaert:2025avl} is a continuous parameter correspond to those where
\begin{subequations}\label{eq:dilaton solution}
\begin{align}
\Phi_{\rm cont}(\rho)&=\frac{\pi}{2}+\rmi\log(\rho+\rmi\cos \theta)~,\quad \theta\in[0,~\pi]~\label{eq:disc Phi}.\\
    \rme^{-\rmi \Phi_{\rm cont}}\rmd s^2&=g(\rho)\rmd \tau^2+\frac{\rmd \rho^2}{g(\rho)}~,\quad g(\rho)=\rho^2-\sin^2 \theta~.\label{eq:eff geo}
\end{align}
\end{subequations}
where the ETW brane only backreacts along the worldline \eqref{eq:jafferis}.

In the case at hand, \eqref{eq:chord number ETW brane discr} is manifestly of the same form as the continuous energy spectrum case \eqref{eq:length ETW wormhole}; however, there is an additional factor $\rmi$ in the Lorentzian time. We can understand its origin in the bulk as describing a different contour in the complex plane for $\Phi$ parametrised by $\rho$ as
\begin{equation}\label{eq:Phi disc}
    \Phi_{\rm disc}(\rho)=\frac{\pi}{2}+\rmi\log(\rmi\rho+\rmi\cosh \theta_l)~.
\end{equation}
In this way, the Weyl rescaling used in the metric of sine dilaton gravity \eqref{eq:eff geo} now results in
\begin{equation}\label{eq:disc Weyl metric}
    \rme^{-\rmi \Phi_{\rm disc}}\rmd s^2=-g(\rho)\rmd \tau^2+\frac{\rmd \rho^2}{g(\rho)}~,\quad g(\rho)=\rho^2-\sinh^2 \theta_l~,
\end{equation}
where we used $\Phi_h$ in \eqref{eq:Phi h discrete}. Note that the above metric still describes an AdS$_2$ black hole with an ETW brane in the semiclassical geometry at finite temperature. The only difference is that due to the Weyl rescaling \eqref{eq:disc Weyl metric}, there appears an additional factor $\rmi$ in the corresponding metric. From \eqref{eq:chord number ETW brane discr}, one identifies a holographic dictionary in the $n_F=4$ deformed DSSYK model \eqref{eq:H nF4} with sine dilaton gravity with ETW brane parameters \eqref{eq:length ETW wormhole} as,\footnote{Since the dilaton at the would-be horizon becomes discrete and purely imaginary $\Phi_h=\rmi x_l$, this is reminiscent of the complex Liovuille string where the dilaton can be purely imaginary \cite{Collier:2025pbm,Collier:2024kmo}.}
\begin{equation}\label{eq:Phi h discrete}
    m=1+t_1t_2-(t_1+t_2)\cosh \theta_l~,\quad \Phi_h =2\rmi \sinh \theta_l~.
\end{equation}
First note that the brane tension in the AdS$_2$ black hole can be complex valued in sine dilaton gravity (due to $t_i\in\mathbb{C}$), while the black hole radius (which determines its ADM energy) in the effective geometry is discrete. We display the different contours in the complex plane for $\Phi$ in Fig.~\ref{fig:contours}.
\begin{figure}
    \centering
    \includegraphics[width=0.4\linewidth]{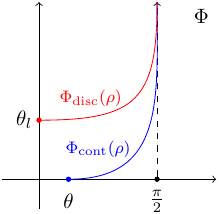}
    \caption{Complex plane of the dilaton $\Phi$ in sine dilaton gravity \eqref{eq:SDG with ETW brane}, where we display the contours for the Weyl rescaling to an AdS$_2$ black hole geometry, where $\Phi=\theta$ is the location of the black hole horizon in the continuous energy spectrum solution, with contour $\Phi_{\rm cont}$ \eqref{eq:disc Phi}, and $\theta_l$ for the discrete one, with $\Phi_{\rm disc}$ in \eqref{eq:Phi disc}.}
    \label{fig:contours}
\end{figure}

We emphasise there is an emergent Wick rotation due to the Weyl rescaling \eqref{eq:disc Weyl metric} with respect to the continuous energy spectrum case \eqref{eq:dilaton solution}, which is triggered by the deformation parameter $t_i$. This connects with Cauchy slice holography \cite{Araujo-Regado:2022gvw,Araujo-Regado:2022jpj,Soni:2024aop,Araujo-Regado:2025elv}, where, by increasing the T$\overline{\text{T}}$ deformation parameter (introduced in \cite{Smirnov:2016lqw,Cavaglia:2016oda}, and reviewed in \cite{Jiang:2019epa,He:2025ppz,Guica:2025jkq}), one can trigger a Euclidean to Lorentzian transition. Similarly, there is a Wick rotation in time when one of the q-Askey deformation parameters $t_i$ is greater than $1$ while obeying the conditions \eqref{eq:dissrete cond 1}, \eqref{eq:dissrete cond 2}. The counterpart of $t_i$ in Cauchy slice holography is the T$\overline{\text{T}}$ deformation parameter. However, in contrast to the T$\overline{\text{T}}$ case, the flow in the energy spectrum with respect to the deformation parameter $t_i$ can be discrete in our case. {In fact, the similarity between q-Askey and $\TT$ deformations is also present in the existence of an energy flow equation obtained by varying the energy spectrum of the deformed theory with respect to any of the deformation parameters $t_i$, seen from \eqref{eq:initial length} with a fixed initial length $\ell_0$
\begin{eqnarray}\label{eq: flow equation q Askey}
    \dv{E}{t_i}=\frac{1}{\sqrt{2\lambda}}\qty(2\dv{b(\ell_0)}{t_i}+\dv{a(\ell_0)}{t_i})~,
\end{eqnarray}
where $E$ is the energy spectrum of either the discrete or continuous energy states. As an example, $n_F=2$ we have that $\dv{b(\ell_0)}{t_i}=0$ and $\dv{a(\ell_0)}{t_i})=\rme^{-2\ell_0}$.
} We present a comparison between the geometric interpretation of q-Askey deformation in sine dilaton gravity and Cauchy slice holography in Fig.~\ref{fig:compare qAskey Cauchy slice}.
\begin{figure}
    \centering
    \subfloat[]{\includegraphics[height=0.4\linewidth]{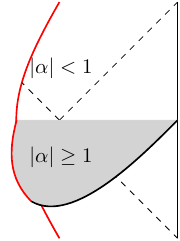}}\hspace{1cm}\subfloat[]{\includegraphics[height=0.455\linewidth]{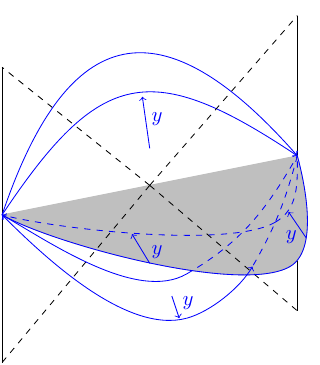}}
    \caption{Bulk interpretation of (a) the discrete spectrum in q-Askey deformations of the DSSYK model and (b) T$\overline{\text{T}}$ deformations in Cauchy slice holography. By increasing the magnitude of the q-Askey deformation parameter $\alpha$ beyond the critical value $\alpha=1$, the effective AdS$_2$ black hole geometry in sine dilaton gravity transitions between Lorentzian and Euclidean bulk geometries with an ETW brane. In Cauchy slice holography, by increasing the T$\overline{\text{T}}$ deformation parameter beyond a critical value, which corresponds to a bulk process where the finite cutoff Euclidean boundary (blue) is ``pushed'' radially inwards until it transitions into a Lorentzian Cauchy slice foliating the spacetime.}
    \label{fig:compare qAskey Cauchy slice}
\end{figure}

While the above description relates Krylov complexity \eqref{eq:chord number ETW brane discr} in the discrete spectrum solution \eqref{eq:discrete spectrum} to sine dilaton gravity, it does not explain the presence of the discreteness in the first place, which comes from analytic continuation of the orthogonality relation \eqref{eq:recurrence discrete} when the conditions in the deformation parameters \eqref{eq:dissrete cond 1} are satisfied. It would be very useful to work on the exact bulk mechanism that gives rise to the discrete spectrum. We discuss about this direction in Sec.~\ref{ssec:open problems}.

\subsection{The \texorpdfstring{$n_F=8$}{} case: Askey-Wilson polynomials}\label{ssec:Krylov nF8}
We again consider the deformed chord Hamiltonian \eqref{eq:H SYK8} in its symmetric form:
{\small
\begin{align}\label{eq:H nF8}
    \hH_{n_F=8}\ket{n}& 
    = \bigg(\frac{\sqrt{[n+1]_{{q^2}}}}{(1-t_1t_2t_3t_4q^{4n})}\sqrt{\frac{\prod_{1\leq i<j\leq 4}(1-t_it_jq^{4n})(1-t_1t_2t_3t_4q^{2(2n-1)})}{(1-t_1t_2t_3t_4q^{2(2n-1)})(1-t_1t_2t_3t_4q^{2(2n+1)})}}\bigg)\ket{{n+1}}
    \;\\[10pt]& 
    +\Biggl(\frac{\sqrt{[n]_{{q^2}}}}{(1-t_1t_2t_3t_4q^{2(2n-2)})}\sqrt{\frac{\prod_{1\leq i<j\leq 4}(1-t_it_jq^{2(n-1)})(1-t_1t_2t_3t_4q^{2(n-2)})}{(1-t_1t_2t_3t_4q^{2(2n-1)})(1-t_1t_2t_3t_4q^{2(2n-3)}) }}\Biggr) \ket{{n-1}}\nonumber
    \\[10pt]& + 
    \bigg(t_1+t_1^{-1}-\frac{(1-t_1t_2q^{2n})(1-t_1t_3q^{2n})(1-t_1t_4q^{2n})(1-t_1t_2t_3t_4 q^{2n-2})}{t_1(1-t_1t_2t_3t_4q^{4n-2})(1-t_1t_2t_3t_4q^{4n})}\nonumber\\& \qquad
    -\frac{t_1(1-q^{2n})(1-t_2t_3q^{2(n-1)})(1-t_2t_4q^{2(n-1)})(1-t_3t_4q^{2(n-1)})}{(1-t_1t_2t_3t_4q^{2(4n-4)})(1-t_1t_2t_3t_4q^{2(4n-2)})}\bigg)\frac{1}{\sqrt{1-{q^2}}}\ket{n}.\nonumber
\end{align}
}
which can be solved by
\begin{flalign}\begin{aligned}\label{eqn:ortn-AW}
    &\bra{n}\ket{\theta}_{n_F=8} =\sqrt{\frac{(1-t_1t_2t_3t_4 q^{4n-2})(t_1t_2t_3t_4q^{-2};q^2)_n}{(1-t_1t_2t_3t_4 q^{-2})\prod_{1\leq i<j\leq 4}(t_it_j;q^2)_n}}\frac{P_n(\cos\theta;t_1,t_2,t_3,t_4|q^2)}{\sqrt{2\pi(q^2;q^2)_n}}~,
\end{aligned}\end{flalign}
when we selected $\bra{0}\ket{\theta}_{n_F=8}=1$, and introduced the Askey-Wilson polynomials as
\begin{flalign}\label{eqn:def-P}
    P_n(\cos\theta;t_1,t_2,t_3,t_4|q^2) = 
    \tfrac{(t_1t_2,t_1t_3,t_1t_4;q^2)_n}{t_1^n} \,_4\phi_3\bigg(\genfrac{}{}{0pt}{}{q^{-2n},t_1t_2t_3t_4q^{2n-2},t_1e^{\pm i\theta}}{ t_1t_2,t_1t_3,t_1t_4};q^2,q^2\bigg)~.
\end{flalign}
We notice that the Lanczos coefficients for the rescaled Hamiltonian in the semiclassical limit are:
\begin{subequations}\label{eq:lanczos nF8}
\begin{align}
    b(\ell)=&\frac{\sqrt{(1-\rme^{-\ell})(1-t_1t_2t_3t_4\rme^{-\ell})\prod_{1\leq i<j\leq 4}(1-t_it_j\rme^{-\ell})}}{(1-t_1t_2t_3t_4\rme^{-2\ell})^2}~,\\
    a(\ell)=&t_1+t_1^{-1}-\frac{(1-t_1t_2\rme^{-\ell})(1-t_1t_3\rme^{-\ell})(1-t_1t_4\rme^{-\ell})(1-t_1t_2t_3t_4 \rme^{-\ell})}{t_1(1-t_1t_2t_3t_4\rme^{-2\ell})^2}\\
    &-\frac{t_1(1-\rme^{-\ell})(1-t_2t_3\rme^{-\ell})(1-t_2t_4\rme^{-\ell})(1-t_3t_4\rme^{-\ell})}{(1-t_1t_2t_3t_4\rme^{-2\ell})^2}~.\nonumber
\end{align}
\end{subequations}
The technical details of the evaluation are presented in App.~\ref{sapp: nF8 def}. We find an explicit solution for the Krylov complexity of the HH state of the $n_F=8$ deformed DSSYK model in the semiclassical limit in terms of several elliptic integral functions, which is difficult to invert in terms of $\mathcal{C}_{\rm S}(t)$. Given that the explicit function $t(\ell)$ is not very illustrative, we report on the numerical evaluation of $\mathcal{C}_{\rm S}(t)$ for the $n_F=8$ case in Fig.~\ref{fig:nF}.

Noticeably, the Krylov spread complexity of the HH state for the $\hH_{n_F=8}$ Hamiltonian \eqref{eq:H nF8} does not correspond to the length between a single ETW brane and an asymptotic boundary in \eqref{eq:length ETW wormhole}. This can be seen from the differential equation of Krylov complexity in the semiclassical limit \eqref{eq:2nd ODE} with the Lanczos coefficients \eqref{eq:lanczos nF8}, which does not take the form of the differential equation corresponding to a single ETW brane in an AdS$_2$ black hole \eqref{eq:ETW brane ODE}. Even though the Schwarzian limit of the $n_F=8$  system in Sec.~\ref{ssec:triple scaling limit} indeed describes ETW branes in the bulk, the result for Krylov state complexity indicates that the bulk dual theory in the semiclassical limit and at finite temperature captures richer dynamics than just that of a single ETW brane in an AdS$_2$ black hole background. Meanwhile, in the very low temperature limit, the system indeed describes a single ETW brane when focusing on the triple-scaling limit, as we showed in Sec.~\ref{ssec:triple scaling limit}. It would be interesting to build an explicit sine dilaton gravity action that captures the properties of the $n_F=8$ deformed DSSYK model at the quantum level, which we leave for future work. Nevertheless, at late enough times, the ETW brane answer is a good approximation, as we will show at the end of this subsection. The reason for this is that the growth of Krylov complexity is determined by the fake temperature \eqref{eq:fake temper}, so that the late time limit corresponds to a very low temperature limit, which approximates to our results on the Schwarzian limit in Sec. \ref{ssec:triple scaling limit}.

\paragraph{Series expansion and convexity}
We can determine more information about the semiclassical solution at finite temperature without explicitly integrating \eqref{eq:2nd special int}. To do this, we go back to the equation of motion \eqref{eq:2nd ODE} with $b(\ell)$ and $a(\ell)$ in \eqref{eq:lanczos nF8} as
\begin{equation}\label{eq:perturbative exp nF=8}
\begin{aligned}
    \dv[2]{\ell}{t}&=\frac{a_1 \rme^{-\ell} \left(\rme^{-4\ell} (t_1t_2t_3t_4)^2+6 \rme^{-2\ell} t_1t_2t_3t_4+1\right)+a_2 \rme^{-2\ell} \left(\rme^{-2\ell} t_1t_2t_3t_4+1\right)}{\left(1-\rme^{-2\ell} t_1t_2t_3t_4\right)^3}\\
    &=a_1\sum_{m=1}^\infty (2m-1)^2(t_1t_2t_3t_4)^{m-1}\rme^{-(2m-1)\ell}+a_2\sum_{m=1}^\infty m^2(t_1t_2t_3t_4)^{m-1}\rme^{-2m\ell}~.
\end{aligned}
\end{equation}
where the coefficients are
\begin{subequations}\label{eq:coefficients nF8}
\begin{align}
    a_1&=2\qty(1+\prod_{i\neq j}t_it_j+\prod_{i=1}^4t_i-\qty(\sum_{i=1}^4t_it_j+\sum_{i\neq j\neq k}t_it_jt_k)\cos\theta)~,\\
    a_2=&8 \cos \theta \left(t_1^2 t_2 t_3 t_4+t_1 \left(t_2^2 t_3
   t_4+t_2 \left(t_3^2 t_4+t_3
   t_4^2+t_3+t_4\right)+t_3 t_4\right)+t_2 t_3
   t_4\right)\\
   &-\left(t_1^2 \left(-t_2^2 (t_3-t_4)^2+2 t_2
   (t_3+t_4) (t_3 t_4+1)-(t_3 t_4-1)^2\right)\right)\nonumber\\
   &-2 t_1
   \left(t_2^2 (t_3+t_4) (t_3 t_4+1)+t_2 \left(t_3^2
   \left(t_4^2+1\right)+12 t_3 t_4+t_4^2+1\right)+(t_3+t_4)
   (t_3 t_4+1)\right)\nonumber\\
   &+t_2^2 t_3^2 t_4^2-2 t_2^2 t_3
   t_4+t_2^2-2 t_2 t_3^2 t_4-2 t_2 t_3 t_4^2-2 t_2
   t_3-2 t_2 t_4+t_3^2-2 t_3 t_4+t_4^2~.\nonumber
\end{align}
\end{subequations}
Note that in contrast to the ETW brane case \eqref{eq:length ETW wormhole} where $a_1>0$ and $a_2>0$, here
\begin{equation}
    \dv[2]{\ell}{t}<0~,\quad{\rm when}~~\rme^{-2\ell} t_1t_2t_3t_4>0,
\end{equation}
which is always satisfied as long as we consider $0\leq t_1t_2t_3t_4\leq 1$. However, the term with the $a_2\in\mathbb{R}$ coefficient is suppressed due to the factors $\rme^{-2\ell}$ at late times. This is the reason that $\mathcal{C}_{\rm S}$ is not a convex function in time, similar to our findings in the $n_F=4$ case discussed in Sec.~\ref{ssec:Krylov nF4}.

Next, we notice from the series expansion \eqref{eq:perturbative exp nF=8} at leading order (i.e. $m=1$), the differential equation takes the form of the second-order differential equations at the beginning of the section \eqref{eq:ETW brane ODE}, and thus we recover a Krylov complexity for the HH state of the form \eqref{eq:Krylov HH state nF} with the coefficients $a_{1}$ and $a_2$ in \eqref{eq:coefficients nF8}. Since $\mathcal{C}_{\rm S}(t)$ is then a monotonically increasing function, the approximation is only valid when $t\gtrsim 2\sin\theta$. This means that for early enough times, the system behaves differently from a single ETW brane. 

\subsection{The adjoint case: continuous q-ultraspherical polynomials}\label{ssec:Krylov nF Adj}
We now express the chord Hamiltonian \eqref{eq:H SYK8} when $t_1=q^{-1}t_2=-t_3=-q^{-1}t_4=\sqrt{\chi}$ in its symmetric form
\begin{flalign}\begin{aligned}\label{eq:H Adj}
    \hH_{\rm Adj}\ket{n}=& 
    \sqrt{[n+1]_{q^2}}\sqrt{\frac{1-\chi^2 q^{2n}}{(1-\chi q^{2n})(1-\chi q^{2n+2})}}\ket{{n+1}}
    \\& 
    + \sqrt{[n]_{q^2}}\sqrt{\frac{1-\chi^2 q^{2n-2}}{(1-\chi q^{2n-2})(1-\chi q^{2n})}}\ket{{n-1}}~,
\end{aligned}\end{flalign}
This relation defines an inner product between the chord number states and the energy eigenstates with initial condition $\bra{0}\ket{\theta}_{\rm Adj.} =1$, which is explicitly solved by
\begin{flalign}\begin{aligned}
    &\bra{0}\ket{\theta}_{n_F=2, adj.} = \sqrt{\frac{1-\chi q^{2n}}{1-\chi}
    \frac{(q^2;q^2)_n}{(\chi^2;q^2)_n}}
    C_n(\cos\theta;\chi | q^2)\label{eqn:zeta_n}
\end{aligned}\end{flalign}
where we denote the continuous q-ultraspherical polynomials as
\begin{equation}\label{eq:ultraspherical}
    C_n(\cos\theta;\chi|q^2) = 
    \frac{(\chi^2 ;q^2)_n}{(q^2;q^2)_n} \chi^{-\frac{n}{2}}\,_4\phi_3\bigg(
    \genfrac{}{}{0pt}{}{q^{-2n},\chi^2q^{2n},\chi^{1/2}e^{\pm i\theta}}{\chi q,-\chi,-\chi q}
    ;q^2,q^2\bigg)~.
\end{equation}
The Lanczos coefficients in the semiclassical limit are
\begin{align}\label{eq: Adj coefficients}
     b(\ell)&=\sqrt{1-\rme^{-\ell}}\frac{\sqrt{1-\chi^2 \rme^{-\ell}}}{1-\chi \rme^{-\ell}} ~,\quad a(\ell)=0~.
\end{align}
Note that in the case of $\chi=1$ we have that $b(\ell)=1$, while when $\chi=0$ we recover the $n_F=0$ case (see App.~\ref{app:Krylov nF0}). The initial condition becomes
\begin{equation}\label{eq:adj initi }
    \rme^{-\ell_0}=\frac{2 \sin ^2\theta}{\sqrt{\left(\chi ^2-\chi  \cos ^2(\theta
   )+1\right)^2-4 \chi ^2 \sin ^2\theta}+\chi ^2-\chi  \cos ^2(\theta
   )+1}~.
\end{equation}
We now evaluate $\mathcal{C}_{\rm S}$ \eqref{eq:spread Complexity} from the integral of motion \eqref{eq:special eq}, which gives
\begin{align}
    &\Biggl[ \tanh ^{-1}\left(\frac{\csc \theta \left(\chi ^2-2 \chi  \cos
   ^2\theta-2 \chi ^2 x \sin ^2\theta+1\right)}{2 \chi  \sqrt{\sin ^2(\theta
   )+\chi ^2 x^2 \sin ^2\theta-x \left(\chi ^2-2 \chi  \cos ^2(\theta
   )+1\right)}}\right)\label{eq:full spread Adj}\\
   &\qquad-\tanh ^{-1}\left(\frac{\csc \theta \left(2
   \sin ^2\theta-x \left(\chi ^2-2 \chi  \cos ^2\theta+1\right)\right)}{2
   \sqrt{\sin ^2\theta+\chi ^2 x^2 \sin ^2\theta+x \left(-\chi ^2+2 \chi  \cos
   ^2\theta-1\right)}}\right)\eval{\Biggr]}_{x=\rme^{-\ell_0}}^{x=\rme^{-\lambda \mathcal{C}_{\rm S}}}=2\sin\theta t~.\nonumber
   \end{align}
This implicit solution is in agreement with the results in Sec.~\ref{ssec:Krylov nF8} when $t_1=-t_3=\sqrt{\chi}$ and $t_2=-t_4=q^2\sqrt{\chi}$ (where we take the limit $\lambda\rightarrow0$). We illustrate the results for $\mathcal{C}_{\rm S}$ \eqref{eq:spread Complexity} in the continuum approximation in Fig.~\ref{fig:nF}. Note that the evolution of Krylov complexity for the adjoint case is more closely related to the $n_F=0$ case than for the other $n_F$ cases that have similar deformation parameters $t_i$. This hints that the bulk dual theory of the adjoint deformed DSSYK model might describe an AdS$_2$ black hole with two asymptotic boundaries, while the other cases are more closely associated with a single one. One hint on that is from the Krylov complexity (interpreted as a geodesic length in the bulk), which in the $n_F=0$ and adjoint cases is greater than the early-time (with respect to the fake temperature \eqref{eq:fake temper}) Krylov complexity for the other cases given the same preparation of state (see Fig.~\ref{fig:nF}).

As consistency check, when $\chi=0$, one recovers\footnote{Meanwhile, the first order correction for $0<\chi\ll1$ becomes
\begin{equation}
    -\frac{\chi  (\cos (2 \theta )+X)}{\sqrt{\sin ^2\theta-X}}+\frac{1}{4} \left(-2
   \csc \theta \tanh ^{-1}\left(\frac{2 \sin \theta-X \csc \theta}{2
   \sqrt{\sin ^2\theta-X}}\right)-i \pi  \csc \theta\right)=t~.
\end{equation}
}
\begin{equation}
    \rme^{-\ell}=\frac{\sin^2\theta}{\cosh^2(\sin\theta t)}~,
\end{equation}
which describes the semiclassical Krylov complexity of the HH state in the DSSYK model \cite{Rabinovici:2023yex}.

\paragraph{ETW brane approximation}
Similar to the $n_F=8$, we will be interested in a series expansion in terms of the variable $\rme^{-\ell(t)}$ which can be interpreted as an expansion around the ETW brane solution at late boundary times.

Using the Lanczos coefficients in \eqref{eq: Adj coefficients}, we find that the expansion of the ODE for Krylov spread complexity of the HH state in the semiclassical limit \eqref{eq:2nd ODE} can be expressed as a series
\begin{equation}
    \dv[2]{\ell}{t}=2\rme^{-\ell}(\chi-1)^2\sum_{m=1}^\infty \chi^{m-1}m^2\rme^{-m\ell}~.
\end{equation}
In the approximation where we keep the $m=1,~2$ terms
\begin{equation}\label{eq:spread adj ETW}
    \mathcal{C}_{\rm S}\simeq\frac{1}{\lambda}\log\frac{(\chi-1)^2+\abs{\chi-1}\sqrt{(\chi-1)^2+8\chi\sin^2\theta}\cosh(2\sin\theta t)}{2\sin^2\theta}~.
\end{equation}
We emphasise that the full non-perturbative solution is given in \eqref{eq:full spread Adj}, while \eqref{eq:spread adj ETW} is an approximation valid only at late times $t\gtrsim 2\sin\theta$.

For the naive limit $\chi = 1$ (which is also the semiclassical limit of $\chi = q^2$), we see that the equation predicts a constant $l$ (by the initial condition), not in agreement with the exact $\chi = q^2$ computation (see Sec.~\ref{ssec:RMT}). This indicates that the semiclassical limit and the $\chi \rightarrow 1$ limit do not commute. 
Interestingly, while the very low temperature limit shows that the adjoint deformation behaves as the Schwarzian theory without the ETW brane, when we allow for finite temperatures, the semiclassical geometry encodes the presence of an ETW brane at late enough times, with respect to the thermal time scale $t\gtrsim 2\sin\theta=\beta_{\rm fake}/2\pi$ which we have expressed in terms of the fake temperature \eqref{eq:fake temper}.

\subsection{Emergent RMT and \texorpdfstring{$\mathcal{N}=2$}{} SU(2) gauge theory}\label{ssec:RMT}
The orthogonality relation of the q-ultraspherical polynomials \eqref{eq:ultraspherical} can be expressed as
\begin{flalign}\begin{aligned}\label{eqn:ort-C-n}
    &\int _0^\pi d\theta \frac{(e^{\pm 2i\theta};q^2)_\infty}{(\chi e^{\pm 2i\theta},\chi q^2 e^{\pm 2i\theta};q^4)_\infty} \frac{(-\chi,q^2;q^2)_m}{(\chi^2,-\chi q^2;q^2)_m} C_n(\cos\theta;\chi | q^2)C_m(\cos\theta;\chi | q^2) =\\&
    \qquad\qquad\qquad\qquad\qquad
    \frac{1-\chi^2}{1-\chi^2 q^{4n}} \frac{2\pi (\chi^2 q^2;q^2)_\infty}{(q^2;q^2)_\infty (\chi^2 q^2;q^4)_\infty^2(-\chi,-\chi q^2;q^2)_\infty} \delta_{n,m}.
\end{aligned}\end{flalign}
This implies the integration measure
\begin{equation}\label{eq:mu Adj}
\begin{aligned}
    \mu_{\rm Adj}(\theta)&=\bra{B_{\rm Adj}}\ket{\theta}\equiv\frac{(e^{\pm 2i\theta};q^2)_\infty}{2\pi}\frac{(\chi^2 q^2;q^4)_\infty^2(q^2,-\chi,-\chi q^2;q^2)_\infty}{(\chi e^{\pm 2i\theta},\chi q^2 e^{\pm 2i\theta};q^4)_\infty(\chi^2q^2;q^2)_\infty}\rightarrow~,\\
    \ket{B_{\rm Adj}}&=\frac{(\chi ^2q^2;q^4)^2_\infty(q^2,-\chi,-\chi q^2;q^2)_\infty}{(\chi ^2q^2;q^2)_\infty(\chi\rme^{\pm2\rmi\theta(\hH)},\chi q^2\rme^{\pm2\rmi\theta(\hH)};q^4)_\infty}\ket{\Omega}~,
\end{aligned}
\end{equation}
which is normalised such that
\begin{equation}
    \int _0^\pi\rmd\theta~\mu_{\rm Adj}(\theta)=1~.
\end{equation}
A special case is $\chi=q^2$ (which becomes $\chi =1$ in the semiclassical limit). In that case, we have the identity
\begin{equation}
    \frac{(q^6;q^4)_{\infty}^2 (q^2,-q^2,-q^4;q^2)_{\infty}}{(q^6;q^2)_{\infty}} =1
\end{equation}
and 
\begin{equation}
    \frac{(e^{\pm 2i \theta};q^2)_{\infty}}{(q^2 e^{\pm 2 i \theta};q^2)_{\infty}} = |1-e^{2i\theta}|^2.
\end{equation}
The measure then collapses to 
\begin{equation}
    \mu(\theta)|_{\chi = q^2} = \frac{2}{\pi} \sin^2 \theta,
\end{equation}
which is exactly the $SU(2)$ Haar measure. In this limit, the $q-$ultraspherical polynomials degenerate to characters of $SU(2)$. 

In that case, we can compute the Krylov complexity exactly, without assuming any semiclassical limit. {Indeed, let us notice that the Lanczos coefficients in this case \eqref{eq:H Adj} are
\begin{equation}
b_n= 1/\sqrt{1-q^2}~,
\end{equation}
just a constant, in agreement with Krylov complexity for RMT \cite{Balasubramanian:2025xkj}. Meanwhile, Krylov complexity itself can be expressed as
\begin{eqnarray}
    \mathcal{C}_{\rm S}=\sum_{k=0}^\infty k\abs{\sum_{n=0}^\infty\frac{(-\rmi t)^n}{n!}\bra{k}\hH_{n_F}^n\ket{\Omega}}^2~.
\end{eqnarray}
We can see that} $\bra{0}H^n \ket{0}$ computes (up to a prefactor $1/\sqrt{1-q^2}^{n}$) the number of singlets in the $n$-fold tensor product of the fundamental representation, and similarly, $\bra{k}H^n \ket{0}$ computes how many spin $k/2$ representations appear in the $n$-fold tensor product of the fundamental. In this special case, the Krylov complexity is simply the speed of the representation-theoretic spread of the $SU(2)$ spin.

Another way to interpret this transfer matrix is a random walk on the natural numbers (which are mapped to the spin of the representation) that starts and ends at zero. This random walk picture has a generalization to different $\chi$ and also to all the other $q$-Askey schemes: there, the endpoint of the random walk is not constrained to be zero; instead, the distribution of possible endpoints is fixed by the integration measure (which can always be decomposed into a sum over characters). For Euclidean temperature $\beta$, we then have the result:
{\small
\begin{equation}
    \begin{split}
        \mathcal{C}_S(t) = & \frac{-\frac{\beta}{\sqrt{1-q^2}}}{ I_1(x+y)}\Bigg[  \left(1+4\frac{t^2}{\beta^2}\right)\left(I_0(x)I_0(y)+I_1(x)I_1(y)\right)\\ - & (1-q^2)^{\frac{1}{2}} i \frac{t}{\beta^2}\left[\left(1+2i\frac{t}{\beta}\right)I_1(x)I_0(y)-\left(1-2i\frac{t}{\beta}\right)I_0(x)I_1(y)\right]
        +  \frac{\sqrt{1-q^2}}{\beta} I_1(x+y) \Bigg],
    \end{split}
\end{equation}
}
with $x=-\frac{\beta + 2 it}{\sqrt{1-q^2}}$, $y=x^*$. At late times, the complexity grows linearly, with coefficient
\begin{equation}
    \mathcal{C}_S(t)=\frac{2}{\pi \beta\; I_1(\frac{2\beta}{\sqrt{1-q^2}})}\left[2\cosh \frac{2\beta}{\sqrt{1-q^2}}-\frac{\sqrt{1-q^2}}{\beta}\sinh \frac{2\beta}{\sqrt{1-q^2}}\right] t + \mathcal{O}(t^0).
\end{equation}

\paragraph{Larger $\chi$}
The theory for $q^2 < \chi < 1$ is still perfectly well-defined. At $\chi = 1$, the $\theta$-dependence of the measure cancels exactly, and we simply get the flat measure on the interval $[0,\pi]$. The orthonormal wavefunctions are plane waves.\footnote{Intriguingly, the transfer matrix naively has the same expression for $\chi=1$ and $\chi=q^2$. However, for $\chi=1$ the action on the state $\ket{1}$ contains the ratio $\sqrt{(1-\chi^2)/(1-\chi)}\rightarrow \sqrt{2}$ which then deforms the transfer matrix, while the $\chi=q^2$ has no such issue. We also see below that the microscopic Hamiltonian takes the same expression for $\chi=1$ and $\chi = q^2$.}
For comparison purposes we illustrate the growth of Krylov complexity in the range $\chi\in[0,1]$ in Fig.~\ref{fig:adj_def}. We note that the growth of Krylov complexity is strongly dependent on $\chi$; as expected, when $\chi$ approaches $q^{2}$ it displays the growth expected in RMT \cite{Camargo:2024rrj}, while it matches to the $n_F=0$ case as $\chi\rightarrow0$, which displays slower growth.
\begin{figure}
    \centering
    \subfloat[]{\includegraphics[width=0.5\linewidth]{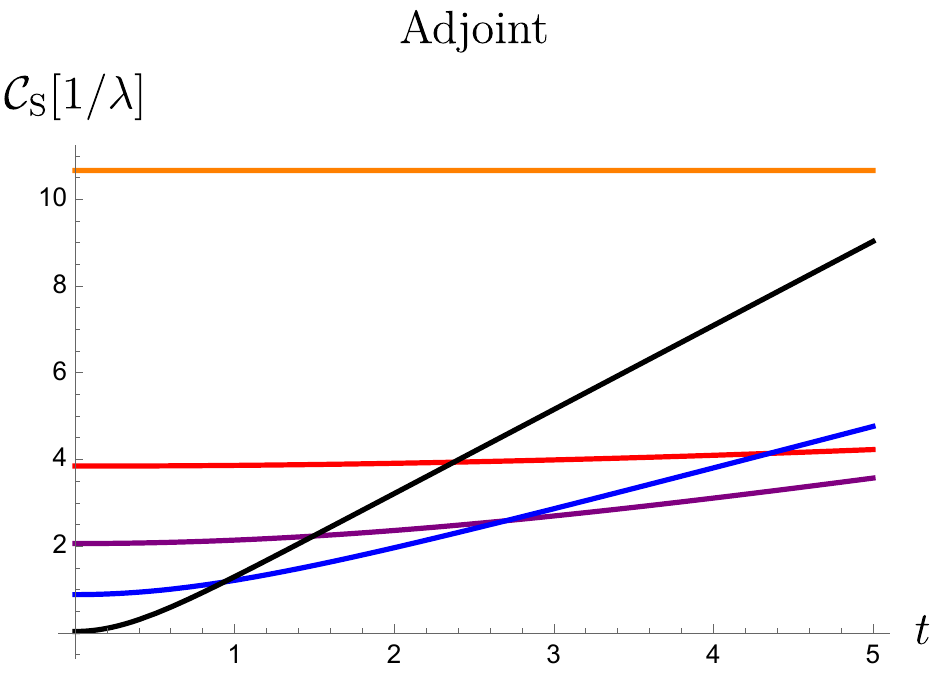}}\subfloat[]{\includegraphics[width=0.5\linewidth]{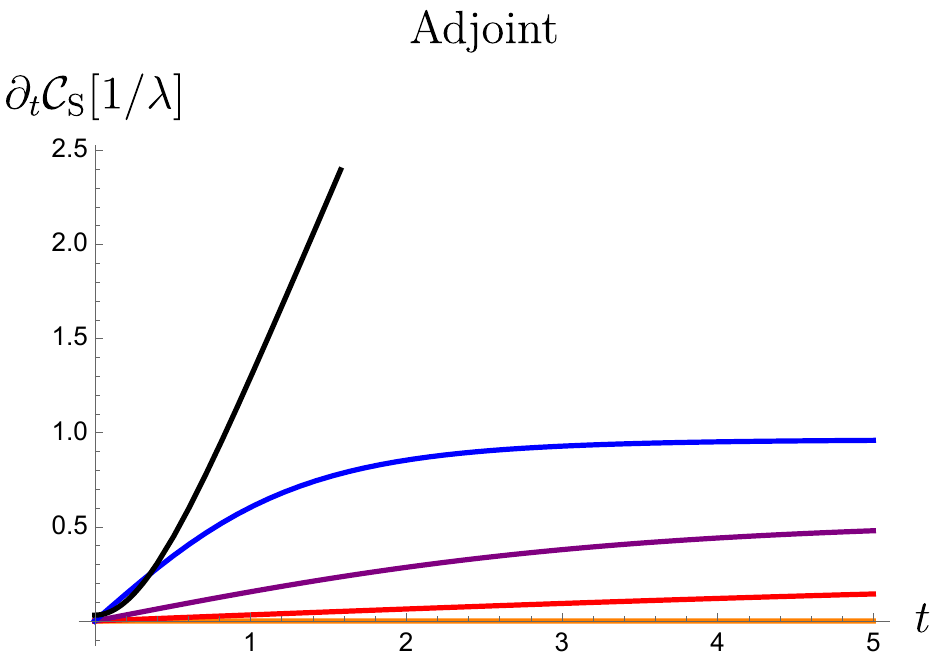}}
    \caption{(a) Krylov complexity \eqref{eq:Krylov HH state nF} and (b) its rate of growth, for the HH state of the adjoint deformation \eqref{eq:H Adj} for different values of $\chi$, $\chi=10^{-2}$ (orange), $0.25$ (red), $0.5$ (purple), $0.75$ (blue), $0.99$ (black). The rate of growth of the complexity increases with $\chi$, while the initial value of Krylov complexity in the HH state for a fixed value $\theta$ is greater when $\chi$ is vanishing.}
    \label{fig:adj_def}
\end{figure}

We can increase $\chi$ further. In that case, one picks up discrete contributions from bound states, as in \cite{Rajgadia:2026ask}. From deforming the integration contour in the thermal partition function appropriately, one finds the relation
\begin{equation}\label{adjoint partition recursion}
    Z(\chi q^{-2};q^2) = Z(\chi;q^2)+ \frac{q(q^2\chi^{-1},\chi q^{-2};q^2)_{\infty}}{\sqrt{\frac{\chi}{q^2}}(\chi^2 q^{-4};q^2)_{\infty}} \cosh \left(\frac{\beta}{2\sqrt{1-q^2}}(\sqrt{\frac{\chi}{q}}+\frac{q}{\sqrt{\chi}})\right). 
\end{equation}
We see that for $\chi >1$, there is a series of bound state contributions (one applies \eqref{adjoint partition recursion} until one gets $\chi <1$). Similarly, one can compute the change of matrix elements $\bra{n} e^{-\beta H} \ket{0}$ to then find the bound state contributions to the Krylov complexity. From \eqref{adjoint partition recursion} (and similar expressions for the matrix elements), one can see that real time evolution only contributes an oscillating phase. The bound state contributions should therefore not affect the asymptotic Krylov complexity. This is intuitive: the spectrum of bound states for each value of $\chi$ gives a finite-dimensional Hilbert space, and the wavefunction can therefore not spread indefinitely in that subspace.  

To summarise the section, Krylov complexity is an advantageous tool to analyse the evolution of the chord number in the DSSYK model and its deformations. It allowed us to understand the holographic dictionary of the boundary theory up to the $n_F=6$ case, and it provides valuable insights for the $n_F=8$ and adjoint deformations. In particular, it displays rich connections with the quantum group of the DSSYK model. While we focused on the semiclassical limit for some of the evaluations, we also found a particular regime in the adjoint case which is fully solvable, and which reveals an emergent SU(2) RMT that has a direct interpretation in terms of the Schur/SYK duality with an $\mathcal{N}=2$ gauge theory with matter hypermultiplets.

\section{Operator algebras and entanglement entropy}\label{sec:OP algebras}

In this section, we employ algebraic methods to study the bulk dual of the q-Askey deformed DSSYK model. To this end, we analyse the algebraic structure and entanglement entropy and extract its bulk interpretation in terms of an RT formula.

We first define the operator algebras for the deformed models. The algebras are constructed based on different operators that include information about the thermal properties of the system, namely the chord Hamiltonians of the deformed theories, and the matter chord operators. Equipped with appropriate tracial states, this gives rise to a set of type II$_1$ factors, which we denote q-Askey double-scaled algebras. 

On the other hand, once we include the total chord number in the algebra, it becomes a type I$_\infty$ factor, which we refer to as a \textit{q-Askey chord algebra}. This algebra has a well-adapted operator, which can recover full information about the purity of the state used to evaluate the expectation values in the microscopic theory, including the PETS in Secs.~\ref{ssec:KM state deformations} and \ref{ssec:explicit microscopic}.

To investigate the q-Askey double-scaled algebras, there are two complementary viewpoints one may take. The first is to consider the double-scaled algebra of the undeformed ($n_F=0$) theory \eqref{eq:algebra}, and relate the ETW brane states to different states (in the algebraic sense) on the same type II$_1$ factor. In this description, the deformation parameters are specified by the relevant ETW brane state. 

The second approach is to consider a deformed algebra $\mathcal{A}_{\rm DS}^{n_F}$, which explicitly depends on the ETW brane parameters $t_i$. One can then attempt to construct a tracial state $\omega_{n_F}$, and inspect the properties of the pair $(\mathcal{A}_{\rm DS}^{n_F},\omega_{n_F})$. This approach has already been studied in \cite{Cao:2025pir} for $n_F=2$, and here we will give a generalization to $n_F>2$. We find that similar to their analysis, there is a no-man's island in the bulk which can only be reconstructed from the boundary, if one includes the wormhole length into the boundary algebra. 

\paragraph{Outline}In Sec.~\ref{ssec:qA DSA} we define the q-Askey double-scaled algebra, which includes the deformed chord Hamiltonians and matter chord operators, resulting in type II$_1$ factors for the algebra and its commutant. In Sec.~\ref{ssec:mod flows} we investigate relative modular flows in the q-Askey double-scaled algebras. In Sec.~\ref{ssec:algebraic EE} we introduce definitions for algebraic entanglement entropy between the type II algebras given a global pure state, which we illustrate in the q-Askey double-scaled algebras. We also define the relative Araki entropy. In Sec.~\ref{ssec:EE from chord number}, we apply the definitions to evaluate algebraic entanglement entropy with respect to the chord number state in the q-Askey deformed DSSYK models. In Sec.~\ref{ssec:bulk AEE} we analyse the bulk interpretation of algebraic entanglement entropy in terms of ETW branes. We conclude the section with Sec.~\ref{ssec:qA chord algebra}, which discusses an extension of the algebras where we include the total chord number. We define this algebra as the q-Askey chord algebra and show that it is type I$_\infty$.

\subsection{q-Askey double-scaled algebra (type II\texorpdfstring{$_1$}{})}\label{ssec:qA DSA}
We study the chord operator algebra that incorporates matter operators acting on the right subsystem in \eqref{eq:algebra} and the deformed Hamiltonians in Sec.~\ref{ssec:def qAskey def}. We define the q-Askey double-scaled algebra as:\footnote{The name is chosen due to its isomorphism to the ``double-scaled'' algebra of DSSYK \cite{Lin:2022rbf}, which we reviewed in Sec.~\ref{ssec:DSSYK review}. 
}
\begin{equation}\label{eq:q-Askey algebra}
\mathcal{A}_{\rm DS}^{n_F}=\langle\hH_{n_F}^{(R)},~\hat{\mathcal{O}}^{(R)}_\Delta\rangle''~,
\end{equation} 
where we included the label $(R)$ in $\hH_{n_F}$ defined in Sec.~\ref{ssec:explicit microscopic} to distinguish it from the chord Hamiltonian in the commutator algebra $\hH^{(L)}_{n_F}${; and we use the same notation as \eqref{eq:part 2}}. 
Note that \eqref{eq:q-Askey algebra} describes thermal operators of the deformed microscopic SYK model. In  App.~\ref{sapp:PETS DS algebra} we show that the algebra \eqref{eq:q-Askey algebra} as well as its commutant algebra $\mathcal{A}_{\rm DS}^{n_F}{}'=\langle\hH^{(L)}_{n_F},~\hmO_\Delta^{(L)}\rangle''$, the latter was defined in \eqref{eq:algebra}, are generated by evaluating correlation functions in PETS of the deformed theories in the double-scaling limit. In the limit $\beta_L\to\infty$, in the construction representing the ETW brane system Fig.~\ref{fig:PETS}, there are no dynamics associated with the left subsystem. Therefore, we will focus on $\mathcal{A}_{\rm DS}^{n_F}$ in the discussion below.

Let us denote the empty chord state in the q-Askey deformed theories in Sec.~\ref{ssec:def qAskey def}, whose Hilbert space appears in \eqref{eq:H space nF}, by $\ket{0}$.
One can show that $\ket{0}$ is a cyclic state in the above algebra following the same arguments as \cite{Xu:2024hoc,Cao:2025pir} (see Lemma 1, therein). One can generate a dense set of states that span the Hilbert space of the system, including matter, by repeated Hamiltonian insertions $\hH^{(R)}_{n_F}$. This can be seen from the construction of the Krylov basis for all the different q-Askey deformations in Sec.~\ref{sec:Krylov Askey} as an explicit construction.

Similar to Definition 1 in \cite{Cao:2025pir}, we introduce a map between the deformed algebra with matter and the one without deformation 
\begin{equation}\label{eq:Td inverse}
\text{Td}^{-1}~:~~\mathcal{A}_{\rm DS}^{n_F}\rightarrow\mathcal{A}_{R}~,
\end{equation}
where $\ket{\Omega}$, defined in \eqref{eq:H space nF=0}, 
corresponds to the cyclic separating tracial state for $\mathcal{A}_{R}$. More explicitly, the map Td$^{-1}$ can be represented as a Cauchy sequence of polynomial functions $f_n$ acting on operators $\tilde{a}\in\mathcal{A}_{\rm DS}^{n_F}$ as
\begin{equation}\label{eq:Td explicit}
    \hat{a}_R=\text{Td}^{-1}(\tilde{a})\equiv\lim_{n\rightarrow\infty}f_n(\hH_{n_F},\hat{\mathcal{O}}^{(R)}_\Delta)~,\quad \hat{a}_R\in\mathcal{A}_R~.
\end{equation}
Using that $\ket{0}$ is a cyclic state for the algebra $\mathcal{A}_{\rm DS}^{n_F}$, we have:
\begin{align}
    &\bra{0}P_1(\hH_{n_F},\hat{\mathcal{O}}^{(R)}_\Delta)\tilde{a}P_2(\hH_{n_F},\hat{\mathcal{O}}^{(R)}_\Delta)\ket{0}\\&
    =\lim_{n\rightarrow\infty}\bra{0}P_1(\hH_{n_F},\hat{\mathcal{O}}^{(R)}_\Delta)f_n(\hH_{n_F},\hat{\mathcal{O}}^{(R)}_\Delta)P_2(\hH_{n_F},\hat{\mathcal{O}}^{(R)}_\Delta)\ket{0}\nonumber\\
    &=\lim_{n\rightarrow\infty}\bra{\Omega}\hat{B}^{(n_F)}_{\vec{t}}P_1(\hH_{R},\hat{\mathcal{O}}^{(R)}_\Delta)f_n(\hH_{R},\hat{\mathcal{O}}^{(R)}_\Delta)P_2(\hH_{R},\hat{\mathcal{O}}^{(R)}_\Delta)\ket{\Omega}~,\label{eq:las for type II}
    \end{align}
where in the last line we applied \eqref{eq:mu nF first}, and $\hH_R$ is the $n_F=0$ chord Hamiltonian with matter.

It follows that $\ket{0}$ is also a separating state for $\mathcal{A}_{\rm DS}^{n_F}$ (analogous to Lemma 3 in \cite{Cao:2025pir}) by simply setting $P_2=1$ above since $\hat{a}_R\ket{\Omega}=0$ for $\hat{a}_R\in\mathcal{A}_R$ is only possible when $\hat{a}_R=0$. We can use the cyclicity of $\ket{\Omega}$ to move $(\hat{B}_{\vec{t}}^{(n_F)})^{1/2}$ in the expectation value \eqref{eq:las for type II} as in \eqref{eq:mu nF first} and write
\begin{equation}
\begin{aligned}
    &\bra{0}P_1(\hH_{n_F},\hat{\mathcal{O}}^{(R)}_\Delta)\tilde{a}P_2(\hH_{n_F},\hat{\mathcal{O}}^{(R)}_\Delta)\ket{0}\\
&=\bra{(B_{\vec{t}}^{(n_F)})^{1/2}}P_1(\hH_{n_F},\hat{\mathcal{O}}^{(R)}_\Delta)\tilde{a}P_2(\hH_{n_F},\hat{\mathcal{O}}^{(R)}_\Delta)\ket{(B_{\vec{t}}^{(n_F)})^{1/2}}~,
\end{aligned}
\end{equation}
therefore $\ket{B_{\vec{t}}^{1/2}}$ acts as a tracial state for $\mathcal{A}_{\rm DS}^{n_F}$. Since $\ket{0}$  
is cyclic and separating, it follows that $\ket{B_{\vec{t}}}$ is a cyclic separating tracial state for $\mathcal{A}_{\rm DS}^{n_F}$. Moreover, one can show that $\mathcal{A}_{\rm DS}^{n_F}$ is a factor from similar arguments as Theorem 5 in \cite{Cao:2025pir}. 

Suppose $\tilde{o}\in\mathcal{A}_R\cap\tilde{A}'$, which denotes the commutant algebra $\hat{a}_R'\in\tilde{A}_R'$ when $[\tilde{a}_R',\tilde{a}_R]=0~,~~\forall\hat{a}_R\in\mathcal{A}_R$. Since $[\tilde{o},\tilde{a}]=0$, then 
\begin{equation}
    \begin{aligned}
        &{\bra{0}P_1(\hH_{n_F},\hat{\mathcal{O}}^{(R)}_\Delta)[\tilde{o},~\tilde{a}]P_2(\hH_{n_F},\hat{\mathcal{O}}^{(R)}_\Delta)\ket{0}}\\
        &=\bra{\Omega}P_1(\hH_{R},\hat{\mathcal{O}}^{(R)}_\Delta)\hat{B}_{\vec{t}}~[\tilde{o},~\tilde{a}]P_2(\hH_{R},\hat{\mathcal{O}}^{(R)}_\Delta)\ket{\Omega}=0~,
    \end{aligned}
\end{equation}
where $\hat{o}={\rm Td}^{-1}(\tilde{o})$ using \eqref{eq:Td explicit}, from which it follows that
\begin{equation}
 [\hat{o},~\hat{a}_{R}]=0~,~~\forall\hat{a}_R\in\mathcal{A}_R   ~,
\end{equation}
and since $\mathcal{A}_R$ is a factor then $\hat{o}\propto \mathbb{1}$, which means that $\mathcal{A}_{\rm DS}^{n_F}$ is also a factor. Since $\mathcal{A}_{\rm DS}^{n_F}$ has a cyclic separating state, it follows that it is a type II$_1$ factor.

\subsection{Relative modular flows from ETW brane states}\label{ssec:mod flows}

As shown in \cite{Berkooz:2025ydg}, a key property of transition amplitudes in the q-Askey deformed models is that they may be expressed as amplitudes in a non-vacuum sector of ordinary DSSYK. This observation implies that the different ETW brane states arising in the deformed theories can be consistently viewed as elements of a single underlying DSSYK chord Hilbert space \eqref{eq:H ffull}. From this point of view, the parameters $t_i$ should be viewed as attributes of the state itself.

Consequently, rather than modifying the operator algebra, we can fix a common von Neumann algebra and vary the state on which we evaluate observables. This naturally leads to a formulation of \textit{relative modular flows} \cite{Araki:1976zv,Liu:2025krl}.

\paragraph{The undeformed double-scaled algebra}
Let us recall the double-scaled algebra of the DSSYK model $\mathcal{A}_{L/R}$ defined in \cite{Lin:2022rbf,Xu:2024hoc}. The construction relies on introducing two commuting left/right copies of the system, which together realise the set of all bounded operators on the chord Hilbert space $\mathcal{H}_{\rm chord}$ \eqref{eq:H ffull}. One then considers the algebra of one subsystem, while the algebra of the other is naturally given by the commutant. We choose:
\begin{flalign}
    &\mathcal{A} =\mathcal{A}_R = \langle\hH_{R}, \hmO_\Delta^{(R)}\rangle''~,\quad \mathcal{A}' = \mathcal{A}_L~,\quad\mathcal{A}_L \vee \mathcal{A}_R = \mathcal{B}(\mathcal{H}_{\rm chord})~,
\end{flalign}
where $\hH_R$ is defined in \eqref{eqn:hH_LR} is the usual DSSYK Hamiltonian and $\hmO_\Delta^{(R)}$ is the matter chord operator. In the remainder of this section, we will use $\hH_R = \hH_{n_F=0}$ to highlight that this is the ordinary DSSYK chord Hamiltonian.

As shown in \cite{Xu:2024hoc}, the empty chord state $\ket{\Omega}$ is cyclic and separating and gives rise to a tracial state\footnote{In the context of von Neumann algebras, given an algebra $\mathcal{M}$, a state is defined as a continuous map $\omega:\mathcal{M}\to\mathbb{C}$, satisfying:
\begin{flalign}\label{eqn:alg-state}
    \omega(aA+bB)=a\omega(A)+b\omega(B), \qquad \omega(A^\dagger) = (\omega(A))^*, \qquad \forall a,b\in\mathbb{C},\, \forall A,B\in \mathcal{M}.
\end{flalign}
Although most generally, not every such map corresponds to a single vector in $\mathcal{H}_{chord}$, for our present purposes, we will interchangeably use ``state'' to mean either the map \eqref{eqn:alg-state}, or the corresponding vector in the chord Hilbert space. The distinction should be clear from context.
} $\omega_0(A)$ on $\mathcal{A}$:
\begin{flalign}
    \omega_0(A) = \Tr(A)= \bra{\Omega}A\ket{\Omega}, \quad \forall A\in \mathcal{A}_{L,R} .
    \label{eqn:trace}
\end{flalign}
The double-scaled algebra $\mathcal{A}$ is a II$_{1}$ factor. By the Tomita-Takesaki theorem, one can introduce a Tomita operator $S_\Omega$ as follows:
\begin{flalign}\label{eq:def tomita}
    &S_\Omega A \ket{\Omega} = A^\dagger\ket{\Omega}~,\quad S_\Omega = J_\Omega \Delta_\Omega^{1/2}~, \quad \forall A \in\mathcal{A}.
\end{flalign}
Here, $\Delta_\Omega$ is commonly referred to as the modular operator and $J_\Omega$ as the modular conjugation operator.

\paragraph{A family of ETW brane states}
We recall the set of ETW states in the chord Hilbert space, defined in \eqref{eq:ETW brane state general ext}-\eqref{eq:ETW op}:
\begin{flalign}
    \begin{aligned}
        &\ket{(B_{\vec{t}}^{(n_F)})^{1/2}} = (B_{\vec{t}}^{(n_F)})^{1/2} \ket{\Omega}, \\
        &\text{ where }
     B_{\vec{t}}^{(n_F)} = \frac{(t_1,..t_{n_F/2};q^2)_\infty}{\prod_{i=1}^{n_F/2} (t_ie^{\pm i\theta(\hH_{n_F=0})};q^2)_\infty}\frac{1}{\prod_{i<j}(t_it_j;q^2)_\infty}~,
    \end{aligned}
\end{flalign}
and $\theta(x) = \arccos(x\sqrt{1-q^2}/2)$. In this section,we will only consider parameters $t_{1\leq i \leq 4}$ satisfying the condition \eqref{eqn:Cond-continuous}, i.e. we will consider the regime where the spectrum does not exhibit discrete energy states.  Since $B_{\vec{t}}^{(n_F)}$ is a bounded function of the chord Hamiltonian $\hat H_{n_F=0}$, it defines an operator in the algebra $\mathcal{A}$. Each vector $\ket{(B_{\vec{t}}^{(n_F)})^{1/2}}$ therefore defines a state $\omega^{(n_F)}:\mathcal{A}\to \mathbb{C}$ as:
\begin{flalign}\begin{aligned}\label{eqn:state-Omega_nF}
    \omega^{(n_F)}(A) &= Z_{n_F}^{-1}\bra{(B_{\vec{t}}^{(n_F)})^{1/2}} A \ket{(B_{\vec{t}}^{(n_F)})^{1/2}} \\&= \bra{\Omega}(Z_{n_F}^{-1} B_{\vec{t}}^{(n_F)})^{1/2}\, A \,(B_{\vec{t}}^{(n_F)}Z_{n_F}^{-1})^{1/2} \ket{\Omega},
\end{aligned}\end{flalign}
where $Z_{n_F} = \bra{(B_{\vec{t}}^{(n_F)})^{1/2}}\ket{(B_{\vec{t}}^{(n_F)})^{1/2}}$ is a normalization factor, included so that $\omega^{(n_F)}(\mathbb{1})=1$.
The states $\omega^{(n_F)}$ are normal and admit a corresponding density operator $\rho^{(n_F)}$. In terms of the trace function \eqref{eqn:trace}, $\omega^{(n_F)}$ can be expressed as:
\begin{flalign}
    \omega^{(n_F)}(A) = \Tr(\rho^{(n_F)}A), \qquad \forall A\in \mathcal{A}.
\end{flalign}
Using the cyclic property of the trace, and the fact that $(B_{\vec{t}}^{(n_F)})^{1/2}, A (B_{\vec{t}}^{(n_F)})^{1/2} \in \mathcal{A}$, we have:
\begin{flalign}
    \bra{\Omega}(B_{\vec{t}}^{(n_F)})^{1/2} A (B_{\vec{t}}^{(n_F)})^{1/2} \ket{\Omega} = \bra{\Omega}B_{\vec{t}}^{(n_F)}\, A  \ket{\Omega}.
\end{flalign}
Hence the canonical density matrix can be expressed as:
\begin{flalign}
\begin{aligned}
    \omega^{(n_F)}(A) &= \bra{\Omega}\bigg( \frac{B_{\vec{t}}^{(n_F)}}{\bra{(B_{\vec{t}}^{(n_F)})^{1/2}}\ket{(B_{\vec{t}}^{(n_F)})^{1/2}}}\bigg) A\ket{\Omega}, \quad \forall A\in\mathcal{A} \\
    \implies
    \rho^{(n_F)} &= \frac{B_{\vec{t}}^{(n_F)}}{\bra{\Omega}B_{\vec{t}}^{(n_F)}\ket{\Omega}}
    = \frac{B_{\vec{t}}^{(n_F)}}{\bra{(B_{\vec{t}}^{(n_F)})^{1/2}}\ket{(B_{\vec{t}}^{(n_F)})^{1/2}}}.
    \label{eqn:DensityMat}
\end{aligned}
\end{flalign}
We now analyse the conditions under which the ETW brane states $\omega^{(n_F)}$ are cyclic and separating with respect to the algebra $\mathcal{A}$. As shown in \cite{Berkooz:2025ydg}, the operators $B_{\vec{t}}^{(n_F)}$ are bounded for values of the parameters $t_i$ satisfying \eqref{eq:dissrete cond 2}.

Under these conditions, the operators $(B_{\vec{t}}^{(n_F)})^{1/2}$ which define the states $\omega^{(n_F)}$ are bounded and invertible. It then follows that the states \eqref{eqn:state-Omega_nF} are cyclic and separating on $\mathcal{A}$. To show this, we rely on the cyclic and separating properties of the tracial state $\ket{\Omega}$.\\[-5pt]

\noindent For any $A\in\mathcal{A}$, we may write:
\begin{flalign}
    A \ket{(B_{\vec{t}}^{(n_F)})^{1/2}} = A (B_{\vec{t}}^{(n_F)})^{1/2} \ket{\Omega} \, .
\end{flalign}
Since $(B_{\vec{t}}^{(n_F)})^{1/2}$ is invertible, the set $\{A (B_{\vec{t}}^{(n_F)})^{1/2} : A\in\mathcal{A}\}$ coincides with $\mathcal{A}$ itself. Cyclicity of $\omega_0$ then implies that $\omega^{(n_F)}$ is \textbf{cyclic}.

If $A \ket{(B_{\vec{t}}^{(n_F)})^{1/2}} = 0$ for some $A\in\mathcal{A}$, then $A (B_{\vec{t}}^{(n_F)})^{1/2} \ket{\Omega} = 0$. Using invertibility of $(B_{\vec{t}}^{(n_F)})^{1/2}$ and the fact that $\ket{\Omega}$ is separating, it follows that $A=0$. Hence $\ket{(B_{\vec{t}}^{(n_F)})^{1/2}}$ is \textbf{separating}.

\paragraph{Relative modular flows} 
Having established the required properties of the ETW brane states \eqref{eqn:state-Omega_nF}, we can now consider relative modular flows induced by pairs of states $\{\omega_0, \omega^{(n_F)}\}$ on the algebra $\mathcal{A}$. We will first state the general definitions, following \cite{Liu:2025krl}. For each pair $\{\omega_0, \omega^{(n_F)}\}$ we can define a relative Tomita operator $S_{B|\Omega}^{(n_F)}$:
\begin{flalign}\label{eq:relative ops}
\begin{aligned}
     &S_{B|\Omega}^{(n_F)} A\ket{\Omega} = A^\dagger Z_{n_F}^{-1/2}\ket{(B_{\vec{t}}^{(n_F)})^{1/2}}, \qquad (S_{B|\Omega}^{(n_F)})^\dagger A'\ket{\Omega} = (A')^\dagger Z_{n_F}^{-1/2}\ket{(B_{\vec{t}}^{(n_F)})^{1/2}},\\
    &\forall A \in\mathcal{A}, \quad \forall A'\in \mathcal{A}'.
\end{aligned}
\end{flalign}
$S_{B|\Omega}^{(n_F)}$ is anti-linear and satisfies $S_{B|\Omega}^{(n_F)} = (S_{\Omega|B}^{(n_F)})^{-1}$.
The Tomita operators admit a polar decomposition from which we can define:
\begin{flalign}
    S_{B|\Omega}^{(n_F)} = J_{B|\Omega}^{(n_F)} (\Delta_{B|\Omega}^{(n_F)})^{1/2}, \qquad
    \Delta_{B|\Omega}^{(n_F)} = (S_{B|\Omega}^{(n_F)})^\dagger S_{B|\Omega}^{(n_F)},
\end{flalign}
where the relative conjugation operator $J_{B|\Omega}^{(n_F)}$ is anti-unitary and the relative modular operator $\Delta_{B|\Omega}^{(n_F)}$ is positive, self-adjoint. We can also define the relative modular Hamiltonians:
\begin{flalign}
    K_{B|\Omega}^{(n_F)} \equiv - \log \Delta_{B|\Omega}^{(n_F)}
\end{flalign}
Let us now establish these quantities in terms of operators in the double-scaled algebra \eqref{eq:algebra}. 
By definition, the relative modular operator is:
\begin{flalign}
    \Delta_{B|\Omega}^{(n_F)}= (S_{B|\Omega}^{(n_F)})^\dagger S_{B|\Omega}^{(n_F)}=Z_{n_F}^{-1}\sqrt{B_{\vec{t}}^{(n_F)}(B_{\vec{t}}^{(n_F)})^\dagger}. \label{eqn:SB}
\end{flalign}
$(B_{\vec{t}}^{(n_F)})^{1/2}$ has a polar decomposition as $(B_{\vec{t}}^{(n_F)})^{1/2}=u_{B|\Omega}^{(n_F)}\sqrt{|B_{\vec{t}}^{(n_F)}|}$, where $u_{B|\Omega}^{(n_F)}$ is some unitary operator. We now have:
\begin{flalign}
    (\Delta_{B|\Omega}^{(n_F)} )^{1/2} &= Z_{n_F}^{-1/2}\qty(B_{\vec{t}}^{(n_F)}(B_{\vec{t}}^{(n_F)})^\dagger)^{1/4} = Z_{n_F}^{-1/2} u_{B|\Omega}^{(n_F)}\qty(B_{\vec{t}}^{(n_F)})^{1/2}(u_{B|\Omega}^{(n_F)})^\dagger\\
    S_{B|\Omega}^{(n_F)} &= J_{\Omega} Z_{n_F}^{-1/2} \sqrt{(B_{\vec{t}}^{(n_F)})^\dagger} = J_\Omega Z_{n_F}^{-1/2}(u_{B|\Omega}^{(n_F)})^\dagger (\Delta_{B|\Omega}^{(n_F)} )^{1/2} \label{eqn:SJ},
\end{flalign}
which implies $J_{B|\Omega}^{(n_F)} = J_\Omega (u_{B|\Omega}^{(n_F)})^\dagger$. Since $B_{\vec{t}}^{(n_F)}$ is positive definite, it follows from \eqref{eqn:SB} and \eqref{eqn:SJ} that:
\begin{flalign}
    J_{B|\Omega}^{(n_F)} = J_{\Omega}, \qquad \Delta_{B|\Omega}^{(n_F)} = Z_{n_F}^{-1} B_{\vec{t}}^{(n_F)} = \frac{B_{\vec{t}}^{(n_F)}}{\bra{(B_{\vec{t}}^{(n_F)})^{1/2}}\ket{(B_{\vec{t}}^{(n_F)})^{1/2}}},
\end{flalign}
and we also have $\Delta_{\Omega|B}^{(n_F)} = (\Delta_{B|\Omega}^{(n_F)})^{-1} = Z_{n_F}(B_{\vec{t}}^{(n_F)})^{-1}$. Therefore, the relative modular Hamiltonian is given by:
\begin{flalign}\label{eqn:rModH}
    K^{(n_F)}_{B|\Omega} = -\log \Delta_{B|\Omega}^{(n_F)} = -\log \frac{B_{\vec{t}}^{(n_F)}}{\bra{(B_{\vec{t}}^{(n_F)})^{1/2}}\ket{(B_{\vec{t}}^{(n_F)})^{1/2}}}.
\end{flalign}
Operators $A\in\mathcal{A}$ and $B\in\mathcal{A}'$ experience modular evolution governed by $B_{\vec{t}}^{(n_F)}$, in modular time $s$:
\begin{flalign}
    &\sigma_s^{(n_F)}(A) = (\Delta_{B|\Omega}^{(n_F)})^{-is} A (\Delta_{B|\Omega}^{(n_F)})^{is} = (B_{\vec{t}}^{(n_F)})^{-is/2}A(B_{\vec{t}}^{(n_F)})^{is/2} \\&
    {\sigma'}_s^{(n_F)}(B) = (\Delta_{\Omega|B}^{(n_F)})^{-is} B (\Delta_{\Omega|B}^{(n_F)})^{is} = (B_{\vec{t}}^{(n_F)})^{is/2}A(B_{\vec{t}}^{(n_F)})^{-is/2} .
\end{flalign}
The relative modular operator between $\omega^{(n_F)}$ and $\omega_0$ may be expressed using the corresponding density operators as:
\begin{equation}
\Delta_{B|\Omega}^{(n_F)}=\rho^{(n_F)} \rho_\Omega^{-1}.
\end{equation}
$\ket{\Omega}$ is the tracial state, therefore $\rho_{\Omega}=\mathbb{1}$. We thus recover the same expression as in \eqref{eqn:rModH}.

It would be interesting to examine the relative modular Hamiltonian in the triple-scaling limit. As discussed in Sec.~\ref{ssec:triple scaling limit}, in the triple-scaling limit, the bulk Hamiltonian becomes that of a particle in a Morse potential:
\begin{equation}\label{eq:H nF mod}
\hH_{n_F}=\lambda^{2}(\hat{P}^2 + U(\hat{L}))~,
\end{equation}
where $U(L)=\nu\rme^{-L}+(1/4)\rme^{-2L}$, which in the ordinary DSSYK case ($n_F=0$) reduces to the Liouville potential \eqref{eq:triple scaling IR}. 

In the triple-scaling limit, the relative modular Hamiltonian \eqref{eqn:rModH} does not generate ordinary Schwarzian time evolution. Since it is expressed via the spectral dressing factor $B_{\vec{t}}^{(n_F)}$, in its leading semiclassical form we thus expect the relative modular Hamiltonian to encode the deformation from the Liouville potential $U(\hat{L})\vert_{n_F=0}$ to the Morse potential $U(\hat{L})$ associated with the corresponding ETW brane state. We leave this as a topic of future study.

\paragraph{Entropy of the ETW brane states}
The relative modular structure also naturally defines a notion of entropy for the ETW brane states, relative to the the tracial state $\omega_0$. This is the Araki relative entropy:
\begin{flalign}\label{eq:araki special}
    S_{\mathcal{A}}\bigg((Z_{n_F}^{-1} B_{\vec{t}}^{(n_F)})^{1/2}\;||\;\Omega\bigg) = Z_{n_F}^{-1}\bra{(B_{\vec{t}}^{(n_F)})^{1/2}}\log\Delta_{B|\Omega}^{(n_F)}\ket{(B_{\vec{t}}^{(n_F)})^{1/2}}.
\end{flalign}
By using properties of the tracial state $\ket{\Omega}$, the Araki relative entropy can be rewritten as:
\begin{flalign}
    S_{\mathcal{A}}\bigg((Z_{n_F}^{-1} B_{\vec{t}}^{(n_F)})^{1/2}\;||\;\Omega\bigg) &= 
    \bra{\Omega} \bigg(\frac{B_{\vec{t}}^{(n_F)}}{Z_{n_F}}\bigg) \log\bigg(\frac{B_{\vec{t}}^{(n_F)}}{Z_{n_F}}\bigg)\ket{\Omega} .
\end{flalign}
The entropy \eqref{eq:araki special} is known to have properties like positivity and convexity under restrictions to subalgebras of $\mathcal{A}$. Positivity directly follows from the inequality $x\log x\geq x-1$, implying:
\begin{flalign}
    S_{\mathcal{A}}\bigg((Z_{n_F}^{-1} B_{\vec{t}}^{(n_F)})^{1/2}\;||\;\Omega\bigg) \geq \bra{(Z_{n_F}^{-1}B_{\vec{t}}^{(n_F)})^{1/2}}\ket{(Z_{n_F}^{-1}B_{\vec{t}}^{(n_F)})^{1/2}} - \bra{\Omega}\ket{\Omega} = 0,
\end{flalign}
where the last equality holds, since $\ket{\Omega}$ and $\ket{(Z_{n_F}^{-1}B_{\vec{t}}^{(n_F)})^{1/2}}$ are normalised so that $\omega_0(\mathbb{1})=1$ and $\omega^{(n_F)}(\mathbb{1})=1$.

\subsection{Algebraic entanglement entropy}\label{ssec:algebraic EE}

We are interested in evaluating the entanglement entropy between the q-Askey double-scaled algebra \eqref{eq:q-Askey algebra} and its commutant algebra for a given pure state. We discuss the notion of algebraic entanglement entropy in general before specializing to our case of interest. Consider two generic algebras $\widehat{\mathcal{A}}_{L/R}$ which are commutants of each other over the GNS Hilbert space, i.e., $\widehat{\mathcal{A}}_L \vee \widehat{\mathcal{A}}_R = B(\mathcal{H})$ and:
\begin{equation}
    [\hat{w}_L,\,\hat{w}_R]=0~,\quad\forall\hat{w}_{L/R}\in\widehat{\mathcal{A}}_{L/R}~.
\end{equation}
We are interested in type II$_1$ algebras which incorporate a tracial state, induced by the vector $\ket{\omega}$, for the left/right algebras\footnote{Note that the algebra trace is defined to recover finite expectation values for trace class operators in the algebra, while the Hilbert space trace would lead to a generically diverging answer \cite{Chandrasekaran:2022cip,Witten:2021unn}.}
\begin{equation}\label{eq:algebra trace}
    \Tr_{{L/R}}[\hat{w}_{L/R}]=\bra{\omega}\hat{w}_{L/R}\ket{\omega}~,
\end{equation}
where $\Tr_{{L/R}}(\hat{w}_{L/R}^\dagger \hat{w}_{L/R})\geq0$, recalling that the definition of the trace is \emph{unique up to a non-zero constant rescaling,} corresponding to a different normalization for $\ket{\omega}$.

Based on the above properties, one can define a density matrix $\hrho_{L/R}\in\widehat{\mathcal{A}}_{L/R}$ associated to a state in the GNS Hilbert space $\ket{\Psi}\in\mathcal{H}$ \eqref{eq:H ffull}, namely:
\begin{equation}\label{eq:rho from state definition}
    \bra{\Psi}\hat{w}_{L/R}\ket{\Psi}=\bra{\omega}\hat{\rho}_{L/R}\hat{w}_{L/R}\ket{\omega}~,\quad\forall\hat{w}_{L/R}\in\widehat{\mathcal{A}}_{L/R}~.
\end{equation}
After suppressing indices, the von Neumann entropy of the density matrices $\hat{\rho}\in\widehat{\mathcal{A}}$, is defined using the algebraic trace \eqref{eq:algebra trace} as
\begin{equation}\label{eq:algebraic entropy}
\begin{aligned}
        S(\hrho)&\equiv\log\Tr\hrho-\frac{\Tr(\hrho\log\hrho)}{\Tr\hrho}
        =\log\bra{\omega}\hrho\ket{\omega}-\frac{\bra{\omega}\hrho\log\hrho\ket{\omega}}{\bra{\omega}\hrho\ket{\omega}}~,
\end{aligned}
\end{equation}
where $\hrho$ is not necessarily normalised; one can choose to normalise it ${\bra{\omega}\hrho\ket{\omega}}=1$ to simplify the above expression as $\Tr(\hrho\log\hrho)$. 

Given that $\ket{\Psi}$ \eqref{eq:rho from state definition} is a pure global state, it follows that $S(\hrho_L)=S(\hrho_R)$. This means that \eqref{eq:algebraic entropy} corresponds to entanglement entropy between the algebras $\widehat{\mathcal{A}}_L$ and $\widehat{\mathcal{A}}_R$ for a given chord state. For this reason, we will denote \eqref{eq:algebraic entropy} as \emph{algebraic entanglement entropy}. The above arguments are model-independent; \eqref{eq:algebraic entropy} corresponds to entanglement entropy between algebras as long as $\hrho_{L/R}\in\widehat{\mathcal{A}}_{L/R}$ with $\widehat{\mathcal{A}}_{L}=(\widehat{\mathcal{A}}_{R})'$ is obtained from a pure global state $\ket{\Psi}$ through \eqref{eq:rho from state definition}.
From now on, we will suppress indices $L/R$ unless explicitly stated. 

To be more specific about $\hrho$; any state $\ket{\Psi}\in\mathcal{H}$ can be constructed from the cyclic separating state $\ket{\omega}$ through the GNS construction by an iterative application of the operators in either of the algebras $\widehat{\mathcal{A}}_{L/R}$. For illustration, let us then denote the state \begin{equation}
    \ket{\Psi}=f(\hat{a}_i)\ket{\omega}\in\mathcal{H}_{\rm chord}~,
\end{equation}
where $\hat{a}_i\in\widehat{\mathcal{A}}$, and $f$ is a polynomial function. Then, we can identify the corresponding reduced density matrix from \eqref{eq:rho from state definition} as
\begin{equation}
    \hrho_{L/R}=f(\hat{a}_i)f(\hat{a}_i)^\dagger~,
\end{equation}
and its von Neumann entropy is just \eqref{eq:algebraic entropy}.

\subsection{Entanglement from the chord number state}\label{ssec:EE from chord number}
Let us specialise to the double-scaled q-Askey algebra $\mathcal{A}_{\rm DS}^{n_F}$ \eqref{eq:q-Askey algebra} and its commutant $\mathcal{A}_{\rm DS}^{n_F}{}'$ described in Sec.~\ref{ssec:qA DSA} to evaluate the algebraic entanglement entropy for a given pure state in the full chord Hilbert space, and compare it to a geometric quantity in Sec.~\ref{ssec:bulk AEE}. We now study the IR/UV triple-scaling limits of the entanglement entropy between the algebras associated to a chord number state in the ETW brane system (described in \cite{Aguilar-Gutierrez:2025hty}) in the corresponding Krylov basis
\begin{equation}
    \ket{\Psi}\equiv\ket{n}=f_n\qty(\sqrt{1-q^2}\hH_{n_F})\ket{\Omega}~,\quad \hrho=f_n\qty(\sqrt{1-q^2}\hH_{n_F})^2
\end{equation}
where $f_n(\hH)$ obeys the recurrence relation \eqref{eq:lanczos new}.
\begin{equation}
    b_{n+1}f_{n+1}(x)=\hH_{n_F}f_n(x)-b_nf_{n-1}(x)-a_nf_n(x)~.
\end{equation}
Applying the resolution of the identity in the energy basis \eqref{eq:recurrence discrete} results in the following representation of algebraic entanglement entropy \eqref{eq:algebraic entropy},
\begin{equation}\label{eq:algebraic simplified entropy}
\begin{aligned}
    S&=\begin{cases}
     -\int \rmd\theta p_n(\theta)\log\frac{
    p_n(\theta)}{\mu_{n_F}(\theta)}~,&{\rm Cond_1=True}\\
    -\int \rmd\theta p_n(\theta)\log\frac{
    p_n(\theta)}{\mu_{n_F}(\theta)}+\sum_{l=0}^{l_{\rm max}}p_n(\theta_l)\log\frac{p_n(\theta_l)}{\tilde{w}_l}~,&{\rm Cond_2=True}
    \end{cases}\\
    p_n&\equiv \mu_{n_F}(\theta)f_n(2\cos\theta)^2.
    \end{aligned}
\end{equation}
where $\tilde{w}_l=w_l/h_0$ as in \eqref{full partition function}.

To evaluate the integral \eqref{eq:algebraic simplified entropy} we use the triple-scaling limit of the $n_F=2,4,6,8$ chord Hamiltonians, i.e.
\begin{equation}\label{eq:IR Hamiltonian}
    \hH_{\rm IR}\equiv\eval{\frac{\sqrt{1-q^2}\hH_{n_F}-2}{\lambda^2}}_{\mathcal{O}(1)}={\hP^2}+U(\hat{L})~,
\end{equation}
where in the $n_F=2,4,6,8$ triple-scaled Hamiltonians, the potential has the form
\begin{equation}\label{eq:U ell}
    U(L)\equiv \nu\rme^{-L}+\frac{1}{4}~\rme^{-2L}~,
\end{equation}
where $L$ and $\nu$ appear in \eqref{eq:ETW brane dictionary substitution} in terms of the deformation parameters. The explicit form of the potential is irrelevant for now.

\paragraph{Continuous spectrum contribution}
We first focus on the continuous energy spectrum of the deformed theories, i.e.~when the condition \eqref{eqn:Cond-continuous} is satisfied. In the limit \eqref{eq:IR Hamiltonian}, then the algebraic entropy \eqref{eq:algebraic simplified entropy} becomes
\begin{equation}\label{eq:S to the IR}
   S\underset{IR}{\rightarrow}2\lambda\int _0^\infty\rmd k~p(L,k)\log{\frac{p(L,k)}{\mu_{n_F}(\lambda k)}}~,
\end{equation}
where
\begin{equation}\label{eq:constructin pell k}
    p_n(\theta)\underset{IR}{\rightarrow} p(L,k)~,\quad \int _0^\infty\rmd k~p(L,k)=1~,
\end{equation}
and we are using $k\equiv\theta/\lambda$ as in past sections. In this approximation,
\begin{equation}\label{eq:eigen k val}
    \hH_{\rm IR}\ket{\theta=\lambda k}= k^2 \ket{\theta=\lambda k}~.
\end{equation}
{The wave equation \eqref{eq:eigen k val} expressed as
\begin{equation}
(-\partial_L^2+U(L))\psi(L,k)=k^2\psi(L,k)~,
\end{equation}
are Whittaker functions $\psi(L,k)\propto\rme^{-L}W_{-\nu,\rmi k}(\rme^{-L})^2$, where $p(L,k)\equiv \psi(L,k)$ with normalization such that \eqref{eq:constructin pell k} is satisfied.}

Since $p(\ell,k)\sim\mathcal{O}(1)$ as $\lambda\rightarrow0$ in \eqref{eq:constructin pell k}, while
\begin{equation}\label{eq:mu_nF}
    \lambda\log\mu_{n_F}(\lambda k)=-\frac{1}{2}\pi^2+\sum_{i=1}^{4}\qty(2{\rm Li}_{2}\qty(t_i)+\sum_{j=1}^4{\rm Li}_2(t_it_j))-{\rm Li}_2(t_1t_2t_3t_4)+\mathcal{O}(\lambda\log\lambda)~,
\end{equation}
where we have expanded the energy measure $\mu_{n_F}(\theta)$ in \eqref{eq:mu nF first} using
\begin{equation}\label{eq:entropy contribution from ETW brane}
\begin{aligned}
    &\frac{(t_1t_2t_3t_4;q^2)_\infty}{(t_1\rme^{\pm\rmi\theta},\dots, t_{4}\rme^{\pm\rmi\theta};q^2)_\infty\prod_{i\leq j\leq 4}(t_it_j;q^2)_\infty}\\
    &\underset{\lambda\rightarrow0}{=}\exp\qty(\frac{1}{\lambda}\qty(\sum_{i=1}^{4}\qty(\sum_{\epsilon=\pm}{\rm Li}_{2}\qty(t_i\rme^{\epsilon\rmi\theta})+\sum_{j=1}^4{\rm Li}_2(t_it_j))-{\rm Li}_2(t_1t_2t_3t_4)))~,
\end{aligned}
\end{equation}
with ${\rm Li}_2(x)$ as the dilogarithm function, and we assumed that $t_i=a_iq^\Delta_i$ with $a_i$ and $\Delta_i\sim\mathcal{O}(1)$ as $\lambda\rightarrow0$.

Then, \eqref{eq:S to the IR} gives
\begin{equation}\label{eq:interm S}
    S\underset{IR}{\rightarrow}2\int _0^\infty\rmd k~\qty(-\frac{\pi^2}{2}+\qty(\sum_{i=1}^{4}\qty(2{\rm Li}_{2}\qty(t_i)-\sum_{j=i}^4{\rm Li}_2(t_it_j))+{\rm Li}_2(t_1t_2t_3t_4)))p(L,k)+\mathcal{O}(\lambda\log\lambda)~.
\end{equation}
Using the WKB approximation ($P^2=0$), there is a critical value of the wavenumber $k$ in \eqref{eq:eigen k val}:
\begin{equation}
    k_{\rm crit}=\sqrt{U(L)}~.
\end{equation}
The probability distribution in this approximation is
\begin{equation}\label{eq:approx p}
    p(\ell,k)=\begin{cases}
        A\exp\qty(-2\int\rmd L~\tilde{P})~,&k\lesssim k_{\rm crit}\\
        B\sin^2(\int\rmd L~ P)~,&k>k_{\rm crit}~,
    \end{cases}
\end{equation}
where $A/B$ are normalization constants such that \eqref{eq:constructin pell k} is respected, namely
\begin{equation}
    \begin{aligned}
        A&=\frac{2}{\pi}\sqrt{\frac{k^2}{k^2-U(L)}}~,\quad B=\frac{1}{2\pi}\sqrt{\frac{k^2}{U(L)-k^2}}~,
    \end{aligned}
\end{equation}
and $P$ and $\tilde{P}$ the momenta in the allowed and disallowed regions of the corresponding Schrödinger equations
\begin{equation}
    P\equiv\sqrt{k^2-U(L)}~,\quad \tilde{P}\equiv \sqrt{U(L)-k^2}~.
\end{equation}
In the WKB approximation, we consider that the wavenumber obeys a hierarchy with respect to the critical value
\begin{equation}\label{eq:condition vanishing}
\begin{aligned}
    U(L)&\gg k^2~,\quad k\lesssim k_{\rm crit}~,\\
    U(L)&\ll k^2~,\quad k> k_{\rm crit}~.
\end{aligned}
\end{equation}
We then recover the probability distribution \eqref{eq:approx p} as
\begin{equation}
    p(L,k)\sim\begin{cases}
    0&k\lesssim k_{\rm crit}~,\\
       \frac{2}{\pi}\sin^2(L k)~,&k>k_{\rm crit}~.
    \end{cases}
\end{equation}
where the vanishing of the distribution for the interval $k\in[0,~k_{\rm crit}]$ follows from the exponential suppression \eqref{eq:approx p} when the potential energy term obeys the condition \eqref{eq:condition vanishing} in the leading order analysis of the WKB approximation.

The algebraic entanglement entropy between the algebras for the chord number state in the Krylov basis in the triple-scaling limit \eqref{eq:interm S} becomes:
\begin{equation}
\begin{aligned}
    S&=S_0+\qty(\pi-\frac{2}{\pi}\qty(\sum_{i=1}^{4}\qty(2{\rm Li}_{2}\qty(t_i)-\sum_{j=i}^4{\rm Li}_2(t_it_j))+{\rm Li}_2(t_1t_2t_3t_4)))\int _0^{k_{\rm crit}}~\rmd k~,\\
    S_0&\equiv\qty(\frac{2}{\pi}\qty(\sum_{i=1}^{4}\qty(2{\rm Li}_{2}\qty(t_i)-\sum_{j=i}^4{\rm Li}_2(t_it_j))+{\rm Li}_2(t_1t_2t_3t_4))-\pi)\int _0^{\infty}\rmd k~.
\end{aligned}
\end{equation}
Here, $S_0$ corresponds to a renormalization of the thermodynamic entropy in the IR limit. Note that since $t_i$ is an overall constant factor evaluated in the triple-scaling limit, and traces are defined up to a non-zero overall rescaling, then we can recast the entropy difference
\begin{equation}\label{eq:S difference}
    S-S_0\underset{IR}{\rightarrow}\frac{2\pi}{\lambda}\sqrt{U(L)}\propto \sqrt{\frac{1}{4}\rme^{-2L}+\nu ~\rme^{-L}}~,
\end{equation}
where we used \eqref{eq:U ell} and the overall rescaling in the trace is selected based on the relation between $\lambda$ in the DSSYK model and $G_N$ in sine dilaton gravity \cite{Blommaert:2024ymv} or JT gravity \cite{Lin:2022rbf}, where $\lambda=8\pi G_N$. 

There is a similar behaviour in the adjoint case: in the triple-scaling \eqref{eq:adj H opt 2}, we have that
\begin{equation}
    S-S_0\propto\rme^{-L/2}~,
\end{equation}
which is the holographic entanglement entropy in the JT gravity \cite{Tang:2024xgg}, as can be seen from  \eqref{eq:S difference} in the tensionless limit.

\paragraph{Including discrete spectrum}
The equation of motion for the discrete spectrum
\begin{equation}
\qty(-\partial_L^2+U(L))\psi_{k_l}(L)=-k_l^2\psi_{k_l}(L)~,
\end{equation}
where we included the Morse potential $U(L)$ in \ref{eq:morse potential} and $k_l^2=(\nu+l+1/2)$. The solutions $\psi_{k_l}(L)$, can be used to write a probability distribution $p_{k_l}(L)\equiv\abs{\psi_{k_l}(L)}^2$, with normalization
\begin{equation}
\int _{-\infty}^{\infty}\rmd L~p_{k_{l}}(L)=1~,
\end{equation}
We can then evaluate the discrete part of the algebraic entanglement entropy, which in the triple-scaling limit becomes
\begin{equation}\label{eq:final discrete piece}
S_{\rm disc}=\sum_{l=0}^{l_{\rm max}}p_{k_l}(L)\log\frac{p_{k_l}(L)}{w_{k_l}(\nu)}~,
\end{equation}
where $l_{\rm max}$ \eqref{eq:ceiling} in the triple-scaling limit becomes \cite{Rajgadia:2026ask} $l_{\rm max}\rightarrow\lfloor-\nu-1/2\rfloor$, and the weight $\tilde{w}_l$ in \eqref{eq:algebraic simplified entropy} in the Schwarzian regime becomes $w_{k_l}(\nu)\equiv-(2(\nu+l)+1)/(l!\Gamma(-2\nu-l))$.

We note that \eqref{eq:final discrete piece} gives an increase in the algebraic entanglement entropy with respect to the continuous contribution, indicating an increase in entropy once \eqref{eq:dissrete cond 1} is satisfied. This can be understood in our analysis of Sec.~\ref{ssec:sine dilaton gravity int}, where the discrete contributions are associated with black holes in sine dilaton gravity with an ETW brane. They contribute to the total black hole entropy for a given geodesic length $L$ in the bulk. However, as we will find below, a sharp bulk interpretation of algebraic entanglement entropy is mostly manifest in the $\nu\rightarrow0$ limit, i.e. when there are no discrete energy levels, we will focus on the present bulk/boundary comparison for the continuous part of the energy spectrum.

\subsection{The bulk picture of algebraic entanglement entropy}\label{ssec:bulk AEE}
At last, we seek to match the algebraic entanglement entropy \eqref{eq:algebraic entropy} to the dilaton in JT gravity evaluated on an extremal surface from the RT formula,
\begin{equation}\label{eq:bulk entropy}
    S_{\rm bulk}=\frac{\Phi(\gamma)}{4G_N}~,
\end{equation}
with $\gamma$ the extremal surface that minimises the dilaton. Note there is a modification of the RT formula when the RT surface touches ETW branes \eqref{eq:bulk entropy}.

The dilaton in global AdS coordinates is shown in \eqref{eq:metric AdS}. The extremization tells us that the RT surface is located at
\begin{equation}
    \partial_\sigma\Phi=0\implies\sigma=\pi/2~.
\end{equation}
Meanwhile, the temporal location of the asymptotic boundary can be described from the wormhole length \eqref{eq:length ETW wormhole} where \cite{Blommaert:2025avl}
\begin{equation}
    \rme^{L-L_0}=\frac{m+\sqrt{m^2+\Phi_{h}^2}\sec T}{m+\sqrt{m^2+\Phi_{h}^2}}~.
\end{equation}
This means \eqref{eq:bulk entropy} transforms into
\begin{equation}
    S_{\rm bulk}=\frac{1}{4G_N}\frac{\sqrt{m^2+\Phi_h^2}}{\qty(m+\sqrt{m^2+\Phi_h^2})\rme^{L-L_0}-m}~.
\end{equation}
Thus, we see that the evaluation of the algebraic entanglement entropy between the algebras $\mathcal{A}_{L/R}$ \eqref{eq:algebra} in the chord number state matches the codimension-two area (measured by the dilaton in JT gravity) \eqref{eq:metric AdS} when
\begin{equation}\label{eq:special regime}
    \rme^{L-L_0}\gg1~,
\end{equation}
which to leading order is
\begin{equation}
    S_{\rm bulk}\simeq\frac{\rme^{L_0}}{8G_N}\rme^{-L}~.
\end{equation}
If we assume $a~q^n\ll b~q^{2n}$, then 
\begin{equation}\label{eq:restriction general}
    S-S_0=\lambda^{-1}\rme^{-\tilde{L}}~.
\end{equation}
which corresponds to the tensionless limit of the brane.  

The reason for the apparent mismatch between the algebraic entanglement entropy and the bulk calculation for finite brane tension is that the homology constraint in the RT formula is trivialised in the AdS$_2$/CFT$_1$ limit with an ETW brane in the asymptotic boundary. This can be understood from the Lewkowycz-Maldacena \cite{Lewkowycz:2013nqa} argument when the size of the boundary subregion in a one-sided black hole vanishes, as illustrated in Fig.~\ref{fig:ETW_LM}. This contribution corresponds to the thermal surface in the BTZ description. Given an anchoring surface at the asymptotic boundary of spatial size $\abs{A}$, the holographic entanglement entropy in the BTZ description is
\begin{equation}
S^{(\rm th)}=\frac{1}{2G_3}\log\frac{2}{r_h\epsilon}\sinh\frac{r_h\abs{A}}{2}~, 
\end{equation}
where we set the AdS scale to $1$; $r_h$ the black hole horizon, and $\epsilon$ a regulator.

Thus, by the homology constraint, the extremization of the dilaton is restricted to only the asymptotic boundary, due to the zero size of the boundary region in the s-wave limit, $S^{(\rm th)}=0$. Meanwhile, when the anchoring surface has a large enough spatial extension, there is an additional RT surface connecting the ETW brane to the anchoring region \cite{Lee:2022efh}; which, however, is not present in the s-wave reduction. On the other hand, when we consider the tensionless limit (\eqref{eq:restriction general}) there exists a non-trivial RT surface which indeed matches the algebraic entanglement answer \eqref{eq:S difference} for $\nu\rightarrow0$. Thus, we find precise agreement between our previous evaluations from the bulk and boundary perspectives where $S\sim \rme^{-\tilde{L}}$. The homology constraint with the brane in this setting is thus in agreement with the s-wave reduction in the higher-dimensional picture in AdS$_3$/CFT$_2$.
\begin{figure}
    \centering
    \subfloat[]{\includegraphics[height=0.23\linewidth]{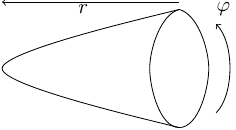}}\hspace{0.1cm}\subfloat[]{\includegraphics[height=0.3\linewidth]{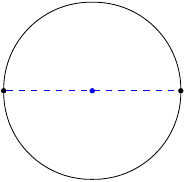}}\hspace{0.1cm}\subfloat[]{\includegraphics[height=0.3\linewidth]{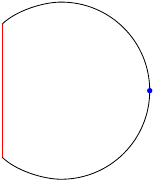}}
    \caption{Representation of the RT surfaces in Euclidean AdS$_2$. (a) As in the description by \cite{Lewkowycz:2013nqa}, in higher-dimensional AdS/CFT, we consider a boundary subregion which regenerates a conical deficit in the bulk. (b) In the lower-dimensional version, we perform the s-wave reductions, consisting of taking the boundary subregion to have zero size. This leads to a nontrivial RT surface as long as we consider two boundary anchor points that are connected through the RT surface. (c) However, once we include the ETW brane, the asymmetric point in the asymptotic boundary to anchor the RT surface stops existing, which trivialises the homology constraint, leading to a trivial RT surface in the pure JT gravity setting.}
    \label{fig:ETW_LM}
\end{figure}

\subsection{The q-Askey chord algebra (type I\texorpdfstring{$_\infty$}{})}\label{ssec:qA chord algebra}
A simple extension is to incorporate a bounded function of the total chord number operator $q^{2\hat{N}}$ in the q-Askey double-scaled algebra \eqref{eq:q-Askey algebra}. We define the resulting algebra as the q-Askey chord algebra,\footnote{The name is chosen due to its isomorphism to the ``chord algebra'' of DSSYK \cite{Lin:2023trc}.}
\begin{equation}\label{eq:def q Askey chord algebra}
    \mathcal{A}^{n_F}_{\rm chord}\equiv\langle\hH_{n_F},~\hat{\mathcal{O}}^{(R)}_\Delta,~q^{\hat{N}}\rangle''~.
\end{equation}
One can show it is a type $I_\infty$ algebra. It has been proved (see Lemma 9 and Theorem 10 in \cite{Gao:2021uro}) that
\begin{equation}
    \langle\hH_{R},~\hat{\mathcal{O}}^{(R)}_\Delta,q^{\hat{N}}\rangle=\mathcal{B}(\mathcal{H}_{\rm chord})~,
\end{equation}
with $\hat{N}$ the total chord number operator \eqref{eq:total chord number} which counts the number of operator insertions, and $\mathcal{H}_{\rm chord}$ the full chord Hilbert space \eqref{eq:H ffull}.

We can use the inverse of the isomorphism in \eqref{eq:Td inverse} to map elements in $\mathcal{A}_R$ to $\mathcal{A}_{\rm DS}^{n_F}$ by a Cauchy sequence function
\begin{equation}
    \tilde{\hat{a}}=\text{Td}(\hat{a}_R)\equiv\lim_{n\rightarrow\infty}f_n(\hH_{R},\hat{\mathcal{O}}^{(R)}_\Delta)~.
\end{equation}
This implies that \eqref{eq:def q Askey chord algebra} can be expressed as
\begin{equation}\label{eq:A chord new}
    \mathcal{A}^{n_F}_{\rm chord}=\langle\hH_R,~\hat{\mathcal{O}}^{(R)}_\Delta,~q^{\hat{N}}\rangle''~.
\end{equation}
Since the right-hand side is the chord algebra of the undeformed DSSYK model, which is a type I$_\infty$ algebra \cite{Lin:2023trc}, the same follows for $\mathcal{A}^{n_F}_{\rm chord}$.

We can understand this as follows: as the result of gaining access to the purity of the state used to define the expectation values in the microscopic theory \eqref{eq:Hamiltonia moments}, the algebra now incorporates the total chord number operator $q^{2\hat{N}}$. It is consistent with our findings on Krylov complexity as measuring a bulk length connecting the asymptotic boundary to the ETW brane in the bulk. This allows us to gain access to the full quantum state defined in the corresponding Cauchy slice. Meanwhile, the type II$_1$ algebra corresponds to having access to thermal operators in the microscopic theory, for which the system is not distinguishable from a thermal mixed state. Similar findings have been recently pointed out in other works \cite{Rajgadia:2026ask,Cao:2025pir}, which emphasise the appearance of the ETW brane theories with the KM states. Meanwhile, our work provides a different perspective, since the microscopic state is a PETS one builds from the infinite temperature TFD (presented in Sec.~\ref{sec:micro models}) by taking the zero temperature limit in one of the boundaries to generate the system describing an ETW brane (previously envisioned in \cite{Goel:2018ubv}). This perspective shows that the operator size (or equivalently $\hH_{\rm int'}$ \eqref{eq:H int}) encodes the purity of the microscopic PETS construction. This serves as an explicit example of an operator adapted to the purity of the underlying state used to define expectation values. It is also interesting to note that this is valid for a family of PETS in Sec.~\ref{ssec:KM state deformations}. The chord number is a seemingly universal feature for the different deformations of the SYK model studied in this work. This hints that the emergent wormhole length might be associated with the double-scaling limit taken in all the cases we considered.

An alternative to our discussion is to incorporate the dressed matter operators $\hat{M}_{R}$, $\hat{M}_{R}^\dagger$ of Sec.~\ref{ssec:DSSYK review} together with the chord Hamiltonian of the deformed theory, by taking its double-commutant
\begin{equation}\label{eq;chord prime algebra}
    \langle\hH^{(R)}_{n_F},\hat{M}_R,\hat{M}_R\rangle''~.
\end{equation}
This type of algebra has been recently discussed by \cite{Rajgadia:2026ask} in the undeformed DSSYK model, where one would replace $\hH_{n_F}$ in \eqref{eq;chord prime algebra} by $\hH_{n_F=0}$ \eqref{eq:H nF0}. We study this explicitly in Sec~\ref{app:PETS extension algebra}. As a result, the algebra indeed remains type I$_\infty$ and $\hat{M}_R$, $\hat{M}^\dagger_R$ take the role of $q^{2\Delta\hat{N}}$ to probe the purity of the global quantum state of the system.

\section{Discussion}\label{sec:disc}
To summarise, we uncovered new types of deformations of the SYK model which have tractable analytic properties in the double-scaling limit of the theory, and which retain a bulk hologram incorporating an ETW brane with different parameters determined by those in the boundary deformation. The microscopic theories were constructed such that they can be described by an auxiliary quantum system whose transfer matrix is solved in terms of basic hypergeometric orthogonal polynomials in the different hierarchies of the q-Askey scheme. The auxiliary quantum mechanical model is determined by requiring that annealed ensemble-averaged expectation values of SYK operators in a PETS state equal those of chord operators acting on the auxiliary Hilbert space of each deformed theory. We discovered that the different deformed auxiliary theories admit a Schwarzian limit, corresponding to a semiclassical regime at very low temperature in the canonical ensemble corresponding to the low energy sector of the theory. In the two sided picture Fig.~\ref{fig:PETS}, this means that we take $\beta_L\rightarrow\infty$ while $\beta_R\simeq\mathcal{O}(\lambda)$ as $\lambda\rightarrow0$. The chord Hamiltonians in this limit are isomorphically dual to the ADM Hamiltonians of JT gravity with an ETW brane, except for the adjoint deformation \eqref{eq:H Adj}, which is described by the Liouville Hamiltonian of JT gravity, indicating the presence of two asymptotic boundaries in the bulk dual geometry. We demonstrated that several microscopic deformations of the SYK model lead to equivalent chord Hamiltonians in the q-Askey scheme. The details of the microscopic construction, such as the PETS used to evaluate expectation, values lead to modifications in the observables of the theory.

To gain a deeper understanding of the semiclassical limit of the deformed theories at finite temperatures, we investigated the Krylov complexity of the HH state for a given deformed theory. We discovered that the convexity of Krylov complexity in time is sensitive to the deformation parameters, indicating that the chaotic behaviour in the chord theories is different from that of the undeformed DSSYK model. In many of these cases, we matched the Krylov complexity of the deformed models to the lengths of an Einstein-Rosen bridge connecting the asymptotic boundary of an AdS$_2$ black hole to an ETW brane. This permitted the decoding of a precise holographic dictionary. However, there are families of deformations not described by simple ETW branes; their precise bulk interpretation in the semiclassical limit and at finite temperatures remains an open problem. Nevertheless, they admit a late-time approximation related to ETW branes. Krylov complexity can also be phrased in terms of the quantum group of the DSSYK model and its deformations. In particular, we found that the radial displacement of a particle moving on a quantum disk with a magnetic field is described by the Krylov spread complexity of the HH state. There is an emergent SU(2) RMT described by the adjoint deformation in a specific parametric regime. In the context of Schur/SYK duality \cite{Gaiotto:2024kze,Berkooz:2025ydg,Lewis:2025qjq}, Krylov complexity describes the rate of growth of the SU(2) spin in the index of a four-dimensional $\mathcal{N}=2$ gauge theory with matter hypermultiplets. \footnote{An interesting observation is the following: we have observed that the $n_F=2,4,6$ cases describe an ETW brane in the bulk, the $n_F=8$ and adjoint case warrant a more sophisticated bulk interpretation. Via the SYK-Schur duality \cite{Gaiotto:2024kze,Berkooz:2025ydg}, there is a parallel to the $\mathcal{N}=2$ $SU(2)$ gauge theories: $n_F=2,4,6$ are asymptotically free, while $n_F=8$ and adjoint are conformal. We did not find a satisfying explanation for this match.}

Crucially, each family of the q-Askey deformed DSSYK models can describe discrete energy spectrum solutions after one of the deformation parameters reaches a critical value. These solutions describe bound-state excitations. We approached its bulk interpretation in sine dilaton gravity through Krylov complexity, which describes the length of an Einstein-Rosen bridge connecting an asymptotic boundary to the corresponding ETW brane in the bulk. In contrast to the solutions with continuous spectrum, the length encodes the quantised ADM energy of an AdS$_2$ black hole in sine dilaton gravity where the time coordinate is Wick rotated with respect to the continuous spectrum. Moreover, the discrete spectrum of the deformed theories provides a vanishing contribution to the total partition function in the semiclassical limit; their effects are non-perturbative in the parameter $\lambda$. {As expected, the discrete energy levels do not contribute to the semiclassical theory.}\footnote{{It would be interesting to cross-check our results using a diagnostic of non-classicality for the discrete spectrum contribution, such Wigner negativity \cite{Kenfack:2004ges}.}} In connection to Cauchy slice holography, by controlling the deformation parameter one obtains a transition between Lorentzian to Euclidean geometries with important differences regarding the flow in the energy spectrum in terms of the ETW brane parameters. 

We also inquired about the algebraic properties of the deformed theories. We defined the q-Askey double-scaled algebra, consisting of the deformed chord Hamiltonians and matter chord operators in the deformed theories. We show that this algebra is a type II$_1$ factor with a tracial state associated to the ETW brane in the bulk. This algebra only contains thermal information about the system. We studied the relative modular flows and entanglement entropy in the corresponding algebras. This provided new insights into the bulk theory. We evaluated the algebraic entanglement entropy with respect to the chord number state in the Schwarzian regime of the deformed theories, and we recovered a match to an extremal area in the bulk in the tensionless limit of the brane. We argued that the match is consistent with the s-wave reduction of a BTZ black hole with an ETW brane in terms of JT gravity. This serves as a new example of holographic entanglement entropy from algebraic methods. At last, we defined an additional algebra, the q-Askey chord algebra, which incorporates the total chord number. We showed that it is type I$_\infty$, and thus, it encodes the full quantum state used to construct the ensemble-averaged observables that define each auxiliary chord theory in the q-Askey deformations. This is consistent with the interpretation of the chord number operator being dual to a geodesic length, which probes the full quantum state on the Cauchy slices of the bulk dual geometry, and it serves as an example of a state-adapted operator discussed in \cite{Rajgadia:2026ask}.

We now comment on some concrete future directions.

\subsection{Open problems}\label{ssec:open problems}
\paragraph{Geometric interpretation of $n_F=8$ and beyond}
As found in Sec.~\ref{ssec:Krylov nF6}, the $n_F=6$ deformation of the DSSYK model has a bulk dual description in terms of a single ETW brane in an AdS$_2$ black hole at the semiclassical level with finite temperature. We would like to develop a modification of the sine dilaton gravity action \eqref{eq:SDG with ETW brane} which incorporates an additional ETW brane tension parameter, which can be expressed in terms of the q-Askey deformation parameters $t_{1\leq i\leq 3}$. Next, we found in Sec.~\ref{ssec:Krylov nF8} that the semiclassical limit of the $n_F=8$ and particularly the adjoint theory (Sec.~\ref{ssec:Krylov nF Adj}) needs to be described (at the semiclassical limit with finite temperature) by a geometry that goes beyond simply an ETW brane in an AdS$_2$ black hole. It would be very interesting to derive an appropriate modification of sine dilaton gravity that, after canonical quantization, leads to an ADM Hamiltonian that is isomorphically dual to the chord Hamiltonian of the deformed boundary theory, and particularly to learn what type of geometry it describes in the bulk. The same type of question can be stated for the adjoint case, which seems to describe a geometry with two asymptotic AdS$_2$ boundaries, as our findings in the Schwarzian regime point towards. We investigate this in App.~\ref{app:Ent branes}, which suggests that the geometry might be more complicated than a pair of entangled branes. It is also natural to wonder what the geometric interpretation of the q-Askey classification stopping at $n_F=8$ could be. Is there a bulk argument for why one cannot engineer a generalization in sine dilaton gravity with more ETW brane parameters? A possible expectation is that by doing so, one might generate an AdS$_2$ geometry with cusps by modifying the action.  We wish to investigate this further in future work.

It would be interesting to explore further our formulation of the q-Askey deformations based on the PETS construction in Sec.~\ref{ssec:KM state deformations} and \ref{ssec:explicit microscopic}.
{There are $2^{\mathbb{k}N/2}$ microstates given $N$ Majorana fermions, with $\mathbb{k}\in[0,~1]$, where $\mathbb{k}$ parametrizes the bipartition of the fermions. Each of the PETS corresponds to a different bulk state describing a black hole with an ETW brane in our construction after taking the double scaling limit with annealed ensemble average. Similarly, the Penington-Shenker-Stanford-Yang model \cite{Penington:2019kki} considers several numbers of ETW brane states in JT gravity coupled to a thermal bath to collect Hawking radiation. It would be interesting to use the q-Askey deformations of the SYK model to construct a toy model of black hole evaporation with a very precise boundary theoretic interpretation.} Moreover, there is a richer algebra of observables when one includes the $\hat{M}_{L/R}$ and $\hat{M}_{L/R}^\dagger$ operators. It would be interesting to extend our discussion to include the interpretation of these operators in the bulk dual.

\paragraph{Multitrace correlators and geometric wormholes}
We would like to perform further checks on the correspondence between the q-Askey deformation of the SYK model with sine dilaton gravity. In particular, it was explained in \cite{Blommaert:2025avl} that one can evaluate a double trumpet amplitude in sine dilaton gravity (which we illustrate in Fig.~\ref{fig:DT}) 
\begin{figure}
    \centering
    \subfloat[]{\includegraphics[width=0.48\linewidth]{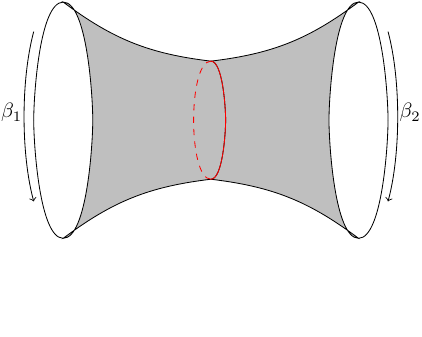}}\hfill\subfloat[]{\includegraphics[width=0.48\linewidth]{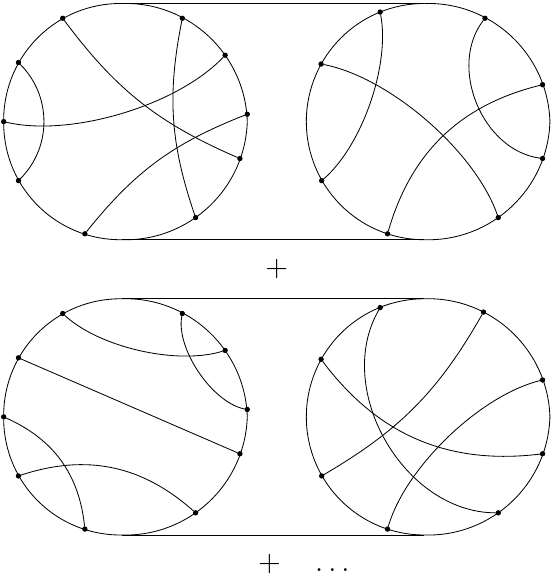}}
    \caption{(a)Double-trumpet amplitude \eqref{eq:Z beta 1 beta 2} in sine dilaton gravity, where the red solid line indicates an ETW brane with a fixed size used to glue to single trumpet amplitudes. (b) Chord diagram representation of the connected multitrace correlators between two pairs of Wick contracted deformed SYK Hamiltonians in \eqref{eq:new Z beta 12}. We suspect that in the double-scaling limit \eqref{eq:Z beta 1 beta 2} = \eqref{eq:new Z beta 12} based on the results in \cite{Blommaert:2025avl}.}
    \label{fig:DT}
\end{figure}
using the ETH matrix model of Jafferis-Kolchmeyer-Mukhametzhanov-Sonner \cite{Jafferis:2022uhu,Jafferis:2022wez}, which takes the form
\begin{equation}\label{eq:Z beta 1 beta 2}
    \tilde{Z}(\beta_1,\beta_2)\equiv\expval{\Tr(\rme^{-\beta_1\hH_{\rm ETH}})\Tr(\rme^{-\beta_2\hH_{\rm ETH}})}_{\rm conn}~,
\end{equation}
where $\hH_{\rm ETH}$ represents the Hamiltonian matrix used in \cite{Jafferis:2022uhu,Jafferis:2022wez} (see also \cite{Okuyama:2023aup,Okuyama:2023kdo,Okuyama:2023yat,Okuyama:2024eyf,Okuyama:2025hsd,Miyaji:2025ucp}) as one of the two types of matrices in the model, and the subindex `conn' represents connected contributions of two pairs of Hamiltonians in the two different traces inside the expectation value. However, App.~B in \cite{Blommaert:2025avl}  contains an alternative derivation that reproduces the same answer as \eqref{eq:Z beta 1 beta 2} which has no direct connection with the ETH matrix model; instead it relies on the canonically quantised ADM Hamiltonian of sine dilaton gravity with two ETW brane parameters, which is isomorphically dual to the chord Hamiltonian for the $n_F=4$ q-Askey deformed DSSYK model in \eqref{eq:H nF4}. However, to derive \eqref{eq:Z beta 1 beta 2}, one has to start with a different amplitude with divergent properties and extract a finite answer by performing a Fourier transform for the corresponding single trumpet amplitude, which can then be glued together along the would-be ETW brane to form the double-trumpet amplitude, as explained in App.~B in \cite{Blommaert:2025avl}. This procedure is arguably unnatural, at least from the boundary perspective. On the other hand, the natural parallel of \eqref{eq:Z beta 1 beta 2} from the microscopic $n_F=4$ deformation of the SYK model is
\begin{equation}\label{eq:new Z beta 12}
    Z(\beta_1,\beta_2)\equiv\expval{\Tr(\rme^{-\beta_1\hH_{\rm SYK}^{n_F}})\Tr(\rme^{-\beta_2\hH_{\rm SYK}^{n_F}})}_{\rm conn}~,
\end{equation}
where $\hH_{\rm SYK}^{(n_F)}$ is the corresponding deformed Hamiltonian (whose most general form is \eqref{eq:micro H nF8}), where the subindex conn has the same meaning as in \eqref{eq:new Z beta 12}. Since we find that the annealed averaged expectation values of $\hH_{\rm SYK}^{(n_F)}$ correspond to expectation values of $\hH_{n_F}$ \eqref{eq:H SYK8}, and that the $n_F=4$ case corresponds to the ADM Hamiltonian in sine dilaton gravity with ETW branes, it would be interesting to explicitly evaluate the multitrace correlator in \eqref{eq:new Z beta 12} in the double-scaling limit. In case of the ordinary SYK model (i.e.~$n_F=0$), \eqref{eq:new Z beta 12} is known to leading order in a $\begin{pmatrix}
    N\\p
\end{pmatrix}^{-1}$ expansion \cite{Berkooz:2020fvm}. We could employ similar methods as \cite{Berkooz:2020fvm,Raz:2025wjw,Rajgadia:2026ask} to perform the evaluation in the $n_F=4$ case and to perform the evaluation of \eqref{eq:new Z beta 12} in the double-scaling limit to verify if it agrees with the one in sine dilaton gravity \cite{Blommaert:2025avl}. This means that there are technical complications since the corresponding microscopic Hamiltonian also depends on the operator size, which mixes left/right fermions. This would permit connecting the two Euclidean boundaries in the double trumpet amplitude, corresponding to a wormhole geometry in the bulk. Furthermore, we have argued that even the $n_F\geq6$ deformations have a semiclassical bulk description (at finite or infinite temperatures, discussed in Secs.~\ref{ssec:triple scaling limit} and \ref{sec:Krylov Askey} respectively). It is thus important to carry out the evaluation of \eqref{eq:new Z beta 12} beyond the $n_F=4$ case, which could give hints at the expected double-trumpet amplitudes in an extension of sine dilaton gravity with ETW brane of \cite{Blommaert:2025avl}.

\paragraph{Closed baby universes from q-Askey deformations}
The Morse potential of the ETW brane theories has interesting applications to generate baby universes in JT gravity, which has been mostly explored by coupling a pair of SYK models with a Maldacena-Qi interaction term and by adding matter. 

For instance, the nucleation of baby universes has been realised through a Hawking-Page transition from a (pair of) black holes into thermal AdS$_2$ entangled with the closed baby universe in \cite{Sasieta:2025vck,Sontag:2026iiu}, while \cite{AALO} considers a dynamical process (involving additional matter) to an eternal traversable wormhole. In the latter, the role of discrete energy levels and scattering states in the Morse potential takes a more central role, like in our work. Another related work \cite{Sontag:2026iiu} recovers a chord diagram description of a system of coupled SYK models with matter and the Maldacena-Qi interaction term, which produces the baby universe through a Hawking-Page transition. The bulk interpretation relies on the low energy limit and does not incorporate the generalised Maldacena-Qi interaction term \eqref{eq:H int} that leads to the discrete spectrum \eqref{eq:discrete spectrum}.

It would be interesting to extend some of the lessons in the previous literature beyond the low energy limit, even in the semiclassical limit in the bulk, where new effects arise, as seen for instance from the bulk interpretation of the discrete spectrum states of the q-Askey Hamiltonians $\hH_{n_F}$. In particular, it would be interesting to investigate the interpretation of the discrete spectrum from the Schur half index of the SU(2) picture in \cite{Berkooz:2025ydg}, and in the semi-open channel of sine dilaton gravity with ETW branes \cite{Blommaert:2025avl}. For instance, the discrete states, seen from the Morse potential in JT gravity, correspond to negative energy states generating a traversable wormhole \cite{AALO}, one could try building a similar setting in sine dilaton gravity.

The simplest setting to analyse the discrete energy spectrum \eqref{eq:discrete spectrum} in more detail is the $n_F=2$ case \eqref{eq:H nF2}, where the deformation \eqref{eq:H int} acts as an interaction term coupling the two SYK Hamiltonians. Each one of them is associated with the generator of time translation in a two-sided black hole. Meanwhile, those presented in Sec.~\ref{ssec:explicit microscopic} correspond to the deformation for a single SYK Hamiltonian, and there is no phase transition in the single deformed SYK model, which can be modified by accounting for a second one.

Starting from the $n_F=2$ microscopic deformation of the SYK model \eqref{eq:micro nF2}, it is natural to include both the $L$ and $R$ Hamiltonians in \eqref{eq:HLR SYK}
\begin{eqnarray}\label{eq:microscopic n_F revisited}
    \hH_{\rm tot}\equiv\hH_{\rm SYK}^{(L)}+\hH_{\rm SYK}^{(R)}+\frac{t_1}{\sqrt{1-q^2}}\hH_{\rm int'}~,
\end{eqnarray}
where $\hH_{\rm int'}$ is the deformation parameter \eqref{eq:H int} which involves both left and right fermions, so that it acts as a coupling term in \eqref{eq:microscopic n_F revisited}. $\hH_{\rm tot}$ is precisely the same microscopic model proposed in \cite{Sasieta:2025vck} to investigate the transition between the black hole and closed baby universe saddle points in the gravitational path integral of the bulk dual, seen as JT gravity in the large $N$ limit. This occurs due to a Hawking-Page phase transition between the AdS$_2$ black hole with matter insertion to thermal AdS$_2$ space coupled to a closed baby universe \cite{Sasieta:2025vck}. We are interested in investigating its double-scaling limit, which captures finite $q$ effects beyond JT gravity, and then take its Schwarzian limit. In the double-scaling limit of observables  involving $\hH_{\rm tot}$ with respect to the PETS: $\ket{\hat{O}_{\rm SYK}}\equiv\hat{O}_{\rm SYK}^{(L)}\ket{\rm TFD_{\infty}}$, as described in \eqref{eq:Hamiltonia moments} and \eqref{eq:relation with chord number}, one recovers
\begin{subequations}
    \begin{align}
        \overline{\bra{\hat{O}_{\rm SYK}}P(\hH_{\rm tot})\ket{\hat{O}_{\rm SYK}}}&=\bra{\Omega}P\qty(\hH^{(L)}_{n_F=2}+\hH^{(R)}_{n_F=2})\ket{\Omega}~,\label{eq:OSYK exp val}\\
    \hH^{(L/R)}_{n_F=2}&\equiv\hat{a}_{L/R}^\dagger+\hat{a}_{L/R}+\frac{t_1 q^{\Delta_s\hat{N}}}{2\sqrt{1-q^2}}~.\label{eq:new H nF2}
    \end{align}
\end{subequations}
We find that \eqref{eq:new H nF2} under the condition $\theta_L=\theta_R$ (parametrising the spectrum \eqref{eq:cont spectrum}) and in the Schwarzian limit precisely agree with the JT gravity model presented in \cite{Sasieta:2025vck}. The $\hH_{\rm int'}$ term in \eqref{eq:microscopic n_F revisited} acts as a generalised Maldacena-Qi \cite{Maldacena:2018lmt} coupling connecting the two otherwise decoupled SYK models, generating the Morse potential. By increasing the conformal dimension of the operator $\hmO_\Delta$ in \eqref{eq:OSYK exp val} the correlation between the SYK models increases, which contributes to Morse potential. In a related work \cite{Sontag:2026iiu} a Maldacena-Qi interaction term decreases the range of propagation of the correlation due to condensation of certain chords, which results in the emission of a baby universe. The $n_F=2$ deformation, seen as an interaction term between the SYK models, would be closer to the setting presented in \cite{Sasieta:2025vck}. The $n_F\geq 4$ might lead to baby universes with more parameters, in a similar way as the ETW brane interpretation presented in Sec.~\ref{sec:Krylov Askey}.

\paragraph{Representation-theoretic interpretation}
{We now comment on concrete developments on the representation theory of quantum groups associated with the DSSYK model that might relate to our findings.}

{Recently, \cite{Belaey:2025ijg} studied gravitational amplitudes describing ETW brane states in the context of the DSSYK model. The strategy is to insert characters of specific representations of $\mathcal{U}_q(\text{sl}(2,\mathbf{R}))$ in the partition function of the trumpet amplitude.}

{Later, \cite{vanderHeijden:2025zkr} derived a map, an upgrade of the chord intertwiner in \cite{Aguilar-Gutierrez:2025mxf}, to decompose chord states with multiple particle insertions into the irreducible representations of $\mathcal{U}_q(\text{su}(1,1))$, where the coefficients in the decomposition are expressed in terms of Askey-Wilson polynomials. This might indicate a connection with the ETW brane interpretation of the deformations of the DSSYK model in our work, which is a specific type of matter (as indicated by $\hH_{\rm int'}$).}

{Meanwhile, \cite{Schouten:2025tvn} found that the Casimir operator of the $\mathcal{U}_q(\text{su}(1,2))$ quantum group in the Gauss decomposition leads to recurrence relations of the Askey-Wilson polynomials, which under appropriate limits correspond to the q-Hermite wavefunctions of the DSSYK model. In particular, the corresponding harmonic analysis of the von Neumann algebraic quantum group $\mathcal{A}_q(SU(1,1) \rtimes \mathbb{Z}_2)$ leads to an argument about the positivity of the chord number. They suggested a natural notion of ``Askey-Wilson DSSYK’’ chord theory, which is seemingly realised by our results.}

{All the above procedures point towards a natural connection between Askey-Wilson polynomials with quantum groups based on $\mathcal{U}_q(sl(2))$ with different structures, reflecting different putative bulk geometries.
It would be very interesting to connect our approach based on a microscopic model in the double-scaling limit with $\mathcal{U}_q(\text{su}(1,1))$. This quantum group and $\mathcal{U}_q(\text{sl}(2,\mathbf{R}))$ are expected to describe distinct but complementary properties of the DSSYK model \cite{Belaey:2025ijg,vanderHeijden:2025zkr}.
}

\paragraph{Quantum chaos}
{In Sec.~\ref{sec:Krylov Askey}, we uncovered different conditions where the Krylov state complexity of the HH state for the deformed chord theories is not a convex function in time. Meanwhile, it has convex growth in chaotic systems at infinite $N$ \cite{Alishahiha:2024vbf,Nandy:2024evd,Rabinovici:2025otw,Baiguera:2025dkc}. It is therefore expected that the deformations that we introduced in the microscopic models modify the chaotic properties of the SYK. In particular, the deformations presented in Sec.~\ref{ssec:explicit microscopic} involve the rescaled operator size \eqref{eq:rescaled op size}, which is bilinear in the fermion operators in the left/right subsystems, which can be used in integrable deformations of the SYK model \cite{Huh:2024ytz,Baggioli:2024wbz}. It would then be interesting to investigate the spectral statistical properties of the microscopic Hamiltonians. In particular, one could determine the spectral form factor \cite{Cotler:2016fpe}, and the distribution that the energy eigenvalue level spacing \cite{Dyson:1962es,Dyson:1962oir,wigner1993characteristic} for the deformed models obeys. In particular, we found that in the adjoint deformation with the specific value $\chi=q^{2}$ there is an emergent SU(2) RMT description encoded in the partition function of the theory, which we relate to the Shur half-index of a $\mathcal{N}=2$ gauge theory with matter hypermultiplets. It is thus natural to expect that the corresponding microscopic theory in Sec.~\ref{ssec:explicit microscopic} will also encode properties expected from RMT. It would be beneficial to perform a careful analysis of its spectral statistics to confirm our expectations. It is however, not necessary that the microscopic model obeys a RMT behaviour for the auxiliary chord system to display emergent RMT, given that they are different quantum mechanical systems.\footnote{In a similar way, there are emergent integrable chord operators resulting from dressed matter operators in the DSSYK model \cite{Rajgadia:2026ask}, even though the physical system is unrelated to the commuting SYK model \cite{Gao:2023gta} where such operators naturally arise in the double-scaling limit and with ensemble averaging \cite{Berkooz:2024evs,Berkooz:2024ofm,Almheiri:2024xtw,Gao:2024lem}.}. Another natural calculation to understand scrambling properties of the deformed models would be to evaluate OTOCs and to analyse under what conditions in terms of the deformation parameters one can obtain the Lyapunov exponent.}

\paragraph{Combining finite cutoff holography and q-Askey deformations}
Another interesting class of deformation with a natural interpretation in the holographic correspondence are T$\overline{\text{T}}$ deformations (defined in \cite{Smirnov:2016lqw,Cavaglia:2016oda}). In particular, in higher dimensions, this allows one to understand holography beyond the constraints in AdS/CFT  \cite{Jiang:2019epa,He:2025ppz,Guica:2025jkq}; for instance the deformation parameters introduces an effective length scale that changes the CFT into a more generic QFT, which results in a finite cutoff in the bulk dual theory \cite{McGough:2016lol} or in a modification of the asymptotic boundary conditions \cite{Guica:2019nzm} depending on the sign of the deformation parameter. Finite cutoff holography in lower dimensions has some advantages. In particular, when probing the high-energy spectrum of the DSSYK model, one can realise very explicitly a precise connection between static patch holography and dS$_2$/CFT$_1$ \cite{Aguilar-Gutierrez:2024oea,Aguilar-Gutierrez:2026ogo}. This procedure may also allow us to understand the holographic duality with other asymptotic boundary conditions in the bulk \cite{Aguilar-Gutierrez:2026ogo,Parvizi:2025shq,Parvizi:2025wsg,Sheikh-Jabbari:2025kjd}. Thus, it is interesting to try combining both the q-Askey deformations and those in finite cutoff holography, where a lower-dimensional version of T$\overline{\text{T}}$ is a special case. In particular, given that the deformation modifies the chord Hamiltonian \cite{Aguilar-Gutierrez:2026ogo} it will also lead to a modification of the Krylov basis and Krylov complexity of the q-Askey deformed theories in this work. It would be particularly interesting to investigate what effects the deformations in finite cutoff holography might lead to for the discrete energy spectrum solutions in Sec.~\ref{ssec:discrete spectrum}. A simple, but useful approach to this problem would be to compute geodesic lengths in AdS$_2$ space with the ETW brane and a finite cutoff in the remaining asymptotic AdS$_2$ boundary. This could be useful intuition to then evaluate the Krylov spread complexity in the deformed boundary theory, which obeys a flow equation of the form \cite{Aguilar-Gutierrez:2026ogo}
\begin{equation}
    \dv{\hH_{n_F}}{y}=\frac{(\hH_{n_F})^2+(1-\eta) y^{-2}}{2(1-y \hH_{n_F})}~,
\end{equation}
where $y$, $\eta$ are deformation parameters, while $\hH_{n_F}$ is the q-Askey chord Hamiltonian \eqref{eq:H nF8}. {Note that there is a similar flow equation that depends on the q-Askey deformation parameters $t_i$ in \eqref{eq: flow equation q Askey}.} The flow parametrised by $y$ and $\eta$ above leads to modifications in  $\hH_{n_F}$ which recover those in this work when $\eta=+1$ and $y=0$. The Krylov basis will therefore depend on the deformation parameters. We hope to investigate this in the future.

\paragraph{Extensions of generalised q-coherent states}
Previous literature on deformations of the DSSYK chord Hamiltonian found that the $\ket{B_{\vec{t}}^{(n_F)}}$ \eqref{eq:ETW brane state general ext} is a q-coherent and generalised q-coherent states when $n_F=2$ \cite{Okuyama:2023byh} and $4$ \cite{Watanabe:2025rwp} respectively. This means that the ETW brane state is an eigenstate of a linear combination of the creation and annihilation operators of the zero-particle chord Hamiltonian, where the coefficients are determined by the $t_i$ deformation parameters. It is then natural to ask if there are extended notions of the generalised q-coherent states applicable for $n_F=6,~8$ and adjoint. In App. \ref{app:extended q coherent}, we provide a criterion for extending the definition to the most general case. This leads to a recursion relation associated with the action of the creation and annihilation operators on the ETW brane state. Developing this extension of generalised q-coherent states in more detail could be useful to identify a new set of natural ETW brane states with additional matter contributions, similar to \cite{Cao:2025pir}, and to understand the precise algebraic properties of the ETW brane state $\ket{B^{(n_F)}_{\vec{t}}}$ in terms of the q-oscillator algebra.

\paragraph{Symmetry sectors}
In recent work \cite{Aguilar-Gutierrez:2025hty}, it was shown that ETW brane theories in sine dilaton gravity can be described in terms of matter chord operators in the DSSYK model without deformations by implementing constraint quantisation \cite{Henneaux:1994lbw,dirac2013lectures}. In App.~\ref{app:symmetry sectors}, we show that one might understand the cases $n_F=2,~4$ in terms of symmetry sectors in the DSSYK model with matter \cite{Aguilar-Gutierrez:2025hty}. However, the procedure becomes more involved for the $n_F\geq 6$ cases. For this reason, we view the chord Hamiltonian in all cases as the auxiliary system describing deformations of the microscopic SYK model after annealed ensemble averaging in the double-scaling limit. One recovers the same Hamiltonian for the auxiliary theory by restricting the set of states in the chord space. {This can be understood from the fact that the Hamiltonian has different representations depending on the set of states in the Hilbert space where it acts. In this way, the families of q-Askey deformed chord Hamiltonians are compatible with the undeformed DSSYK chord Hamiltonian in the extended Hilbert space with matter proposed in \cite{Lin:2022rbf}.} It would be interesting to establish a connection between the explicit microscopic SYK deformations uncovered in Sec.~\ref{ssec:explicit microscopic} with the symmetry sectors in the $n_F=2,~4$ cases, and to realise the $n_F=6,~8$ cases from constraint quantization.

\section*{Acknowledgments}
We thank Ofer Aharony, Goncalo Araujo-Regado, Xuchen Cao, Chuanxin Cui, Pawel Caputa, Gabriele Di Ubaldo, Ping Gao, Henry Lin, Takato Mori, Shoy Ouseph, Jarod Tall, Thomas Tappeiner, Herman Verlinde, Masataka Watanabe for very useful discussions; especially Micha Berkooz, Andreas Blommaert, Mikhail Isachenkov, and Jiuci Xu. We acknowledge Harshit Rajgadia and Jiuci Xu for sharing a draft of \cite{Rajgadia:2026ask}, which inspired us to study some parts of this work. SEAG thanks the high-energy group at the Weizmann Institute of Science for their hospitality and travel support, which enabled the start of this collaboration. SEAG and JS gratefully acknowledge support from the Simons Center for Geometry and Physics, Stony Brook University at which some of the research for this paper was performed during the workshop ``Double Scaled Sachdev-Ye-Kitaev Model: From Gravity to Many-Body Quantum Chaos'' and where this work was presented. SEAG acknowledges the Yukawa Institute for Theoretical Physics at Kyoto University, where part of this work was developed and presented during the ``Holographic Universe'' conference. The authors are grateful to the Mainz Institute for Theoretical Physics (MITP) of the Cluster of Excellence PRISMA+ (Project ID 390831469), for its hospitality and its partial support during the completion of this work, where it was presented during the workshop ``Quantum Chaos and Holography''. SEAG is supported by the Okinawa Institute of Science and Technology Graduate University.
The work of TK and JS is supported in part by the Israel Science Foundation grant no. 2159/22, by the Minerva foundation, and by a German-Israeli Project Cooperation (DIP) grant "Holography and the Swampland". The work of TK is supported by a research grant from Martin Eisenstein. The work of JS is supported by a Dean Award of Excellence of the Weizmann School of Science and by a Weizmann internal grant. This work was made possible through the support of the WOST, WithOut SpaceTime project (\hyperlink{https://withoutspacetime.org}{https://withoutspacetime.org}), supported by Grant ID\# 63683 from the John Templeton Foundation (JTF), and ID\#62312 grant from the JTF, as part of the ‘The Quantum Information Structure of Spacetime’ Project (QISS), as well as Grant ID\# 62423 from the JTF. The opinions expressed in this work are those of the author(s) and do not necessarily reflect the views of the John Templeton Foundation. 

\appendix

\section{Key nomenclature}\label{app:nomenclature}
\begin{itemize}[noitemsep]
    \item $p$, $N$: Number of all-to-all interactions, total number of Majorana fermions $\psi_i$, respectively.
\item $q^2\equiv\rme^{2\lambda}$, $\lambda\equiv p^2/N$: SYK Hamiltonian penalty factor.
\item $\hH_{\rm SYK}$ \eqref{eqn:H-SYK}: Undeformed SYK Hamiltonian.
\item $\Psi_I$ \eqref{eq:string Psi}: String of p-Majorana fermions
\item $\hmO_{\rm SYK}$ \eqref{eq:def SYK matter ops}: SYK matter operator.
\item DS \eqref{eq:DS limit}: Double-scaling limit, $p^2/N=$fixed.
\item $\ket{\Omega}$ \eqref{eq:H space nF=0}, $\ket{0}$ \eqref{eq:Hamiltonia moments}: Empty chord state in the undeformed and deformed theories respectively.
\item $\hH_{n_F=0}$ \eqref{eq:H SYK0}: Undeformed DSSYK chord Hamiltonian.
\item $\hat{n}$ \eqref{eq: Oscillators 1}, $\hat{N}$ \eqref{eq:total chord number}: Zero-particle sector chord number and total chord number with matter respectively.
\item $\mathcal{A}_{L/R}$ \eqref{eq:algebra}: Double-scaled algebras of the undeformed DSSYK model.
\item $\hmO_\Delta^{(L/R)}$ \eqref{eq:reference pets}: Matter chord operators in the extended chord Hilbert space.
\item $\mathcal{H}_{\rm chord}$ \eqref{eq:H ffull}: Chord Hilbert space with matter.
\item $\ket{\Delta_1,\dots,\Delta_m,n_0,\dots,n_m}$ \eqref{eq:states notation matter}: Chord number basis with matter.
\item $\hH_{L/R}$ \eqref{eqn:hH_LR}: Two-sided DSSYK chord Hamiltonian acting in the Hilbert space with matter operators.
\item $\hat{a}_{L/R}$, $\hat{a}_{L/R}^\dagger$ \eqref{eq:two_sided_DSSYK_H}: Creation and annihilation operators for $\hH_{L/R}$. If the labels L/R are omitted, it’s for $\hH$.
\item $\hat{N}$ \eqref{eq:total chord number}: total chord number operator.
\item $\hH_{\rm SYK}^{(L)}\equiv\hH_{\rm SYK}\otimes \mathbb{1}$, $\hH_{\rm SYK}^{(L/R)}\equiv\mathbb{1}\otimes \hH_{\rm SYK}$ \eqref{eq:HLR SYK}: SYK Hamiltonian acting on L/R subsystems.
\item $\varepsilon=L/R$: Label used in different algebras.
\item $n_F$ (Tab.~\ref{tab:qaskey}): Label for the family of basic hypergeometric orthogonal polynomials. $n_F=$Adj.~indicates continuous q-ultraspherical polynomials.
\item $\ket{\rm TFD_{\infty}}$ \eqref{eq:inf temp TFD}: Infinite temperature TFD state.
\item $\mathcal{H}_{n_F}$, $\ket{n}$ \eqref{eq:H space nF}: Chord Hilbert space and chord number basis of deformed DSSYK models.
\item $E(\theta)=2\cos\theta/\sqrt{1-q^2}$ \eqref{eq:cont spectrum}: Continuous spectrum of deformed DSSYK chord Hamiltonian for $0\leq\theta\leq\pi$.
\item $t_i$ \eqref{eq:dissrete cond 1}: Deformation parameters in the SYK Hamiltonian for $n_F=2,\dots,~8$, corresponding to parameters in the q-Askey scheme of polynomials. 
\item $\mu_{n_F}(\theta)$ \eqref{eq:mu nF first}: Integration measure in continuous energy basis.
\item $\ket{B^{(n_F)}_{\vec{t}}}\equiv \hat{B}^{(n_F)}_{\vec{t}}\ket{\Omega}$, $\ket{(B^{(n_F)}_{\vec{t}})^{1/2}}\equiv \qty(\hat{B}^{(n_F)}_{\vec{t}})^{1/2}\ket{\Omega}$ \eqref{eq:ETW brane state general ext}: ETW brane states.
\item $\hH_{\rm int}$ \eqref{seq:H int} and $\hH_{\rm int'}$ \eqref{seq:H intp}: Microscopic operators associated to $q^{2 \hat{n}}$ in the deformed chord Hamiltonians \eqref{eq:micro nF8} using KM and TFD states respectively.
\item $\hH_{\rm def}$ \eqref{seq:H def} and $\hH_{\rm def'}$ \eqref{seq:H defp}: Microscopic operators associated to $2\ha^\dagger$ in the deformed chord Hamiltonians \eqref{eq:micro nF8} using KM and TFD states respectively.
\item $\hat{s}$ \eqref{eq:rescaled op size}: Rescaled operator size.
\item $\hH_{\rm SYK}^{n_F}$ \eqref{eq:micro nF8} and $\hH_{\rm SYK'}^{n_F}$ \eqref{eq:micro H nF8}: q-Askey deformed SYK Hamiltonians formulated in terms of KM and TFD states respectively.
\item $\hH_{n_F}$ \eqref{eq:micro nF8}: Deformed chord Hamiltonian (i.e.~after ensemble averaging).
\item $\chi$ \eqref{eq:H SYKAdj}: Deformation parameter for the Adj.~case.
\item $L=2\lambda n+\log(2\lambda)$ \eqref{eq:def IR UV triple scaling limit}: Emergent length variable in the  Schwarzian (i.e.~ $L$ fixed as $\lambda\rightarrow0$) limits in the deformed theories.
\item $\hH_{\rm IR}$ \eqref{eq:IR H one}: Schwarzian limit of $\hH_{n_F}$.
\item $U(L)$ \eqref{eq:morse potential}: Potential in the Schrodinger equation.
\item $\nu$ \eqref{eq:morse potential}: ETW brane tension in the Morse potential
\item $\ket{s}_{\rm PETS}$ \eqref{eq:def PETS}: PETS.
\item $\hat{S}_i$ \eqref{eq:spin op}: Spin operators.
\item $\hat{M}_{I''}$, $\hat{M}^\dagger_{I''}$ \eqref{eq:def M Mdagger}: Dressed operators.
\item $C_{I''}$, $C_{I''}^\dagger$ \eqref{eq:C upper}: String of $w$ creation or annihilation spin flips.
\item $a_n$, $b_n$ \eqref{eq:lanczos special}: Lanczos coefficients.
\item $\mathcal{C}_{\rm S}$ \eqref{eq:spread Complexity}: Krylov state complexity.
\item $a(\ell)$, $b(\ell)$ \eqref{eq:classical H}: semiclassical approximation of the Lanczos coefficients.
\item $\ell\equiv \lambda\mathcal{C}_{\rm S} $ \eqref{eq:2nd ODE}: Canonical length variable used to define the semiclassical limit, i.e.~when $\ell$ fixed as $\lambda\rightarrow0$.
\item $\beta_{\rm fake}$ \eqref{eq:fake temper}: Fake temperature.
\item $E(\theta_l)=2\cosh \theta_l/\sqrt{1-q^2}$ \eqref{eq:discrete spectrum}: Discrete spectrum of deformed DSSYK chord Hamiltonian for $l\in\mathbb{Z}_{\geq0}$.
\item $\Phi$ \eqref{eq:SDG with ETW brane}: Dilaton in sine dilaton gravity or JT gravity with an ETW brane. 
\item $\mathcal{A}_{\rm DS}^{n_F}$ \eqref{eq:q-Askey algebra}: q-Askey double-scaled algebra (type II$_1$ factor).
\item $S_{\Omega}$, $J_\Omega$, $\Delta_{\Omega}$ \eqref{eq:def tomita}: Tomita operator, its modular conjugation operator and the modular operator respectively. Subindices $B|\Omega$ indicate relative operators \eqref{eq:relative ops}.
\item $K_{B|\Omega}^{(n_F)}\equiv -\log(\Delta^{(n_F)}_{B|\Omega})$ \eqref{eqn:rModH}: Relative modular Hamiltonian.
\item $\hat{\rho}_{L/R}$ \eqref{eq:rho from state definition}: Density matrices in the algebras $\mathcal{A}_{L/R}$ constructed a pure global state.
\item $S(\hat{\rho})$ \eqref{eq:algebraic entropy}: Algebraic entanglement entropy.
\item $\mathcal{A}_{\rm chord}^{n_F}$ \eqref{eq:A chord new}: q-Askey double-scaled algebra (type I$_\infty$ factor).
\item $n_{\rm tot}\equiv  \mathbb{k} N$ \eqref{eq:def K}: Number of fermions adjacent even and odd fermion pairings.
\item $\mathfrak{A}_{\rm DS}^{(\varepsilon)}$ \eqref{eq:specific U type II1}: $*$-algebras associate to the q-Askey double-scaled algebra \eqref{eq:q-Askey algebra}.
\end{itemize}

\section{More on the microscopic models}\label{app:more micro models}
Complementing our discussion of Sec.~\ref{ssec:explicit microscopic}, we state the microscopic models for the $n_F=4$ and $6$ deformations, which simply follows from \eqref{eq:micro H nF8} when $t_3=t_4=0$ and $t_4=0$ respectively.

\paragraph{\texorpdfstring{The $n_F=4$}{} deformation}
Consider the theory
\begin{align}\label{eq:H micro nF4}
    \begin{aligned}
\hH_{\rm SYK'}^{(n_F=4)}=&\frac{1}{2}\qty(1-\sum_{n=1}^\infty\frac{(2n-3)!!}{2^n\,n!}\qty(t_1t_2q^{-4}\hH_{\rm int})^m)\hH_{\rm def'}+\frac{t_1+t_2}{\sqrt{1-q^2}}\hH_{\rm int'}\\
&+\frac{1}{2}\hH_{\rm def'}^\dagger\qty(1-\sum_{n=1}^\infty\frac{(2n-3)!!}{2^n\,n!}\qty(t_1t_2q^{-4}\hH_{\rm int})^m)~,
\end{aligned}
\end{align}
where $\hH_{\rm def'}$ appears in \eqref{eq:micro H nF8}, $\hH_{\rm int'}$ in \eqref{eq:H int}. Based on the double-scaling limit in \eqref{eq:relation with chord number}, the auxiliary system in \eqref{eq:Hamiltonia moments} is:
\begin{equation}\label{eq:nF4 case 1}
    {\hH_{n_F=4}}=\sqrt{1-t_1t_2 q^{2\hat{N}-2}}\ha_R^\dagger+\ha_R\sqrt{1-t_1t_2 q^{2\hat{N}-2}}+\frac{t_1+t_2}{\sqrt{1-q^2}}q^{2\hat{N}}~.
\end{equation}
In the $\Delta\rightarrow0$ limit we have:
\begin{equation}
    \eval{\hH_{n_F=4}}_{\Delta=0}=\sqrt{1-t_1t_2 q^{2\hat{n}-2}}\ha^\dagger+\ha\sqrt{1-t_1t_2 q^{2\hat{n}-2}}+\frac{t_1+t_2}{\sqrt{1-q^2}}q^{2\hat{n}}~.
\end{equation}
This is just the $n_F=4$ Hamiltonian of \cite{Berkooz:2025ydg}, and one can verify that the eigenvalue problem of this Hamiltonian can indeed be solved in terms of Al Salam-Chihara polynomials.

\paragraph{\texorpdfstring{The $n_F=6$}{} deformation}\label{app: moremicroscopic}
We propose the following microscopic deformation of the SYK model:
\begin{equation}
\begin{aligned}
\hH_{\rm SYK'}^{(n_F=6)}=&\frac{1}{2}\prod_{1\leq i\leq j\leq3}\qty(1-\sum_{n=1}^\infty\frac{(2n-3)!!}{2^n\,n!}\qty(t_it_jq^{-2}\hH_{\rm int})^m)\hH_{\rm def'}+\frac{t_1+t_2+t_3}{\sqrt{1-q^2}}\hH_{\rm int'}\\
&+\frac{1}{2}\prod_{1\leq i\leq j\leq3}\hH_{\rm def'}^\dagger\qty(1-\sum_{n=1}^\infty\frac{(2n-3)!!}{2^n\,n!}\qty(t_it_jq^{-2}\hH_{\rm int})^m)\\
&+\frac{t_1t_2t_3}{{\sqrt{1-q^2}}}\qty(q^{-2}\hH_{\rm int'}-(1+q^{-2})\hH_{\rm int'}^2)~,
\end{aligned}
\end{equation}
such that in the double-scaling limit and after ensemble averaging, we recover
\begin{equation}\label{eq:H SYK6}
\begin{aligned}
    \hH_{n_F=6}=&\prod_{1\leq i<j\leq 3}\sqrt{1-t_it_jq^{2\hat{N}-2}}\hat{a}^\dagger_R+\hat{a}_R\prod_{1\leq i<j\leq 3}\sqrt{1-t_it_jq^{2\hat{N}-2}} \\
    &+\frac{t_1+t_2+t_3}{\sqrt{1-q^2}}q^{2\hat{N}}+\frac{t_1t_2t_3}{\sqrt{1-q^2}}(q^{2\hat{N}-2}-q^{4\hat{N}}-q^{4\hat{N}-2}) ~.
\end{aligned}
\end{equation}
When $\Delta=0$, we obtain:
\begin{equation}
\begin{aligned}
    \eval{\hH_{n_F=6}}_{\Delta=0}=&\prod_{1\leq i<j\leq 3}\sqrt{1-t_it_jq^{2\hat{n}-2}}\ha^\dagger+\ha\sqrt{1-t_it_jq^{2\hat{n}-2}}+\frac{t_1+t_2+t_3}{\sqrt{1-q^2}}q^{2\hat{n}}\\
    &+\frac{t_1t_2t_3}{\sqrt{1-q^2}}(q^{2\hat{n}-2}-q^{4\hat{n}}-q^{4\hat{n}-2}) ~,
\end{aligned}
\end{equation}
which encodes the recurrence relation for the continuous dual q-Hahn polynomials, discussed in Sec.~\ref{ssec:Krylov nF6}.

\section{Double-scaling limit of partially entangled thermal states}\label{app:more PETS}
In this appendix, we complement our discussions of PETS in the SYK model and its q-Askey deformations of Sec.~\ref{ssec:KM state deformations} with some technical details. {In particular, in App.~\ref{sapp:H int H def DS} we prove that the terms $\hH_{\rm int}$ \eqref{seq:H int} and $\hH_{\rm def}$ \eqref{seq:H def} in the ensemble averaged theory after the double-scaling limit correspond to $q^{\Delta_w\hat{n}}$ and $2\hat{a}^\dagger$ respectively, which gives rise to the Askey-Wilson transfer matrix \eqref{eq:micro nF8}.}

We can carry out the same analysis as in \cite{Rajgadia:2026ask} to deduce the chord rules in the PETS of Sec~\ref{ssec:KM state deformations}. For intermediate values of $\mathbb{k}$, we incorporate the operators $\hat{M}_{L/R},~\hat{M}_{L/R}^\dagger$ into  different polynomials of the kind \eqref{eq:full polynomial expression}. 
Consider a normalised PETS $\ket{\Phi}$ which satisfies,\footnote{The explicit representation of $\ket{\Phi}$ in the energy basis of the two-copy SYK model,
\begin{equation}
    \ket{\Phi}=\sum_{nm}\Phi_{nm}\ket{E_n}\otimes\ket{E_m}~,
\end{equation}
with $\Phi_{nm}$ some coefficients, is not required for the evaluations in this subsection. We only require that $\ket{\Phi}$ satisfies \eqref{eq:spin op}.}
\begin{equation}\label{eq:def Phi state}
    \hat{S}_k\ket{\Phi}=-\ket{\Phi}~,\quad 0\leq k\leq  n_{\rm tot}~,
\end{equation}
which we can use to evaluate expectation values. {Note that while one can evaluate chord rules for matrix elements involving combinations of operators acting on the L/R subsystems, we are interested in the emergent algebras in the double-scaling limit resulting from operators acting on a single subsystem.} Similar to \cite{Rajgadia:2026ask}, one can show that the only matrix elements that survive in the double-scaling limit \eqref{eq:double scaling L} with four operator insertions involving $\hHeps$ and $\hMeps$ correspond to,
\begin{subequations}\label{eq:PETS relations}
    \begin{align}
\bra{\Phi}\wick{\c1 \hH_{\rm SYK}^{(\varepsilon)} \c2 \hH_{\rm SYK}^{(\varepsilon)} \c1 \hH_{\rm SYK}^{(\varepsilon)} \c2 \hH_{\rm SYK}^{(\varepsilon)}} \ket{\Phi}&\underset{\rm DS}{\rightarrow} q^2~,\label{eq:1st PETS}\\
    \bra{\Phi}\wick{\c1 \hMeps \c2 \hHeps \c1 \hMeps^\dagger \c2 \hHeps} \ket{\Phi}=\bra{\Phi}\wick{\c1 \hHeps \c2 \hMeps \c1 \hHeps \c2 \hMeps^\dagger} \ket{\Phi}&\underset{\rm DS}{\rightarrow}q^{2\sqrt{\mathbb{k}}\Delta_w}~,\label{eq:2nd PETS}\\
    \bra{\Phi}\wick{\c1 \hMeps \c2 \hHeps \c2 \hHeps \c1 \hMeps^\dagger} \ket{\Phi}&\underset{\rm DS}{\rightarrow}q^{2\sqrt{\mathbb{k}}\Delta_w}~, \label{eq:3rd PETS}\\
    \bra{\Phi}\wick{\c2 \hMeps \c1 \hMeps \c1 \hMeps^\dagger \c2 \hMeps^\dagger} \ket{\Phi}=\bra{\Phi}\wick{\c1 \hMeps \c2 \hMeps \c1 \hMeps^\dagger \c2 \hMeps^\dagger} \ket{\Phi}&\underset{\rm DS}{\rightarrow}q^{2\mathbb{k}\Delta_w^2}~,\label{eq:4th PETS}
    \end{align}
\end{subequations}
where in our notation, the Wick contraction includes the ensemble average over random couplings to shorten the notation.

The relation \eqref{eq:1st PETS}, follows from expanding the expectation value as 
\begin{equation}
\bra{\Phi}\wick{\c1 \hH_{\rm SYK}^{(\varepsilon)} \c2 \hH_{\rm SYK}^{(\varepsilon)} \c1 \hH_{\rm SYK}^{(\varepsilon)} \c2 \hH_{\rm SYK}^{(\varepsilon)}} \ket{\Phi}=\begin{pmatrix}
        2N\\p
    \end{pmatrix}^{-2}\sum_{IJ}\bra{\Phi}\Psi^{(\varepsilon)}_I\Psi^{(\varepsilon)}_J\Psi^{(\varepsilon)}_I\Psi^{(\varepsilon)}_J\ket{\Phi}    
\end{equation}
and identifying that $\Psi^{(\varepsilon)}_I\Psi^{(\varepsilon)}_J\Psi^{(\varepsilon)}_I\Psi^{(\varepsilon)}_J\propto\delta_{IJ}$, so that the correlator is proportional to a trace, which gives the $q$ penalty factor due to H-chord Hamiltonian intersections. Similarly, if we included matter SYK operators \eqref{eq:def SYK matter ops} on either L/R, we recover the same chord rules as \cite{Berkooz:2018jqr}. Meanwhile, the case \eqref{eq:2nd PETS} comes from expanding 
\begin{equation}
    \bra{\Phi}\wick{\c1 \hHeps \c2 \hMeps \c1 \hHeps \c2 \hMeps^\dagger} \ket{\Phi}=\begin{pmatrix}
        n_{\rm tot}\\w
    \end{pmatrix}^{-1}\begin{pmatrix}
        2N\\p
    \end{pmatrix}^{-1}\bra{\Phi}\Psi^{(\varepsilon)}_IC^{(\varepsilon)}_I\Psi^{(\varepsilon)}_I{C^{(\varepsilon)}_I}^\dagger\ket{\Phi}~,
\end{equation}
where the leading contribution in the sum comes from $\Psi^{(\varepsilon)}_I$ that commutes with $C_I$ \cite{Rajgadia:2026ask} and resulting in a factor $\begin{pmatrix}
    2N-2w\\p-N+n_{\rm tot}
\end{pmatrix}$, which then leads to (\ref{eq:2nd PETS}) when combined with other factors. Similarly for \eqref{eq:3rd PETS} when $\Psi^{(\varepsilon)}_I$ does not common fermions with $\mathcal{C}_I$. The last relations \eqref{eq:4th PETS} similarly follow from applying the anticommutation relations to order $C^{(\varepsilon)}_I$ and its hermitian conjugate. As noticed in \cite{Rajgadia:2026ask} this product of fermions contains $2w$ annihilation operators appearing $\begin{pmatrix}
    2w\\w
\end{pmatrix}$ times. This gives an overall factor
\begin{equation}\label{eq:def K}
    \begin{pmatrix}
        n_{\rm tot}&\\
        w
    \end{pmatrix}^{-2}\begin{pmatrix}
        n_{\rm tot}&\\
        2w
    \end{pmatrix}\begin{pmatrix}
        2w\\
        w
    \end{pmatrix}\underset{\rm DS}{\rightarrow} q^{2\mathbb{k}\Delta_w^2}~.
\end{equation}
In all cases, we find that the penalty factor for intersections between $\hH_{L/R}$ with $\hat{M}_{L/R}$ and $\hat{M}_{L/R}^\dagger$ is $\mathbb{k}$ dependent, such as $q^{2\sqrt{\mathbb{k}}\Delta_w}$ 
with $\Delta_w\equiv w/p$. Note that this contribution is trivial in the TFD case $\mathbb{k}=0$, and finite otherwise.

\subsection{Double-scaling limit of \texorpdfstring{$\hH_{\rm int}$ \eqref{seq:H int} and $\hH_{\rm def}$ \eqref{seq:H def}}{}}\label{sapp:H int H def DS}
{In this subappendix, we justify that the double-scaling limit of the microscopic deformation terms  $\hH_{\rm int}$ \eqref{seq:H int} and $\hH_{\rm def}$ \eqref{seq:H def} in the q-Askey SYK models \eqref{eq:def_1} corresponds to \eqref{eq:micro nF8}. We start from the known relations in \cite{Rajgadia:2026ask} in our notation
\begin{equation}\label{eq:starting comm relation}
\lim_{\rm DS}\overline{\bra{s}P(\hH_{\rm SYK},\hM,\hM^\dagger)\ket{s}}=\bra{\Omega}P(\hat{a}+\hat{a}^\dagger,q^{\frac{\Delta_w}{2}\hat{N}}\hat{e},\hat{e}^\dagger q^{\frac{\Delta_w}{2}\hat{N}})\ket{\Omega}
\end{equation}
where the above operators obey the algebras 
\begin{subequations}\label{eq:commrelations integrable piece}
\begin{align}
[\hat{e},~\hat{e}^\dagger]=1~,\quad[\hat{e},~q^{\frac{\Delta_w}{2}\hat{N}}]_{q^{\frac{\Delta_w^2}{2}}}=0~,\quad[q^{\frac{\Delta_w}{2}\hat{N}},\hat{e}^\dagger]_{q^{\frac{\Delta_w^2}{2}}}=0~,
\end{align}
\end{subequations}
We seek to reproduce the term $q^{\Delta_w\hat{n}}$ in the q-Askey deformed chord theories \eqref{eq:micro nF8} from the double-scaling limit of the microscopic one. Combining \eqref{eq:starting comm relation} in \eqref{eq:commrelations integrable piece} we notice that
\begin{equation}\label{eq:double scaled H int}
\begin{aligned}
    \lim_{\rm DS}\overline{\bra{s}P(\hH_{\rm SYK},~\hM\hM^\dagger)\ket{s}}&=\bra{\Omega}P(\ha+\ha^\dagger,q^{\frac{\Delta_w}{2}\hat{N}}\hat{e}\hat{e}^\dagger q^{\frac{\Delta_w}{2}\hat{N}}\ket{\Omega}\\
    &=\bra{\Omega}P(\ha+\ha^\dagger,~q^{\Delta_w\hat{n}})\ket{\Omega}~,
\end{aligned}
\end{equation}
where $\hat{n}$ is the zero-particle chord number, in contrast to the total chord number $\hat{N}$ (which includes M-chords). Note that the evaluation crucially relies on the fact that $\hM\hM^\dagger$ only acts onto states in $\mathcal{H}_{0}$ \eqref{eq:zero particle H} (spanned by the basis $\ket{n}$) which only involves H-chords due to the action of $\ha+\ha^\dagger$ terms in the polynomial $P$ onto $\ket{\Omega}$. Given that $\hat{e}\ket{n}=0$, the commutation relation \eqref{eq:double scaled H int} means that $\hat{e}\hat{e}^\dagger\ket{n}=\ket{n}$, which leads to the last relation in \eqref{eq:double scaled H int}. Thus, we can add $\hM\hM^\dagger$ to generate the q-Askey chord Hamiltonians \eqref{eq:micro nF8}. However, $\hM\hM^\dagger$ is not a Hermitian operator. One can simply consider the combination $\hM\hM^\dagger+\hM^\dagger\hM$, which is Hermitian, since in the double scaling limit
\begin{eqnarray}
    \begin{aligned}
\lim_{\rm DS}\overline{\bra{s}P(\hH_{\rm SYK},~\hM\hM^\dagger)\ket{s}}&=\bra{\Omega}P(\ha+\ha^\dagger,\hat{e}^\dagger q^{\Delta_w\hat{N}}\hat{e}\ket{\Omega}=\bra{\Omega}P(\ha+\ha^\dagger,0)\ket{\Omega}
    \end{aligned}
\end{eqnarray}
due to $\hat{e}\ket{n}=0$. On the other hand, if we had considered a more general correlation function than \eqref{eq:double scaled H int} involving M-chords that do not simply appear in combinations $\hM\hM^\dagger$ and $\hM^\dagger\hM$, such as those in \eqref{eq:PETS relations}, then we would be forced to use an operator $[\hat{M},\hM^\dagger]_{q^{\Delta^2}}$ instead of $\qty{\hM,\hM^\dagger}$ to recover a term $q^{\Delta_w\hat{N}}$ in the double-scaling limit, i.e.
\begin{eqnarray}
    &\lim_{\rm DS}\overline{\bra{s}P(\hH_{\rm SYK},~[\hM,\hM^\dagger]_{q^{\Delta^2}},\hM,\hM^\dagger)\ket{s}}\\
    &=\bra{\Omega}P(\ha+\ha^\dagger,q^{\Delta_w\hat{N}},q^{\frac{\Delta_w}{2}\hat{N}}\hat{e},\hat{e}^\dagger q^{\frac{\Delta_w}{2}\hat{N}})\ket{\Omega}
\end{eqnarray}
However, a disadvantage of using $[\hM,\hM^\dagger]_{q^{\Delta^2}}$ instead of $\hH_{\rm int}\equiv\qty{\hM,\hM^\dagger}$ to define the deformed Hamiltonians $H_{\rm SYK}^{n_F}$ \eqref{eq:def_1} (or equivalently \eqref{eq:H nf8 again}) is that $\hH_{\rm int}$ is not Hermitian at finite $N$, even though its counterpart in the auxiliary quantum theory ($q^{\Delta_w\hat{n}}$) itself is Hermitian in the DS limit.\footnote{{We thank Jiuci Xu for very useful correspondence about this point.}}}

{Next, let us show that $\hH_{\rm def}$ \eqref{seq:H def} reproduces the term $2\hat{a}$ in the ensemble averaged theory  \eqref{eq:micro nF8}. First, we analyze the terms inside the commutators in \eqref{seq:H def}. We know from \eqref{eq:double scaled H int} that $\hH_{\rm SYK}$ corresponds to $\ha+\ha^\dagger$ in the ensemble averaged theory. Meanwhile, as we just showed, the term $\hM\hM^\dagger$ restricted to the zero-particle space $\mathcal{H}_0$ \eqref{eq:zero particle H} corresponds to $q^{\Delta_w\hat{n}}$. This implies
\begin{eqnarray}
    \lim_{\rm DS}\overline{\bra{s}P\qty(\hH_{\rm SYK},\sum_{I''_1I''_2}R_{I''_1}R_{I''_2}C_{I_1''}C_{I_2''}^\dagger)\ket{s}}=\bra{\Omega}P\qty(\ha+\ha^\dagger,~q^{\Delta_w\hat{n}})\ket{\Omega}~.
\end{eqnarray}
In particular in the $w=1$ limit, where $\lambda\Delta_w\rightarrow p/N$, we have that
\begin{eqnarray}\label{eq:impt eq}
\begin{aligned}
    \lim_{\rm DS}\overline{\bra{s}P\qty(\hH_{\rm SYK},~\sum_{j=1}^{N}\frac{1}{N}\qty(\psi_{2j-1}+\rmi \psi_{2j})\qty(\psi_{2j-1}-\rmi\psi_{2j}))\ket{s}}&\\
    =\bra{\Omega}P\qty(\ha+\ha^\dagger,~1-\frac{p\hat{n}}{N}+\mathcal{O}(N^{-2}))\ket{\Omega}&~,
\end{aligned}
\end{eqnarray}
where we used $\overline{R_{I_1''}R^*_{I_2''}}=N^{-1}\delta_{I_1''I_2''}$ \eqref{eq:gaussian dist R} when $w=1$. To complete the proof, we follow a similar strategy as \cite{Lin:2022rbf} (57), which relates the rescaled operator size \eqref{eq:rescaled op size} to the chord number operator. Namely, we analyze the $\mathcal{O}(N^{-1})$ terms in \eqref{eq:impt eq}
\begin{eqnarray}\label{eq:towards chord number}
    \lim_{\rm DS}\overline{\bra{s}P\qty(\hH_{\rm SYK},\sum_{j=1}^{N}\frac{1}{p}\qty(1+\frac{\rmi}{4}\qty([\psi_{2j-1},\psi_{2j}]-2)))\ket{s}}=\bra{\Omega}P\qty(\ha+\ha^\dagger,~\hat{n})\ket{\Omega}~,
\end{eqnarray}
where we have used $\qty{\psi_i,\psi_j}=2\delta_{ij}$. Our result is indeed consistent with \cite{Rajgadia:2026ask} (3.9). Now, using \eqref{eq:towards chord number} and \eqref{eq:starting comm relation}, we can now evaluate the double-scaling limit of 
\begin{eqnarray}
    \hH_{\rm def}\equiv \hH_{\rm SYK}+\frac{-\rmi}{4p}\sum_{j=1}^N\qty[\hH_{\rm SYK},[\psi_{2j},\psi_{2j-1}]]
\end{eqnarray}
in the auxiliary system as
\begin{eqnarray}\label{eq:Double scaling H def}
\begin{aligned}
    \lim_{\rm DS}\overline{\bra{s}P\qty(\hH_{\rm SYK},\hH_{\rm def})\ket{s}}&=\bra{\Omega}P\qty(\ha+\ha^\dagger,~\ha+\ha^\dagger+[\ha+\ha^\dagger,\hn])\ket{\Omega}\\
    &=\bra{\Omega}P\qty(\ha+\ha^\dagger,~2\ha^\dagger)\ket{\Omega}~,
\end{aligned}
\end{eqnarray}
where we have used \eqref{eq:property n}. Thus, $2\ha^\dagger$ is the chord operator corresponding to $\hH_{\rm def}$ \eqref{seq:H def} in the microscopic theory. While the operator $\hM\hM^\dagger$ itself is not Hermitian, the combination that appears in the deformation term $\hH_{\rm def}$ itself is Hermitian.
}

\section{Extension of the q-Askey algebras}\label{app:PETS extension algebra}
In this appendix, we show that the q-Askey double-scaled algebra (type II$_1$) and the q-Askey chord algebra (type I$_\infty$) in Sec.~\ref{sec:OP algebras} can be formulated from a limiting sequence of correlation functions evaluated on PETS of the deformed SYK models in the double-scaling limit in Sec.~\ref{ssec:KM state deformations}{ssec:explicit microscopic}. This generalises recent work \cite{Rajgadia:2026ask}, which was based on the KM state for the undeformed DSSYK model. 

\subsection{q-Askey double-scaled algebra}\label{sapp:PETS DS algebra}
In the following, we construct a pair of $*$-algebras generated by two sets of symbols $\hH^{(\varepsilon)}_{n_F}$ and $\hat{\mathcal{O}}^{(\varepsilon)}_\Delta$ where $\varepsilon=L/R$,
\begin{equation}\label{eq:specific U type II1}
\mathfrak{A}^{(\varepsilon)}_{\rm DS}\equiv\left\langle\hH_{n_F}^{(\varepsilon)},~\hmO^{(\varepsilon)}_\Delta\right\rangle_{\mathbb{C}}~,
\end{equation}
where the $\langle,~\rangle_{\mathbb{C}}$ indicates polynomials of the algebra generators with complex coefficients. In the case of physical interest where $\hH_{n_F}^{(\varepsilon)}$ and $\hmO^{(\varepsilon)}_\Delta$ are identified with a deformed chord Hamiltonian and a matter chord operator respectively,\footnote{The dual bulk algebra would be generated by the ADM Hamiltonians for the left/right boundaries, and dressed matter operators (the non-minimally coupled scalar in sine dilaton gravity in \cite{Blommaert:2024ymv,Bossi:2024ffa,Cui:2025sgy}) with respect to the boundaries. Similar to out boundary analysis, we anticipate that the expectation values would be evaluated in a sequence of quantum gravity states with finite $G_N$ corrections in sine dilaton. It would be interesting to develop this explicitly following the same arguments in this appendix.} we will assume the generators in \eqref{eq:specific U type II1} are self adjoint, i.e.~the involution operation ($^*$) acts leaves them invariant, $\hH_{n_F}^*=\hH_{n_F}$, $\qty(\hmO^{(\varepsilon)}_\Delta)^*=\hmO^{(\varepsilon)}_\Delta$. We will extend the involution to act anti-linearly and anti-multiplicatively on the generators, i.e. given $\hat{A}$ and $\hat{B}$ in a $*$-algebra, then the involution reverses scalar multiplication for a linear combination, and it reverses the order of products respectively as
\begin{equation}\label{eq:def products}
(a \hat{A}+b\hat{B})^*=a^* \hat{A}^*+b^*\hat{B}^*~,\quad (\hat{A}\hat{B})^*= \hat{B}^*\hat{A}^*~.
\end{equation}
Next, we define a linear functional $\omega$ which evaluates correlation functions from the generators of the algebras based on the operators $\hH_{\rm SYK}^{n_F(\varepsilon)}$, $\hmO_{\rm SYK}^{(\varepsilon)}$ acting on a representative PETS $\ket{\Phi_N}$ \eqref{eq:def Phi state} in the microscopic theory with $N$ finite as,
\begin{subequations}\label{eq:omega def}
    \begin{align}
\omega~&:~~\mathfrak{A}^{(\varepsilon)}_{\rm DS}\rightarrow\mathbb{C}~,\\
\omega(P)&{\equiv}\lim_{\rm DS}{\overline{\bra{\Phi}P\qty(\hH_{\rm SYK}^{n_F(\varepsilon)},\hmO_{\rm SYK}^{(\varepsilon)})\ket{\Phi}}}~,
\end{align}
\end{subequations}
{where DS denotes the double-scaling limit of the sequence of microscopic PETS,} and we introduced the polynomial function $P(\hH_{n_F},\hmO_\Delta)\in\mathfrak{A}$; and DS in the definition indicates that we evaluate the expectation value in the $N,~p\rightarrow\infty$ limit with $p^2/N$ fixed \eqref{eq:DS limit}. Note that
\begin{equation}\label{eq:DS expval}
{\omega(P^*P)}=\lim_{\rm DS}{\rightarrow}{\overline{\bra{\Phi_N}P\qty(\hH_{\rm SYK}^{n_F(\varepsilon)},\hmO_{\rm SYK}^{(\varepsilon)})^\dagger P\qty(\hH_{\rm SYK}^{n_F(\varepsilon)},\hmO_{\rm SYK}^{(\varepsilon)})\ket{\Phi_N}}}\geq0~,
\end{equation}
which follows from the non-negativity the same expectation value at finite $N$ version when the generators are self-adjoint
\begin{align}
\overline{\bra{\Phi}P\qty(\hH_{\rm SYK}^{n_F(\varepsilon)},\hmO_{\rm SYK}^{(\varepsilon)})^\dagger P\qty(\hH_{\rm SYK}^{n_F(\varepsilon)},\hmO_{\rm SYK}^{(\varepsilon)})\ket{\Phi}}\geq0~.
\end{align}
Then, it follows from \eqref{eq:def Phi state} that in the double-scaling limit, \eqref{eq:DS expval} is a positive linear functional. Next, we construct a left null ideal, which is defined as the set of polynomials $P\in\mathfrak{A}$ such that 
\begin{align}
\mathcal{I}\equiv \qty{P\qty(\hH_{n_F}^{(\varepsilon)},~\hmO^{(\varepsilon)}_\Delta)\in\mathfrak{A}_{\rm DS}^{(\varepsilon)}~:~\omega(P^*P)=0}~.
\end{align}
We construct an inner product on $\mathfrak{A}_{\rm DS}^{n_F}/\mathcal{I}$ from the linear functional in \eqref{eq:omega def}
\begin{align}\label{eq:A B boxed}
\omega(\hat{A}^*\hat{B})\equiv \lim_{\rm DS}{\overline{\bra{\Phi_N}P\qty(\hat{A}+\mathcal{I})^\dagger P\qty(\hat{B}+\mathcal{I})\ket{\Phi_N}}}
\end{align}
where $\hat{A}$, $\hat{B}\in\mathfrak{A}_{\rm DS}^{n_F}$. We can thus complete $\mathfrak{A}_{\rm DS}^{n_F}/\mathcal{I}$ with respect to the norm
\begin{eqnarray}
    \norm{\hat{A}+\mathcal{I}}\equiv \omega(\hat{A}^*\hat{A})~,
\end{eqnarray}
We can now define a representation $\pi_{\rm GNS}$ for operators $P\in\mathfrak{A}$ acting on the normal state of the algebra $\ket{0}_{\rm GNS}$, defined from \eqref{eq:A B boxed} when $\hat{A}=\mathbb{1}$, $\hat{B}=\mathbb{1}$. This generate the corresponding Hilbert space $\mathcal{H}_{\rm GNS}$, as
\begin{align}\label{eq:GNS H}
\omega(P)&={}_{\rm GNS}\bra{0}\pi_{\rm GNS}\qty(P\qty(\hH_{n_F}^{(\varepsilon)},~\hmO^{(\varepsilon)}_\Delta))\ket{0}_{\rm GNS}~,\\
\pi_\omega:&~\mathfrak{A}_{\rm DS}^{(\varepsilon)}\rightarrow\mathcal{B}(\mathcal{H}_{\rm GNS})~.
\end{align}
We can thus define a von Neumann algebra
\begin{align}\label{eq:VN new}
\mathcal{A}_{\rm DS}^{(\varepsilon)}\equiv\pi_{\rm GNS}\qty(\mathfrak{A}_{\rm DS}^{(\varepsilon)})''~,
\end{align}
where ${}''$ is the double-commutant in the set of bounded operators $\mathcal{B}(\mathcal{H}_{\rm GNS})$, as described in \eqref{eq:part 2}.

We now choose the representations in the GNS algebra $\pi_{\rm GNS}$ such that the states in $\mathcal{H}_{\rm GNS}$ are identified with those in the chord Hilbert space of the deformed chord theory \eqref{eq:H space nF}, i.e.
\begin{equation}\label{eq:const GNS}
\pi_{\rm GNS}\qty(P\qty(\hH^{(\varepsilon)}_{n_F},~\hat{\mathcal{O}}^{(\varepsilon)}_\Delta))\ket{0}_{\rm GNS} =P\qty(\hH^{(\varepsilon)}_{n_F},~\hat{\mathcal{O}}^{(\varepsilon)}_\Delta)\ket{0}~,
\end{equation}
where $\ket{0}$ is the zero-chord state of the $n_F$-deformed DSSYK Hilbert space \eqref{eq:H space nF}. \eqref{eq:const GNS} says that the states in $\mathcal{H}_{\rm GNS}$ are constructed in \eqref{eq:GNS H} by applying $\hH_{n_F}^{(\varepsilon)}$ and $\hat{\mathcal{O}}^{(\varepsilon)}_\Delta$.

In this way, the GNS Hilbert space and the cyclic state can be respectively identified with
\begin{equation}
\mathcal{H}_{\rm GNS}=\mathcal{H}_{n_F}~,\quad \ket{0}_{\rm GNS}=\ket{0}~.
\end{equation}
Therefore, in the GNS construction, \eqref{eq:VN new} is isomorphic to the q-Askey double-scaled algebra in Sec.~\ref{ssec:qA DSA} (i.e.~$\mathcal{A}_{\rm DS}^{(R)}=\mathcal{A}^{n_F}_{\rm DS}$ and $\mathcal{A}_{\rm DS}^{(L)}=\mathcal{A}^{n_F}_{\rm DS}{'}$), which means that the the operator algebra constructed by a sequence of PETS $\mathcal{A}_{\rm DS}^{(\varepsilon)}$ \eqref{eq:VN new} is a type II$_1$ factor.

\subsection{q-Askey chord algebra}\label{sapp:PETS chord algebra}
Next, we construct a different $*$-algebra involving the following generators
\begin{equation}\label{eq:new chord algebra}
\mathfrak{A}_{\rm chord}\equiv \left\langle\hH^{(R)}_{n_F},~\hat{M}_R,~\hat{M}_R^\dagger\right\rangle_{\mathbb{C}}~,
\end{equation}
where $\hat{M}_R$ and $\hat{M}_R^\dagger$ are independent operators at this level, {which will later be identified with those in \eqref{eq:def M Mdagger}}, and $\langle,~\rangle_{\mathbb{C}}$ is defined in \eqref{eq:specific U type II1}. When we identify the generators in \eqref{eq:new chord algebra} to the chord operators in Secs.~\ref{ssec:DSSYK review}, we have to require that the involution of the $*$-algebra acts as 
\begin{equation}
\hH_{n_F}^{(R)}{}^*=\hH_{n_F}^{(R)}~,\quad \hat{M}_{R}^*=\hat{M}_{R}^\dagger~,\quad \qty(\hat{M}_{R}^\dagger)^*=\hat{M}_{R}~,
\end{equation}
where we again apply the involution extension in \eqref{eq:def products}. We apply the GNS construction of the previous subsection, where we choose the representation of the generators $\pi_{\rm GNS}$ to be that of the correlation functions in the deformed DSSYK model \eqref{eq:PETS relations} such that the corresponding GNS Hilbert space can again be identified with \eqref{eq:H space nF}, with $\ket{0}$ being the cyclic state. We then construct a von Neumann algebra from \eqref{eq:new chord algebra} in the same way as \eqref{eq:VN new}, namely
\begin{equation}\label{eq:new chord enw}
\mathcal{A}_{\rm chord}\equiv \pi_{\rm GNS}(\mathfrak{A}_{\rm chord})''~.
\end{equation}
One can then follow the same steps as Sec.~4.2 in \cite{Rajgadia:2026ask} to show that \eqref{eq:new chord algebra} is a type I$_\infty$ factor, by showing that $\mathcal{A}_{\rm chord}'=c\mathbb{1}$, with $c$ a constant. The main difference with respect to \cite{Rajgadia:2026ask} is that the chord rules involve a factor $\mathbb{k}$ shown in \eqref{eq:PETS relations}. Therefore, following the previous subsection, we identify the GNS Hilbert space with the chord Hilbert space of the deformed theory \eqref{eq:H space nF}, and the corresponding von Neumann algebra in \eqref{eq:new chord enw} becomes in a type I$_\infty$, just as the one in Sec.~\ref{ssec:qA chord algebra} where the dressed operators $\hat{M}_R$ and $\hat{M}_R^\dagger$ take a similar role to that of $q^{2\hat{N}}$ in \eqref{eq:A chord new}.

\section{Two-point functions in the quantum regime}\label{app:more two-point functions}
In this appendix, we elaborate on the two-point correlation functions of Sec.~\ref{ssec:two-point functions} beyond the semiclassical level. This discussion relies on Fourier kernels for the polynomials in the q-Askey scheme. The mathematical results for the continuous energy spectrum are well known in the literature \cite{askey1996general}. We discuss the discrete spectrum solutions, which result in significant simplifications. For that, we study correlation functions in the energy basis, given by
\begin{subequations}
    \begin{align}
G_2(\tau_1,\tau_2)&=\bra{0}\rme^{-\tau_1\hH_{n_F}}q^{2\Delta\hat{n}}\rme^{-\tau_2\hH_{n_F}}\ket{0}~,\label{eq:G2 cont}\\
&=\int _0^\pi\rmd\theta\mu_{n_F}(\theta)\rmd\phi\mu_{n_F}(\phi)\rme^{-\tau_1E(\theta)-\tau_2E(\phi)}\bra{\theta}q^{2\Delta\hat{n}}\ket{\phi}~,\nonumber\\
\bra{\theta}q^{2\Delta\hat{n}}\ket{\phi}&=\sum_{n=0}^\infty q^{2\Delta n}\phi_n(\theta)\phi_n(\phi)~,\label{eq:theta part}
\end{align}
\end{subequations}
where $\hat{n}$ is the chord number operator \eqref{eq:total chord number}, and $\phi_n(\theta)\equiv \bra{n}\ket{\theta}_{n_F}$ are the rescaled Askey-Wilson polynomials \eqref{eqn:ortn-AW}. Using the explicit special functions \eqref{eqn:def-P}, the above integration can be performed exactly. In fact, some of these expressions are available in the form of Fourier kernels in \cite{askey1996general}. While general expressions exist, it is not particularly useful to illustrate them for our purposes. Below, we consider a tractable case, when $n_F=4$. The symmetric Fourier kernel used to evaluate \eqref{eq:theta part} in this case is \cite{Berkooz:2018jqr}
\begin{align}
    \label{eq:kernel-Q}
&\sum_{n=0}^{\infty} \frac{Q_n\left(\cos\theta ; t_1, t_2 \mid q^2\right) Q_n\left(\cos\phi ; t_1, t_2 \mid q^2\right)}{\left(t_1t_2, q^2 ; q^2\right)_n} z^n   \\
&=\frac{\left(t_2 z / t_1, t_1 z e^{ \pm i \theta}, t_1 z e^{ \pm i \phi} ; q^2\right)_{\infty}}{\left(t_1^2z, z e^{ \pm i \theta \pm i \phi} ; q^2\right)_{\infty}} \sum_{r=0}^{\infty}\frac{1-t_1^{2}zq^{2r-1}}{1-t_1^{2}zq^{-1}}\frac{\left(t_1^{2}zq^{-1},\frac{t_1z}{t_2},t_1e^{\pm i\theta},t_1e^{\pm i\phi};q^2\right)_{r}}{\left(t_1t_2,t_1ze^{\pm i\theta},t_1ze^{\pm i\phi},q^2;q^2\right)_{r}}\left(\frac{t_2z}{t_1}\right)^{r},\nonumber
\end{align}
where $z\equiv q^{2\Delta}$. As noticed in the $n_F=2$ case by \cite{Rajgadia:2026ask}, there are important simplifications in the Fourier kernel once we focus on the solutions with a discrete energy spectrum where 
\begin{equation}\label{eq:int ti}
    \rme^{x_l}=t_i q^{2l}~,\quad l\in \mathbb{Z}_{\geq0}~,
\end{equation}
since 
\begin{equation}
    (q^{-2l};q^2)_n=0~,\quad n>l~.
\end{equation}
and where $t_i$ in \eqref{eq:int ti} is the deformation parameter satisfying $\abs{t_i}\geq 1$. In the following, we consider that $\abs{t_1}\geq 1$ is the deformation parameter leading to the discrete spectrum. 

After applying the continuation $\theta\rightarrow\rmi x_{l_1}$ and $\phi\rightarrow \rmi x_{l_2}$ \eqref{eq:kernel-Q} can be put into the form
\begin{align}
      \bra{x_{l_1}}z^{\hat{n}}\ket{x_{l_2}}=&\frac{\left(t_2 z / t_1, t_1^2 z q^{2l_1}, z e^{-2l_1}, t^2_1 z q^{2l_2},  z q^{-2l_2} ; q^2\right)_{\infty}}{\left(t_1^2z, 
        zt_1^2 q^{2l_1+2l_2}, 
        z e^{2l_1-2l_2}, 
        z e^{2l_2-2l_1}, 
        zt_1^{-2} q^{-2(l_1+l_2)} ; q^2\right)_{\infty}}\cdot\\
        &\cdot \sum_{r=0}^{\rm min(l_1,l_2)}\frac{1-t_1^{2}zq^{2r-1}}{1-t_1^{2}zq^{-1}}\frac{\left(t_1^{2}zq^{-1},\frac{t_1z}{t_2},t_1^2q^{2l_1},q^{-2l_1},t_1^2q^{2l_2},q^{-2l_2};q^2\right)_{r}}{\left(t_1t_2,t_1^2zq^{2l_1},zq^{-2l_1},t_1^2zq^{2l_2},zq^{-2l_2},q^2;q^2\right)_{r}}\left(\frac{t_2z}{t_1}\right)^{r}~.\nonumber
\end{align}
This truncation can then be used to evaluate the Euclidean two-point correlation function \eqref{eq:G2 cont} for the discrete spectrum solution based on the orthogonality relation {\eqref{eq:recurrence discrete}}. It takes the form
\begin{equation}
    G_2(\tau_1,\tau_2)=\sum_{l_1,l_2}\rme^{-\tau_1 E(x_{l_1})-\tau_2 E(x_{l_2})}\bra{x_{l_1}}z^{\hat{n}}\ket{x_{l_2}}~.
\end{equation}
The above expressions allow one to evaluate the two-point function in the quantum regime of the theory, $\forall q\in[0,1)$. The $q\rightarrow1^-$ limit at finite temperature is analysed in Sec.~\ref{ssec:Krylov nF4}. 

\section{More on Krylov complexity}\label{app:Krylov nF0}
In this appendix, to complement our discussion in Sec.~\ref{sec:Krylov Askey}, we discuss the evaluation of Krylov complexity for the HH state as the reference state in the Lanczos algorithm for (i) the DSSYK model without deformations in Sec.~\ref{sapp: nF0}, and (ii) the $n_F=8$ deformed DSSYK model in Sec.~\ref{sapp: nF8 def}.

\subsection{Undeformed DSSYK model}\label{sapp: nF0}
We start from the DSSYK chord Hamiltonian \eqref{eq:H SYK0} where we apply a basis transformation, such that it takes its symmetric form:
\begin{equation}\label{eq:H nF0}
    \hH_{n_F=0}\ket{n}=\sqrt{[n+1]_{q^2}}\ket{{n+1}}+\sqrt{[n]_{q^2}}\ket{{n-1}}~,
\end{equation}
This expression determines the inner product between the energy basis and the chord number basis as
\begin{equation}
    \bra{n}\ket{\theta}=\frac{H_n(\cos\theta|q^2)}{\sqrt{(q^2;q^2)_n}}~.
\end{equation}
where we assumed as the initial condition in the recurrence relation that $\bra{\Omega}\ket{\theta}=1$, and we introduced the q-Hermite polynomials
\begin{equation}
    H_n(\cos\theta|q^2) \equiv\sum_{k=0}^n \frac{(q^2;q^2)_n}{(q^2;q^2)_k(q^2;q^2)_{n-k}}e^{i(n-2k)\theta}~.
\end{equation}
After rescaling the Hamiltonian \eqref{eq:H nF0} by a factor $\sqrt{1-q^2}$, one then identifies the Lanczos coefficients in the semiclassical limit as:
\begin{equation}
    b(\ell)=\sqrt{1-\rme^{-\ell}}~,\quad a(\ell)=0~.
\end{equation}
Solving the equation of motion \eqref{eq:EOM krylov basis def} with initial condition \eqref{eq:initial length}:
\begin{equation}
    \ell_0=-2\log(\sin\theta)~,
\end{equation}
one finds \eqref{eq:spread Complexity} in the continuum approximation as (see Fig.~\ref{fig:nF}):
\begin{equation}
    \mathcal{C}_{\rm S}(t)=\frac{\ell(t)}{\lambda}=\frac{2}{\lambda}\log\frac{\cosh(\sin\theta~t)}{\sin\theta}~.
\end{equation}
This corresponds to the regularised geodesic length between the asymptotic boundaries of an AdS$_2$ black hole with black hole radius $\sin\theta$ \cite{Harlow:2018tqv,Blommaert:2024ymv}, see Fig.~\ref{fig:CS_nF0}.  Numerical evaluations of Krylov spread complexity for the HH state \eqref{eq:liouv} in \cite{Rabinovici:2023yex} show great agreement with the saddle point answer when $\mathcal{C}_{\rm S}\sim \mathcal{O}(1/\lambda)$.

\begin{figure}
    \centering
    \subfloat[]{\includegraphics[height=0.33\linewidth]{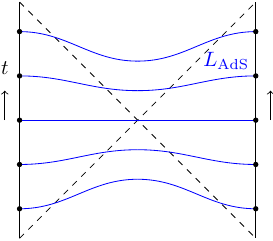}}\hspace{0.8cm}\subfloat[]{\includegraphics[height=0.33\linewidth]{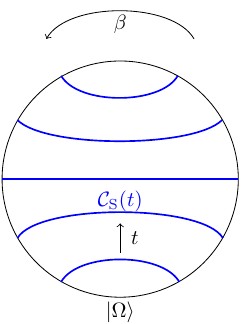}}
    \caption{(a) Bulk interpretation of $\hH_{n_F=0}$ \eqref{eq:H nF0} and $\mathcal{C}_{\rm S}$, corresponding to minimal geodesics (blue) joining the asymptotic anchoring point (black dots) of an (effective) AdS$_2$ black hole in the bulk. (b) `Chord diagram' representation (suppressing the chords) where $n(t)=Z(\beta)^{-1}\bra{\Omega}\rme^{-\tau^*\hat{H}_{n_F=0}}\hat{n}{\rm e}^{-\tau\hat{H}_{n_F=0}}\ket{\Omega}$ with $\tau=\beta/2+\rmi t$.}
    \label{fig:CS_nF0}
    \end{figure}

\subsection{\texorpdfstring{$n_F=8$}{} deformed DSSYK model}\label{sapp: nF8 def}
We now provide details about evaluation in Sec.~\ref{ssec:Krylov nF8}. Let us solve the equation of motion that determines Krylov complexity in the semiclassical limit \eqref{eq:special eq} by introducing a substitution $X(t)=\rme^{-\lambda\mathcal{C}_{\rm S}(t)}$ in the explicit Lanczos coefficients in the semiclassical limit $b(\ell)$ and $a(\ell)$ \eqref{eq:lanczos nF8}. It takes the form:
{\small
\begin{equation}\label{eq:2nd special int}
    2t=\int _{X_0}^{X(t)}\frac{(X')^{-1}(1-z(X')^2)^2\rmd X'}{\sqrt{\sin^2\theta(1-z^2(X')^4)^2+a'_1X'(1-z (X')^2-z^2 (X')^4+z^3(X')^6)+a'_2(X')^2(1-z(X')^2)^2}}~,
\end{equation}
}
where $z\equiv t_1t_2t_3t_4$,
\begin{subequations}
\begin{align}
    a'_1&\equiv\cos\theta\qty(\prod_{i=1}^4t_i+\sum_{i\neq j\neq k}t_it_jt_k)-1-\prod_{i\neq j}t_it_j~,\\
    4a'_2&\equiv t_1^2 \left(-t_2^2 (t_3-t_4)^2+2 t_2 (t_3+t_4) (t_3
   t_4+1)-(t_3 t_4-1)^2\right)+8 t_1 t_2 t_3 t_4 \cos
   2 \theta\\
   &-8 \cos \theta \left(t_1^2 t_2 t_3
   t_4+t_1 \left(t_2 t_3 t_4^2+t_4 (t_2+t_3) (t_2
   t_3+1)+t_2 t_3\right)+t_2 t_3 t_4\right)\nonumber\\
   &+2 t_1 \left(t_4
   \left(t_2^2 \left(t_3^2+1\right)+8 t_2 t_3+t_3^2\right)+t_4^2
   (t_2+t_3) (t_2 t_3+1)+t_2 t_3
   (t_2+t_3)+t_2+t_3+t_4\right)\nonumber\\
   &+2 t_4 (t_2+t_3) (t_2
   t_3+1)-(t_2-t_3)^2-t_4^2 (t_2 t_3-1)^2~,\nonumber
\end{align}
\end{subequations}
and the initial condition $X_0=\rme^{-\ell_0}$ is determined by $\ell_0$ in \eqref{eq:initial length}. We emphasise that in the evaluation of \eqref{eq:2nd special int} we do not pick poles neither for the continuous not discrete solutions (where one sends $\cos\theta\rightarrow\cosh x_\ell$), since $t_1t_2t_3t_4<1$, for both the continuous and discrete spectrum solutions due to the conditions \eqref{eq:dissrete cond 1}, \eqref{eq:dissrete cond 2}, and since $\rme^{-\ell}\leq 1$ for $\ell\in\mathbb{R}^+_0$.

\section{Semiclassical partition function}\label{app:semiclassical Z}
In this appendix, we provide details about the evaluation of the semiclassical partition function in Sec.~\ref{eq:partitil func}.

We start from the continuous contribution \eqref{eq:mu nF first}. We write the partition function in the energy basis as \cite{Aguilar-Gutierrez:2025hty}
\begin{eqnarray}\label{eq:Z cont }
   Z_{\rm cont}(\beta)\equiv \int _0^\pi\rmd\theta \rme^{S_{n_F}(\theta)-\beta E(\theta)}~,
\end{eqnarray}
where $S_{n_F}(\theta)\equiv\log \mu_{n_F}(\theta)$. We are interested in the leading order behaviour of $S_{n_F}(\theta)$ as $\lambda\rightarrow0$. Let $x$ be a fixed parameter as $\lambda\rightarrow0$. We have that $-2\lambda\log(x;q^2)_\infty={\rm Li}_2(x)$ \cite{Goel:2023svz} where Li$_2(x)\equiv\sum_{k=1}^\infty k^{-2}x^k$ is the dilogarithm function. Using the explicit form of $\mu_{n_F}$ in \eqref{eq:mu nF first}, and choosing $\alpha\in t_i$ in the same way as in the condition \eqref{eq:dissrete cond 1} ($\abs{\alpha}>1$), one then finds
\begin{equation}\label{eq:thermal ETW}
    \begin{aligned}
        S_{n_F}(\theta)=&\frac{2\pi\theta}{\lambda}\qty(1-\frac{\theta}{\pi})\\
    &+\frac{1}{\lambda}\qty(\frac{\pi^2}{6}+{\rm Li}_2(t_1t_2t_3t_4)+\sum_{i=1}^4\qty[\sum_{\epsilon=\pm1}{\rm Li}_2(t_j\rme^{\rmi\epsilon\theta})-\sum_{j\neq i}^4{\rm Li}_{2}(t_it_j)]+\mathcal{O}(\lambda))~.
    \end{aligned}
\end{equation}
In the following, we define the semiclassical limit of the $n_F$ deform theories where we select one of the deformation parameters to scale as $\alpha=q^{2\nu+1}\in t_i$ and $\nu$ and the other parameters $t_{j}$ remain fixed as $\lambda\rightarrow0$. This definition of semiclassical limit is consistent with the scaling of the parameters $t_i$ in the triple-scaling limit \eqref{eq:triple scaling}. Namely, we simply do not need to restrict the analysis to low energies (where $\theta$ would not be fixed). 

Under the above semiclassical limit, $S_{n_F}(\theta)$ reduces to
\begin{eqnarray}
\begin{aligned}
    S_{n_F}(\theta)\simeq\frac{1}{\lambda}\Biggl[&-\frac{(\pi-3\theta)^2}{6}+{\rm Li}_2\qty(\prod_{t_i\neq\alpha}t_i)\\&+\sum_{t_i\neq\alpha}\Biggl[\sum_{\epsilon=\pm1}{\rm Li}_2\qty(t_i\rme^{\rmi\epsilon\theta})-{\rm Li}_2(t_i)-\sum_{j\neq i,~t_j\neq \alpha}{\rm Li}_{2}(t_it_j)\Biggr]\Biggr]~.
\end{aligned}
\end{eqnarray}
Using $1/\lambda$ dependence in the entropy and $1/\sqrt{1-q^2}$ in the energy, it follows that we can evaluate \eqref{eq:Z cont } in the saddle point approximation when $\lambda\rightarrow0$ as
\begin{eqnarray}
    Z_{\rm cont}(\beta)=\rme^{S_{n_F}(\theta)-\beta E(\theta)}~,\quad \beta(\theta)=\pdv{S_{n_F}(\theta)}{E(\theta)}~.
\end{eqnarray}
as in \eqref{eq:semiclassical disc S2}.

Next, we evaluate the contribution of the discrete spectrum in \eqref{full partition function}, namely
\begin{eqnarray}\label{eq:disc Z}
    Z_{\rm disc}(\beta)\equiv\sum_{l=0}^{l_{\rm max}}\rme^{S_l-\beta x_l}~,\quad S_{l}\equiv\log \tilde{w}_l~,
\end{eqnarray}
can be carried out using the definition for the semiclassical limit above, and (7.30, 31) in \cite{Rajgadia:2026ask}, which in our notation corresponds, with $\alpha\equiv q^{2\nu+1}$ as $\lambda\rightarrow0$, to
\begin{equation}
\frac{1-\alpha^2q^{4l}}{1-\alpha^2}\rightarrow\frac{2\nu+2l+1}{2\nu+1}~,\quad \frac{(\alpha^2;q^2)_l}{(q^2;q^2)_l}\rightarrow\frac{(2\nu+1)_l}{l!}~,\quad \frac{(\alpha^{-2};q^2)_\infty}{(q^2;q^2)_\infty}\rightarrow\frac{\lambda^{4\nu+2}}{\Gamma(-2\nu-1)}
\end{equation}
where $(x)_n$ is the Pochhammer symbol. Collecting the contributions in \eqref{eq:disc Z}, we find
\begin{equation}\label{eq:semiclassical disc S}
\begin{aligned}
    w_l&=\frac{(-1)^l(2\nu+2l+1)(2\nu+1)!\lambda^{4\nu+2}}{(2\nu+1)l!~\Gamma(-\nu-1)}+\mathcal{O}(\lambda)~,\\
    S_l&=\frac{1}{\lambda}\qty({\rm Li}_2\qty(\prod_{t_{i}\neq\alpha}t_i)+\sum_{t_i\neq\alpha}\qty({\rm Li}_2(t_i)-\sum_{j\neq i,~t_j\neq\alpha}{\rm Li}_2(t_it_j))+\frac{\pi^2}{6})+\mathcal{O}(\log\lambda)~.
\end{aligned}
\end{equation}
At leading order as $\lambda\rightarrow0$, the entropy is independent of the index $l$ over the sum $l$, while the next order contribution to the sum which depends on $l$ is
\begin{eqnarray}\label{eq:leading order conti}
   \eval{\rme^{S_l}}_{\mathcal{O}(\lambda^0)}=(2\nu+2l+1)\frac{(2\nu+1)_l}{l!}\qty(-1)^l
\end{eqnarray}
up to an overall $l$-independent coefficient.

This semiclassical entropy can be introduced back in \eqref{eq:disc Z}, which gives the term in \eqref{eq:semiclassical disc S2}. Importantly, by evaluating the series with the leading order contribution \eqref{eq:leading order conti}, we find that it vanishes since $E(\theta_l)\simeq(\alpha+1/\alpha)/2$ at order $\mathcal{O}(\lambda^0)$ as $\lambda\rightarrow0$. If one allows the first correction $\mathcal{O}(\lambda^{1/2})$ (alternatively, rescale the Hamiltonians entering in the partition function by a factor $1/\sqrt{\lambda}$), then there is a non-vanishing contribution to the partition function from the discrete energy spectrum, which will be discussed elsewhere.

\section{Algebraic entanglement between ETW branes}\label{app:Ent branes}
In \cite{Berkooz:2025ydg}, it was pointed out that the adjoint case resembles an entangled pair of ETW brane reservoirs.  We illustrate this in Fig.~\ref{fig:Ent_branes}. 
In this appendix, we comment on this resemblance using the operator algebras for each of the different systems in the q-Askey scheme.
\begin{figure}
    \centering
    \subfloat[]{\includegraphics[height=0.32\linewidth]{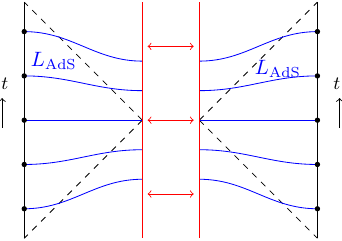}\label{sfig:CS4_2}}\hfill\subfloat[]{\includegraphics[height=0.32\linewidth]{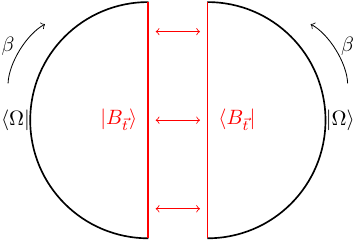}\label{sfig:ETW4_2}}
    \caption{A pair of $n_F=2$ ETW branes which are entangled with each other, $L_{\rm AdS}$ is the distance between the asymptotic boundary and the ETW brane at a given asymptotic boundary time $t$. (a) Lorentzian, (b) Euclidean perspectives.}
    \label{fig:Ent_branes}
\end{figure}
For these purposes, let us first describe the entanglement entropy between the double-scaled algebras given a chord state dual to a black hole spacetime with an ETW brane. ETW branes in sine dilaton gravity have been discussed in \cite{Blommaert:2025avl}. It was shown in \cite{Aguilar-Gutierrez:2025hty} that the corresponding Hamiltonians can be recovered by gauging specific symmetries of the theory. For this reason, we can identify the ETW branes in the bulk with matter in the DSSYK under specific limits. The algebra of observables remains unchanged since we study operators in the double-scaled algebras in particular limits. 

Consider a single-sided black hole with an ETW brane, as displayed in Fig.~\ref{fig:CS_nF2}. The respective black hole background is in a pure state, even though the dual CFT is in a thermal state. This has been previously discussed in the context of PETS \cite{Goel:2018ubv}; arguing that the ETW brane acts as a purifier of the combined system, motivated by other work \cite{Takayanagi:2011zk}. To test this proposal, we investigate the entanglement structure of DSSYK with ETW brane states. We reiterate that the ETW brane corresponds to taking a limit in the chord Hilbert space itself (i.e.~after ensemble averaging) in terms of chord number, which is expected to have a microscopic counterpart in the physical SYK model \cite{Goel:2018ubv}.

We are interested in evaluating the von Neumann entropy for the density matrix in the double-scaled algebra encoded in the partition \eqref{eq:mu nF first} using the ETW brane operator in \eqref{eq:ETW brane state general ext}, namely
\begin{equation}
    \Tr\hrho=\bra{\Omega}B_{t_1,t_2,t_3,t_4}(\hH)\rme^{-\beta\hH_{n_F=0}}\ket{\Omega}~, \quad \hrho=B_{t_1,t_2,t_3,t_4}(\hH)\rme^{-\beta\hH_{n_F=0}},
\end{equation}
where we used that $\ket{\Omega}$ is the tracial state in the algebra.

The corresponding entanglement entropy \eqref{eq:algebraic entropy} between the double-scaled algebras for $\hrho$ above is simply the thermal result \eqref{eq:mu nF first}, which we can be expressed as
\begin{equation}\label{eq:thermal entropy ETW}
        S_{n_F}= \qty(1-\beta\pdv{\beta})\log \bra{0}\rme^{-\beta\hH_{n_F=8}}\ket{0}~.
    \end{equation}
We now seek to compare the proposal for the entropy associated with the ETW brane in a pure one-sided black hole in JT gravity by \cite{Goel:2018ubv} (5.4), namely $S=2\pi\Delta$ with $\Delta$ being the conformal dimension of the operator associated with the ETW brane. Note that \eqref{eq:thermal ETW} is an algebraic entanglement entropy, and it should not be thought of as a measure of entanglement between a CFT on one of the AdS$_2$ boundaries and the ETW brane, since we work at $N\rightarrow\infty$. Nevertheless, we seek to test if \eqref{eq:thermal ETW} shows quantitative similarities to the conjecture in \cite{Goel:2018ubv}. In terms of operators $\hmO_\Delta$, one-sided ETW branes, alike those considered by \cite{Goel:2018ubv}, correspond to \cite{Aguilar-Gutierrez:2025hty}
\begin{equation}
    t_1=q^{2\Delta}~,\quad t_2=t_3=t_4=0~.
\end{equation}
To describe JT gravity, we are interested in the IR triple-scaling limit, where $\theta/\lambda$ and $\Delta$ are fixed as $\lambda\rightarrow0$; so that the semiclassical entanglement entropy in \eqref{eq:thermal ETW} becomes
\begin{equation}\label{eq:important entropy}
\begin{aligned}
     S_{n_F}(t_1=q^\Delta,t_2=0)&=\frac{2\pi\theta}{\lambda}\qty(1-\frac{\theta}{\pi})+\frac{1}{\lambda}\qty(\frac{\pi^2}{6}+\sum_{\epsilon=\pm1}\qty({\rm Li}_2(t_1\rme^{\rmi\epsilon\theta}))+\mathcal{O}(\lambda))\\
     &\rightarrow \frac{2\pi\theta}{\lambda}+\frac{\pi^2}{2\lambda}+2\Delta\log\lambda+\mathcal{O}(1)~,
\end{aligned}
\end{equation}
where the arrow indicates the IR triple-scaling limit, and we use the series expansion
\begin{equation}
    {\rm Li}_2(\rme^{-x})=\frac{\pi^2}{6}-x(1-\log(x))+\mathcal{O}(x^2)~.
\end{equation}
Thus, the leading order correction to the black hole entropy in the IR triple-scaling limit is $2\Delta\log\lambda$ (where $\lambda=8\pi G_N$ \cite{Blommaert:2025eps}),
 which then shows quantitative similarities with the conjectured formula for the entropy associated with an ETW brane in JT gravity by \cite{Goel:2018ubv}.

In particular, let us apply the above result for the adjoint case. We use the corresponding density of states in \eqref{eq:mu Adj}
\begin{equation}
   Z_{\rm Adj}(\chi)=\int _0^\pi\rmd\theta~\mu(\theta)\frac{(\chi ^2q^2;q^4)^2_\infty(-\chi,-\chi q^2;q^2)_\infty}{(\chi ^2q^2;q^2)_\infty(\chi\rme^{\pm2\rmi\theta},\chi q^2\rme^{\pm2\rmi\theta};q^4)_\infty}~.
\end{equation}
We can then evaluate the thermal entropy \eqref{eq:thermal ETW}
\begin{equation}\label{eq:adj entropy}
    S_{\rm Adj}=\frac{1}{\lambda}\qty(\frac{\pi^2}{6}+2\pi\theta\qty(1-\frac{\theta}{\pi})-2{\rm Li}_2\qty(-\chi)+2\sum_{\epsilon=\pm1}{\rm Li}_2(\chi\rme^{\epsilon2\rmi\theta})+\mathcal{O}(\lambda))~.
\end{equation}
where we take $\chi\sim\mathcal{O}(1)$ as $\lambda\rightarrow0$. It can be seen that \eqref{eq:adj entropy} takes a very similar form to \eqref{eq:important entropy} in the triple-scaling limit where $\theta\sim 0+\mathcal{O}(\lambda)$; $\chi$ and $t_1$ can be related as overall additional constants. 

It was expected in \cite{Berkooz:2025ydg} that the adjoint case describes a pair of entangled brane reservoirs. If the reservoirs were at equal temperatures in an overall thermal state, then we would have
\begin{equation}
\begin{aligned}
   S_{\rm total}=\qty(1-\beta\partial_\beta) Z_{\rm total}=\qty(1-\beta\partial_\beta)Z_{n_F}(\beta)^2\underset{\lambda\rightarrow0}{=}2S_{n_F}~,
\end{aligned}
\end{equation}
where $S_{n_F}$ appears in \eqref{eq:thermal ETW}. Beyond the triple-scaling limit, there is thus a difference between the adjoint case and a pair of entangled brane reservoirs associated with the $n_F$ cases.

This indicates that although the adjoint system has properties very similar to those of ETW brane systems, as we inquired in Sec.~\ref{ssec:Krylov nF Adj}, it might describe a more complex configuration than just a pair of entangled ETW branes. It would be interesting to study this problem more explicitly; also, its sine dilaton gravity description if one were able to find the semiclassical bulk dual geometry describing the $n_F=8$ deformed theory. Nevertheless, our findings from Sec.\ref{ssec:semiclassical limit} indicate that the bulk theory should have at least two asymptotic boundaries. It is then natural to expect that a realization of the bulk picture in \cite{Berkooz:2025ydg} could be made more precise with sine dilaton gravity.

\section{Extension of q-coherent states}\label{app:extended q coherent}
In this appendix, we review the construction of generalised q-coherent states, which are relevant to describing the $n_F=4$ ETW brane state $\ket{B_{\vec{t}}^{(n_F)}}$, and we define an extension of this definition for the $n_F=8$ case. 

It is known \cite{Okuyama:2023byh,Watanabe:2025rwp,Aguilar-Gutierrez:2025hty} that the brane state $\ket{B_{t_1,t_2}}\equiv \ket{B^{(n_F=2)}_{\vec{t}}}$ \eqref{eq:ETW brane state general ext} (with $t_3=t_4=0$) is a generalised q-coherent state (in the zero-particle DSSYK chord Hilbert space), as it satisfies the relation
\begin{equation}\label{eq:genealized q-coherent}
     (\hat{a}+t_1t_2\hat{a}^\dagger)\ket{B_{t_1,t_2}}\equiv\frac{t_1+t_2}{\sqrt{1-q^2}}\ket{B_{t_1,t_2}}~,
 \end{equation}
with $\ha$, $\ha^\dagger$ being annihilation and creation operators in the zero-particle chord Hilbert space \eqref{eq:zero particle H}, and $\ket{B_{t_1,t_2}}$ is defined by \eqref{eq:ETW brane state general ext}, i.e.
\begin{subequations}
    \begin{align}
        \ha^\dagger\ket{n}&=\sqrt{[n+1]_{q^2}}\ket{n+1}~,\quad&\ha\ket{n}&=\sqrt{[n]_{q^2}}\ket{n-1}~,\quad\hn\ket{n}=n\ket{n}~,\label{eq:q-oscillator algebra}\\
    \ket{B_{t_1,t_2}}&=\hat{B}_{t_1,t_2}\ket{\Omega}~,\quad &\hat{B}_{t_1,t_2}&\equiv\frac{(t_1t_2;q^2)_\infty}{(t_1\rme^{\pm\rmi\theta(\hH)},t_2\rme^{\pm\rmi\theta(\hH)};q^2)_\infty}~.\label{eq:ETW brane state}
    \end{align}
\end{subequations}
Note that we use the Krylov orthogonal basis in App.~\ref{app:Krylov nF0}.

The generalised q-coherent brane states can then be expanded as
\begin{equation}\label{eq:Bt1t2}
    \ket{B_{t_1,t_2}}=\sum_nc_n(t_1,t_2)\ket{n}_{n_F=0}~,
\end{equation}
where
\begin{equation}\label{eq:coefficients t1 t2}
    \frac{t_1+t_2}{\sqrt{1-q^2}}c_n=\sqrt{[n+1]_{q^2}}~c_{n+1}+t_1t_2\sqrt{[n]_{q^2}}~c_{n-1}~.
\end{equation}
which is solved by
\begin{equation}
    c_n(t_1,t_2)=\frac{(-t_1t_2)^n}{\sqrt{(q^2;q^2)_n}}H_n\qty(\eval{\frac{t_1+t_2}{2\sqrt{t_1t_2}}}q^2)~.
\end{equation}
In particular, when $t_2=0$ in \eqref{eq:genealized q-coherent} we recover the usual q-coherent state
\begin{equation}\label{eq:Bt1}
    \ket{B_{t_1}}\equiv \ket{B_{t_1,t_2=0}}~,\quad \hat{a}\ket{B_{t_1}}=\frac{t_1}{\sqrt{1-q^2}}\ket{B_{t_1}}~,
\end{equation}
describing the ETW brane state in \cite{Okuyama:2023byh} (see Fig.~\ref{sfig:ETW_nF2}).

In fact, one can express the generalised q-coherent state in terms of a product of q-coherent states. One can see this from the definition
\begin{equation}\label{eq:eq gen q coherent}
    \begin{aligned}
        \bra{B_{t_1,t_2}}\rme^{-\beta\hH_{n_F=0}}\ket{\Omega}&=\int _0^\pi\rmd\theta\frac{\rme^{-\beta E(\theta)}\mu(\theta)}{(t_1\rme^{\pm\rmi\theta},t_2\rme^{\pm\rmi\theta};q^2)_\infty}=\bra{B_{t_1}}\int _0^\pi\rmd\theta\mu(\theta)\rme^{-\beta E(\theta)}\ket{\theta}\bra{\theta}\ket{B_{t_2}}\\
    &=\bra{B_{t_1}}\rme^{-\beta\hH_{n_F=0}}\ket{B_{t_2}}~,
    \end{aligned}
\end{equation}
where $\bra{\theta}\ket{B_{t_1}}\equiv1/(t_1\rme^{\pm\rmi\theta};q^2)_\infty$. One can similarly write the measure in the partition function of the $n_F\geq2$ case as the transition amplitudes between other $n_F'\leq n_F$ states \cite{Berkooz:2025ydg}. 

As discussed in Sec.~\ref{ssec:open problems}, it is natural to verify if there is some extension of the generalised q-coherent state in \eqref{eq:genealized q-coherent} which incorporates all $t_{1\leq i\leq 4}$ parameters associated to the ETW brane states \eqref{eq:ETW brane state general ext}
\begin{equation}\label{eq:def ext q coherent}
    (b_0(\hat{a})^2+b_1\hat{a}+b_2\hat{a}^\dagger+b_3(\hat{a}^\dagger)^2)\ket{B^{(n_F)}_{\vec{t}}}=b_4\ket{B^{(n_F)}_{\vec{t}}}
\end{equation}
where $b_{0}$, \dots $b_4$ are functions of the deformation parameters $t_{1\leq i\leq4}$ that need to be determined based on \eqref{eq:ETW brane state general ext}. Note that we include additional powers of the creation and annihilation operators with respect to \eqref{eq:genealized q-coherent} since it contains twice the number of free parameters $t_{i}$ obeying either \eqref{eq:cont spectrum} or \eqref{eq:dissrete cond 1}.

Similar to \eqref{eq:coefficients t1 t2}, the definition \eqref{eq:def ext q coherent} requires identifying the recurrence relation obeyed by the set of constants in the expansion of the brane state \eqref{eq:ETW brane state general ext},
\begin{subequations}
    \begin{align}
        \ket{B^{(n_F)}_{\vec{t}}}&=\sum_{n=0}c_n\ket{n}~,\\
        c_n&=\int _0^\pi\rmd\theta\mu(\theta)\frac{H_n(\cos\theta|q^2)}{\sqrt{(q^2;q^2)_n}}\frac{\prod_{i<j<4}(t_it_j;q^2)_\infty}{(t_1\rme^{\pm \rmi \theta},\dots,t_{4}\rme^{\pm\rmi \theta};q^2)_\infty(t_1t_2t_3t_4;q^2)_\infty}~.\label{eq:new cn}
    \end{align}
\end{subequations}
One should be able to identify the recurrence relation for the coefficients $c_n$ in \eqref{eq:new cn} based on the definition of extended q-coherent states in \eqref{eq:def ext q coherent} and the recurrence relation for the q-Hermite polynomials \eqref{eq:H nF0}, at least in an asymptotic expansion for $n\gg1$.\footnote{We thank Masataka Watanabe for valuable discussions regarding this point.} We leave this for future work.

\section{Symmetry sectors in the chord Hilbert space}\label{app:symmetry sectors}
In this appendix, we provide technical details about our discussion on symmetry sectors in the chord Hilbert space discussed in Sec.~\ref{ssec:open problems}.

By a symmetry sector we refer to a set of states that remain invariant under a transformation, i.e.
\begin{equation}
    \hat{U}_i\ket{\psi}=\ket{\psi}\in\mathcal{H}_{\rm const}~,
\end{equation}
which is implemented by a set of constraints (i.e.~the generators of the transformation)
\begin{equation}
    \hat{U}_i=\rme^{\alpha_i\hat{\varphi}_i}~,\quad \hat{\varphi}_i\ket{\psi}=0~.
\end{equation}
Below, we illustrate this procedure in connection with the q-Askey deformation of the DSSYK model, where we consider the (left) DSSYK chord Hamiltonian \eqref{eq:HLmultiple}, and the states in the extended Hilbert space \eqref{eq:H ffull}.

We start with $m=1$ particle insertion in the chord Hamiltonian \eqref{eqn:hH_LR}, which can be described by
\begin{equation}\label{eq:pair DSSYK Hamiltonians 1 particle}
\begin{aligned}
    \hH_{L/R}=\frac{1}{\sqrt{1-q^2}}\qty(\rme^{-\rmi \hat{P}_{L/R}}+\rme^{\rmi \hat{P}_{L/R}}[\hat{n}_{L/R}]+q^{2\Delta}\rme^{\rmi \hat{P}_{R/L}}q^{\hat{n}_{L/R}}[\hat{n}_{R/L}])~,
    \end{aligned}
\end{equation}
which acts on the one-particle chord Hilbert space (\eqref{eq:states notation matter} with $m=1$), where
\begin{equation}
\begin{aligned}
    \rme^{-\rmi \hP_{L}^\dagger}\ket{\Delta;n_L,n_R}&=\sqrt{1-q^2}\ket{\Delta;n_L+1,n_R}~,\quad  \rme^{-\rmi \hP_{R}^\dagger}\ket{\Delta;n_L,n_R}&&=\sqrt{1-q^2}\ket{\Delta;n_L,n_R+1}\\
    \rme^{\rmi \hP_{L}^\dagger}\ket{\Delta;n_L,n_R}&=\frac{1}{\sqrt{1-q^2}}\ket{\Delta;n_L-1,n_R}~,\quad  \rme^{\rmi \hP_{R}^\dagger}\ket{\Delta;n_L,n_R}&&=\frac{1}{\sqrt{1-q^2}}\ket{\Delta;n_L,n_R-1}~,\\
    \hat{n}_{L/R}\ket{\Delta;n_L,n_R}&={n}_{L/R}\ket{\Delta;n_L,n_R}~.
\end{aligned}
\end{equation}
One then recovers the $n_F=2$ chord Hamiltonian \eqref{eq:H nF2} for e.g. $\hH_L\rightarrow\hH_{n_F=2}$ by imposing a constraint
\begin{equation}\label{eq:important constraint}
\varphi_A\equiv\hat{n}_L-n_{1}~,\quad \varphi_B\equiv \hat{P}_L~,
\end{equation}
and relabeling $\hn_R\equiv \hn$, $\hP_R=\hP$ (similar to \cite{Aguilar-Gutierrez:2025hty}), where one identifies
\begin{equation}
    t_1=q^{2\Delta}(1-q^{2n_1})~.
\end{equation}
This implies that the case studied by \cite{Okuyama:2023byh} has a DSSYK embedding where $n_L=n_1$ is fixed, and $p_L=0$. As a consequence, $\theta_L$ is also fixed (see further details in \cite{Aguilar-Gutierrez:2025hty}).

Similarly, for the $m=2$ case, where, focusing on
\begin{equation}
\begin{aligned}
  {\sqrt{1-q^2}}\hH_L=&\rme^{-\rmi \hP_0}+\rme^{\rmi \hP_0}\qty({1-q^{2\hn_0}})+\rme^{\rmi \hP_1}\qty({1-q^{2\hn_1}})q^{2\hn_0}q^{\Delta_1}\\
     &+\rme^{\rmi \hP_2}\qty({1-q^{2\hn_2}})q^{2(\hn_0+\hn_1)}q^{2(\Delta_1+\Delta_2)}~,
\end{aligned}
\end{equation}
one can implement the same constraint \eqref{eq:important constraint} and find that $\hH_L\rightarrow\hH_{n_F=4}$ \eqref{eq:H nF4}, where
\begin{equation}\label{eq:m=2 case}
    q^{2\Delta_1}(1-q^{2n_1})=t_1+t_2~,\quad q^{2(\Delta_1+\Delta_2+n_1)}=-\frac{t_1t_2}{q^2}~.
\end{equation}
This means that, generically, one can describe the $n_F=4$ ETW brane system in terms of the DSSYK model with two particle insertions with the same constraint as the $n_F=2$ case. The one-particle chord Hamiltonian of the DSSYK model can also be used to derive other $n_F=4$ deformed Hamiltonians with specific $t_1$ and $t_2$ parameters \cite{Aguilar-Gutierrez:2025hty}.

\bibliographystyle{JHEP}
\bibliography{references.bib}
\end{document}